\documentclass[reqno,11pt,nosumlimits]{article}
\usepackage{amssymb,amsfonts,amscd}
\usepackage[mathscr]{eucal}
\usepackage{amsmath}
\usepackage{graphics}
\usepackage{color}

\usepackage{ulem}
\usepackage{graphicx}

\allowdisplaybreaks[1]
\numberwithin{equation}{section}

\newtheorem{theorem}{Theorem}[section]

\newtheorem{proposition}[theorem]{Proposition}

\newtheorem{condition}{Condition}

\newtheorem{definition}[theorem]{Definition}

\newtheorem{remark}[theorem]{Remark}
\newtheorem{example}[theorem]{Example}
\newtheorem{observation}[theorem]{Observation}
\newtheorem{convention}[theorem]{Convention}

\DeclareMathOperator{\Image}{Image}
\DeclareMathOperator{\arc}{Image}
 \DeclareMathOperator{\Tr}{Tr}
\DeclareMathOperator{\GL}{GL}
\DeclareMathOperator{\gl}{gl}

 \DeclareMathOperator{\Mat}{Mat}
\DeclareMathOperator{\supp}{supp}
 \DeclareMathOperator{\Det}{Det}
\DeclareMathOperator{\dom}{dom}
\DeclareMathOperator{\codim}{codim}

 \DeclareMathOperator{\End}{End}
\DeclareMathOperator{\Hom}{Hom}
\DeclareMathOperator{\Ad}{Ad}
\DeclareMathOperator{\ad}{ad}
\DeclareMathOperator{\Aut}{Aut}

\DeclareMathOperator{\wind}{wind}

\DeclareMathOperator{\sgn}{sgn}

\DeclareMathOperator{\Ker}{ker}
\DeclareMathOperator{\gleam}{gleam}

\DeclareMathOperator{\sing}{sg}
\DeclareMathOperator{\aff}{aff}

\DeclareMathOperator{\Hol}{Hol}

\DeclareMathOperator{\contr}{contr}

\DeclareMathOperator{\col}{col}

\newcommand{\bN}{{\mathbb N}}
\newcommand{\bQ}{{\mathbb Q}}
\newcommand{\bZ}{{\mathbb Z}}
\newcommand{\bR}{{\mathbb R}}
\newcommand{\bC}{{\mathbb C}}

\newcommand{\bK}{{\mathbb K}}

\newcommand{\ct}{{\mathfrak t}}
\newcommand{\cG}{{\mathfrak g}}

\newcommand{\ck}{{\mathfrak k}}
\newcommand{\cg}{c_{\cG}}

\newcommand{\id}{{ \rm id}}

\newcommand{\CW}{{\mathcal C}}

\newcommand{\G}{{\mathcal G}}

\newcommand{\cl}{{\mathrm h}}
\newcommand{\reg}{reg}
\newcommand{\orth}{\perp}
\newcommand{\eps}{\epsilon}

\newcommand{\cA}{{\mathcal A}}
\newcommand{\cB}{{\mathcal B}}
\newcommand{\cC}{{\mathcal C}}
\newcommand{\cD}{{\mathcal D}}

\newcommand{\cF}{{\mathcal F}}

\newcommand{\cI}{{\mathcal I}}
\newcommand{\cJ}{{\mathcal J}}
\newcommand{\cK}{{\mathcal K}}

\newcommand{\cN}{{\mathcal N}}
\newcommand{\cO}{{\mathcal O}}
\newcommand{\cP}{{\mathcal P}}

\newcommand{\cR}{{\mathcal R}}

\newcommand{\cS}{{\mathscr S}}

\newcommand{\cT}{{\mathcal T}}

\newcommand{\cW}{{\mathcal W}}

\begin{document}

\title{The non-Abelian Chern-Simons path integral on $M=\Sigma \times S^1$ in the torus gauge: a review}

\date{\today}

\maketitle

\begin{center} \large
Atle Hahn
\end{center}

\begin{center}
\it   \large  Group of Mathematical Physics \\
Faculty of Science of the University of Lisbon \\
Campo Grande, Edificio C6\\
PT-1749-016 Lisboa, Portugal\\
Email: atle.hahn@gmail.com
  \end{center}

\begin{abstract}
In the present paper we review the main results of a series of recent papers
on the non-Abelian Chern-Simons path integral on $M=\Sigma \times S^1$
in the so-called ``torus gauge''.
More precisely, we study the torus gauge fixed version
 of the Chern-Simons path integral expressions
$Z(\Sigma \times S^1,L)$ associated to  $G$ and $k \in \bN$
where  $\Sigma$ is a compact, connected, oriented surface,  $L$ is a  framed,
colored link in $\Sigma \times S^1$, and
 $G$ is a  simple, simply-connected, compact Lie group.\par

We demonstrate that  the torus gauge approach
allows a rather quick explicit evaluation of
$Z(\Sigma \times S^1,L)$. Moreover,
 we  verify in several special cases
that the explicit values obtained for
 $Z(\Sigma \times S^1,L)$ agree with the values of the corresponding Reshetikhin-Turaev invariant.  Finally, we  sketch three different approaches for obtaining
 a  rigorous realization of the torus gauge fixed CS path integral. \par

 It remains to be seen  whether also for general $L$ the  explicit values obtained for
 $Z(\Sigma \times S^1,L)$ agree with those of the corresponding Reshetikhin-Turaev invariant.
 If this is indeed the case then this could lead to progress towards the solution of several open questions in Quantum Topology.
\end{abstract}





\medskip

\section{Introduction}
\label{sec1}

Let $M$ be a  compact, connected, oriented 3-manifold, let $k \in \bN$, and let
 $G$ be a simple, simply-connected, compact Lie group.
 At a heuristic level, the map  which associates to every (colored) link $L$ in $M$ the (informal) Chern-Simons path integral  $Z(M,L)$ corresponding to $(G,k)$
is then a link invariant\footnote{This was first observed in \cite{Schw}
where it was also suggested that $Z(M,L)$ might be related to the Jones polynomial.}.
In a celebrated paper, cf. \cite{Wi},   Witten  succeeded in evaluating $Z(M,L)$
and $Z(M):= Z(M,\emptyset)$ explicitly using arguments from Conformal Field Theory.
    Later, Reshetikhin and  Turaev found  rigorous versions $RT(M)$ and  $RT(M,L)$ of Witten's  invariants    $Z(M)$, $Z(M,L)$ using the representation theory of quantum groups, cf. \cite{ReTu1,ReTu2} and  \cite{turaev}.  \par

At present  it is not  clear if/how one can derive the algebraic expressions
$RT(M,L)$ directly from  the (informal) path integral expressions $Z(M,L)$
and whether it is possible  to find a  rigorous path integral realization $Z_{rig}(M,L)$ of  $Z(M,L)$ (or, alternatively, of a gauge fixed version of $Z(M,L)$).
These two open problems, ``Problem (P1)'' and ``Problem (P2)'', are important  by themselves, cf. \cite{Kup,Saw99}.
Moreover,   if Problems (P1) and (P2) can be solved this could  lead to  progress
 towards the solution of several other  open problems in the field of  3-manifold quantum topology, cf.  Sec. \ref{subsec5.2} below.  \par

In the present paper we will  restrict our attention  to the special base manifolds $M$ of the form $M=\Sigma \times S^1$ for which   ``torus gauge fixing'' is available.
This gauge fixing was first applied  to the Chern-Simons path integral by  Blau and Thompson in \cite{BlTh1}
where it was shown (cf. Remark \ref{rm_BlTh_Lob} in Sec. \ref{subsec3.1} below)  that this allows a remarkably quick and  simple (informal)  evaluation of $Z(M)$ and of $Z(M,L)$  in the special case where $L$  is a ``fiber link'' in $M= \Sigma \times S^1$,
 i.e. a link consisting only of loops which are ``parallel to the $S^1$-component''
  (or, more precisely, a link consisting only of loops each of which is contained completely
    in some  $S^1$-fiber of  $M= \Sigma \times S^1$). \par

In more recent work Blau and Thompson generalized torus gauge fixing first to non-trivial $S^1$-bundles $M$ (cf. \cite{BlTh4}) and then to Seifert fibered spaces $M$ (cf. \cite{BlTh5}) and used this
to evaluate   $Z(M)$ and $Z(M,L)$ for ``fiber links'' $L$ in $M$.
By doing so they recovered the explicit expressions obtained earlier in \cite{BeaWi,Bea}
where non-Abelian localization was applied to the CS path integral (cf. Remark \ref{rm_3.3.9_II} in Sec. \ref{subsec3.3} below). \par

Instead of trying to generalize the torus gauge fixing approach to the CS path integral
to more general manifolds like in \cite{BlTh4,BlTh5}  one can also try to generalize this approach to general (colored) links $L$ in the original (trivial) $S^1$-bundles $M = \Sigma \times S^1$. This question is studied in the series of papers \cite{Ha3b,Ha3c,Ha4,HaHa,Ha7a,Ha7b,Ha6b,Ha9} where apart from the explicit
evaluation of $Z( \Sigma \times S^1,L)$ for general $L$   we also  consider
the issue of finding a  rigorous realization of $Z( \Sigma \times S^1,L)$.
The short term goal of the program initiated in \cite{Ha3b,Ha3c,Ha4,HaHa,Ha7a,Ha7b,Ha6b,Ha9}
 is to obtain a complete solution of the aforementioned problems  (P1) and (P2)
 for manifolds $M$ of the form $M = \Sigma \times S^1$ (cf. Sec. \ref{subsubsec3.5.2} and Sec. \ref{sec4} below).
 The medium term goal is to solve  problems  (P1) and (P2)
 for general links in all those manifolds $M$  considered in \cite{BlTh4,BlTh5}
 (by combining the ideas/methods in the present paper with those in \cite{BlTh4,BlTh5},
  cf. Sec.  \ref{subsec5.1} below).
  The long term goal is to
exploit this for making progress regarding some of the open problems in Quantum Topology
hinted at  above (and described in more detail in Sec.  \ref{subsec5.2} below).

\smallskip

 The present paper  reviews and extends\footnote{We have included some new material, cf. Sec. \ref{subsubsec3.2.3},
  Sec. \ref{subsec4.3}, Sec. \ref{subsec5.2},
  Appendix \ref{appA.2}, Appendix \ref{appA.6}, and Appendix \ref{appD}.
  Moreover,
  we have streamlined the presentation in \cite{Ha3b,Ha3c,Ha4,HaHa,Ha7a,Ha7b,Ha6b,Ha9},
  see, in particular, Sec. \ref{subsec2.2}, Sec. \ref{subsec2.3}, Sec. \ref{subsec3.2},  and Appendices \ref{appA.3}--\ref{appA.5}. }
  the results of \cite{Ha3b,Ha3c,Ha4,HaHa,Ha7a,Ha7b,Ha6b,Ha9}.
The emphasis is on the explicit evaluation of $Z( \Sigma \times S^1,L)$.
 The  rigorous realization of (the torus gauge fixed version of)
 $Z( \Sigma \times S^1,L)$ is only outlined, cf. Sec. \ref{sec4} below.

 \smallskip

 The present paper is organized as follows: \par

In Sec. \ref{sec2} we give a self-contained
rederivation of the  formula, found in \cite{Ha7a}, for the general\footnote{i.e. for general, colored links $L$ in $M=\Sigma \times S^1$} (informal) Chern-Simons path integral on $M= \Sigma \times S^1$ in the torus gauge,   see  Eq. \eqref{eq2.48_pre} below. (Eq. \eqref{eq2.48_pre} below is later rewritten in a suitable way,  cf.  Eq. \eqref{eq2.48} and cf. also Eq. \eqref{eq2.48_reg} in Sec. \ref{subsec3.2} below).\par

In Sec. \ref{sec3} we evaluate $Z(\Sigma \times S^1,L)$ explicitly in several situations.
First we give a complete evaluation of  $Z(\Sigma \times S^1,L)$ in three interesting special cases\footnote{\label{ft_sec1_3}From the knot theoretic point of view the most interesting of these special cases is the case considered
in Sec. \ref{subsec3.3} where $L$ belongs to a large class
of colored torus (ribbon) knots in  $S^2 \times S^1$.
The explicit formula for $Z(\Sigma \times S^1,L)$ in this special case
(cf. Eq. \eqref{eq_Step6_Ende} below)
can be  generalized in a straightforward way (cf. Eq. \eqref{eq_Step7_1}
and the rewritten version Eq. \eqref{eq_RT_WLO_spec} below).
 By combining Eq. \eqref{eq_RT_WLO_spec} with a special case of Witten's surgery formula
 (cf. Eq. \eqref{eq_surgery_formula1}) we then arrive (for all $G$) at the so-called ``Rosso-Jones formula''
for colored torus knots in $S^3$ (cf. Eq. \eqref{eq_my_RossoJones3} below).} (cf. Secs
\ref{subsec3.1}, \ref{subsec3.3}, and \ref{subsec3.4})
and then we  sketch the evaluation of $Z(\Sigma \times S^1,L)$ in the case of general $L$
(cf. Sec. \ref{subsec3.5} and cf. also Appendix \ref{appD}).
We refer to the beginning of Sec. \ref{sec3} for a more detailed summary of the
content of Sec. \ref{sec3}. \par

In Sec. \ref{sec4} we summarize and sketch
the various approaches studied in \cite{Ha3b,Ha3c,Ha4,HaHa,Ha7a,Ha7b,Ha6b,Ha9}
for obtaining a rigorous realization of  $Z(\Sigma \times S^1,L)$
and of the computations in Sec. \ref{sec3}. \par
In Sec. \ref{sec5} we conclude the main part of the present paper with a
 short outlook explaining in more detail  the medium and long term goals mentioned above.

\smallskip

The present paper has four appendices. \par

In Appendix \ref{appB} we list the Lie theoretic and quantum algebraic notation used in the paper.\par

In Appendix \ref{appA} we fill in some technical details omitted in Sec. \ref{sec2}. \par

In Appendix \ref{appC} we recall the definition of Turaev's shadow invariant in the special case relevant for us and discuss its relation with the Reshetikhin-Turaev invariant. \par
Appendix \ref{appD} is a supplement to Sec. \ref{subsubsec3.5.2}.

\section{The  Chern-Simons  path integral  in the torus gauge}
\label{sec2}

\subsection{The original Chern-Simons path integral}
\label{subsec2.1}

Let $M$ be a compact, connected, oriented 3-manifold
and let $G$ be a  simple,  simply-connected, compact Lie group.
We denote the Lie algebra of $G$ by $\cG$ and we set\footnote{Here and in the following  $\Omega^p(N,V)$ denotes, for every finite-dimensional real vector space $V$,  the space of $V$-valued  $p$-forms on a smooth manifold $N$.}
\begin{align} \cA & :=  \Omega^1(M,\cG), \\
    \G & :=C^{\infty}(M,G)
\end{align}

Let $k \in \bN$ and let $\langle \cdot , \cdot \rangle$
be the unique $\Ad$-invariant scalar product
on $\cG$ normalized such that $\langle \Check{\alpha} , \Check{\alpha} \rangle = 2$
for every short real coroot $\Check{\alpha}$ (w.r.t. to any fixed Cartan subalgebra of $\cG$). \par

The ``Chern-Simons action function''
associated to $M$, $G$, and the ``level''
  $k$ is the function $S_{CS}: \cA \to \bR$ given by
    \begin{equation} \label{eq2.2}
 S_{CS}(A) = - k \pi \int_M \langle A \wedge dA \rangle
   + \tfrac{1}{3} \langle A\wedge [A \wedge A]\rangle \quad \forall A \in \cA
 \end{equation}
where $[\cdot \wedge \cdot]$  denotes the wedge  product associated to the
Lie bracket $[\cdot,\cdot] : \cG \times \cG \to \cG$
and    $\langle \cdot \wedge  \cdot \rangle$  the wedge product  associated to the
 scalar product $\langle \cdot , \cdot \rangle : \cG \times \cG \to \bR$.
The normalization of the scalar product $\langle \cdot,  \cdot \rangle$ chosen above
ensures that
$$\cA \ni A \mapsto \exp(i S_{CS}(A)) \in \bC$$
 is ``gauge invariant'', i.e.  invariant under the standard right-action of the  group  $\G$  on $\cA$ (cf. Eq. \eqref{eq_gauge_trans} below), see Sec. 1 in \cite{Wi} for the case $G=SU(N)$ and, e.g., \cite{Roz} for the case of general $G$.

\smallskip

A (smooth) knot in $M$ is a smooth embedding $K:S^1 \to M$.
Note that, using the surjection $i_{S^1}:[0,1] \ni s \mapsto e^{2 \pi i s} \in \{ z \in \bC \mid |z| =1 \}  \cong S^1$, we can  consider each  knot $K$ in $M$ as a (smooth) loop $l:[0,1] \to M$, $l(0) = l(1)$, in the obvious way. \par
In the following let us fix an (ordered) ``link''  in $M$, i.e.
a finite tuple $L = (l_1, l_2, \ldots, l_m)$, $m \in \bN$, of pairwise non-intersecting
knots $l_i$ . We equip  each $l_i$ with a ``color'', i.e. an irreducible,
 finite-dimensional, complex representation $\rho_i$ of $G$.
 By doing so we obtain a ``colored link''  $((l_1,l_2,\ldots,l_m), (\rho_1,\rho_2,\ldots,\rho_m))$, which will also be denoted by ``$L$'' in the following.\par

The ``Chern-Simons path integral associated to $(M,G,k)$ and $L$''  is the informal integral expression given by
(cf. Sec. 1 in \cite{Wi})
\begin{equation} \label{eq_WLO_orig}
Z(M,L) := \int_{\cA} \left( \prod_{i=1}^m  \Tr_{\rho_i}(\Hol_{l_i}(A)) \right) \exp( i S_{CS}(A)) DA
\end{equation}
 where $DA$ is the (ill-defined) ``Lebesgue measure'' on the
infinite-dimensional space $\cA$ and where\footnote{In the physics literature the notation $P \exp(\int_l A)$
is often
used instead of $\Hol_{l}(A)$.} $\Hol_{l}(A) \in G$ is the holonomy of $A \in \cA$ around the knot $l = l_i$, $i \le m$
(considered as a smooth loop), cf. Eqs \eqref{eq_Hol_definition} below.
It will sometimes be convenient to work with
 the  ``normalization'' $\langle L \rangle$ of $Z(M,L)$ given by
\begin{equation} \label{eq_sec2.1_bracket}
\langle L \rangle:= \frac{Z(M,L)}{Z(M)}
\end{equation}
where
\begin{equation}Z(M):= \int_{\cA}  \exp( i S_{CS}(A)) DA.
\end{equation}

\subsubsection{Restriction to the case of matrix Lie groups}
\label{subsubsec2.1.1}

Since every compact Lie group
is isomorphic to a matrix Lie group
we can (and will) assume, without loss of generality, that
 $G \subset \GL(N,\bR)$ (and hence $\cG \subset \Mat(N,\bR)$) for some fixed $N \in \bN$.
(This will be very convenient in Sec. \ref{subsec2.3}
and in parts \ref{appA.3}--\ref{appA.5} of the appendix below.)
We can then rewrite Eq. \eqref{eq2.2} as
  \begin{equation} \label{eq2.2'} S_{CS}(A) =  k \pi \int_M \Tr(A \wedge dA
   + \tfrac{2}{3}  A \wedge A \wedge A)
 \end{equation}
 where  $\wedge$  is the wedge product
for $\Mat(N,\bR)$-valued forms and
where  $\Tr:\Mat(N,\bR) \to \bR$ is the trace functional normalized such that
\begin{equation} \label{eq_scal_prod_normalization}
 \Tr(C  D) = - \langle C, D \rangle \quad \text{ for all } C, D \in \cG \subset  \Mat(N,\bR).
 \end{equation}
(This is always possible because,  by assumption, $G$ is simple.)\par

For two $\Mat(N,\bR)$-valued forms $\alpha$ and  $\beta$
we will simply write  $\alpha \beta$ instead of $\alpha \wedge \beta$
if $\alpha$ or $\beta$ is a $0$-form.
The standard right-action of $\G$ on  $\cA$ mentioned above can then be written as
\begin{equation} \label{eq_gauge_trans}
A \cdot \Omega = \Omega^{-1}A\Omega + \Omega^{-1} d\Omega \quad
\forall A \in \cA, \Omega \in \G
\end{equation}

Moreover, for every $A \in \cA$ and every smooth loop $l:[0,1] \to M$ we then have
\begin{subequations} \label{eq_Hol_definition}
\begin{equation}\Hol_{l}(A) = P_1(A)
\end{equation}
 where
$P(A) = (P_s(A))_{s \in [0,1]}$ is the unique smooth map $[0,1] \to \Mat(N,\bR)$ such that
\begin{equation}\forall s \in [0,1]: \ \tfrac{d}{ds} P_s(A) =  P_s(A) \cdot A(l'(s)), \quad P_0(A)=1
\end{equation}
\end{subequations}
where ``$\cdot$'' is the multiplication of $\Mat(N,\bR)$.

\subsection{Torus gauge fixing}
\label{subsec2.2}

As mentioned in Sec. \ref{sec1} ``torus gauge fixing'' was introduced in \cite{BlTh1}
 and used for the (informal) evaluation of $Z(\Sigma \times S^1)$ and
 $Z(\Sigma \times S^1,L)$ for ``fiber links'' $L$ in  $M = \Sigma \times S^1$.
  In \cite{Ha3c,Ha4,Ha7a} the formula (7.9) in \cite{BlTh1} was
generalized\footnote{cf. part (ii) of Remark \ref{rm2.2} in Sec. \ref{subsubsec2.2.2} below
and Remark \ref{rm_non_equiv} in Sec. \ref{subsec3.1}
for more comments regarding the relation between Eq. (7.9) in \cite{BlTh1} and Eq. (6.8)
  in \cite{BlTh3}
on the one hand and our Eq. \eqref{eq2.24}, Eq. \eqref{eq2.48_pre}, and Eq. \eqref{eq2.48} on the other hand.}
to arbitrary links $L$ in $M = \Sigma \times S^1$, cf.  Eq. \eqref{eq2.24},
 Eq. \eqref{eq2.48_pre}, and Eq. \eqref{eq2.48} below.
In Sec. \ref{subsec2.2},  Sec. \ref{subsec2.3}, and Appendix \ref{appA}
 we give a shortened (but self-contained) rederivation  of  these three equations.

\medskip

In order to motivate the derivation of Eq.  \eqref{eq2.24} in Sec. \ref{subsubsec2.2.2} below we will first derive an analogous formula
for the  manifold $S^1$.

\subsubsection{Motivation: Torus gauge fixing for the manifold $S^1$}
\label{subsubsec2.2.1}

We will now fix, for the rest of this paper, a maximal torus  $T$ of $G$ and denote
by $\ct$  the Lie algebra of $T$ and by $\ck$  the  $\langle \cdot,\cdot \rangle$-orthogonal
complement of $\ct$ in $\cG$.

\medskip

By  $\frac{\partial}{\partial t}$ we will denote the vector field
  on  $S^1$ which is induced by the map  $i_{S^1}: [0,1]  \ni s \mapsto  \exp(2\pi i s) \in  \{ z \in \bC \mid |z| =1 \} \cong S^1$ and by $dt$ we denote the 1-form on $S^1$ which is dual to  $\frac{\partial}{\partial t}$. \par

Moreover, we set
\begin{subequations}
\begin{align}
\cA_{S^1}& := \Omega^1(S^1,\cG), \\
\G_{S^1} & := C^{\infty}(S^1,G).
\end{align}
\end{subequations}
The group $\G_{S^1}$ acts on $\cA_{S^1}$ from the right by the obvious analogue of Eq. \eqref{eq_gauge_trans} in Sec. \ref{subsec2.1} above.

\medskip

In the following we want to show, informally, that for every continuous\footnote{\label{ft_no_top}Here we
assume that $\cA_{S^1}$ is equipped with a suitable topology, which we do not specify
since the derivation of Eq. \eqref{eq_AppA1_lang} is informal anyway.}
$\G_{S^1}$-invariant function $\chi: \cA_{S^1} \to \bC$ we have
\begin{align} \label{eq_AppA1_lang}
 \int_{\cA_{S^1}} \chi(A) DA  \sim \int_{\ct} \chi(b \ dt)
   \det(1_{\ck} - \exp(\ad(b))_{|\ck}) db
  \end{align}
  where $\sim$ denotes equality up to a multiplicative constant
independent of $\chi$ and where   $1_{\ck}$  is the identity on $\ck$.
Moreover, $DA$ is the (informal) Lebesgue measure on $\cA_{S^1}$ and
 $db$ denotes the  normalized Lebesgue measure on $\ct = (\ct,\langle\cdot,\cdot\rangle)$.
 (Here  $\int_{\ct} \cdots db$ and   $\int \cdots DA$
  are sloppy notations
  for the improper integrals  $\int^{\sim}_{\ct} \cdots db$ and
  $\int^{\sim} \cdots DA$
   appearing in Proposition \ref{prop2.1} and
   Remark \ref{rm_sec2.2.1}   below.)

\medskip

Eq. \eqref{eq_AppA1_lang}  can be  derived using  a standard Faddeev-Popov determinant computation.  In the present section we will
  give an alternative derivation of Eq. \eqref{eq_AppA1_lang}
which is based on a corollary  of the Weyl integral formula (cf. Proposition \ref{prop2.1} and  Appendix \ref{appA.1} below) and which has at least the following two advantages: Firstly, it is probably more accessible for mathematicians
and secondly, and more importantly,
 it can  be extended successfully to the situation in Sec. \ref{subsubsec2.2.2} below.
 [By contrast, the argument using the  Faddeev-Popov determinant computations
 leads to certain difficulties when applied to $M= \Sigma \times S^1$
 if the surface $\Sigma$  is compact, cf. the second paragraph after Eq. \eqref{eq2.18} below.]

 \begin{proposition} \label{prop2.1}
For every continuous conjugation invariant function
$f:G \to \bC$  we have
\begin{align} \label{eq_WeylInt}  \int_G f(g) dg  & \sim  \int^{\sim}_{\ct} f(\exp(b)) \det(1_{\ck} - \exp(\ad(b))_{|\ck}) db
\end{align}
where  $dg$ is the normalized  Haar measure  on $G$
and  where $\int^{\sim}_{\ct} \phi(b)  db$ is a suitably defined\footnote{\label{ft_improper_int}For example,
we  can use the definition  $\int^{\sim}_{\ct} \phi(b)  db := \lim_{R \to \infty} \tfrac{1}{vol(R \cdot Q)} \int_{R \cdot Q} \phi(b) db$
where  $Q$ is, e.g., a unit hypercube centered around $0 \in \ct$
or the unit  ball  around $0$ or we can make the ansatz
$\int^{\sim}_{\ct} \phi(b)  db := \lim_{\eps \to 0} \int \phi(b) d\mu_{\eps}(b)$
where $ d\mu_{\eps}(b) := e^{-\eps\|b\|^2} db /\int e^{-\eps\|b\|^2}  db$.}
improper integral which extracts the
 ``mean value'' of a periodic\footnote{Note that $
 \det(1_{\ck} - \exp(\ad(b))_{|\ck}) = \det(1_{\ck} - \Ad(\exp(b))_{|\ck})$
 so  $\phi(b) =  f(\exp(b)) \det(1_{\ck} - \exp(\ad(b))_{|\ck})$
 is periodic.} function $\phi$  on $\ct$.
\end{proposition}

Using Proposition \ref{prop2.1} we can now derive Eq. \eqref{eq_AppA1_lang} above as follows:

\smallskip

Let $ \tilde{\G}_{S^1}  := \{\Omega \in \G_{S^1} \mid \Omega(1) = 1\} $.
It is not difficult to see  that\footnote{The surjectivity of $\psi$ follows from the fact that, since $G$ was assumed to be compact and connected, the exponential map
$\exp:\cG \to G$ is surjective. The
 injectivity follows from a short explicit calculation.}
$\psi: \cA_{S^1}/\tilde{\G}_{S^1}  \ni [A] \mapsto \Hol_{i_{S^1}}(A) \in G$
is a well-defined bijection. From the bijectivity of $\psi$
it follows that for every  $\tilde{\G}_{S^1}$-invariant
 $\chi:\cA_{S^1} \to \bC$ there is a
$\bar{\chi}:  G \to \bC$ such that
$$\chi = \bar{\chi} \circ p$$
   where
$$p:\cA_{S^1} \ni A \mapsto \Hol_{i_{S^1}}(A) \in G.$$

Accordingly,  we obtain, informally, for every $\tilde{\G}_{S^1}$-invariant function $\chi:\cA_{S^1} \to \bC$
\begin{subequations}
\begin{equation} \label{eq_sec2.2.1a}
 \int_{\cA_{S^1}} \chi(A) DA =  \int_{\cA_{S^1}} \bar{\chi}(p(A)) DA    \overset{(*)}{\sim} \int_{G} \bar{\chi}(g) dg
  \end{equation}
where  step $(*)$ is justified in Remark \ref{rm_sec2.2.1}  below.\par
Next observe that as $\chi$ was not only $\tilde{\G}_{S^1}$-invariant  but even ${\G}_{S^1}$-invariant
the function $\bar{\chi}$ will be conjugation invariant.
Accordingly,  we now obtain from Proposition \ref{prop2.1} above
and the relation $ \exp(b) = \Hol_{i_{S_1}}(b dt) = p(b dt)$,  $b \in \ct$,
\begin{align}  \label{eq_sec2.2.1b}
  \int_{G} \bar{\chi}(g) dg & \sim \int_{\ct}^{\sim} \bar{\chi}(\exp(b))    \det(1_{\ck} - \exp(\ad(b)))_{|\ck}) db \nonumber \\
&  \sim \int_{\ct}^{\sim} \chi(b \ dt)
   \det(1_{\ck} - \exp(\ad(b)))_{|\ck}) db
  \end{align}
  \end{subequations}
By combining Eq. \eqref{eq_sec2.2.1a} with Eq. \eqref{eq_sec2.2.1b} we arrive at Eq. \eqref{eq_AppA1_lang} above.

\begin{remark} \label{rm_sec2.2.1}
In order to justify step $(*)$ in Eq. \eqref{eq_sec2.2.1a} above
we will now (re)interpret the informal integral  $\int \chi(A) DA$
appearing above as the (informal) improper  integral
 $\int^{\sim} \chi(A) DA :=  \lim_{\eps \to 0} \int \chi(A) d\mu_{\eps}(A)$
where $ d\mu_{\eps}$ is the informal Gaussian measure on $\cA_{S^1}$
given by $ d\mu_{\eps}(A)  := e^{-\eps\|A\|_2^2} DA /\int e^{-\eps\|A\|_2^2} DA$
where $\|\cdot\|_2$ is the $L^2$-norm on $\cA_{S^1}$ associated to any fixed Riemannian metric on $S^1$.  Step $(*)$ in Eq. \eqref{eq_sec2.2.1a} above then follows at an informal level
  provided that  we can argue that $p_*(d\mu_{\eps}) \to dg$
   weakly as $\eps \to 0$  where $p_*(d\mu_{\eps})$ is the pushforward of $d\mu_{\eps}$ under $p$.    By using standard techniques in probability theory  one can obtain rigorous versions of this informal result.
 (One such rigorous version will be included in an additional part of the appendix
  in the next version of the present paper.)
 \end{remark}

\subsubsection{Torus gauge fixing for $M=\Sigma \times S^1$}
\label{subsubsec2.2.2}

Let $M$ be a smooth 3-manifold  of the form $M = \Sigma \times S^1$ where $\Sigma$ is a
 connected, orientable surface.
As in Sec. \ref{subsec2.1}
we use the notation
 $\cA  =  \Omega^1(\Sigma \times S^1,\cG)$ and $\G  = C^{\infty}(\Sigma \times S^1,G)$.

\smallskip

 Recall that at the beginning of Sec. \ref{subsubsec2.2.1}
 we introduced a vector field $\frac{\partial}{\partial t}$ and a 1-form  $dt$
 on $S^1$.   The obvious ``lift''/pullback of $\frac{\partial}{\partial t}$ and $dt$
 to  $M = \Sigma \times S^1$
will also be denoted by $\frac{\partial}{\partial t}$ and $dt$ in the following.
Observe that we have
\begin{equation} \label{eq_cA_decomp}
\cA = \cA^{\orth} \oplus \cA^{||}
\end{equation}
where we have set
\begin{subequations} \label{eq_basic_spaces_cont}
\begin{align}
\cA^{\orth} & :=  \{ A \in \cA \mid A(\partial/\partial t) = 0\}, \\
\cA^{||} & := \{ A_0 dt \mid A_0 \in C^{\infty}(\Sigma \times S^1,\cG) \}.
 \end{align}
\end{subequations}

In the following we set
\begin{equation}
\cB  := C^{\infty}(\Sigma,\ct)
\end{equation}
and we will make  the identification
\begin{equation}
\cB \cong \{A_0 \in  C^{\infty}(\Sigma \times S^1,\ct) \mid
 \forall \sigma \in \Sigma: A_0(\sigma, \cdot) \text{ is constant }\}
\end{equation}

Now let  $\chi:\cA \to \bC$ be a   $\G$-invariant function,
which we assume to be continuous w.r.t. a suitable topology on $\cA$
 (cf. Footnote \ref{ft_no_top} above).
By applying (in a naive way)  a standard Faddeev-Popov determinant argument  to the situation at hand one arrives, informally, at the following analogue of Eq. \eqref{eq_AppA1_lang} above
   \begin{equation} \label{eq2.18}
 \int_{\cA} \chi(A) DA
 \sim  \int_{\cB} \biggl[ \int_{\cA^{\orth}} \chi(A^{\orth} + B dt)
   DA^{\orth}  \biggr]   \det\bigl(1_{\ck}-\exp(\ad(B))_{| \ck}\bigr) DB
  \end{equation}
  where $\sim$ denotes equality up to a multiplicative constant
independent of $\chi$.
 Moreover,  $DB$ denotes the (informal) ``Lebesgue measure'' on $\cB$,
    $1_{\ck}-\exp(\ad(B))_{|\ck} $ is the
  linear operator on $C^{\infty}(\Sigma,\ck)$  given by
 $(1_{\ck}-\exp(\ad(B))_{|\ck} \cdot f)(\sigma) = (1_{\ck}-\exp(\ad(B(\sigma)))_{|\ck}) \cdot f(\sigma)$ for all
 $\sigma \in \Sigma$, $f \in   C^{\infty}(\Sigma,\ck)$
 and  $\det\bigl(1_{\ck}-\exp(\ad(B))_{| \ck}\bigr)$ is its (informal) determinant.
 (See  Sec. \ref{subsubsec2.3.2} below for a rigorous realization\footnote{More precisely, in Sec. \ref{subsubsec2.3.2} below we will make rigorous sense of the informal
 expression which we obtain after combining $\det\bigl(1_{\ck}-\exp(\ad(B))_{| \ck}\bigr)$
  with another factor, cf.  Eq. \eqref{eq2.49b}  and Eq. \eqref{eq_def_det(B)} in Sec. \ref{subsubsec2.3.1}.} of $\det\bigl(1_{\ck}-\exp(\ad(B))_{| \ck}\bigr)$.)

\smallskip

However, a more careful analysis of ``torus gauge fixing'' and its properties
shows that, as a result of  certain topological
obstructions, Eq. \eqref{eq2.18} is not correct if $\Sigma$ is compact,
 cf.  Sec. 6 in \cite{BlTh3} and Sec. 2.2 in \cite{Ha7a}.
(It will not be necessary to repeat
 the analysis in \cite{BlTh3} or \cite{Ha7a} here. The origin of the aforementioned
 topological  obstructions  will get obvious during our derivation of Eq. \eqref{eq2.24} below in Appendix \ref{appA.2}.)

 \smallskip

 Before we state the corrected version of  Eq. \eqref{eq2.18}
   we need some notation. Let
\begin{equation}\G_{\Sigma} := C^{\infty}(\Sigma,G).
\end{equation}
The group $\G_{\Sigma}$
 acts on  $C^{\infty}(\Sigma,G/T)$ from the left by
$$\Omega \cdot \bar{g}  = \Omega\bar{g}  \quad \text{  for all $\bar{g} \in
C^{\infty}(\Sigma,G/T)$ and $\Omega \in \G_{\Sigma}$ } $$
where $\Omega\bar{g} \in C^{\infty}(\Sigma,G/T)$ is given by
$(\Omega\bar{g})(\sigma) = \Omega(\sigma) \bar{g}(\sigma)$
for all $\sigma \in \Sigma$.
The corresponding orbit space will be denoted by\footnote{The standard
notation for this orbit space would be $\G_{\Sigma} \backslash C^{\infty}(\Sigma,G/T)$ but in order to be consistent with the notation in \cite{Ha3c,Ha7a} where we worked with  a right-action  of $\G_{\Sigma}$ on $C^{\infty}(\Sigma,G/T)$
 we will use the notation $C^{\infty}(\Sigma,G/T)/\G_{\Sigma}$ in the following.}
\begin{equation}C^{\infty}(\Sigma,G/T)/\G_{\Sigma}.
\end{equation}
Moreover,  we set
\begin{equation}\cB_{reg}:= C^{\infty}(\Sigma,\ct_{reg})
\end{equation}
where $\ct_{reg} \subset \ct$ is the union of the Weyl alcoves of $\ct$,
cf.  Appendix \ref{appB.1} below.  \par

Finally, we  set
$ \bar{g} b \bar{g}^{-1} := g b g^{-1} \in \cG$ for each  $\bar{g} \in G/T$ and $b \in \ct$
where $g$ is an arbitrary  element of $G$ fulfilling $gT = \bar{g}$.

\smallskip

In Appendix \ref{appA.2} below we will derive (at an informal level)
the following corrected version of Eq. \eqref{eq2.18}
\begin{multline} \label{eq2.24}
\int_{\cA} \chi(A) DA \sim \sum_{\cl \in C^{\infty}(\Sigma,G/T)/\G_{\Sigma}}
 \int_{\cB_{reg}} \biggl[ \int_{\cA^{\orth}} \chi(A^{\orth} +
  \bar{g}_{\cl} B  \bar{g}_{\cl}^{-1} dt)  DA^{\orth} \biggr] \\
  \times   \det\bigl(1_{\ck}-\exp(\ad(B))_{| \ck}\bigr) DB
\end{multline}
where
$(\bar{g}_{\cl})_{\cl \in  C^{\infty}(\Sigma,G/T)/ \G_{\Sigma}}$ is any fixed system
 of representatives   of $C^{\infty}(\Sigma,G/T)/ \G_{\Sigma}$
   and where $\bar{g}_{\cl} B \bar{g}_{\cl}^{-1} \in C^{\infty}(\Sigma,\cG)$,
   for  $\bar{g}_{\cl} \in C^{\infty}(\Sigma,G/T)$ and $B \in \cB = C^{\infty}(\Sigma,\ct)$,   is given by
$(\bar{g}_{\cl}B \bar{g}_{\cl}^{-1})(\sigma):= \bar{g}_{\cl}(\sigma)
B(\sigma) \bar{g}_{\cl}(\sigma)^{-1}$ for all $\sigma \in \Sigma$.

\smallskip

See Remark \ref{rm_appA.2} in Appendix \ref{appA.2}
for a quick explanation regarding the origin of the
 differences between the RHS of Eq. \eqref{eq2.24}
and the RHS of Eq. \eqref{eq2.18}, and see also part (ii) of Remark \ref{rm2.2} below
for a comparison between our Eq. \eqref{eq2.24} and
 formula (6.8) in \cite{BlTh3}.

 \begin{remark} \label{rm2.1}
(i) From the assumption that $\dim(\Sigma)=2$ and that $G$ is simply-connected it follows that
  two maps $\bar{g}_1, \bar{g}_2 \in  C^{\infty}(\Sigma,G/T)$
  are in the same $\G_{\Sigma}$-orbit iff they are homotopic (cf. Proposition 3.2 in \cite{Ha3c}).
  Accordingly, we can identify     $C^{\infty}(\Sigma,G/T)/\G_{\Sigma}$  with the set $[\Sigma,G/T]$
    of  homotopy classes of  (smooth or continuous) maps $\Sigma \to G/T$. \par

 On the other hand,  every\footnote{That this indeed holds for every non-compact
  (connected, orientable) surface $\Sigma'$
 is a rather deep result in low-dimensional topology.
 It follows, e.g., from the result by Behnke and Stein (1948)
 that every non-compact, connected Riemann surface is a Stein manifold.
 I emphasize that in the main part of the present paper we need
 this result only in the special case  $\Sigma' = \Sigma \backslash \{\sigma_0\}$
   where $\Sigma$ is  a compact, connected,  orientable surface and $\sigma_0 \in \Sigma$ a fixed point.  Using the classification theorem for compact, connected, orientable surfaces
 it is not difficult to show directly that  $\Sigma \backslash \{\sigma_0\}$ is
 indeed homotopy equivalent to a 1-dimensional CW-complex.} non-compact (connected, orientable) surface is homotopy equivalent to a 1-dimensional CW-complex.
Since  $G/T$ is simply-connected (cf. Prop. 7.6 in Chap. V in \cite{Br_tD})
this means that every  continuous map $\bar{g}:\Sigma \to G/T$ is null-homotopic.
  This implies that for non-compact $\Sigma$ we have $[\Sigma, G/T] = \{[1_T]\}$ where   $1_T:\Sigma \to G/T$ is the constant map taking only the value $T \in G/T$.  Accordingly, Eq. \eqref{eq2.24} then reduces to Eq.  \eqref{eq2.18} (with $\cB$
  replaced by  $\cB_{reg}$).

  \smallskip

(ii) In part (i) of the present remark we observed  that if $\Sigma$ is non-compact then every continuous map $\bar{g}:\Sigma \to G/T$  is
null-homotopic. Since $\pi_{G/T}: G \to G/T$ is a  fiber bundle and therefore possesses   the homotopy lifting property  (cf., e.g.,  \cite{Hu})  we conclude that if $\Sigma$ is non-compact then every continuous map $\bar{g}:\Sigma \to G/T$  can   be lifted to a
continuous map $\Omega: \Sigma \to G$.
Moreover, if $\bar{g}$ is smooth then $\Omega$ can be chosen to be smooth as well.
This observation will play an important role in Sec. \ref{subsec2.3} below,
 cf. the paragraph after Eq. \eqref{eq_sigma0_cond}  below.
\end{remark}

\begin{remark} \label{rm2.2}   (i) In Sec. \ref{subsec2.3} below we will apply
 Eq. \eqref{eq2.24} to the situation relevant for the CS path integral, i.e. we
 take $\chi = \chi_L$ where  $\chi_L$ is given by Eq. \eqref{eq2.31} below.
 Using a suitable change of variable argument
we then arrive at the very simple  equation Eq. \eqref{eq2.48_pre} below (which is later rewritten as  Eq. \eqref{eq2.48_Ha7a} and Eq. \eqref{eq2.48}).

 \smallskip

  (ii)  In the special case where  $L$ is either  the ``empty link'' or a  ``fiber link'',
 which is the only case treated in \cite{BlTh1} (cf. Eq. (7.9) in \cite{BlTh1} and  Remark \ref{rm_non_equiv} in Sec. \ref{sec3} below),    Eq. \eqref{eq2.48} mentioned in part (i) simplifies considerably leading to Eq. \eqref{eq2.48_simpl}  below.
In order to compare Eq. \eqref{eq2.48_simpl} below with Eq. (7.9) in \cite{BlTh1} observe first that
 according to Remark \ref{rm2.1} there is a natural 1-1-correspondence
between the elements of  the orbit space $C^{\infty}(\Sigma,G/T)/\G_{\Sigma}$ and the elements of  $[\Sigma,G/T]$.
 On the other hand, since $G$ is path-connected and simply-connected
 the $T$-bundle $G \to G/T$ is ``2-universal'' (cf. Sec. 16  in \cite{Borel}) so
 there is a  1-1-correspondence between the elements of $[\Sigma,G/T]$  and  the set of equivalence classes of $T$-bundles over $\Sigma$ (cf. Theorem 16.1 in \cite{Borel}).
Accordingly, it is clear that Eq. \eqref{eq2.48_simpl} below and Eq. (7.9) in \cite{BlTh1}
are very closely related. (Note that the derivation of Eq. (7.9)  in \cite{BlTh1}  involves a sum over equivalence classes of $T$-bundles over $\Sigma$ and so does  Eq. (6.8) in \cite{BlTh3}.)
And indeed, the explicit evaluation of Eq. (7.9) in \cite{BlTh1} gives rise to the same
concrete values as the explicit evaluation of Eq.  \eqref{eq2.48_simpl} below, so from a computational point of view both equations are equivalent.
On the other hand, from a conceptual point of view Eq. (7.9) in \cite{BlTh1} does not seem
to be equivalent to Eq. \eqref{eq2.48_simpl} below, cf. Remark \ref{rm_non_equiv} in Sec. \ref{sec3} below.
\end{remark}

\begin{remark} \label{rm_eigentlich_average}
(i)  In Eq. \eqref{eq2.24}  we can replace the space $\cB_{reg}$ by $C^{\infty}(\Sigma,P)$
(for any fixed Weyl alcove $P$)
   or by each of the two spaces $\cB^{ess}_{reg}$ or   $\{B \in \cB_{reg}^{ess} \mid B(\sigma_0) \in P\}$
    introduced in Sec. \ref{subsubsec3.2.3} and Appendix \ref{appA.6} below.
  In the present section we chose to work with $\cB_{reg}$ for stylistic reasons.
  In   Sec. \ref{subsec3.1} it will be convenient to work with the space $C^{\infty}(\Sigma,P)$ (cf. the last paragraph before Sec. \ref{subsubsec3.1.1}).
   From Sec. \ref{subsubsec3.2.6} on we will work with the space
   $\cB^{ess}_{reg}$, for reasons explained in Remark \ref{rm_why_B_ess_reg} in Sec. \ref{subsubsec3.2.3} below.

\smallskip

 (ii) Observe that, similarly to the situation in Eq. \eqref{eq_AppA1_lang} in Sec. \ref{subsubsec2.2.1}
above, the integral $\int \cdots DB$ on the RHS of Eq. \eqref{eq2.24} should be interpreted as a suitable improper integral $\int^{\sim} \cdots DB$, cf. Remark \ref{rm_sec2.2.1} above. (The same applies if we use the space $\cB^{ess}_{reg}$
instead of $\cB_{reg}$.) However, since the integral on the RHS of Eq. \eqref{eq2.24}
is informal anyway we do not use a notation like $\int^{\sim} \cdots DB$.
\end{remark}

\subsection{Torus gauge fixing applied to the Chern-Simons path integral}
\label{subsec2.3}

Let us now go back to Eq. \eqref{eq_WLO_orig} in Sec. \ref{subsec2.1}.
We will consider the special case of Eq. \eqref{eq_WLO_orig}
where $M$ is of the form
 $M=\Sigma \times S^1$ where $\Sigma$ is a compact, connected, oriented surface  and where
      $L= ((l_1, l_2, \ldots, l_m),(\rho_1,\rho_2,\ldots,\rho_m))$  is a colored link in $M= \Sigma \times S^1$.
      Obviously, the RHS of Eq. \eqref{eq_WLO_orig} then agrees with the LHS of Eq. \eqref{eq2.24}
      if we choose $\chi$ to be the $\G$-invariant function $\chi_L: \cA \to \bC$ given by
\begin{equation} \label{eq2.31}  \chi_L(A)= \left( \prod_{i=1}^m  \Tr_{\rho_i}\bigl(\Hol_{l_i}(A)\bigr) \right)
 \exp(iS_{CS}(A)) \quad \forall A \in \cA
 \end{equation}
 From  Eq. \eqref{eq_WLO_orig} and  Eq. \eqref{eq2.24}   we therefore obtain
\begin{multline} \label{eq_2.24_WLO} Z(\Sigma \times S^1,L)
 \sim \sum_{\cl \in C^{\infty}(\Sigma,G/T)/\G_{\Sigma}}
 \int_{\cB_{reg}} \biggl[ \int_{\cA^{\orth}}
   \left( \prod_{i=1}^m  \Tr_{\rho_i}(\Hol_{l_i}(A^{\orth} +
  \bar{g}_{\cl} B  \bar{g}_{\cl}^{-1} dt)) \right) \\
  \times    \exp(iS_{CS}(A^{\orth} +
  \bar{g}_{\cl} B  \bar{g}_{\cl}^{-1} dt))   DA^{\orth} \biggr]
    \det\bigl(1_{\ck}-\exp(\ad(B))_{| \ck}\bigr) DB
\end{multline}
where ``$\sim$'' denotes equality up to a multiplicative constant independent
of the colored link $L$ and the level $k$ and
where $(\bar{g}_{\cl})_{\cl \in  C^{\infty}(\Sigma,G/T)/ \G_{\Sigma}}$,
$\bar{g}_{\cl} \in C^{\infty}(\Sigma,G/T)$,
 is a fixed system of representatives   of $C^{\infty}(\Sigma,G/T)/ \G_{\Sigma}$. \par

For each fixed  $B \in \cB$ and $\cl \in C^{\infty}(\Sigma,G/T)/\G_{\Sigma}$
we will now simplify the integral
$$\int_{\cA^{\orth}}
   \left( \prod_{i=1}^m  \Tr_{\rho_i}(\Hol_{l_i}(A^{\orth} +
  \bar{g}_{\cl} B  \bar{g}_{\cl}^{-1} dt)) \right)     \exp(iS_{CS}(A^{\orth} +
  \bar{g}_{\cl} B  \bar{g}_{\cl}^{-1} dt))   DA^{\orth}$$
by applying a  suitable change of variable (and by exploiting the special properties of the
functions  $\Tr_{\rho_i}(\Hol_{l_i}(\cdot))$ and $S_{CS}$). \par

\smallskip

As a preparation let us set
$l^i_{\Sigma}:= \pi_{\Sigma} \circ l_i$ for each loop $l_i$ appearing in the link $L$
where $\pi_{\Sigma}: \Sigma \times S^1 \to \Sigma$ is the canonical projection and let us
 fix  $\sigma_0 \in \Sigma$ such that
\begin{equation} \label{eq_sigma0_cond} \sigma_0 \notin  \bigcup_{i=1}^m \arc(l^i_{\Sigma})
\end{equation}

Since the (connected, oriented) surface $\Sigma \backslash \{\sigma_0\}$ is non-compact,
according to Remark \ref{rm2.1} in Sec. \ref{subsec2.2} above
the map  $(\bar{g}_{\cl})_{|\Sigma \backslash
  \{\sigma_0\}} \in C^{\infty}(\Sigma \backslash \{\sigma_0\},G/T)$
  is null-homotopic and can therefore be lifted (in the fiber bundle $\pi:G \to G/T$)
 to a map   $\Omega_{\cl} \in   \G_{\Sigma \backslash \{\sigma_0\}} := C^{\infty}(\Sigma \backslash \{\sigma_0\},G)$.   We will keep  $\Omega_{\cl}$ fixed in the following.

 \medskip

Since   $\Tr_{\rho_i}(\Hol_{l_i}(\cdot))$ is $\G$-invariant we have,  in particular,
  \begin{equation} \label{eq_TrHol} \Tr_{\rho_i}(\Hol_{l_i}(A^{\orth} +   \Omega B \Omega^{-1}  dt))   = \Tr_{\rho_i}(\Hol_{l_i}(A^{\orth} \cdot \Omega +   B dt))
 \end{equation}
 for all $\Omega \in \G_{\Sigma} \subset \G$. Now observe that
 as $\sigma_0$  was chosen to that  condition \eqref{eq_sigma0_cond} above is fulfilled,
  Eq. \eqref{eq_TrHol} also holds for all
 $\Omega \in \G_{\Sigma \backslash \{\sigma_0\}} = C^{\infty}(\Sigma \backslash \{\sigma_0\},G)$
 if the  expressions
  $\Hol_{l_i}(A^{\orth} \cdot \Omega +   B dt)$
  and $\Hol_{l_i}(A^{\orth} +  \Omega B  \Omega^{-1} dt)$ are   defined in the obvious
  way. Applying this to the special case $\Omega = \Omega_{\cl}$ we then obtain
 \begin{multline} \label{eq_Tr_Hol_change_of_gauge} \Tr_{\rho_i}(\Hol_{l_i}(A^{\orth} +
  \bar{g}_{\cl} B  \bar{g}_{\cl}^{-1} dt)) \\
   = \Tr_{\rho_i}(\Hol_{l_i}(A^{\orth} +  \Omega_{\cl} B  \Omega_{\cl}^{-1} dt))   = \Tr_{\rho_i}(\Hol_{l_i}(A^{\orth} \cdot \Omega_{\cl} +   B dt)
  \end{multline}

Let us now derive a similar formula for the factor  $\exp(iS_{CS}(A^{\orth} +
  \bar{g}_{\cl} B  \bar{g}_{\cl}^{-1} dt))$ in Eq. \eqref{eq_2.24_WLO}.
  Recall from Sec. \ref{subsubsec2.1.1} above that we have been assuming (without loss of generality)
that $G$ is a matrix Lie group, i.e.  $G \subset \GL(N,\bR) \subset \Mat(N,\bR)$ for some $N \in \bN$. From  Eq. \eqref{eq2.2'}  in Sec. \ref{subsubsec2.1.1} above we then obtain
\begin{align} \label{eq_S_CS_torus_gauge}
& S_{CS}(A^{\orth} + B dt ) \nonumber \\
& =  k\pi  \int_M \bigl[ \Tr(A^{\orth} \wedge dA^{\orth} ) +
2 \Tr( A^{\orth} \wedge  B dt \wedge A^{\orth}) + 2  \Tr(A^{\orth}
\wedge dB \wedge dt) \bigr] \nonumber \\
& = - k\pi  \int_{S^1} \biggl[ \int_{\Sigma}  \Tr\bigl(A^{\orth}(t) \wedge \bigl(\partial/\partial t + \ad(B) \bigr) \cdot
 A^{\orth}(t)\bigr)  -  2  \Tr(A^{\orth}(t) \wedge dB)  \biggr] dt
\end{align}
where in the last expression we used the identification (cf. Sec. 2.3.1 in \cite{Ha7a})
\begin{equation}
\cA^{\orth}  \cong C^{\infty}(S^1,\cA_{\Sigma})
\end{equation}
where \begin{equation}
 \cA_{\Sigma}   :=  \Omega^1(\Sigma,\cG)
\end{equation}
and where $C^{\infty}(S^1,\cA_{\Sigma})$ is the space of maps
$f:S^1 \to \cA_{\Sigma}$ which are ``smooth'' in the sense that
$\Sigma \times S^1 \ni (\sigma,t) \mapsto (f(t))(X_{\sigma}) \in \cG$
is smooth for every smooth vector field $X$ on $\Sigma$.
The operator  $\partial/\partial t:C^{\infty}(S^1,\cA_{\Sigma}) \to C^{\infty}(S^1,\cA_{\Sigma})$
is defined in the obvious way.  \par

For reasons which become clear later we will rewrite Eq. \eqref{eq_S_CS_torus_gauge}
using a suitable  improper integral, as
\begin{multline} \label{eq_S_CS_improper} S_{CS}(A^{\orth} + B dt ) \\
=   - k\pi  \int_{S^1} \lim_{\eps \to 0} \biggl[  \int_{\Sigma \backslash  B_{\eps}(\sigma_0)}  \Tr\bigl(A^{\orth}(t) \wedge \bigl(\partial/\partial t + \ad(B) \bigr) \cdot
 A^{\orth}(t)\bigr)  -  2  \Tr(A^{\orth}(t) \wedge dB)  \biggr] dt
\end{multline}
 where  $ B_{\eps}(\sigma_0)$, $\eps >0$, denotes the   closed $\eps$-ball around $\sigma_0$
 w.r.t.  to an arbitrary fixed   Riemannian metric ${\mathbf g}_{\Sigma}$ on $\Sigma$.
In Appendix \ref{appA.3} below we show that
  \begin{multline} \label{eq_S_CS_change_of_gauge} \exp(iS_{CS}(A^{\orth} +
  \bar{g}_{\cl} B  \bar{g}_{\cl}^{-1} dt))
  =  \exp(iS_{CS}(A^{\orth} + \Omega_{\cl} B  \Omega_{\cl}^{-1} dt)) \\
    =  \exp(i S_{CS}(A^{\orth} \cdot \Omega_{\cl} +   B dt) \times  \exp\bigl( - 2 \pi i k \ \langle  n(\cl), B(\sigma_0)\rangle \bigr)
  \end{multline}
  where the  expressions $S_{CS}(A^{\orth} + \Omega_{\cl} B  \Omega_{\cl}^{-1} dt)$ and
$ S_{CS}(A^{\orth} \cdot \Omega_{\cl} +   B dt)$ are defined by the obvious analogues of
the RHS of Eq. \eqref{eq_S_CS_improper} and where we set
\begin{equation} \label{eq_def_ncl}  n(\cl)
 := \lim_{\eps \to 0} \int_{\partial B_{\eps}(\sigma_0)}  \pi_{\ct}\bigl( \Omega_{\cl}^{-1} d\Omega_{\cl}\bigr) \in \ct
\end{equation}
 where  $\pi_{\ct}$ is  the   $\langle \cdot,\cdot \rangle$-orthogonal projection $\cG \to \ct$ and where the orientation of $\partial B_{\eps}(\sigma_0)$ is opposite
 to the  orientation induced by $\overline{B_{\eps}(\sigma_0)}$ (cf. Footnote \ref{ft_opposite_orientation} in Appendix \ref{appA.3}).

 \smallskip

 The following proposition will be proven in Appendix \ref{appA.4} below.
 \begin{proposition} \label{prop_n(cl)}
 \begin{enumerate}
 \item  The limit in Eq. \eqref{eq_def_ncl} exists.

\item  $n(\cl)$ is  independent of the special choice of  $\bar{g}_{\cl}$ and $\Omega_{\cl}$
above\footnote{Part (ii) of Proposition \ref{prop_n(cl)} is used in the proof of part (iii) in
Appendix \ref{appA.4} below. Note that, not surprisingly,
 $n(\cl)$ is also independent of the special choice of    ${\mathbf g}_{\Sigma}$ and  $\sigma_0$ above
   but this plays no role in the proof of part (iii).}.

 \item The map $C^{\infty}(\Sigma,G/T)/\G_{\Sigma} \ni \cl \mapsto n(\cl) \in \ct$
 is injective and its image is $I:= \ker(\exp_{| \ct})  \subset \ct$.

 \end{enumerate}
 \end{proposition}

From  Eq. \eqref{eq_Tr_Hol_change_of_gauge} and Eq. \eqref{eq_S_CS_change_of_gauge}  we obtain
\begin{align} \label{eq_change_of_variable}
& \int_{\cA^{\orth}}
   \left( \prod_{i=1}^m  \Tr_{\rho_i}(\Hol_{l_i}(A^{\orth} +
  \bar{g}_{\cl} B  \bar{g}_{\cl}^{-1} dt)) \right)   \exp(iS_{CS}(A^{\orth} +
  \bar{g}_{\cl} B  \bar{g}_{\cl}^{-1} dt))   DA^{\orth}\nonumber \\
& = \bigl( \exp\bigl( - 2 \pi i k \ \langle  n(\cl), B(\sigma_0)\rangle \bigr) \bigr)\nonumber \\
& \quad \times \int_{\cA^{\orth}}
   \left( \prod_{i=1}^m  \Tr_{\rho_i}(\Hol_{l_i}(A^{\orth} \cdot \Omega_{\cl} +
 B dt)) \right)   \exp(iS_{CS}(A^{\orth} \cdot  \Omega_{\cl} +
 B dt)) DA^{\orth} \nonumber \\
 & \overset{(*)}{=} \bigl( \exp\bigl( - 2 \pi i k \ \langle  n(\cl), B(\sigma_0)\rangle \bigr) \bigr)  \nonumber \\
 & \quad \times \int_{\cA^{\orth}}   \left( \prod_{i=1}^m  \Tr_{\rho_i}(\Hol_{l_i}(A^{\orth} +
 B dt)) \right)    \exp(iS_{CS}(A^{\orth} +B dt))   DA^{\orth}
\end{align}
where in step $(*)$ we have applied the change of variable
$A^{\orth} \to A^{\orth} \cdot \Omega_{\cl}^{-1} = \Omega_{\cl}A^{\orth} \Omega_{\cl}^{-1} -   d\Omega_{\cl} \Omega^{-1}_{\cl}$
 (and used a similar argument as in step $(*)$ in Eq. \eqref{eq_appA.2_Ende} in Appendix \ref{appA.2} below.)

 \begin{remark} \label{rm_AppA.3}
    Note that because of the discontinuity of $\Omega_{\cl}$
 and the singularity of $d\Omega_{\cl}$  in the point $\sigma_0$
  the change of variable $A^{\orth} \to A^{\orth} \cdot \Omega_{\cl}^{-1}$
  we used above is  not really a transformation of the
 space $\cA^{\orth}$ and we can  not be sure  whether  this change of variable will lead to the
 correct results. In Appendix \ref{appA.5}
 below we will therefore give a careful justification for the change of variable
 $A^{\orth} \to A^{\orth} \cdot \Omega_{\cl}^{-1}$  in Eq. \eqref{eq_change_of_variable}   above.
 That such a justification is necessary also becomes clear
  from the following observation: If we rewrite some of the formulas above using the expression
 $S'_{CS}(A^{\orth} \cdot  \Omega_{\cl} +  B dt)$  introduced in Eq. \eqref{eq_A1_0} in Appendix \ref{appA.3} below and then perform the  change of variable $A^{\orth} \to A^{\orth} \cdot \Omega_{\cl}^{-1}$
in  a naive\footnote{On  the other hand, as explained in Remark \ref{rm_comm1} in Appendix \ref{appA.5} below,
 also when working with $S'_{CS}(A^{\orth} \cdot  \Omega_{\cl} +  B dt)$
we arrive at the correct result if we replace the argument based on the aforementioned
naive change of variable by a more careful argument.} way we would arrive at an incorrect result, cf.  Remark \ref{rm_comm1} in Appendix \ref{appA.5} below.
 \end{remark}

 By applying Eq. \eqref{eq_change_of_variable} for each fixed $B \in \cB$ and $\cl \in C^{\infty}(\Sigma,G/T)/\G_{\Sigma}$
 and by taking  into account part (iii) of Proposition \ref{prop_n(cl)} above we finally  obtain from Eq. \eqref{eq_2.24_WLO}
\begin{multline} \label{eq2.48_pre}  Z(\Sigma \times S^1,L)
 \sim \sum_{y \in I}  \int_{\cB_{reg}}  \biggl[ \int_{\cA^{\orth}} \prod_i  \Tr_{\rho_i}\bigl(
 \Hol_{l_i}(A^{\orth} + Bdt)\bigr)
  \exp(i  S_{CS}( A^{\orth} + B dt)) DA^{\orth} \biggr] \\
 \times \exp\bigl( -  2\pi i k  \langle y,  B(\sigma_0)\rangle \bigr) \bigr)
  \det\bigl(1_{\ck}-\exp(\ad(B))_{|\ck}\bigr) DB
\end{multline}

\subsubsection{Rewriting  Eq. \eqref{eq2.48_pre}}
 \label{subsubsec2.3.1}

It will be convenient  to rewrite  the RHS of
 Eq. \eqref{eq2.48_pre} as an iterated ``Gauss-type'' integral  (cf. Remark \ref{rm_2.7} below).
In order to do so let us set
\begin{subequations}
\begin{align}
\cA_{\Sigma,\ct} & := \Omega^1(\Sigma,\ct),\\
 \cA_{\Sigma,\ck} & := \Omega^1(\Sigma,\ck)\\
\label{eq_part_f}
\Check{\cA}^{\orth} & := \{ A^{\orth} \in \cA^{\orth} \mid \int A^{\orth}(t) dt \in \cA_{\Sigma,\ck} \} \\
\label{eq_part_g}   \cA^{\orth}_c & := \{ A^{\orth} \in \cA^{\orth} \mid \text{ $A^{\orth}$ is constant and
 $\cA_{\Sigma,\ct}$-valued}\} \cong \cA_{\Sigma,\ct}
 \end{align}
\end{subequations}
(Recall that we made the identification
$\cA^{\orth}  \cong C^{\infty}(S^1,\cA_{\Sigma})$
where $C^{\infty}(S^1,\cA_{\Sigma})$ is as in the paragraph before Eq. \eqref{eq_S_CS_improper} above).
Observe that
\begin{equation} \label{eq_cAorth_decomp}
\cA^{\orth} = \Check{\cA}^{\orth} \oplus  \cA^{\orth}_c
\end{equation}

From Eq. \eqref{eq_S_CS_torus_gauge} it follows easily that
\begin{equation}
 S_{CS}(\Check{A}^{\orth} + A^{\orth}_c + B dt)
= S_{CS}(\Check{A}^{\orth}  + B dt) + S_{CS}(A^{\orth}_c + B dt)
\end{equation}
for $\Check{A}^{\orth} \in \Check{\cA}^{\orth}$, $A^{\orth}_c \in \cA^{\orth}_c $, and $B \in \cB$.
Taking this into account we can rewrite Eq. \eqref{eq2.48_pre} as
\begin{multline}  \label{eq2.48_Ha7a} Z(\Sigma \times S^1,L)
 \sim \sum_{y \in I}  \int_{\cA^{\orth}_c \times \cB} \biggl\{ \Det_{FP}(B) 1_{\cB_{reg}}(B)\\
 \times   \biggl[ \int_{\Check{\cA}^{\orth}} \prod_i  \Tr_{\rho_i}\bigl(
 \Hol_{l_i}(\Check{A}^{\orth} + A^{\orth}_c ,B)\bigr)
  \exp(i  S_{CS}( \Check{A}^{\orth}, B)) D\Check{A}^{\orth} \biggr] \\
 \times \exp\bigl( - 2\pi i k  \langle y,  B(\sigma_0)\rangle \bigr) \biggr\}
 \exp(i S_{CS}(A^{\orth}_c, B)) (DA^{\orth}_c \otimes DB)
\end{multline}
where $D\Check{A}^{\orth}$,  $DA^{\orth}_c$, and $DB$ are the informal ``Lebesgue measures''
on $\Check{\cA}^{\orth}$, $\cA^{\orth}_c$, and $\cB$,
 where  we have introduced the
short notation
\begin{subequations} \label{eq2.49}
\begin{align}
S_{CS}(A^{\orth},B) & := S_{CS}(A^{\orth} + B dt ) \\
\label{eq2.49b} \Det_{FP}(B) &:=    \det\bigl(1_{\ck}-\exp(\ad(B))_{|
\ck}\bigr)\\
\label{eq2.49c} \Hol_{l_i}(A^{\orth},   B) & := \Hol_{l_i}(A^{\orth} +  B dt)
\end{align}
\end{subequations}
and where $1_{\cB_{reg}}$ is the indicator function of the subset $\cB_{reg}$ of $\cB$.

\begin{remark} \label{rm_2.7}
 For any (fixed) Riemannian metric  ${\mathbf g}_{\Sigma}$ on $\Sigma$ we  have
 \begin{align} \label{eq_SCS_expl}
S_{CS}(\Check{A}^{\orth},B) & =  \pi k  \ll \Check{A}^{\orth},
\star  \bigl(\tfrac{\partial}{\partial t} + \ad(B) \bigr) \Check{A}^{\orth} \gg_{\cA^{\orth}}   \\
\label{eq_SCS_expl2}
 S_{CS}(A^{\orth}_c,B) & =  - 2 \pi k  \ll   A^{\orth}_c,  \star dB \gg_{\cA^{\orth}}
\end{align}
for  $B \in \cB$, $\Check{A}^{\orth} \in \Check{\cA}^{\orth}$, and $A^{\orth}_c \in \cA^{\orth}_c$
where   $\star$ is the  Hodge star operator\footnote{More precisely, $\star$ denotes
  both the Hodge star operator   $\star: \cA_{\Sigma} \to \cA_{\Sigma}$
   and  the linear automorphism
 $\star: C^{\infty}(S^1, \cA_{\Sigma}) \to C^{\infty}(S^1, \cA_{\Sigma})$ given
 by $(\star A^{\orth})(t) = \star (A^{\orth}(t))$ for all $A^{\orth} \in \cA^{\orth}$ and $t \in S^1$.}
  associated to ${\mathbf g}_{\Sigma}$ and    $\ll \cdot , \cdot \gg_{\cA_{\Sigma}}$ and
    $\ll \cdot , \cdot \gg_{\cA^{\orth}}$  are the scalar products on $\cA_{\Sigma}$ and
  $\cA^{\orth} \cong C^{\infty}(S^1, \cA_{\Sigma})$   which are induced by ${\mathbf g}_{\Sigma}$.   \par

 In view of Eq. \eqref{eq_SCS_expl} and Eq. \eqref{eq_SCS_expl2} it is clear that
 both measures on the RHS of Eq. \eqref{eq2.48_Ha7a} are (complex) ``Gauss-type''  measures.
 This greatly simplifies the tasks of finding
 a rigorous realization of the RHS of Eq. \eqref{eq2.48_Ha7a}, cf. Sec. \ref{sec4} below.
It also simplifies the explicit evaluation of the  RHS of Eq. \eqref{eq2.48_Ha7a}
which will be the topic in Sec. \ref{sec3} below.
We would like to point out, however, that in an informal evaluation of the RHS of Eq. \eqref{eq2.48_Ha7a}
it is useful to rewrite the outer ``Gauss-type'' integral
$\int \cdots  \exp(i S_{CS}(A^{\orth}_c, B)) (DA^{\orth}_c \otimes DB)$ as an iterated integral $\int \bigl[ \int \cdots  \exp(i S_{CS}(A^{\orth}_c, B)) DA^{\orth}_c \bigr] DB$,
cf. Sec. \ref{subsubsec3.1.1}, Sec. \ref{subsubsec3.3.3}, and Sec. \ref{subsubsec3.4.1} below.
\end{remark}

Eq. \eqref{eq2.48_Ha7a} is not yet our final  formula for  $Z(\Sigma \times S^1,L)$.
There are three more things to do before we obtain our final formula, i.e.
Eq. \eqref{eq2.48_reg}  in Sec. \ref{subsubsec3.2.6} below:

\begin{enumerate}

\item[1.]  Recall that our goal is to find a rigorous realization of Witten's CS path integral expressions
 which reproduces the Reshetikhin-Turaev invariants (in the special situation described in the Introduction).
 Since the Reshetikhin-Turaev invariants are defined for ribbon links (or, equivalently\footnote{From the knot theory point of view   the framed link picture and the ribbon link picture are equivalent.
 However,  the ribbon picture seems to be better suited for the
 study of the  Chern-Simons path integral in the torus gauge.},
  for framed links) we will later introduce  a ribbon analogue of Eq. \eqref{eq2.48_Ha7a}, cf.  Sec. \ref{subsubsec3.2.1} below.

\item [2.] For reasons explained in Sec. \ref{subsubsec3.2.3} below we will replace the subspace $\cB_{reg}$ of $\cB$ appearing  in Eq. \eqref{eq2.48_Ha7a} by the slightly larger subspace $\cB^{ess}_{reg}$     of $\cB$, cf. Eq. \eqref{eq_def_B^ess} below.

\item [3.] We will set,  for each $B \in \cB_{reg}$ (and later for $B \in \cB^{ess}_{reg}$),
\begin{align}  \label{eq_def_Z_B}
\Check{Z}(B) & := \int \exp(i  S_{CS}( \Check{A}^{\orth}, B)) D\Check{A}^{\orth},\\
 \label{eq_def_mu_B}
d\mu^{\orth}_B & := \tfrac{1}{\Check{Z}(B)} \exp(i  S_{CS}( \Check{A}^{\orth}, B)) D\Check{A}^{\orth}
\end{align}
 and
\begin{equation} \label{eq_def_det(B)}
\Det(B) := \Det_{FP}(B) \Check{Z}(B)
\end{equation}
and we will then find a rigorous realization $\Det_{rig}(B)$ of $\Det(B)$.
\end{enumerate}

Since neither ribbons nor the space $\cB^{ess}_{reg}$
are  necessary for the simple situation of ``fiber links'' treated in Sec. \ref{subsec3.1} below  we will, for now, only incorporate point 3 above.
Doing so we obtain from Eq. \eqref{eq2.48_Ha7a}
\begin{multline}  \label{eq2.48} Z(\Sigma \times S^1,L)
 \sim \sum_{y \in I}  \int_{\cA^{\orth}_c \times \cB} \biggl\{ 1_{\cB_{reg}}(B) \Det_{rig}(B) \\
 \times   \biggl[ \int_{\Check{\cA}^{\orth}} \left( \prod_i  \Tr_{\rho_i}\bigl(
 \Hol_{l_i}(\Check{A}^{\orth} + A^{\orth}_c, B)\bigr) \right)
d\mu^{\orth}_B(\Check{A}^{\orth}) \biggr] \\
 \times \exp\bigl( - 2\pi i k  \langle y, B(\sigma_0) \rangle \bigr) \biggr\}
 \exp(i S_{CS}(A^{\orth}_c, B)) (DA^{\orth}_c \otimes DB)
\end{multline}
with $\Det_{rig}(B)$ as defined in Sec. \ref{subsubsec2.3.2} below.

\subsubsection{Rigorous realization $\Det_{rig}(B)$ of $\Det(B)$}
\label{subsubsec2.3.2}

We will now introduce a rigorous realization $\Det_{rig}(B)$ of $\Det(B)$.
In order to do so we will use a standard $\zeta$-function regularization argument
 and a variant  of the heat kernel regularization method used in Sec. 6 in  \cite{BlTh1},
cf.  Remark \ref{rm_3.1_2} in Sec. \ref{subsubsec3.1.1} below.

 \smallskip

 As  above let us fix an auxiliary  Riemannian metric ${\mathbf g}_{\Sigma}$ on $\Sigma$
 and let $\star$ and $ \ll \cdot, \cdot \gg_{\cA^{\orth}}$ be as in Remark \ref{rm_2.7} above.
 Then we obtain, informally, for  $B \in \cB_{reg}= C^{\infty}(\Sigma,\ct_{reg})$
 \begin{align} \label{eq_sec2.5_1_3}
\Check{Z}(B) & = \int \exp(i  S_{CS}( \Check{A}^{\orth}, B)) D\Check{A}^{\orth} \nonumber \\
& = \int  \exp(i  \pi k  \ll \Check{A}^{\orth}, \star  \bigl(\tfrac{\partial}{\partial t} + \ad(B) \bigr) \Check{A}^{\orth} \gg_{\cA^{\orth}}) D\Check{A}^{\orth} \nonumber \\
& \sim \det\bigl(\tfrac{\partial}{\partial t} + \ad(B)\bigr)^{ -1/2}
 \end{align}
 where  ``$\sim$'' denotes equality up to a multiplicative constant independent of $B$. \par

In order to make sense of the RHS of Eq. \eqref{eq_sec2.5_1_3} we first consider the analogous problem of making sense of the  determinant
of the linear operator  $\tfrac{\partial}{\partial t} + \ad(b): C^{\infty}(S^1,\ck) \to  C^{\infty}(S^1,\ck)$ with $b \in \ct_{reg}$.
This problem can be solved using a  standard  $\zeta$-function regularization argument.
Using this one arrives at
\begin{equation} \label{eq2.58}
 \det(\tfrac{\partial}{\partial t} + \ad(b)\bigr) \sim \det\bigl(\bigl(1_{\ck}-\exp(\ad(b))_{|\ck}\bigr)\bigr) \quad \forall b \in \ct_{reg}
\end{equation}

Now let $(1_{\ck}-\exp(\ad(B))_{|\ck})^{(p)}$,
for  $p \in \{0,1,2\}$, denote the linear operator   on $\Omega^p(\Sigma,\ck)$
given by\footnote{Note that under the identification $C^{\infty}(\Sigma,\ck) \cong \Omega^0(\Sigma,\ck)$
the operator  $(1_{\ck}-\exp(\ad(B))_{|\ck})^{(0)}$ coincides with what above we call
 $1_{\ck}-\exp(\ad(B))_{|\ck}$.}
\begin{equation} \label{eq_def_det(p)}
 \bigl( \bigl(1_{\ck}-\exp(\ad(B))_{|\ck}\bigr)^{(p)} \cdot \alpha\bigr)(X_{\sigma}) = (1_{\ck}-\exp(\ad(B(\sigma))_{|\ck}) \cdot \alpha(X_{\sigma})
\end{equation}
for all $\alpha \in  \Omega^p(\Sigma,\ck)$, $\sigma \in \Sigma$, $X_{\sigma} \in \wedge^p
 T_{\sigma} \Sigma$. \par

In view of Eq. \eqref{eq2.58} we then have, informally, for $B \in \cB_{reg}$
\begin{equation}
 \det\bigl(\tfrac{\partial}{\partial t} + \ad(B)\bigr) \sim
 \det\bigl(\bigl(1_{\ck}-\exp(\ad(B))_{|\ck}\bigr)^{(1)}\bigr)
 \end{equation}
 and combining this with  Eq. \eqref{eq_sec2.5_1_3} and  the (informal) definition of $\Det(B)$ (cf. Eq. \eqref{eq_def_det(B)} and Eq. \eqref{eq2.49b} above) we therefore obtain for $B \in \cB_{reg}$
 \begin{equation} \label{eq_Det(B)_rewrite-2}
\Det(B)  =  \det\bigl(\bigl(1_{\ck}-\exp(\ad(B))_{|\ck}\bigr)^{(0)}\bigr)
   \det\bigl(\bigl(1_{\ck}-\exp(\ad(B))_{|\ck}\bigr)^{(1)}\bigr)^{-1/2}
\end{equation}

It will be convenient to rewrite Eq. \eqref{eq_Det(B)_rewrite-2} in a suitable way.
In order to do so let us denote by  $M^{(p)}_f$,
for  $p =0,1,2$ and $\bK \in \{\bR,\bC\}$,
  the multiplication operator $\Omega^p(\Sigma,\bK) \to \Omega^p(\Sigma,\bK)$
 obtained by multiplication
  with a smooth function $f:\Sigma \to \bK$.  For every $B \in \cB_{reg}$ and $p=0,1,2$
  we then obtain, informally,
\begin{multline} \label{eq_multoperators}
  \det\bigl(\bigl(1_{\ck}-\exp(\ad(B))_{|\ck}\bigr)^{(p)}\bigr) =
   \det\nolimits_{\bC}\bigl(\bigl(1_{\ck}-\exp(\ad(B))_{|\ck}\bigr)^{(p)} \otimes_{\bR} \bC\bigr)
  \overset{(*)}{=}   \prod_{\alpha \in \cR} \det\nolimits_{\bC}\bigl(M^{(p)}_{1-e^{2 \pi i \alpha(B(\cdot))}}\bigr)\\
 = \prod_{\alpha \in \cR_+} \det\nolimits_{\bC} \bigl(M^{(p)}_{4  \sin(\pi  \alpha(B(\cdot)))^2}\bigr)
  = \prod_{\alpha \in \cR_+} \det\bigl(M^{(p)}_{2 | \sin(\pi  \alpha(B(\cdot))) |}\bigr)^2
 \end{multline}
 where we use the notation of Appendix \ref{appB.1}
 and  where in step $(*)$ we applied  a suitable diagonalization argument.
(Note that since $B \in \cB_{reg}$
  the function  $|\sin(\pi  \alpha(B(\cdot)))|: \Sigma \ni \sigma \mapsto  |\sin(\pi  \alpha(B(\sigma)))| \in \bR $  is smooth.)  Setting
 \begin{equation} O^{(p)}_{\alpha}(B) := M^{(p)}_{2 | \sin(\pi  \alpha(B(\cdot))) |}
 \end{equation}
   we can therefore rewrite Eq. \eqref{eq_Det(B)_rewrite-2}
   as\footnote{The reader may wonder we we do not rewrite  Eq. \eqref{eq_Det(B)_rewrite-1}
   in terms of the (informal) determinants of the multiplication operators
$M^{(p)}_{4  \sin^2(\pi  \alpha(B(\cdot)))}$.
  The advantage of using  Eq. \eqref{eq_Det(B)_rewrite-1}
 will become clear in Sec. \ref{subsubsec3.2.5} below where
  we will generalize Eq. \eqref{eq_Det(B)_rewrite-1} to the case
  where $B \in \cB^{ess}_{reg}$.}
\begin{align} \label{eq_Det(B)_rewrite-1}
\Det(B) & =  \prod_{\alpha \in \cR_+} \det(O^{(0)}_{\alpha}(B))^2 \det(O^{(1)}_{\alpha}(B))^{-1}
\end{align}

\medskip

We will now use Eq. \eqref{eq_Det(B)_rewrite-1} as the starting point
for obtaining a rigorous realization $\Det_{rig}(B)$ of $\Det(B)$
for  $B \in \cB_{reg}$ (by means of a suitable ``heat kernel regularization'' argument).

\smallskip

Recall that above we have fixed an auxiliary  Riemannian metric ${\mathbf g}_{\Sigma}$ on $\Sigma$.
Let us now equip the two spaces $\Omega^i(\Sigma,\bR)$, $i =0,1$, with
the scalar product which is induced by ${\mathbf g}_{\Sigma}$.
 By $\overline{\Omega^i(\Sigma,\bR)}$ we will denote the
completion of the pre-Hilbert space $\Omega^i(\Sigma,\bR)$, $i =0,1$,
and by  $\triangle_i$  the (closure of the) Hodge Laplacian on $\overline{\Omega^i(\Sigma,\bR)}$.
\begin{definition}
In view of  Eq. \eqref{eq_Det(B)_rewrite-1} above we now define
\begin{subequations} \label{eq_Det_rig_B}
\begin{equation} \label{eq_Det_rig_B1} \Det_{rig}(B):=  \prod_{\alpha \in \cR_+} \Det_{rig,\alpha}(B)
\end{equation}
with
\begin{equation}\label{eq_Det_rig_B2}  \Det_{rig,\alpha}(B):= \lim_{\eps \to 0} \bigl[ \det\nolimits_{\eps}(O^{(0)}_{\alpha}(B))^2 \det\nolimits_{\eps}(O^{(1)}_{\alpha}(B))^{-1} \bigr]
\end{equation}
where for $i=0,1$ we have set\footnote{This ansatz is, of course, motivated by the
rigorous formula $\det(A) = \exp(\Tr(\log(A)))$ which holds for every strictly positive
(self-adjoint) operator $A$ on a finite-dimensional Hilbert-space.}
\begin{equation} \label{eq_Det_rig_B3}
\det\nolimits_{\eps}(O^{(i)}_{\alpha}(B)) :=
 \exp\bigl( \Tr\bigl( e^{- \eps \triangle_i} \log(O^{(i)}_{\alpha}(B))\bigr) \bigr)
\end{equation}
\end{subequations}
\end{definition}

Note that each of the  operators
$O^{(i)}_{\alpha}(B)$, $i = 0,1$, is a  symmetric, bounded,  positive operator
whose spectrum is bounded away from zero.
Hence also  $\log(O^{(i)}_{\alpha}(B))$ is a well-defined symmetric, bounded operator.
Moreover,  $e^{- \eps \triangle_i} $ is trace-class
 so the product $e^{- \eps \triangle_i} \log(O^{(i)}_{\alpha}(B))$ is also trace-class and
the expression $\Tr( e^{- \eps \triangle_i} \log(O^{(i)}_{\alpha}(B)))$ in Eq. \eqref{eq_Det_rig_B3} is  well-defined.
In Sec. \ref{subsubsec3.2.5}  we will show that the $\eps \to 0$ limit on the RHS of Eq. \eqref{eq_Det_rig_B2}
 exists for all $B \in \cB_{reg}$   and  $\Det_{rig,\alpha}(B)$ and
 $\Det_{rig}(B)$ are therefore well-defined.

\subsection{A remark on the relation between $Z(M,L)$ and the Reshetikhin-Turaev invariant $RT(M,L)$}
\label{subsec2.4}

Let   $\cG_{\bC}$ be a simple complex Lie algebra,
and $q \in U(1)$ a root of unity (of sufficiently high order).
Moreover, let $M$ be a compact, connected, oriented
 3-manifold and $L$ a  framed, colored link in $M$.
The  Reshetikhin-Turaev invariant $RT(M,L)$ associated to the quantum group $U_q(\cG_{\bC})$
is believed to be equivalent to  Witten's informal path integral expression $Z(M,L)$
   based on the Chern-Simons action function     associated to $(G,k)$  where
 $G$ is the simply connected, compact Lie group corresponding to the compact real form
  $\cG$ of $\cG_{\bC}$   and $k \in \bN$ is chosen suitably.
 It it often assumed that this relationship between $q$ and $k \in \bN$ is given by
 \begin{equation} \label{eq_q_k_rel1} q = e^{2 \pi i/(k+\cg)}
 \end{equation}
 where $\cg$ is the dual Coxeter number of $\cG$.
The appearance of  $k + \cg$ instead of $k$
is the famous ``shift of the level'' $k$.
However,  several  authors (cf., e.g., \cite{GMM2}) have argued  that the occurrence
(and magnitude) of such a shift in the level depends on the regularization procedure and
renormalization prescription which is used for making sense of the informal path integral.
Accordingly, it should not be surprising that there  are several papers (cf. the references in \cite{GMM2}) where the shift $k \to k + \cg$ is not observed and one is therefore led to
 the following relationship  between  $q$ and
 $k \in \bN$ with\footnote{In view of the definition of the set $\Lambda_+^k$
   in Appendix \ref{appB.2} below it is clear that the situation $k \le \cg$ is not interesting}  $k > \cg$
 \begin{equation}\label{eq_q_k_rel2}
q = e^{2 \pi i/k}
 \end{equation}
This is also the case in \cite{Ha7a,Ha7b,Ha9,Ha6b} and the present paper.
By contrast, in \cite{BlTh1,BlTh4,BlTh5} (and also in \cite{HaHa}) it is assumed that
$q$ and $k$ are related by  \eqref{eq_q_k_rel1}, cf. Remark \ref{rm_3.1_2} below.

\begin{remark} In Sec. \ref{subsec3.3} below we will compare   $Z(\Sigma \times S^1,L)$
directly with the Reshetikhin-Turaev invariant $RT(\Sigma \times S^1,L)$. By contrast, in  Secs \ref{subsec3.4}
and \ref{subsec3.5}  below we will compare   $Z(\Sigma \times S^1,L)$ with the reformulation of
 $RT(\Sigma \times S^1,L)$ in terms of   Turaev's shadow invariant $|L|$,
 cf. Eq. \eqref{eq_Z(M,L) sim_shadow} at the beginning of Sec. \ref{subsec3.4} and
 Eq. \eqref{eq_RT_vs_|L|} in  Appendix \ref{appC}  below.
\end{remark}

\section{Explicit evaluation of  $Z(\Sigma \times S^1,L)$}
\label{sec3}

We will now  evaluate $Z(\Sigma \times S^1,L)$
explicitly (at an informal level), first in several special cases
and then, in Sec. \ref{subsec3.5} below, we will consider the case of
general (``admissible'') $L$.

\smallskip

We begin in Sec. \ref{subsec3.1} with the special case of ``fiber links''
 $L$, which is the only class of links considered  in \cite{BlTh1}, cf.
 Remark \ref{rm_non_equiv} below.
Even though from a knot theoretic point of view
fiber links are trivial the study of such links is still very interesting,
due to the relationship to the Verlinde formula  for the WZW model,
 cf. Remark \ref{rm_BlTh_Lob} below. \par

Sec. \ref{subsec3.2} is a preparation for Secs \ref{subsec3.3}--\ref{subsec3.5}. In Sec. \ref{subsec3.2} we will derive our final formula  for  $Z(\Sigma \times S^1,L)$, Eq. \eqref{eq2.48_reg} below.
(We do this by  incorporating into Eq. \eqref{eq2.48_Ha7a} above
the first two points of the list appearing in Sec. \ref{subsubsec2.3.1},
cf. the beginning of Sec. \ref{subsec3.2} for more details.)\par

In Sec. \ref{subsec3.3} we will then study an interesting class of non-trivial knots in $S^2 \times S^1$,
namely the class of all  torus knots in $S^2 \times S^1$ of ``standard type''
(cf. Definition \ref{def_3.3_1} and Definition \ref{def_3.3_2} below).
We will first derive an $S^2 \times S^1$-analogue of the so-called Rosso-Jones formula (cf. Footnote \ref{ft_sec1_3} in Sec. \ref{sec1}) and we will then show how  a simple argument based on Witten's surgery formula allows us to derive
for arbitrary (simple, simply-connected, compact) $G$
the original version of the Rosso-Jones formula, which is concerned with  torus knots in $S^3$.\par

In Sec. \ref{subsec3.4} we then study  links in $\Sigma \times S^1$ without ``double points''.
Even though such links are  not very interesting
from a knot theoretic point of view (although they are not trivial)
they are interesting in so far as they allow us to see
how major building blocks of the shadow invariant $|L|$ arise. \par

Finally, in Sec. \ref{subsec3.5}
we consider the case of general (``strictly admissible'') $L$
and sketch the  strategy for evaluating $Z(\Sigma \times S^1,L)$.
I want to emphasize that with some extra work
we can expect to obtain an explicit formula for $Z(\Sigma \times S^1,L)$
 also for general (strictly admissible) $L$, cf. the paragraph before  Remark \ref{rm_KiRe} in Sec. \ref{subsubsec3.5.2} below.
The difference with respect to the treatment of  the three special cases mentioned
above is that in the present paper we do not verify
 that the explicit expressions
 obtained for $Z(\Sigma \times S^1,L)$ for general $L$
 agree with those in $RT(\Sigma \times S^1,L)$ (even though we do give some
plausibility arguments later, cf. Appendix \ref{appD} below.)

\medskip

{\it Note:} In Sec. \ref{subsec3.1} we essentially\footnote{\label{ft_BT_difference}There are two differences, though: we begin our computations  with Eq. \eqref{eq2.48} above as the starting point
rather than with Eq. (7.9) in \cite{BlTh1}, cf. Remark \ref{rm_non_equiv} below.
The second difference is described in Remark \ref{rm_3.1_2} below.} give a rederivation of  the main result of \cite{BlTh1}.
Sec. \ref{subsec3.2} is based on \cite{Ha6b} (with the exception of
Sec. \ref{subsubsec3.2.3},  which is new).
In Sec. \ref{subsec3.3} we have rewritten
the rigorous, ``simplicial'' treatment in  \cite{Ha9} using the continuum setting
introduced in Sec. \ref{subsec3.2} below.
Similarly, in  Sec. \ref{subsec3.4} we have rewritten
  the informal computations in  \cite{Ha4,HaHa}
  within the continuum setting of Sec. \ref{subsec3.2}.
 Finally, in Sec. \ref{subsec3.5} we have modified and generalized
the treatment in Sec. 5.3 in \cite{Ha4}.

\subsection{Special case I.  ``Fiber links'' in $M = \Sigma \times S^1$}
\label{subsec3.1}

Let us now consider the special case where $L = (l_1,l_2, \ldots, l_m)$ consists only of ``fiber  loops'', i.e. loops which are ``parallel'' to the $S^1$-component of $\Sigma \times S^1$. (Note that we could also work with ribbons, cf. Sec. \ref{subsubsec3.2.1}  below,   but in the present section
it is sufficient  to work with loops.)
  More precisely, we assume that each $l_i$, $i \le m$, is given by
  $l_i(s) = (\sigma_i, i_{S^1}(s))$ for all $s \in [0,1]$ for some fixed point $\sigma_i$ in $\Sigma$.
(This special case was already  treated in \cite{BlTh1}, cf. Remark \ref{rm_non_equiv} below for a comparison).
Observe that in this special case   we have $\Hol_{l_i}(A^{\orth} +  B dt) = \exp(B(\sigma_i))$
and therefore
$$ \int_{\Check{\cA}^{\orth}} \left( \prod_i  \Tr_{\rho_i}\bigl(
 \Hol_{l_i}(\Check{A}^{\orth} + A^{\orth}_c, B)\bigr) \right)
d\mu^{\orth}_B(\Check{A}^{\orth}) =  \prod_{i=1}^m  \Tr_{\rho_i}\bigl(\exp(B(\sigma_i))\bigr)$$
so  Eq. \eqref{eq2.48}  reduces to
 \begin{multline} \label{eq2.48_simpl}
Z(\Sigma \times S^1,(\sigma_i)_i,(\rho_i)_i) \sim \sum_{y \in I} \int_{\cB} \biggl[ \int_{\cA^{\orth}_c} \biggl\{
 1_{\cB_{reg}}(B)  \Det_{rig}(B)  \bigl( \prod_{i=1}^m  \Tr_{\rho_i}\bigl(\exp(B(\sigma_i))\bigr) \bigr) \\
 \times   \exp\bigl( - 2\pi i k  \langle y,  B(\sigma_0)\rangle \bigr) \biggr\}
 \exp\bigl(i S_{CS}(A^{\orth}_c, B)\bigr) DA^{\orth}_c \biggr] DB
 \end{multline}
where we have written
 $Z(\Sigma \times S^1,(\sigma_i)_i,(\rho_i)_i)$ instead of $Z(\Sigma \times S^1,L)$
 and where $\sim$ denotes equality up to a multiplicative constant which is independent
 of $(\sigma_i)_i$ and $(\rho_i)_i$.

\begin{remark} \label{rm_non_equiv}
 In view of  the relation  $S_{CS}(A^{\orth}_c, B) = 2 \pi k  \int_{\Sigma }   \Tr\bigl(B \cdot dA_c^{\orth}  \bigr)$
  it is clear  that  Eq. \eqref{eq2.48_simpl} is closely related
         to the  formula (7.9) in \cite{BlTh1},  or rather, the obvious generalization/modification of (7.9) in \cite{BlTh1}
which one obtains after including the analogue of the factor $\prod_{i=1}^m  \Tr_{\rho_i}\bigl(\exp(B(\sigma_i))\bigr)$ appearing above (cf. Sec. 7.6 in \cite{BlTh1}), replacing\footnote{Here $h$ is the notation in \cite{BlTh1} for the
 the dual Coxeter number of $\cG$ (which we denote by $\cg$).
 We refer to Sec. \ref{subsec2.4} above and Remark \ref{rm_3.1_2} below for a comment regarding the  replacement $k+h \to k$.}
  ``$k+h$'' by $k$ and replacing the group $SU(n)$ by a general simple, simply-connected, compact Lie group $G$.
 Both formulas turn out to be ``computationally'' equivalent
  in the sense that their explicit evaluation  of their RHS   leads
 to the same explicit expressions, cf.  Eq. \eqref{eq_Gen_Verlinde} below. \par

 On the other hand, from a conceptual point of view these two formulas do not seem to be equivalent.  Observe that in Eq. \eqref{eq2.48_simpl} we have a sum $\sum_{y \in I}$,
   a  factor $ \exp(- 2\pi i k  \langle y, B(\sigma_0)\rangle)$, and the integration
   $\int \cdots DA^{\orth}_c$. By
   contrast in  (the generalization of) formula (7.9) in  \cite{BlTh1}
 we have  a sum $\sum_{\lambda \in \Lambda}$ over the weight lattice $\Lambda$,
 a factor ``$ \exp\bigl( - i \int_{\Sigma} tr(\lambda \ F)\bigr)$'',
 and the integration    $\int \cdots DF$ where $F$  runs over the space of all  2-forms
 on $\Sigma$.  As a result of the appearance of arbitrary  2-forms,
  formula (7.9) in  \cite{BlTh1} does not seem to have a natural generalization
         to the situation of general links $L$ (while Eq. \eqref{eq2.48_simpl} above obviously has
         such a generalization, namely  Eq. \eqref{eq2.48} above).
\end{remark}

Instead of working with the original version
of Eq. \eqref{eq2.48_simpl}
let us now switch, for simplicity\footnote{The evaluation of the RHS of  Eq. \eqref{eq2.48_simpl} (which leads
to the same expression) is somewhat more involved since we then have to use similar arguments as in
the second half of Sec. \ref{subsubsec3.3.5} below.}, to its $1_{C^{\infty}(\Sigma,P)}$-analogue (cf. Remark \ref{rm_eigentlich_average}  in Sec. \ref{subsubsec2.2.2} above), i.e. to
 \begin{multline} \label{eq2.48_simpl0}
Z(\Sigma \times S^1,(\sigma_i)_i,(\rho_i)_i) \sim \sum_{y \in I} \int_{\cB} \biggl[ \int_{\cA^{\orth}_c} \biggl\{
 1_{C^{\infty}(\Sigma,P)}(B)  \Det_{rig}(B)  \bigl( \prod_{i=1}^m  \Tr_{\rho_i}\bigl(\exp(B(\sigma_i))\bigr) \bigr) \\
 \times   \exp\bigl( - 2\pi i k  \langle y,  B(\sigma_0)\rangle \bigr) \biggr\}
 \exp\bigl(i S_{CS}(A^{\orth}_c, B)\bigr) DA^{\orth}_c \biggr] DB
 \end{multline}

\subsubsection{Explicit evaluation of the RHS of Eq. \eqref{eq2.48_simpl0}}
\label{subsubsec3.1.1}

In order to  evaluate the RHS of Eq. \eqref{eq2.48_simpl0}
 we first integrate out the variable $A^{\orth}_c$. Since the only term in Eq. \eqref{eq2.48_simpl0}
 depending on $A^{\orth}_c$ is the factor
 $$ \exp\bigl(i S_{CS}(A^{\orth}_c, B)\bigr) =  \exp\bigl( - 2 \pi i k  \ll   A^{\orth}_c, \star dB \gg_{\cA^{\orth}}
 \bigr) $$
 (cf. Remark \ref{rm_2.7}   above)  we obtain, informally, a  delta function expression $\delta(dB)$.
 In view of this delta-function the $\int \cdots DB$-integral  can be replaced
 by an integral over the subspace $\cB_{c} := \{ B \in \cB \mid B \text{ is constant}\} \cong \ct$.
From Eq. \eqref{eq2.48_simpl0}  we therefore obtain
  \begin{multline} \label{eq2.48_simpl_b}
Z(\Sigma \times S^1,(\sigma_i)_i,(\rho_i)_i) \sim \sum_{y \in I}  \int_{\ct} \bigl\{ 1_{P}(b)  \Det_{rig}(b)   \\ \times \bigl( \prod_{i=1}^m  \Tr_{\rho_i}\bigl(\exp(b)\bigr) \bigr)  \exp\bigl( - 2\pi i k  \langle y, b \rangle \bigr) \bigr\} db
 \end{multline}

In order to evaluate the expression $\Det_{rig}(b)$,
  $b \in  \ct_{reg}$, given by Eqs \eqref{eq_Det_rig_B1}--\eqref{eq_Det_rig_B3} above
 observe first that
\begin{multline} \label{eq_Euler_char}
\lim_{\eps \to 0} \bigl( 2 \Tr\bigl( e^{- \eps \triangle_0}) -  \Tr\bigl( e^{- \eps \triangle_1}) \bigr)     \overset{(*)}{=}
 2 \dim(\ker(\triangle_0)) - \dim(\ker(\triangle_1)) \\
   \overset{(**)}{=} 2 \dim(H^0(\Sigma,\bR)) -   \dim(H^1(\Sigma,\bR))
    =   \chi(\Sigma)
 \end{multline}
where $\chi(\Sigma)$ is the Euler characteristic of $\Sigma$.
    Here step $(*)$ in Eq. \eqref{eq_Euler_char}
    follows from a well-known argument by McKean \& Singer (cf. \cite{McSin})
    and  step $(**)$ in Eq. \eqref{eq_Euler_char} follows because according to the Hodge theorem we have $\Ker(\triangle_i) \cong H^i(\Sigma,\bR)$.
  From Eq. \eqref{eq_Det_rig_B2}, Eq. \eqref{eq_Det_rig_B3}, and Eq. \eqref{eq_Euler_char} we obtain
  \begin{equation} \label{eq_det_rig_alpha_b}
  \Det_{rig,\alpha}(b) =  (2 |\sin(\pi  \alpha(b))|)^{\chi(\Sigma)} =
  (2 \sin(\pi  \alpha(b)))^{\chi(\Sigma)}
 \end{equation}
 Combining this with Eq. \eqref{eq_Det_rig_B1} and
 Eq. \eqref{eq_appB_det1/2} in Appendix \ref{appB} we arrive at
  \begin{equation} \label{eq_RaySinger} \Det_{rig}(b) = \det(1_{\ck}-\exp(\ad(b))_{| \ck})^{\chi(\Sigma)/2}
 \end{equation}
 (In particular, the value of $\Det_{rig}(b)$
is independent of the auxiliary Riemannian metric  ${\mathbf g}_{\Sigma}$.)

\smallskip

From Eq. \eqref{eq2.48_simpl_b}, Eq. \eqref{eq_RaySinger}, and  the Poisson summation formula (at an informal level\footnote{If one wants a
 rigorous version of this argument one has to regularize the integrand in a suitable way,
 cf. Secs 3.6 and 5.4 in \cite{Ha7b} where the rigorous framework (F1) described in Sec. \ref{subsec4.1} below
 is used.}) we therefore obtain
 \begin{multline} \label{eq_sec3.1_Ende}
Z(\Sigma \times S^1,(\sigma_i)_i,(\rho_i)_i) \\
\sim \sum_{\lambda \in \Lambda} 1_{P}(\lambda/k) \bigl\{ \det(1_{\ck}-\exp(\ad(\lambda/k))_{| \ck})^{\chi(\Sigma)/2}   \bigl( \prod_{i=1}^m  \Tr_{\rho_i}\bigl(\exp(\lambda/k)\bigr) \bigr)  \bigr\}
 \end{multline}
 where $\Lambda$ is the lattice dual to $I$,
 i.e. the (real) weight lattice of $(\cG,\ct)$, cf. Appendix \ref{appB.1} below.

\begin{remark}  \label{rm_3.1_2}
  As mentioned in Sec. \ref{subsubsec2.3.2} above, the  approach for obtaining
  a rigorous realization $\Det_{rig}(B)$ of $\Det(B)$ used in the present paper
 is a  variant of the approach in Sec. 6 in \cite{BlTh1}.
 In the present paper  we use the exponentials  $e^{- \eps \triangle_i}$
of the original (=``plain'') Hodge Laplacians $\triangle_i$ for defining  $\Det_{rig}(B)$
(and we later use the classical Gauss-Bonnet theorem for proving the well-definedness
and for the explicit evaluation of $\Det_{rig}(B)$, cf. Sec. \ref{subsubsec3.2.5} below).
By contrast  in Sec. 6 in \cite{BlTh1}
  ``covariant Hodge Laplacians'' are used (and instead of the  Gauss-Bonnet theorem
 the index theorem for the Dolbeault operator is applied).
This leads to an additional term containing the  dual Coxeter number $\cg$ of $\cG$. The overall effect in the present situation where $L$ is a ``fiber link'' is precisely the  ``shift'' $k \to k+\cg$ mentioned in Sec. \ref{subsec2.4}.
Since in the present paper we are using  the plain Hodge Laplacian
we  do not obtain such a shift, but  according to  Sec. \ref{subsec2.4}
this is not a problem.  Anyway, it would be interesting to study whether
also in the case of general links $L$ the use of the covariant Hodge Laplacian can  produce the shift $k \to k+\cg$  in all places where this is necessary.
(Note that such a shift must also appear in the expressions  $T^{\eps}_{cl}(A^{\orth}_c,B)$, $cl \in Cl_2(L,D)$ appearing in Sec. \ref{subsec3.5} below.)
\end{remark}

 \subsubsection{Rewriting Eq. \eqref{eq_sec3.1_Ende} using quantum algebraic notation}
\label{subsubsec3.1.2}

First we apply  the change of variable $\lambda \to \lambda - \rho$
where $\rho \in \Lambda$ is the   half sum of positive roots of  $(\cG,\ct)$. We then obtain
   \begin{multline} \label{eq_sec3.1_Ende_mod}
Z(\Sigma \times S^1,(\sigma_i)_i,(\rho_i)_i) \\
\sim \sum_{\lambda \in  \Lambda \cap (k P - \rho)} \det(1_{\ck}-\exp(\ad((\lambda+\rho)/k))_{| \ck})^{\chi(\Sigma)/2}   \bigl( \prod_{i=1}^m  \Tr_{\rho_i}\bigl(\exp((\lambda+\rho)/k)\bigr) \bigr)
 \end{multline}
 Without loss of generality we can assume that $P$ is the fundamental Weyl alcove,
 cf. Appendix \ref{appB.1}.
 Then, using  the  notation of Appendix \ref{appB.2} below we have
$$ \Lambda \cap (k P - \rho) = \Lambda_+^k$$

 Let  $\mu_i \in \Lambda_+$ be the highest weight of the representation $\rho_i$.
 In the following we will restrict our attention to the special case where
 $\mu_i \in \Lambda_+^k$.
 According to Eq. \eqref{eq_Weyl_char} in Appendix \ref{appB} we then have  for all $\lambda \in  \Lambda_+^k$
 \begin{equation}\Tr_{\rho_i}(\exp((\lambda + \rho)/k)) = \frac{S_{\lambda \mu_i}}{S_{\lambda 0}}
 \end{equation}
where  $(S_{\mu \nu})_{\mu, \nu \in \Lambda_+^k}$ is the  ``$S$-matrix''
 defined by Eq. \eqref{eq_def_C+S}  in Appendix \ref{appB} (cf. Remark \ref{rm_mod_cat} in Appendix \ref{appB.2}).
Moreover,  we have for every $\lambda \in \Lambda_+^k$  (cf. Eq. \eqref{eq_S_lamba0})
 \begin{equation}\det(1_{\ck}-\exp(\ad((\lambda+\rho)/k))_{| \ck}) \sim (S_{\lambda 0})^2
 \end{equation}
Using this we can rewrite Eq. \eqref{eq_sec3.1_Ende_mod} above as
 \begin{equation}
Z(\Sigma \times S^1,(\sigma_i)_i,(\rho_i)_i) \sim \sum_{\lambda \in \Lambda_+^k} \bigl(\prod_{i=1}^m \tfrac{ S_{\lambda \mu_i}}{S_{\lambda0}}\bigr) (S_{\lambda0})^{\chi(\Sigma)}
\end{equation}
In other words, we have
 \begin{equation} \label{eq_Gen_Verlinde}
Z(\Sigma \times S^1,(\sigma_i)_i,(\rho_i)_i) = C(\Sigma,G,k) \sum_{\lambda \in \Lambda_+^k} \bigl(\prod_{i=1}^m \tfrac{ S_{\lambda \mu_i}}{S_{\lambda0}}\bigr) (S_{\lambda0})^{\chi(\Sigma)}
\end{equation}
where $C(\Sigma,G,k)$ is a constant depending only on (the homeomorphism class of)  $\Sigma$,
$G$, and $k$.

\begin{remark} \label{rm_BlTh_Lob}
(i) Let us now first consider the special case  $\Sigma \cong S^2$.
Let $N_{\mu_1 \mu_2\mu_3}$ be the  dimension  of the vector space
$V_{\bar{k},(\rho_1, \rho_2,\rho_3)}:= V_{S^2,G,\bar{k},(\rho_1, \rho_2,\rho_3)}$
of conformal blocks of the WZW model  with group $G$ at level $\bar{k} \in \bN$
on the punctured surface $\Sigma = S^2$ with three punctures
at $(\sigma_1,\sigma_2,\sigma_3)$
with colors $(\rho_1,\rho_2,\rho_3)$.  According to \cite{Wi} we have
\begin{equation} \label{eq_Z_vs_N}
 N_{\mu_1 \mu_2\mu_3} = Z(S^2 \times S^1,(\sigma_1,\sigma_2,\sigma_3), (\rho_1,\rho_2,\rho_3))
\end{equation}
if   $Z(S^2 \times S^1,(\sigma_1,\sigma_2,\sigma_3), (\rho_1,\rho_2,\rho_3))$ is evaluated explicitly using the method in \cite{Wi} for  $k = \bar{k}$.  \par
Now recall from  Sec. \ref{subsec2.4} above
 that  we expect that, for general $L$,  the explicit values for $Z(\Sigma \times S^1,L)$ which we obtain by using the  method in the present paper  coincide with the values for $Z(\Sigma \times S^1,L)$ obtained in \cite{Wi} up to a  ``shift'' $k \to k+\cg$.
In particular,  in the present special case where $L$ is the fiber link described above
we expect to obtain Eq. \eqref{eq_Z_vs_N}
  provided that we choose $k = \bar{k} + \cg$.
 By  taking into account that $N_{0 0 0} = 1$ or,
equivalently, $Z(S^2 \times S^1) =1$ (cf. Sec. 4.4 in \cite{Wi})
   we obtain\footnote{In view of Eq. \eqref{eq_S2=C} in Appendix \ref{appB.2} and $C_{00} = \delta_{00}$ the relation $Z(S^2 \times S^1) =1$ implies $C(S^2,G,k)= 1$.}     from Eq. \eqref{eq_Z_vs_N} and  Eq. \eqref{eq_Gen_Verlinde} above  (applied to the special case $m=3$)
 \begin{equation} \label{eq_Verlinde_conj}
N_{\mu_1 \mu_2\mu_3} = \sum_{\lambda \in \Lambda_+^k}  \tfrac{ S_{\lambda \mu_1} S_{\lambda \mu_2} S_{\lambda \mu_3} }{S_{\lambda0}}
\end{equation}
where  $\Lambda_+^k$ and $(S_{\mu \nu})_{\mu \nu}$ are defined as above with $k = \bar{k} + \cg$ (cf. Remark \ref{rm_mod_cat} in Appendix \ref{appB.2}).
 Eq. \eqref{eq_Verlinde_conj}  is called the ``fusion rules''\footnote{In Sec. 4.5 \cite{Wi} this formula is called  the ``Verlinde formula''
  but in the present paper we restrict the use of the term ``Verlinde formula'' to  Eq. \eqref{eq_Verlinde_formula} below.}   in \cite{BlTh1}.

\smallskip

(ii)  The second special case we consider is the case $m=0$ (for general $\Sigma$).
     According to \cite{Wi} we have for $k = \bar{k}$
     \begin{equation} \label{eq_dimV_vs_Z} \dim(V_{\Sigma,\bar{k}}) = Z(\Sigma \times S^1)
     \end{equation}  where
     $V_{\Sigma,\bar{k}}:= V_{\Sigma,G,\bar{k}}$ is the the vector space of conformal blocks
     of the WZW model on $\Sigma$ with group $G$ at level $\bar{k}$.
    From what we said in part (i)   of the present remark (cf. again Sec. \ref{subsec2.4})
     we expect Eq. \eqref{eq_dimV_vs_Z} to hold (for our value of $Z(\Sigma \times S^1)$) if we choose $k = \bar{k} + \cg$.
  Combining Eq. \eqref{eq_dimV_vs_Z} with  Eq. \eqref{eq_Gen_Verlinde}
  we obtain\footnote{Of course,  Eq. \eqref{eq_Verlinde_formula0}  does not contain any information as long as we know nothing about $C(\Sigma,G,k)$. Still it is interesting to see how the expression $\sum_{\lambda \in \Lambda_+^{k}} (S_{\lambda0})^{2-2g}$  on the RHS of Eq. \eqref{eq_Verlinde_formula} below appears automatically
  in the torus gauge approach to the CS path integral. Moreover,
     as is sketched in Sec. 7.5 in \cite{BlTh1} it seems to be possible in principle
     to obtain the full Eq. \eqref{eq_Verlinde_formula}  by using suitable additional informal arguments.}
  \begin{equation} \label{eq_Verlinde_formula0}
\dim(V_{\Sigma,\bar{k}}) = C(\Sigma,G,k) \ \sum_{\lambda \in \Lambda_+^{k}} (S_{\lambda0})^{2-2g}
\end{equation}
where $g$ is the genus of $\Sigma$.
The correct value of  $C(\Sigma,G,k)$ turns out to be $1$.
Accordingly, we arrive at the ``Verlinde formula''
\begin{equation} \label{eq_Verlinde_formula}
\dim(V_{\Sigma,\bar{k}}) = \sum_{\lambda \in \Lambda_+^{k}} (S_{\lambda0})^{2-2g}
\end{equation}
Observe that in view of the Weyl denominator formula and
Eq. \eqref{eq_def_S} in Appendix \ref{appB} below (and the paragraph after Eq. \eqref{eq_def_S})  Eq. \eqref{eq_Verlinde_formula}
is indeed equivalent to  Eq. (1.2) in \cite{BlTh1}.

\end{remark}

\subsection{Preparations for Secs \ref{subsec3.3}--\ref{subsec3.5}}
\label{subsec3.2}

As mentioned in Sec. \ref{subsubsec2.3.1} above, Eq. \eqref{eq2.48}
is not yet our final  formula for $Z(\Sigma \times S^1,L)$. Before we arrive at the final  formula
(cf. Eq. \eqref{eq2.48_reg} below)
we will need to incorporate the first two points  appearing in the list
after Remark \ref{rm_2.7} in Sec. \ref{subsubsec2.3.1} above,
which is what we will do in the present section.
In particular, we will  introduce ribbon holonomies (and later regularized ribbon holonomies), we will give the definition of the space $\cB_{reg}^{ess}$ mentioned above,
we will show the existence of $\Det_{rig}(B)$ for each $B \in \cB_{reg}$
(and generalize its definition to all $B \in \cB^{ess}_{reg}$),
and for a suitable choice of the auxiliary Riemannian metric ${\mathbf g}_{\Sigma}$ on $\Sigma$ we will give an
explicit evaluation of $\Det_{rig}(B)$ for those $B$ which will be relevant in Secs \ref{subsec3.3}--\ref{subsec3.5}.

\subsubsection{Closed ribbons and ribbon holonomies}
\label{subsubsec3.2.1}

\begin{definition}
(i)  A closed ribbon  in $M = \Sigma \times S^1$ is a  smooth embedding
$R:S^1 \times [0,1] \to \Sigma \times S^1$.

\smallskip

(ii) A ribbon link in $M = \Sigma \times S^1$ is
 a finite tuple $L=(R_1,R_2, \ldots, R_m)$, $m \in \bN$, of non-intersecting closed ribbons $R_i$ in  $M = \Sigma \times S^1$.

\smallskip

(iii) A colored ribbon link in $M = \Sigma \times S^1$ is
 a pair $L=((R_1,R_2, \ldots, R_m),(\rho_1,\rho_2, \ldots, \rho_m))$, $m \in \bN$,
where $(R_1,R_2, \ldots, R_m)$ is a ribbon link in $M = \Sigma \times S^1$
and each $\rho_i$, $i \le m$, is an irreducible,
 finite-dimensional, complex representation  of $G$.
\end{definition}

\begin{definition} \label{def_3.2_1}
A closed ribbon $R$ in $M = \Sigma \times S^1$ is called
``horizontal''
iff  every $t \in S^1$ has an open neighborhood $U$  such that the restriction
 of $R_{\Sigma} := \pi_{\Sigma} \circ R$
 to $U \times [0,1] \to \Sigma$ is a smooth embedding.
A ribbon link $L=(R_1,R_2, \ldots, R_m)$, $m \in \bN$, in $M=\Sigma \times S^1$
is called horizontal iff each $R_i$, $i \le m$, is horizontal.
\end{definition}

\begin{definition} \label{def_3.2_2}
Let $L=(R_1,R_2, \ldots, R_m)$, $m \in \bN$, be a ribbon link in $M=\Sigma \times S^1$.
Then we will denote by $L^0$ the proper link in  $M=\Sigma \times S^1$ given by
$L^0=(l_1,l_2, \ldots, l_m)$ where $l_i=R_i(\cdot,1/2)$ for each $i \le m$.
Note that each $R_i$ induces a framing of the loop $l_i$ in a natural way.
If $R_i$ is horizontal then the  framing on $l_i$ induced by $R_i$ will
also be called ``horizontal''.
\end{definition}

In the following we will assume that the
(proper) link $L =(l_1,l_2, \ldots, l_m)$ in $M=\Sigma \times S^1$
which we fixed in Sec. \ref{subsec2.3} above
has the property that\footnote{Note that this will always be the case if $L$ is admissible in the sense of Definition \ref{def_3.5_0} in Sec.\ref{subsec3.5} below.
See also Remark \ref{rm_adm_links_gen}.} $L= L^0_{ribb}$
for some horizontal ribbon link $L_{ribb} = (R_1,R_2, \ldots, R_m)$ in $M=\Sigma \times S^1$.
We will keep  $L_{ribb}$ fixed in the following
and we will usually write  $L$ instead of $L_{ribb}$.

\smallskip

From now on we will assume that  $\sigma_0 \in \Sigma$ was chosen such that
\begin{equation} \label{eq_sec3.2_sigma0_cond}
\sigma_0 \notin  \bigcup_{i=1}^m \Image(R^i_{\Sigma})
\end{equation}
where we have set
$R^i_{\Sigma} := (R_i)_{\Sigma} := \pi_{\Sigma} \circ R_i$.
For every $R \in \{R_1, R_2, \ldots, R_m\}$
we define
\begin{subequations}
\begin{equation} \label{eq_Hol_R_def} \Hol_{R}(A) := P_1(A)
\end{equation}
 where
$(P_s(A))_{s \in [0,1]}$ is the unique solution of\footnote{More precisely,
$P(A) = (P_s(A))_{s \in [0,1]}$ is the unique smooth map $[0,1] \to \Mat(N,\bR)$
fulfilling Eq. \eqref{eq3.16}.}
\begin{equation} \label{eq3.16}
\tfrac{d}{ds} P_s(A) =  P_s(A) \cdot \biggl( \int_0^1 A(l'_u(s)) du \biggr), \quad P_0(A)=1
\end{equation}
\end{subequations}
  where $l_u$, $u \in [0,1]$, is the loop
   $[0,1] \to  \Sigma  \times S^1$ associated to the knot
    $K_u:=R(\cdot,u)$ in $\Sigma \times S^1$, cf. Sec. \ref{subsec2.1}.
  (In other words: $l_u$ is given by
    $l_u(s)=  K_u(i_{S^1}(s)) = R(i_{S^1}(s),u)$ for all $s \in [0,1]$.)

\medskip

From Eq. \eqref{eq2.48} we now obtain, after replacing each $l_i$, $ i \le m$, by $R_i$,
\begin{multline}  \label{eq2.48_ribbon} Z(\Sigma \times S^1,L)
 \sim \sum_{y \in I}  \int_{\cA^{\orth}_c \times \cB} \biggl\{ 1_{\cB_{reg}}(B) \Det_{rig}(B) \\
 \times   \biggl[ \int_{\Check{\cA}^{\orth}} \left( \prod_i  \Tr_{\rho_i}\bigl(
 \Hol_{R_i}(\Check{A}^{\orth} + A^{\orth}_c, B)\bigr) \right)
d\mu^{\orth}_B(\Check{A}^{\orth}) \biggr] \\
 \times \exp\bigl( - 2\pi i k  \langle y, B(\sigma_0) \rangle \bigr) \biggr\}
 \exp(i S_{CS}(A^{\orth}_c, B)) (DA^{\orth}_c \otimes DB)
\end{multline}

Observe that when working with ribbon holonomies instead of the usual loop holonomies
there are two ``complications'':
\begin{itemize}
\item While the original holonomy $\Hol_l(A)$ is invariant
 under a reparametrization of the loop $l$, the ribbon holonomy $\Hol_R(A)$ is not invariant under a  reparametrization of $R$. More precisely, if $R'$ is a reparametrization of $R$
   (i.e.  $R' = R \circ \phi$
   for some diffeomorphism $\phi:S^1 \times [0,1] \to S^1 \times [0,1]$) then in general we will have
   $\Hol_R(A) \neq \Hol_{R'}(A)$.

\item While the functions $\cA \ni A \mapsto \Tr_{\rho}(\Hol_l(A)) \in \bC$
    are $\G$-invariant, the functions $\cA \ni A \mapsto \Tr_{\rho}(\Hol_R(A)) \in \bC$
    are not. This is one reason why we did not introduce a ribbon analogue of
     Eq. \eqref{eq_WLO_orig} above.
    Instead we postponed the introduction of  (closed) ribbons until now,
    that is, after the gauge has been fixed\footnote{One should note that this is totally analogous to what is done in the Lorenz gauge     approach to the CS path integral mentioned in Sec. \ref{subsec5.2} below
    where one introduces a framing of the link components
    after the Lorenz gauge has been fixed.}.
\end{itemize}

We can ``defuse''/bypass these two complications by
``sending the ribbon widths to zero'' in the following sense: \par

 For $s \in (0,1)$ and $i \le m$ let $R^{(s)}_i$  be the (closed) ribbon obtained from $R_i$ by
$$R^{(s)}_i(t,u) := R_i(t,s \cdot ( u  - 1/2) + 1/2) \quad \quad \text{ for all $t \in S^1$ and $u \in [0,1]$}$$
(Observe that each $R^{(s)}_i$ is a rescaling of the restriction of $R_i$
onto $S^1 \times [1/2 - s/2, 1/2 + s/2]$). \par

By ``sending the ribbon widths to zero'' we mean that
 in  Eq. \eqref{eq2.48_ribbon} above we replace each $R_i$ by $R^{(s)}_i$ and add $\lim_{s \to 0}$ in front of the
  RHS of Eq.  \eqref{eq2.48_ribbon}. Since
\begin{equation} \label{eq_point_split}
 \forall A \in \cA: \quad
\lim_{s \to 0} \Hol_{R^{(s)}_i}(A) = \Hol_{l_i}(A)
\end{equation}
the use of  ribbon holonomies  whose ribbon widths we then send to zero can be considered as a
  ``point-splitting'' regularization   (in the sense of  Sec. 2.1 in \cite{Wi}) of the original loop holonomies
   $\Hol_{l_i}(A)$.   \par

 For reasons explained in the paragraph  before Eq. \eqref{eq3.2_18}  below
 we will introduce in Sec. \ref{subsubsec3.2.2} an additional regularization.

\subsubsection{Regularized ribbon holonomies}
\label{subsubsec3.2.2}

Above we introduced  ribbon holonomies $\Hol_{R}(A)$, $A \in \cA$.
Let us now consider the special situation $A=A^{\orth} + B dt$
where  $A^{\orth} \in \cA^{\orth}$ and $B \in \cB$
and introduce the notation
\begin{equation}
\Hol_{R}(A^{\orth}, B)  :=  \Hol_{R}(A^{\orth} +B dt)
   \end{equation}
According to the definition we have
\begin{subequations}
\begin{equation}\Hol_{R}(A^{\orth},B) = P_1(A^{\orth},B)
\end{equation}
 where $(P_s(A^{\orth},B))_{s \in [0,1]}$ is the unique solution of
\begin{equation} \label{eq3.18}
\tfrac{d}{ds} P_s(A^{\orth},B) =  P_s(A^{\orth},B) \cdot D_s(A^{\orth},B), \quad P_0(A^{\orth},B)=1.
\end{equation}
Here we have set
\begin{equation} \label{eq3.19} D_s(A^{\orth},B) := \int_0^1 A^{\orth}(l'_u(s)) du + \int_0^1 (B  dt)(l'_u(s)) du,
\end{equation}
\end{subequations}
where (as in Sec. \ref{subsubsec3.2.1} above)
 $l_u$, $u \in [0,1]$, is the loop
 $l_u: [0,1] \ni s \mapsto  R(i_{S^1}(s),u) \in \Sigma  \times S^1$.  \par

Let us fix $s, u \in [0,1]$ temporarily. Moreover, let us fix
 for the rest of Sec. \ref{sec3} an arbitrary ortho-normal basis
 $(T_a)_{a \le \dim(\cG)}$  of $\cG$. \par

During the informal calculation below it would be very convenient to be able to write
$A^{\orth}(l'_u(s))$ as a scalar product in a suitable way.
Informally, we have
\begin{equation} \label{eq3.2_18}
A^{\orth}(l'_u(s)) = A^{\orth}((l_u)'_{\Sigma}(s))
  = \text{``$\sum_a  T_a \ll A^{\orth}, T_a (l_u)'_{\Sigma}(s) \delta_{l_u(s)} \gg_{\cA^{\orth}}$''}
\end{equation}
   where  we have set $(l_u)_{\Sigma} := \pi_{\Sigma} \circ l_u$,
    where    $\ll \cdot, \cdot \gg_{\cA^{\orth}}$ is the scalar product on $\cA^{\orth}$
induced by the Riemannian metric ${\mathbf g}:= {\mathbf g}_{\Sigma}$ fixed
in Sec. \ref{subsec2.3} above, and where $\delta_p$ for $p=l_u(s)$ is the  ``Dirac delta function'' in the point $p$. \par
A well-defined version of the last term in Eq. \eqref{eq3.2_18} can be obtained
if,  instead of working, for every fixed $p \in \Sigma \times S^1$,
 with the  ``Dirac delta function'' $\delta_p$ we  work with a suitable ``Dirac family''\footnote{i.e.
  $\delta^{\eps}_{p}$, $\eps \in (0,\eps_0)$, is a non-negative and smooth function  $\Sigma \times S^1 \to \bR$. Moreover,
 $\int \delta^{\eps}_{\sigma} d\mu_{\mathbf g} \otimes dt =1$,
and we have $\delta^{\eps}_{p} \to  \delta_{p}$ weakly as $\eps \to 0$ where
$\delta_{p}$ is the Dirac measure on $\Sigma \times S^1$ in the point $p$.}
 $(\delta^{\eps}_{p})_{\eps < \eps_0}$, $\eps_0 > 0$, w.r.t. the measure  $d\mu_{\mathbf g} \otimes dt$, cf. the second paragraph  after Eq. \eqref{eq3.2_21} below.
 Here $d\mu_{\mathbf g}$ is the volume measure on $\Sigma$ associated to ${\mathbf g} = {\mathbf g}_{\Sigma}$ . \par

After these preparations we will now replace for every  $\eps < \eps_0$ (with $\eps_0 >0$ as in the ``bullet point'' after Eq. \eqref{eq3.2_21} below)
the expression  ${A}^{\orth}(l'_u(s))$ in Eq. \eqref{eq3.19}
  by the ``regularized'' expression\footnote{Here we interpret $T_a {\mathbf X}_{(l_u)'_{\Sigma}(s)} \delta^{\eps}_{l_u(s)}$
  as an element of $\cA^{\orth} \cong C^{\infty}(S^1,\cA_{\Sigma})$
   using the identification $\cA_{\Sigma} \cong \cG \otimes
  VF(\Sigma)$ (induced by ${\mathbf g}_{\Sigma}$)   where $VF(\Sigma)$ is the space of smooth vector fields on $\Sigma$.   }
\begin{equation} \label{eq_Deps_def_prep}
\bigl({A}^{\orth}(l'_u(s))\bigr)^{(\eps)} := \sum_a  T_a \ll A^{\orth}, T_a {\mathbf X}_{(l_u)'_{\Sigma}(s)} \delta^{\eps}_{l_u(s)} \gg_{\cA^{\orth}}
\end{equation}
where for every $v \in T_{\sigma_0} \Sigma$, $\sigma_0 \in \Sigma$,
we denote by ${\mathbf X}_v$  the local vector field on $\Sigma$ around $\sigma_0$
which is obtained by
  parallel transport with the Levi-Civita connection, cf. the paragraph after Eq. \eqref{eq3.2_21} below.
  Next  we replace the expression $D_s(A^{\orth},B)$ in Eq. \eqref{eq3.18} by
\begin{subequations}
\begin{equation} \label{eq_Deps_def}
 D^{\eps}_s(A^{\orth},B) := \int_0^1 \bigl({A}^{\orth}(l'_u(s))\bigr)^{(\eps)} du
 + \int_0^1 (B  dt)(l'_u(s)) du
 \end{equation}
 and arrive at the regularized holonomy
 \begin{equation}\Hol^{\eps}_{R}(A^{\orth},B) := P^{\eps}_1(A^{\orth},B)
 \end{equation}
 where
$(P^{\eps}_s(A^{\orth},B))_{s \in [0,1]}$ is the unique solution of
\begin{equation} \label{eq3.2_21} \tfrac{d}{ds} P^{\eps}_s(A^{\orth},B) =  P^{\eps}_s(A^{\orth},B) \cdot  D^{\eps}_s(A^{\orth},B), \quad P^{\eps}_0(A^{\orth},B)=1
\end{equation}
\end{subequations}

\begin{itemize}

\item The vector field ${\mathbf X}_v$  where $v \in T_{\sigma_0} \Sigma$, $\sigma_0 \in \Sigma$
mentioned above is obtained as follows: \par
Let $d_{\mathbf g}$ be the Riemannian distance function on $\Sigma$ induced by $\mathbf g$.   Since $\Sigma$ is compact there is a $\eps_0 > 0$ such that for all $\sigma_0, \sigma_1 \in \Sigma$ with $d_{\mathbf g}(\sigma_0,\sigma_1) < \eps_0$ there is a unique
(geodesic) segment starting in $\sigma_0$ and ending in $\sigma_1$.
Using parallel transport along this geodesic segment w.r.t. the Levi-Civita connection of $(\Sigma,\mathbf g)$
we can transport every tangent vector $v \in T_{\sigma_0} \Sigma$
to a tangent vector in  $T_{\sigma_1} \Sigma$.
Thus every $v \in T_{\sigma_0} \Sigma$ induces in a natural way a vector field ${\mathbf X}_v$
on the open ball $B_{\eps_0}(\sigma_0) \subset \Sigma$.

\item The Dirac family  $(\delta^{\eps}_{p})_{\eps < \eps_0}$ on $\Sigma \times S^1$, mentioned above,   is obtained as follows: \par
  We first we choose, for each $t \in S^1$,  a Dirac family
  $(\delta^{\eps}_{t})_{\eps < \eps_0}$ around $t$ w.r.t. the measure $dt$ on $S^1$.
  Moreover, we choose  for each $\sigma \in \Sigma$, a  Dirac family $(\delta^{\eps}_{\sigma})_{\eps < \eps_0}$
around $\sigma$ w.r.t. $d\mu_{\mathbf g}$.
For technical reasons  we will  assume that
  for each $\eps$ and $\sigma \in \Sigma$ the support of  $\delta^{\eps}_{\sigma}$
  is contained in the open ball $B_{\eps}(\sigma)$.\par

For every  $p = (\sigma,t) \in  \Sigma \times S^1$ and $\eps < \eps_0$ we define
$\delta^{\eps}_p \in C^{\infty}(\Sigma \times S^1,\bR)$ by
\begin{equation}\delta^{\eps}_p(\sigma',t'):= \delta^{\eps}_{\sigma}(\sigma') \delta^{\eps}_{t}(t') \quad \text{
for all $\sigma' \in \Sigma$ and $t' \in S^1$.}
\end{equation}

\end{itemize}

\begin{remark} \label{rm_gen_notation}
(i)  Instead of  $P^{\eps}_{s}(A^{\orth},B)$ and $D^{\eps}_{s}(A^{\orth},B)$
  we later also  use the notation
      $P^{\eps}_{R,s}(A^{\orth},B)$ and $D^{\eps}_{R,s}(A^{\orth},B)$
 (i.e. in situations where more than one (closed) ribbon $R$ is involved, cf. Sec. \ref{subsec3.4}).

\smallskip

(ii)  Above $R$ was a closed ribbon in $M = \Sigma \times S^1$, i.e. a smooth map $R:S^1 \times [0,1] \to M$,
which can be considered as a map  $R:[0,1] \times [0,1] \to M$ where
$R(0,u)= R(1,u)$ for all $u \in [0,1]$.
The definition of $P^{\eps}_{R,s}(A^{\orth},B)$ and $D^{\eps}_{R,s}(A^{\orth},B)$
can be generalized in an obvious way to all ``ribbons'', i.e. all smooth maps
 $R:[0,1] \times [0,1] \to M$ (where the condition $R(0,u)= R(1,u)$ for all $u \in [0,1]$ need not be fulfilled, in which case we will call $R$ ``open'').
(This  will be useful in Sec. \ref{subsec3.5} below.)
Instead of $P^{\eps}_{R,1}(A^{\orth},B)$ we will simply write $P^{\eps}_{R}(A^{\orth},B)$.
\end{remark}

\subsubsection{The space $\cB^{ess}_{reg}$}
\label{subsubsec3.2.3}

Observe that
$\ct_{reg} = \ct \backslash \bigl(\bigcup_{\alpha \in \cR, k \in \bZ} H_{\alpha,k}\bigr)$
where
$H_{\alpha,k}$, $\alpha \in \cR$, $k \in \bZ$  is the hyperplane in $\ct$
given by  $H_{\alpha,k}:= \alpha^{-1}(k)$, cf. Appendix \ref{appB.1} below.
Accordingly, the space $\cB_{reg}$ defined in Sec. \ref{subsec2.2} above is the space of those $B \in \cB$
whose image does not meet any of these hyperplanes $H_{\alpha,k}$,
$\alpha \in \cR$, $k \in \bZ$. \par

For reasons explained in Remark \ref{rm_why_B_ess_reg}
 below  we will now replace in Eq. \eqref{eq2.48_ribbon} above
 the space $\cB_{reg}$ by the slightly larger space $\cB^{ess}_{reg}$ of those $B \in \cB$ which, whenever their image does meet  a hyperplane $H_{\alpha,k}$, they intersect $H_{\alpha,k}$ ``properly''.
More precisely, we demand that for all $\sigma \in \Sigma$ for which $B(\sigma) \in H_{\alpha,k}$ holds for some $\alpha \in \cR$ and $k \in \bZ$ the differential $d B_\alpha(\sigma)$ of the map $B_{\alpha} := \alpha \circ B:\Sigma \to \bR$ in the point $\sigma$ does not vanish.
More briefly,  $\cB^{ess}_{reg}$ is given by
 \begin{equation} \label{eq_def_B^ess} \cB^{ess}_{reg} := \{ B \in \cB \mid \forall \sigma \in \Sigma:
  \alpha \in \cR: [ B_{\alpha}(\sigma) \in \bZ \ \Rightarrow \ dB_{\alpha}(\sigma) \neq 0 ]\}
 \end{equation}
Observe that  if $B \in \cB_{reg}^{ess}$
then the set $\cN:= \{\sigma \in \Sigma \mid B(\sigma) \notin \ct_{reg} \}$
is a $d\mu_{\mathbf g}$-zero set of $\Sigma$
and the singularities of the functions $\log(\sin(\pi \alpha(B))): \Sigma \to \bC$, $\alpha \in \cR$, are so mild that the RHS of Eq. \eqref{eq_heatkernel1} in   Sec. \ref{subsubsec3.2.5} below exists.
(This will allow us to generalize the definition of $\Det_{rig}(B)$ to
all $B \in \cB^{ess}_{reg}$.)

\begin{remark} \label{rm_why_B_ess_reg}
(i) The  replacement $\cB_{reg} \to \cB^{ess}_{reg}$ will be justified
in Appendix \ref{appA.6} below (cf. also Remark \ref{rm_sec3.2.7} in Sec. \ref{subsubsec3.2.6} below).

\smallskip

(ii) In certain special cases one can actually continue to work with the space
$\cB_{reg}$ rather than having to work with $\cB^{ess}_{reg}$.
For example, this is the case when $L$ is a  vertical link, cf.  Sec. \ref{subsec3.1}.
Moreover, if    $L$ is as in Sec. \ref{subsec3.4}  (and, possibly, also if $L$ is as in Sec. \ref{subsec3.5}) and each   $\rho_i$ is a fundamental representation of $G$
and $G=SU(N)$ (or, possibly, also for general $G$) then\footnote{In order to see this note that in the special case where  each   $\rho_i$ is a fundamental representation of $G$
the step functions $B$ appearing on the RHS of  Eq. \eqref{eq_sec3.4_step_fun} in Sec. \ref{subsec3.4}  below all have small ``step sizes''.
In the special case where $G=SU(N)$ it is easy to show that this implies that
those $B$ for which $1_{\cB_{reg}}(B)=0$ (i.e. those $B$
 whose image is not contained in a single Weyl alcove) will have at least one
``step'' whose (constant) value lies in one of the hyperplanes $H_{\alpha,k}$,
$\alpha \in \cR$, $k \in \bZ$. Accordingly, we then also  have $1_{\cB^{ess}_{reg}}(B) = 0$.  On the other hand, if $1_{\cB_{reg}}(B)=1$ then trivially also $1_{\cB^{ess}_{reg}}(B) = 1$.
Consequently, the value of the RHS of
 Eq. \eqref{eq_sec3.4_step_fun} below does not change if we replace
$1_{\cB^{ess}_{reg}}(B)$ by $1_{\cB_{reg}}(B)$.}  we can also work with the space $\cB_{reg}$ instead of  $\cB^{ess}_{reg}$.
However, in the majority of cases
we have to work with $\cB^{ess}_{reg}$ if we want to have a chance of obtaining
the correct result for the value of $Z(\Sigma \times S^1,L)$.
\end{remark}

\subsubsection{A convenient choice of the auxiliary Riemannian metric ${\mathbf g}_{\Sigma} $ on $\Sigma$}
\label{subsubsec3.2.4}

Let  $L = (R_1,R_2, \ldots, R_m)$  be the (horizontal) ribbon link
in $M=\Sigma \times S^1$ fixed in Sec. \ref{subsubsec3.2.1} above.
Set $R^i_{\Sigma} := \pi_{\Sigma} \circ R_i$, $i \le m$, where $\pi_{\Sigma}: \Sigma \times S^1 \to \Sigma$
is the canonical projection
and let  $S_i$ denote the interior of $ \Image(R^i_{\Sigma})$ in $\Sigma$.

\medskip

Recall that above we have fixed an auxiliary Riemannian metric
${\mathbf g}_{\Sigma}$ on $\Sigma$.
In order to simplify our life  we will assume from now on that ${\mathbf g} := {\mathbf g}_{\Sigma}$ fulfills the following condition:
\begin{condition} \label{ass1}
The auxiliary Riemannian metric
 $\mathbf g $ on $\Sigma$ was chosen such that for each $i \le m$
 the pullback of $\mathbf g_{| S_i}$    via $R^i_{\Sigma}: S^1 \times (0,1) \to \Sigma$
   coincides with the Riemannian product metric
   ${\mathbf g}_{S^1} \otimes {\mathbf g}_{(0,1)}$ on $S^1 \times (0,1)$
   where ${\mathbf g}_{(0,1)}$ is the standard Riemannian metric on $(0,1)$
   and ${\mathbf g}_{S^1}$ is the translation-invariant Riemannian metric on $S^1$, which   we assume to be normalized such that $vol(S^1) = 1$.
(Note that   $\mathbf g $ is uniquely determined    on each $S_i$.)
\end{condition}

There are two reasons why we choose ${\mathbf g}_{\Sigma}$  such that Condition \ref{ass1} is fulfilled:

\begin{itemize}
\item[1.] The evaluation of the inner integral in Eq. \eqref{eq2.48_regc} in Sec. \ref{subsec3.3}
and in Sec. \ref{subsec3.4} as well as the evaluation of the expressions $T^{\eps}_{cl}(A^{\orth}_c, B)$,  $cl  \in Cl_1(L,D)$, in Sec. \ref{subsec3.5} becomes  much easier.
\item[2.] The explicit evaluation of $\Det_{rig}(B)$ in Sec. \ref{subsubsec3.2.5} below
        leads to the correct formula. This may also be the case without Condition  \ref{ass1} (cf. Remark \ref{rm_self_linking} below), however, this point is not yet clarified.
\end{itemize}

\begin{remark} \label{rm_sec2.5_3}
Recall that  the original (informal) path integral expression in Eq. \eqref{eq_WLO_orig} above is  topologically invariant. In particular, it does not involve a  Riemannian metric.
 However,  for technical reasons, we  work with the auxiliary Riemannian metric  ${\mathbf g}_{\Sigma}$
 breaking topological invariance. \par

 Clearly, whenever one introduces an auxiliary object $\cO$ in order to make
 sense of an informal expression in a natural way
  it would be good to have  either (or a combination) of the following:
 \begin{enumerate}
 \item The auxiliary object   $\cO$  can be chosen arbitrarily
  and the final result does not depend on it.

 \item There is a distinguished/canonical choice of $\cO$ and this is the choice which we use.
\end{enumerate}

Condition \ref{ass1} is a combination of these two cases.
The restriction  of ${\mathbf g} = {\mathbf g}_{\Sigma}$ to  $S := \bigcup_{i=1}^m S_i$ is given canonically.
 On the other hand, the restriction  of $\mathbf g$ to  $S^c:= \Sigma \backslash S$
 can  essentially be chosen arbitrarily
 (as long as $\mathbf g_{| S}$ and $\mathbf g_{|S^c}$ ``fit together'' smoothly,
  i.e. induce a smooth Riemannian metric on all of $\Sigma$).
\end{remark}

\begin{remark} \label{rm_sec3.2.2}
 For  the concrete ribbon links $L$ which  we will consider
 in Sec. \ref{subsec3.3} and Sec. \ref{subsec3.4} below
  Condition \ref{ass1} can always be fulfilled and, according to Remark \ref{rm_sec2.5_3}, is natural. \par

   On the other hand, for general ``admissible'' ribbon links  $L$ as defined in Sec. \ref{subsec3.5} below
   Condition \ref{ass1}  can in general only be fulfilled
after replacing each $R_i$ by a suitable  reparametrization\footnote{Observe, for example, that  the two maps $R^i_{\Sigma}$ and $R^j_{\Sigma}$
 will in general induce a different Riemannian metric on $U_{ij} := S_i \cap S_j \neq \emptyset$ if $i \neq j$. On the other hand, if the ribbon link $L$ fixed above is admissible
 in the sense of Definition \ref{def_3.5_3} in Sec. \ref{subsubsec3.5.2} below
 then for every $i, j \le m$ it is always possible to
 reparametrize $R_i$ and $R_j$
 such that the reparametrized versions of $R^i_{\Sigma}$ and $R^j_{\Sigma}$ induce
   the same Riemannian metric on $U_{ij}$.}.
 So in the general case it may be more natural to  use the following
   ``infinitesimal version'' of  Condition \ref{ass1}
   (which is suggested by Observation \ref{obs_ass1} below):
 \begin{quote}
 The auxiliary Riemannian metric
 $\mathbf g $ on $\Sigma$ was chosen such that
   for each $i \le m$
  and  every $p \in C_i := \arc(l^i_{\Sigma})$ with  $l_i := R_i(\cdot, 1/2)$
  the geodesic curvature    $k_{\mathbf g}^{C_i}(p)$ of  $C_i$
 in the point $p$  vanishes.
  \end{quote}
 Observe that this ``infinitesimal version'' of  Condition \ref{ass1}
 can always be fulfilled but it involves considerably more work
 when evaluating  $Z(\Sigma \times S^1,L)$. \par

 Note also that, as mentioned in Remark \ref{rm_self_linking} below, it may be possible that Condition \ref{ass1}
 or its infinitesimal version can be dropped altogether. (Of course, this increases even further   the amount of work we have to do for evaluating  $Z(\Sigma \times S^1,L)$.)
\end{remark}

\subsubsection{Generalization of $\Det_{rig}(B)$ for $B \in \cB^{ess}_{reg}$}
\label{subsubsec3.2.5}

In the present section we will modify (slightly) the definition
of $\Det_{rig}(B)$ for $B \in \cB_{reg}$, given in Sec. \ref{subsubsec2.3.2} above,
and we will then generalize the new definition to the case of all $B \in \cB^{ess}_{reg}$.
Moreover,  for the class of $B$ relevant for us in Sec. \ref{subsec3.3}--\ref{subsec3.5} below
we will give an explicit formula for $\Det_{rig}(B)$
by applying  the  Gauss-Bonnet formula for surfaces with boundary.

\medskip

Recall that for $B \in \cB_{reg}$ we
rewrote $\Det(B)$ informally as (cf. Eq. \eqref{eq_Det(B)_rewrite-1})
\begin{equation} \label{eq_DetB_O-OP}
\Det(B) = \prod_{\alpha \in \cR_+} \det(O^{(0)}_{\alpha}(B))^2 \det(O^{(1)}_{\alpha}(B))^{-1}
\end{equation}
 where  for each fixed $\alpha \in \cR_+$  the  operators
$O^{(i)}_{\alpha}(B): \Omega^i(\Sigma,\bR) \to \Omega^i(\Sigma,\bR)$, $i =0,1$,
 are the multiplication operators obtained by multiplication with the  function
  $ 2 | \sin(\pi  \alpha(B(\cdot)))|$.  Then we used this as the motivation
  to define $\Det_{rig}(B)$ by Eqs. \eqref{eq_Det_rig_B1}--\eqref{eq_Det_rig_B3}
above.

\smallskip

Let us now (re)define $\Det_{rig}(B)$ for   $B \in \cB_{reg}$  again by  Eqs. \eqref{eq_Det_rig_B1}--\eqref{eq_Det_rig_B3}
in Sec. \ref{subsubsec2.3.2} above where now  we  take $O^{(i)}_{\alpha}(B)$ to be the
multiplication operators obtained by multiplication with the  function
 $ 2  \sin(\pi  \alpha(B(\cdot)))$  (rather than $ 2  |\sin(\pi  \alpha(B(\cdot)))|$)
 and we take the ``operator-logarithm'' $\log$ appearing in Eq. \eqref{eq_Det_rig_B3} above
   to come from the  restriction to $\bR \backslash \{0\}$   of the principal branch of the complex logarithm.
(We will give the definition of  $\Det_{rig}(B)$ for general $B \in \cB^{ess}_{reg}$ below.)

\begin{remark} \label{rm_Oalpha_sign}
(i) Observe that when working with the new ansatz for $O^{(i)}_{\alpha}(B)$,
 the RHS of Eq. \eqref{eq_DetB_O-OP}  depends explicitly (via $\cR_+$)
on the Weyl chamber $\cC$ fixed in Appendix \ref{appB} below. However,
one can argue at an informal level that the value of $\Det(B)$
does not depend on $\cC$. (This implies also that if $B \in \cB_{reg}$
 then the value of the expression  Eq. \eqref{eq_DetB_O-OP} is independent of whether we use the original or  the new ansatz for
 the operators  $O^{(i)}_{\alpha}(B)$.)
Moreover, for the special maps $B \in \cB_{reg}^{ess}$ relevant below one can see from
Eq. \eqref{eq_Detrig_step_total} below (by taking into account that $\sum_i \chi(Y_i) = \chi(\Sigma)$ is even)
 that also the value of $\Det_{rig}(B)$ does not depend on the choice of the Weyl chamber $\cC$.

\smallskip

(ii)  Taking $O^{(i)}_{\alpha}(B)$ to be
the multiplication operator with  the function
$2  \sin(\pi  \alpha(B(\cdot)))$  looks natural and,
as we will see below, leads  to the correct values\footnote{By contrast, if we
choose again  $O^{(i)}_{\alpha}(B)$ to be the multiplication operator with  the function $2  | \sin(\pi  \alpha(B(\cdot))) |$ then we will only get the
 the correct values for $Z(\Sigma \times S^2,L)$
 in  the cases mentioned in Remark \ref{rm_why_B_ess_reg} in Sec. \ref{subsubsec3.2.3} above.
 But these are exactly those cases where
 we do not have to work with the space $\cB^{ess}_{reg}$ in the first place
 but can continue to work with the space $\cB_{reg}$.}
 of $Z(\Sigma \times S^2,L)$.
  Anyway,  it would be desirable to obtain a direct justification
for the new ansatz above (i.e. for  taking $O^{(i)}_{\alpha}(B)$ to be
the multiplication operator with  the function
$2  \sin(\pi  \alpha(B(\cdot)))$), for example by using an argument which involves
 the computation of the $\eta$-invariant of a suitable operator.
\end{remark}

If  $B \in \cB_{reg}$ then $\Tr( e^{- \eps \triangle_i} \log(O^{(i)}_{\alpha}(B)))$ is well-defined
(cf. the argument at the end of Sec. \ref{subsubsec2.3.2} above)
and we can  rewrite the RHS of Eq. \eqref{eq_Det_rig_B3} as
\begin{equation} \label{eq_heatkernel1}
\exp(\Tr( e^{- \eps \triangle_i} \log(O^{(i)}_{\alpha}(B)))) = \exp\bigl(\int_{\Sigma} \Tr(K^{(i)}_{\eps}(\sigma,\sigma)) \log(2 \sin(\pi  \alpha(B(\sigma)))) d\mu_{\mathbf g}(\sigma)\bigr)
\end{equation}
 where  $K^{(0)}_{\eps}:\Sigma \times \Sigma \to \bR \cong \End(\bR)$
 is the integral kernel of $e^{- \eps \triangle_0}$
and $K^{(1)}_{\eps}:\Sigma \times \Sigma \to \bigcup_{\sigma_1,\sigma_2 \in \Sigma} \Hom(T_{\sigma_1} \Sigma, T_{\sigma_2} \Sigma)$ is the integral kernel of $e^{- \eps \triangle_1}$ (and where $\log:\bR \backslash \{0\} \to \bC$
  is the restriction to $\bR \backslash \{0\}$   of the principal branch of the complex logarithm).
  Observe that the RHS of Eq. \eqref{eq_heatkernel1} is  well-defined
  not only when $B \in \cB_{reg}$ but even when
   $B \in \cB^{ess}_{reg}$. Accordingly, let us  now set,  for general $B \in \cB^{ess}_{reg}$,
\begin{equation} \label{eq_new_detOi} \det\nolimits_{\eps}(O^{(i)}_{\alpha}(B)):= \exp\bigl(\int_{\Sigma} \Tr(K^{(i)}_{\eps}(\sigma,\sigma)) \log(2 \sin(\pi  \alpha(B(\sigma)))) d\mu_{\mathbf g}(\sigma)\bigr)
\end{equation}
and define  $\Det_{rig,\alpha}(B)$ and $\Det_{rig}(B)$ again by Eqs. \eqref{eq_Det_rig_B1}--\eqref{eq_Det_rig_B2} in Sec. \ref{subsubsec2.3.2} above.

   \smallskip

  According to a well-known result in \cite{McSin}
  the  negative powers of $\eps$ that appear in the asymptotic
   expansion of $K^{(i)}_{\eps}$, $i =0,1$ as $\eps \to 0$ cancel each other
   and we obtain
\begin{equation} \label{eq_heatkernel_expr}
 \bigl[2 \Tr( K^{(0)}_{\eps}(\sigma,\sigma)) -   \Tr(K^{(1)}_{\eps}(\sigma,\sigma))\bigr]
 \to  \tfrac{1}{4 \pi}   R_{\mathbf g} (\sigma) \quad \text{ uniformly in $\sigma$ as $\eps \to 0$  }
 \end{equation}
 where $R_{\mathbf g}$ is the scalar curvature ( = twice the Gaussian curvature) of $(\Sigma,{\mathbf g})$. \par
  From Eqs. \eqref{eq_new_detOi} and \eqref{eq_heatkernel_expr}
  it follows that $\Det_{rig,\alpha}(B)$ is well-defined for all $B \in \cB^{ess}_{reg}$
  (i.e. that  the $\eps \to 0$ limit in Eq. \eqref{eq_Det_rig_B2} really exists)     and that we have
 \begin{equation} \label{eq_explicit_formula_Detreg} \Det_{rig,\alpha}(B) =
 \exp\biggl( \int_{\Sigma} \log(2 \sin(\pi  \alpha(B(\sigma))))  \tfrac{1}{4 \pi}   R_{\mathbf g} (\sigma) d\mu_{\mathbf g}(\sigma)\biggr)
\end{equation}

\smallskip

Let us now evaluate $\Det_{rig}(B)$ explicitly for all $B$ relevant for us.

\smallskip

Recall that in Sec. \ref{subsec3.1} above only the special case   $B \equiv b$  (with $b \in \ct_{reg}$) was relevant. (Observe that we can rederive Eq. \eqref{eq_det_rig_alpha_b} and hence also Eq. \eqref{eq_RaySinger}
in Sec. \ref{subsec3.1} above directly from Eq. \eqref{eq_explicit_formula_Detreg}  above
by applying    the classical  Gauss-Bonnet Theorem
$ 4 \pi \chi(\Sigma) =  \int_{\Sigma}      R_{\mathbf g}  d\mu_{\mathbf g}$.) \par

Now in  Secs \ref{subsec3.3}--\ref{subsec3.5}  below a larger class of maps  $B:\Sigma \to \ct$
 will be relevant for  the explicit evaluation of $Z(\Sigma \times S^1,L)$
(after the $\eps \to 0$-limit on the RHS of Eq. \eqref{eq2.48_regc}
  below have been carried out), namely the class of all those maps $B$
  which are constant on each connected component  $Y_i$, $i \le r$, of
$  \Sigma \backslash \bigcup_{i=1}^m \Image(R^i_{\Sigma})$.

\smallskip

In order to deal with this larger class of maps $B$
we will need the following, more general version\footnote{\label{ft_gen_Gauss_Bonnet}In fact, if we want to have a chance of dealing successfully with the case of general (strictly admissible) $L$ as in   Sec. \ref{subsec3.5} below we
will have to work with yet another generalization
 of   Eq. \eqref{eq_Gauss_bonnet_gen}  where  the boundary $\partial Y$  is only a piecewise smooth (rather than a smooth) submanifold of $\Sigma$ (cf. Remark \ref{rm_appD_Ende} in Appendix \ref{appD}). In the generalized formula there will be an extra term on the RHS involving a sum over the finite number of those points $p$ of $\partial Y$ where $\partial Y$ is not smooth (and containing the corresponding ``angle'' of  $\partial Y$ at $p$).} of  the classical Gauss-Bonnet Theorem mentioned above: \par
Let $Y \subset \Sigma$ be such that the boundary $\partial Y$ is (either empty or) a smooth 1-dimensional submanifold
of $\Sigma$. We equip $\partial Y$ with the Riemannian metric induced by $\mathbf g = \mathbf g_{\Sigma}$
and denote by $ds$ the corresponding ``line element''  on  $\partial Y$.
Then we have
\begin{equation} \label{eq_Gauss_bonnet_gen}
 4 \pi \chi(Y) =  \int_{Y}      R_{\mathbf g}  d\mu_{\mathbf g} + 2 \int_{\partial Y} k^{\partial Y}_{\mathbf g} ds
\end{equation}
where $k^{\partial Y}_{\mathbf g}(p)$ for $p \in \partial Y$ is the geodesic
 curvature of the curve $\partial Y$ in the point $p$.

\begin{observation} \label{obs_ass1}
Condition \ref{ass1} implies that the scalar curvature $R_{\mathbf g}$ vanishes on
$\bigcup_{i=1}^m \Image(R^i_{\Sigma})$.
Moreover, on  each of the curves $C^i_u= \arc((l^i_u)_{\Sigma})$, $i \le m$, $u \in [0,1]$,
 the geodesic curvature  $k^{C^i_u}_{\mathbf g}$ vanishes.
 \end{observation}

Now let $B:\Sigma \to \ct$ be as above\footnote{Observe that on $\bigcup_{i=1}^m \Image(R^i_{\Sigma})$
$B$ does not have to be smooth.}, i.e.
 $B$ is constant on each of  $Y_i$, $i \le r$,
  where $(Y_i)_{i \le r}$ is the family of connected components of
$  \Sigma \backslash \bigcup_{i=1}^m \Image(R^i_{\Sigma})$.
Let $b_i$ denote the unique value of $B$  on $Y_i$.
From Eq. \eqref{eq_explicit_formula_Detreg} and  Observation \ref{obs_ass1}
we conclude that then
\begin{align} \label{eq_Detrig_step}  \Det_{rig,\alpha}(B)
& =  \exp\biggl( \sum_i \int_{Y_i} \log(2 \sin(\pi  \alpha(b_i)))  \tfrac{1}{4 \pi}   R_{\mathbf g} (\sigma) d\mu_{\mathbf g}(\sigma)\biggr) \nonumber \\
& = \prod_i \exp\biggl(  \log(2 \sin(\pi  \alpha(b_i))) \biggl[ \int_{Y_i} \tfrac{1}{4 \pi}   R_{\mathbf g} (\sigma) d\mu_{\mathbf g}(\sigma) \biggr] \biggr) \nonumber \\
& =  \prod_i \exp\biggl(  \log(2 \sin(\pi  \alpha(b_i))) \biggl[ \int_{Y_i} \tfrac{1}{4 \pi}   R_{\mathbf g} (\sigma) d\mu_{\mathbf g}(\sigma) +  \tfrac{1}{2 \pi}  \int_{\partial Y_i}
k^{\partial Y_i}_{\mathbf g} ds  \biggr] \biggr)\nonumber \\
& = \prod_i \exp\biggl(  \log(2 \sin(\pi  \alpha(b_i))) \bigl[ \chi(Y_i) \bigr] \biggr)\nonumber \\
& = \prod_{i} \bigl(2 \sin(\pi \alpha(b_i))\bigr)^{\chi(Y_i)}
\end{align}
From this we  obtain
\begin{equation}
\label{eq_Detrig_step_total}
\Det_{rig}(B) = \prod_{i}  \det\nolimits^{1/2}\bigl(1_{\ck}-\exp(\ad(b_i))_{|\ck}\bigr)^{\chi(Y_i)}
\end{equation}
where $\det^{1/2}\bigl(1_{\ck}-\exp(\ad(\cdot))_{|\ck}\bigr): \ct \to \bR$
 is given by
 $$\det\nolimits^{1/2}(1_{\ck}-\exp(\ad(b))_{|\ck}) =  \prod_{\alpha \in \cR+}  2 \sin(\pi  \alpha(b)) \quad \forall b \in \ct$$

\begin{remark} \label{rm_self_linking}
 Recall from Remark \ref{rm_sec3.2.2} above that
when working with  general (ribbon) links $L$ it may be better
to work with the infinitesimal version of Condition \ref{ass1}  mentioned in
 Remark \ref{rm_sec3.2.2}  above.
It fact, it may even be possible  to avoid the use
of Condition \ref{ass1} (or its infinitesimal version) completely.
This may be surprising since
for the derivation of Eq. \eqref{eq_Detrig_step_total}
above  it was crucial that the geodesic curvature terms
 $k^{\partial Y_i}_{\mathbf g}$ appearing in Eq. \eqref{eq_Detrig_step}  vanish.
However, it is possible that
 during the computations later on these geodesic curvature terms
  appear  when evaluating the inner integral in Eq. \eqref{eq_Ztgf_def}.
  (This is because,  when    Condition \ref{ass1} is not fulfilled,
 the ``covariances'' $\ll \phi^{\eps}_i, C(B) \phi^{\eps}_j \gg$ appearing  Eq. \eqref{eq_self_covariance}  will in general not vanish anymore.)
  \end{remark}

\begin{remark} \label{rm_Det_disc_prep}
One can rewrite Eq. \eqref{eq_DetB_O-OP} in a more symmetric way,
which will be useful in Remark \ref{rm_Det_disc} below. \par

In order to do so  observe first that for $B \in \cB$
and $\alpha \in \cR_+$ we have
 $O^{(0)}_{\alpha}(B) = \star^{-1} \circ O^{(2)}_{\alpha}(B) \circ \star$  where   $\star: \Omega^0(\Sigma,\ck) \to \Omega^2(\Sigma,\ck)$ is the  Hodge star operator induced by any fixed Riemannian metric ${\mathbf g}_{\Sigma}$ on $\Sigma$.
 Thus we obtain,  informally,
\begin{equation} \label{eq_sec2.5_2}
 O^{(0)}_{\alpha}(B) = O^{(2)}_{\alpha}(B)
\end{equation}
and we can rewrite  Eq. \eqref{eq_DetB_O-OP} as
\begin{equation} \label{eq_DetB_O-OP2}
\Det(B) = \prod_{\alpha \in \cR_+} \det(O^{(0)}_{\alpha}(B)) \det(O^{(1)}_{\alpha}(B))^{-1}
\det(O^{(2)}_{\alpha}(B))
\end{equation}

\end{remark}

\begin{remark} \label{rm_Det_disc}  There is  an alternative  method
for making rigorous sense of $\Det(B)$ which is very natural
when using the rigorous frameworks\footnote{It can also be useful with the rigorous
continuum approach (F2) if it is combined with a suitable continuum limit argument.
} (F1) and (F3) described in Sec. \ref{subsec4.1} and Sec. \ref{subsec4.3} below for making sense of $Z(\Sigma \times S^1, L)$. This alternative method consists in
 introducing a ``simplicial analogue'' $\Det^{disc}(B)$  of the RHS
 of    Eq. \eqref{eq_DetB_O-OP2}, cf. Sec. 3.6 in \cite{Ha9}.
 (For the definition of $\Det^{disc}(B)$
 it is crucial to use the RHS of Eq. \eqref{eq_DetB_O-OP2} here.
 If instead   one uses the RHS of  Eq. \eqref{eq_DetB_O-OP} as the starting point for a definition of $\Det^{disc}(B)$ one obtains incorrect results.)
\end{remark}

\subsubsection{The (regularized) torus gauge fixed CS path integral $Z^{t.g.f}(\Sigma \times S^1,L)$}
\label{subsubsec3.2.6}

In order to arrive at our final formula for the  Chern-Simons path integral $Z(\Sigma \times S^1,L)$ we will now incorporate the  constructions in Secs \ref{subsubsec3.2.1}--\ref{subsubsec3.2.5}.
In particular,  we will now replace  the colored (proper) link $L= ((l_1, l_2, \ldots, l_m),(\rho_1,\rho_2,\ldots,\rho_m))$, $i \le m$, in $M= \Sigma \times S^1$ which we fixed in Sec. \ref{subsec2.3} above by  the
colored ribbon link $L_{ribb}=((R_1, R_2, \ldots, R_m),(\rho_1,\rho_2,\ldots,\rho_m))$
chosen in Sec. \ref{subsubsec3.2.1}.
(Recall that we assume that $L_{ribb}$ is horizontal and fulfills $L = L^0_{ribb}$,
 cf. Definition \ref{def_3.2_1}
and Definition \ref{def_3.2_2}). Instead of $L_{ribb}$ we will simply write $L$ in the following and we  set
\begin{subequations}
\begin{multline} \label{eq_Ztgf_def}  Z^{t.g.f}(\Sigma \times S^1,L) :=  \lim_{s \to 0} \lim_{\eps \to 0} \sum_{y \in I}  \int_{\cA^{\orth}_c \times \cB}
  \biggl\{ 1_{\cB^{ess}_{reg}}(B) \Det_{rig}(B) \\
 \times   \biggl[ \int_{\Check{\cA}^{\orth}} \left( \prod_i  \Tr_{\rho_i}\bigl(
 \Hol^{\eps}_{R^{(s)}_i}(\Check{A}^{\orth} + A^{\orth}_c, B)\bigr) \right)
d\mu^{\orth}_B(\Check{A}^{\orth}) \biggr] \\
 \times \exp\bigl( - 2\pi i k  \langle y, B(\sigma_0) \rangle \bigr) \biggr\}
 \exp(i S_{CS}(A^{\orth}_c, B)) (DA^{\orth}_c \otimes DB)
\end{multline}

According to what we said in Secs \ref{subsubsec3.2.1}--\ref{subsubsec3.2.3}
above $Z^{t.g.f}(\Sigma \times S^1,L)$ can be considered
as a regularized and  gauge-fixed version of $Z(\Sigma \times S^1,L)$
and we should therefore have
\begin{equation} \label{eq2.48_reg}
Z(\Sigma \times S^1,L) \sim  Z^{t.g.f}(\Sigma \times S^1,L).
\end{equation}
\end{subequations}
 In  the special situations in Sec. \ref{subsec3.3} and Sec. \ref{subsec3.4} below
(and probably also in the situation of general strictly admissible $L$, cf.  Sec. \ref{subsec3.5})   the $s\to 0$-limit in  Eq. \eqref{eq_Ztgf_def}
 will turn out to be trivial, i.e.  Eq. \eqref{eq2.48_reg}
 will reduce to
 \begin{multline}  \label{eq2.48_regc} Z(\Sigma \times S^1,L)
 \sim  \lim_{\eps \to 0} \sum_{y \in I}  \int_{\cA^{\orth}_c \times \cB}
  \biggl\{ 1_{\cB^{ess}_{reg}}(B) \Det_{rig}(B) \\
 \times   \biggl[ \int_{\Check{\cA}^{\orth}} \left( \prod_i  \Tr_{\rho_i}\bigl(
 \Hol^{\eps}_{R^{(s_0)}_i}(\Check{A}^{\orth} + A^{\orth}_c, B)\bigr) \right)
d\mu^{\orth}_B(\Check{A}^{\orth}) \biggr] \\
 \times \exp\bigl( - 2\pi i k  \langle y, B(\sigma_0) \rangle \bigr) \biggr\}
 \exp(i S_{CS}(A^{\orth}_c, B)) (DA^{\orth}_c \otimes DB)
\end{multline}
for any fixed  $s_0 \in (0,1)$. (We exclude the case $s_0 = 1$
for  technical reasons, cf. the paragraph before Eq. \eqref{eq_self_covariance_0} below.)

\begin{convention} \label{conv_s_0}
In the following we will sometimes write $\bar{R}_i$ instead of $R_i$.
Moreover, we will often write  $R_i$ instead of $R^{(s_0)}_i$.
In other words, $R_i$ can refer both to $\bar{R}_i$ and to $R^{(s_0)}_i$.
\end{convention}

\begin{remark} \label{rm_sec3.2.7}
One point which which needs to be better understood
 is why, in the generalization of the definition of $\Det_{rig}(B)$
 for all $B \in \cB^{ess}_{reg}$ which we gave in  Sec. \ref{subsubsec3.2.5} above,
 we need to define $O^{(i)}_{\alpha}(B)$ to be
the multiplication operator with  the function
$2  \sin(\pi  \alpha(B(\cdot)))$ rather than $2  |\sin(\pi  \alpha(B(\cdot)))|$,
 cf. Remark \ref{rm_Oalpha_sign} above. \par

Moreover, one should try to understand better why,
as remarked in Remark \ref{rm_why_B_ess_reg} above, apart from some special cases we cannot work with the indicator function $1_{\cB_{reg}}(B)$ instead of  $1_{\cB^{ess}_{reg}}(B)$ if we want to obtain the correct values for  $Z(\Sigma \times S^1,L)$,
even though both Eq. \eqref{eq2.48_reg} and the modification of  Eq. \eqref{eq2.48_reg} where
 in Eq. \eqref{eq_Ztgf_def} the factor
 $1_{\cB^{ess}_{reg}}(B)$ is replaced by $1_{\cB_{reg}}(B)$ can be derived/justified at an informal level. \par

Finally, it would  be desirable to check whether  Condition \ref{ass1}
above can be dropped (cf. Remark \ref{rm_self_linking} above),
 i.e. whether in Secs \ref{subsec3.3}--\ref{subsec3.5}
we arrive at the correct values for $Z(\Sigma \times S^1,L)$ for an arbitrary
auxiliary Riemannian metric ${\mathbf g}$ on $\Sigma$.

\end{remark}

\subsection{Special case II. Torus knots in $M= S^2 \times S^1$}
\label{subsec3.3}

We will now evaluate the RHS of Eq. \eqref{eq2.48_regc} in the special case where
$L$ belongs to a large class of (colored) ``torus ribbon knots'' in $M= S^2 \times S^1$
(cf. Definition \ref{def_3.3_2} below).
By doing so we obtain a  $S^2 \times S^1$-analogue
of the Rosso-Jones formula, cf. Eq. \eqref{eq_Step6_Ende} below.
In Sec. \ref{subsubsec3.3.8} we will combine the straightforward generalization
 Eq. \eqref{eq_Step7_1}
 of Eq. \eqref{eq_Step6_Ende} with a short surgery argument and obtain, for arbitrary (simple, simply-connected, compact Lie group) $G$, the original Rosso-Jones formula, which is concerned with arbitrary (colored) torus knots in $S^3$.

\medskip

Recall that a torus knot in $S^3$ is a knot $\tilde{K}:S^1 \to S^3$ whose image is contained in an unknotted torus $\tilde{\cT}  \subset  S^3$.
(Note, for example, that two of the four simplest non-trivial knots in $S^3$ are torus knots,
namely the trefoil knot and the cinquefoil knot.)
We take this as the motivation for the following definition.

\begin{definition} \label{def_3.3_1}
(i) A torus knot  in $S^2 \times S^1$ of standard type is a knot $K: S^1 \to S^2 \times S^1$
whose image is contained in a torus  $\cT$ in $ S^2 \times S^1$  fulfilling
the following condition:
\begin{enumerate}
 \item[(C)] $\cT$ is of the form $\cT = \psi(C_0 \times S^1)$
 where $C_0$ is an embedded circle in $S^2$ and  $\psi:S^2 \times S^1 \to S^2 \times S^1$ is a  diffeomorphism.
\end{enumerate}
In the special case where $\psi = \id_{S^2 \times S^1}$,
i.e. $\cT =  C_0 \times S^1$, we will call $K$ ``canonical''.

\smallskip

\noindent
(ii) For every canonical torus knot $K$ in $S^2 \times S^1$ of standard type
we denote by ${\mathbf p}(K)$ and  ${\mathbf q}(K)$ the two winding numbers of $K$ where we consider $K$  as a continuous map $S^1 \to  C_0 \times S^1 \cong S^1 \times S^1$ in the obvious way.
(As a side remark we mention that ${\mathbf p}(K)$ and  ${\mathbf q}(K)$ will always be coprime.)
\end{definition}

\begin{remark} \label{rm_torus_knots} Note that every unknotted torus $\tilde{\cT}$ in $S^3$ can be obtained
from a  torus  $\cT$ in $ S^2 \times S^1$ fulfilling condition (C) by performing a suitable
Dehn surgery on a separate  knot in  $ S^2 \times S^1$.
Consequently, every torus knot $\tilde{K}$ in $S^3$ can be obtained from a torus knot $K$ in $S^2 \times S^1$ of standard type by performing such a Dehn surgery.
Moreover, every torus knot $\tilde{K}$ in $S^3$ can be obtained up to equivalence
  from some canonical torus knot in $S^2 \times S^1$ of standard type
  by performing a suitable Dehn surgery.
 We will makes use of this observation in Sec. \ref{subsubsec3.3.8} below where we will derive the
aforementioned Rosso-Jones formula for torus knots in $S^3$.
\end{remark}

\begin{definition} \label{def_3.3_2}
(i) A canonical torus ribbon knot in $S^2 \times S^1$ of standard type is a
closed ribbon $R: S^1 \times [0,1] \to S^2 \times S^1$ which is horizontal
(cf. Definition \ref{def_3.2_1} above) and which has the property  that
each of the knots $K_u:=R(\cdot,u)$, $u \in [0,1]$,  is a canonical torus knot  in $S^2 \times S^1$ of standard type.
We set
${\mathbf p}(R):= {\mathbf p}(K_u) \in \bZ$ and  ${\mathbf q}(R):={\mathbf q}(K_u) \in \bZ$
for any $u \in [0,1]$.

\smallskip

\noindent
(ii)  A canonical torus ribbon knot $R$ in $S^2 \times S^1$ of standard type is called
 ``strictly canonical'' iff $\mathbf p:= {\mathbf p}(R) \neq 0$ and
  the following two conditions are fulfilled:
 \begin{itemize}
 \item  $R_{S^2}:[0,1/|{\mathbf p}|[ \times [0,1] \to S^2$ is injective
 \item  $R_{S^2}$ is $1/|\mathbf p|$-periodic in the first component, i.e.
   $R_{S^2}(s,u) = R_{S^2}(s + 1/|\mathbf p| ,u)$ for all $u \in [0,1]$
        and $s \in [0,1- 1/|\mathbf p|]$.
\end{itemize}
 where $R_{S^2}:[0,1] \times [0,1] \to S^2$ is given by
 $R_{S^2}(s,u):=\pi_{S^2}(R(i_{S^1}(s),u))$ for all $s, u \in [0,1]$.
\end{definition}

\begin{remark}
(i) Clearly, if  $R$ is a strictly canonical torus ribbon knot  in $S^2 \times S^1$ of standard type then the knot $K := R(\cdot, 1/2)$  will be a
canonical torus  knot  in $S^2 \times S^1$ of standard type
with ${\mathbf p}(K) \neq 0$.
Conversely, up to equivalence,
every canonical torus  knot $K$  in $S^2 \times S^1$ of standard type
with ${\mathbf p}(K) \neq 0$ can be obtained in this way.
For the derivation of the Rosso-Jones formula in Sec. \ref{subsubsec3.3.8} below (cf.
Remark \ref{rm_torus_knots} above) it will therefore be sufficient to consider only strictly canonical torus ribbon knots  in $S^2 \times S^1$ of standard type.

\smallskip

\noindent
(ii) The advantage of working with  strictly canonical torus ribbon knots in $S^2 \times S^1$ of standard type  is that for them Condition \ref{ass1} above can always be fulfilled.
This simplifies the computations below considerably, cf.  Remark \ref{rm_sec3.2.2} above.
On the other hand, if one is prepared to do the extra work one can
use the  infinitesimal version of Condition \ref{ass1}
 mentioned in Remark \ref{rm_sec3.2.2} above instead of the original Condition \ref{ass1}.
 By doing so  it should be possible to generalize
  the computations in the present section to general canonical torus ribbon knots in $S^2 \times S^1$ of standard type.
\end{remark}

In the following let $ L:=(R_1,\rho_1)$
where  $R_1$  is a strictly canonical torus ribbon knot of standard type in $S^2 \times S^1$ with winding numbers ${\mathbf p} = {\mathbf p}(R_1) \in \bZ \backslash \{0\}$ and ${\mathbf q} = {\mathbf q}(R_1) \in \bZ$  and where
$\rho_1$ is an irreducible, finite-dimensional complex
representation of $G$ with highest  weight $\lambda_1 \in \Lambda_+$.
Without loss of generality we can assume that  ${\mathbf p} > 0$ and that Condition \ref{ass1} above is fulfilled.

\begin{convention} \label{conv_Sigma_statt_S2}
Even though in the present section, i.e. Sec. \ref{subsec3.3}, we have $\Sigma = S^2$
  we will often write $\Sigma$ instead of $S^2$. In particular, we will
   do this whenever $\Sigma = S^2$
appears as a subscript  like, e.g., in $\cA_{\Sigma,\ct}$
or $(R_1)_{\Sigma}$ or ${\mathbf g}_{\Sigma}$.
\end{convention}

\subsubsection{Evaluation of the inner integral in Eq. \eqref{eq2.48_regc}}
\label{subsubsec3.3.1}

We will  first evaluate, for  fixed $A^{\orth}_c \in \cA^{\orth}_c$, $B \in \cB^{ess}_{reg}$, and $\eps \in (0,\eps_0)$ (with $\eps_0$ as in Sec. \ref{subsubsec3.2.2})
the ``inner integral'' in Eq. \eqref{eq2.48_regc}, which in the present special case
where $m=1$ is the integral
\begin{equation} \label{eq_inf_int} \int_{\Check{\cA}^{\orth}}  \Tr_{\rho_1}\bigl(
 \Hol^{\eps}_{R_1}(\Check{A}^{\orth} + A^{\orth}_c, B)\bigr)
d\mu^{\orth}_B(\Check{A}^{\orth}),
\end{equation}
cf. Convention \ref{conv_s_0} above. In order to do so we will exploit an important property
of the (informal) complex measure $d\Check{\mu}^{\orth}_{B}$ introduced in
 Eq. \eqref{eq_def_mu_B}.  Observe that according to Remark \ref{rm_2.7} above $d\Check{\mu}^{\orth}_{B}$   can be written as
 \begin{subequations}
\begin{equation}d\mu^{\orth}_B  = \tfrac{1}{\Check{Z}(B)} \exp\bigl(i  \tfrac{1}{2} \ll \Check{A}^{\orth}, S(B)
 \Check{A}^{\orth} \gg \bigr) D\Check{A}^{\orth}
\end{equation}
where $\ll \cdot, \cdot \gg$ is  the restriction   of the scalar product
 $\ll \cdot, \cdot \gg_{\cA^{\orth}}$ to  $\Check{\cA}^{\orth}$
and where  we have set
\begin{equation} S(B):= 2 \pi k \bigl(\star  \bigl(\tfrac{\partial}{\partial t} + \ad(B) \bigr)\bigr)
\end{equation}
\end{subequations}
Accordingly,  $d\mu^{\orth}_B$ is an
 informal, normalized\footnote{In Sec. \ref{sec4} below we will study rigorous analogues
 of $d\mu^{\orth}_B$, cf. frameworks (F1) and (F3).
There it will be important to check that for those $B$ for which the regularized
version of $1_{\cB_{reg}}(B)$ or  $1_{\cB^{ess}_{reg}}(B)$ does not vanish
the rigorous version of $Z(B)$ is non-zero.
On the other hand, this issue can be circumvented in the rigorous continuum framework (F2) where  we do not try to make sense of $d\mu^{\orth}_B$ itself but only of the associated integral functional $\Phi^{\orth}_B = \int \cdots d\mu^{\orth}_B$.}
 (``Gauss-type'') complex measure   on the pre-Hilbert space $(\Check{\cA}^{\orth}, \ll \cdot, \cdot \gg)$. \par

Since  $B \in \cB^{ess}_{reg}$
  the symmetric, linear operator $S(B)$  on $(\Check{\cA}^{\orth}, \ll \cdot, \cdot \gg)$ is injective and has dense image\footnote{This follows from the observation
that for $b \in \ct_{reg}$ the linear operator
$\partial_t + \ad(b): C^{\infty}(S^1,\ck) \to  C^{\infty}(S^1,\ck)$
is invertible and its inverse  is given  by
$((\partial_t + \ad(b))^{-1}f)(t)  = \bigl((e^{\ad(b)})_{| \ck} - 1_{\ck} \bigr)^{-1} \cdot \int_0^1 e^{s \ad(b)} f(t+ i_{S^1}(s)) ds$ for all $t \in S^1$ and $f \in C^{\infty}(S^1,\ck)$.}.   Accordingly,  its  inverse
\begin{equation} \label{eq_def_CB} C(B):=S(B)^{-1}
\end{equation}
  is  a  densely defined, (symmetric) linear operator on
  $(\Check{\cA}^{\orth}, \ll \cdot, \cdot \gg)$.
  (We remark that if $B \in \cB_{reg}$ then $C(B)$ has full domain and is bounded.)  \par
    Observe however, that $C(B)$  is neither positive nor negative definite, i.e. there are non-zero elements $j \in \dom(C(B)) \subset \Check{\cA}^{\orth}$ such that
$\ll j, C(B) j \gg = 0$.
 This property of $d\mu^{\orth}_B$ has important consequences, which we will exploit
 below. As a preparation let us  first study the following two  examples
 which deal with the analogous properties of suitable  (``Gauss-type'') complex measures
 on  finite-dimensional spaces.

\begin{example} \label{ex1} \rm Let $d\mu$ be the (well-defined)  normalized\footnote{in the sense that $\int_{\sim} 1 d\mu = 1$ with $\int_{\sim} \cdots d\mu$ as below.}
 (``Gauss-type'') complex measure    on $\bR^2$
which is given by
$$d\mu(x) = \tfrac{1}{2 \pi} \exp(i \tfrac{1}{2} \langle x, S x\rangle) dx
\quad \text{ where } S= \left( \begin{matrix} 0 && 1 \\ 1 && 0 \end{matrix} \right)$$
and where $\langle \cdot, \cdot \rangle$ is the standard scalar product on $\bR^2$.
 Clearly, $S$ is symmetric and invertible but its inverse
 $C:= S^{-1}$ $(=S)$ is obviously neither positive-definite nor negative-definite
 since  there are non-zero vectors $v \in \bR^2$
 fulfilling  $\langle v, C v\rangle = 0$,
  for example $v=(1,0)$ or $v=(0,1)$.
For such $v$ we obtain (using a suitable analytic continuation argument and
the well-known formulas for the moments of a genuine Gaussian measure)
$$ \int_{\sim} \langle x,v \rangle^n  d\mu(x) = 0 \quad \forall n \in \bN$$
where we have introduced the regularized integral functional
$$\int_{\sim} \cdots d\mu := \lim_{\eps \to 0} \int \cdots e^{- \eps |x|^2} d\mu(x)$$
Similarly, we have
 $$ \int_{\sim} \Phi(\langle x,v \rangle) d\mu(x) = \Phi(0)$$
 for every polynomial function $\Phi:\bR \to \bR$
 and, more generally, for every  entire analytic function $\Phi:\bR \to \bR$
 whose sequence of Taylor coefficients satisfies a suitable growth condition.
   \end{example}

Example \ref{ex1} can easily be generalized to arbitrary dimension:

\begin{example} \label{ex2} \rm Let $d\mu$ the  (well-defined) normalized
 (``Gauss-type'') complex measure   on $\bR^n$ given by
$$d\mu(x) = \tfrac{1}{Z} \exp(i \tfrac{1}{2} \langle x, S x\rangle) dx$$ where $\langle \cdot, \cdot \rangle$ is the standard scalar product on $\bR^n$,  where $S$ is an arbitrary symmetric, invertible endomorphism on $\bR^n$
and where $Z \in \bC \backslash \{0\}$ is chosen such that\footnote{$Z$ is given explicitly
by $Z =\tfrac{(2 \pi)^{n/2}}{\det^{1/2}(i S)}$
 with $\det^{1/2}(i S):= \prod_{k=1}^n \sqrt{i \lambda_k}$ where $(\lambda_k)_{k \le n}$
are the (real) eigenvalues of $S$ and $\sqrt{\cdot}: \bC \backslash (-\infty,0) \to \bC$
is the standard square root.}
$\int_{\sim} 1 \ d\mu = 1$
where we have again introduced the regularized integral functional
$$\int_{\sim} \cdots d\mu := \lim_{\eps \to 0} \int \cdots e^{- \eps |x|^2} d\mu(x)$$
Moreover, let $(v_i)_{i \le m}$, $m \in \bN$, be a sequence of vectors of $\bR^n$
such that\footnote{Clearly,
this condition is only interesting if  $C$  is neither positive-definite nor negative-definite
since otherwise   Eq. \eqref{eq_toy_model}
can only be fulfilled if  $v_i = 0$ for all $i \le m$.
In the latter case Eq. \eqref{eq_toy_model2} is trivially fulfilled.}
\begin{equation} \label{eq_toy_model} \langle v_i, C v_j \rangle = 0 \quad \forall i,j \le m
\end{equation}
where $C:=S^{-1}$. Then we have for every polynomial function  $\Phi:\bR^n \to \bR$
 (and, more generally, for every  entire analytic function $\Phi:\bR^n \to \bR$
 whose family of Taylor coefficients satisfies a suitable growth condition)
 \begin{equation}\label{eq_toy_model2} \int_{\sim} \Phi((\langle x,v_i \rangle)_{i \le m}) d\mu(x) = \Phi((\langle 0,v_i \rangle)_{i \le m}) = \Phi(0)
 \end{equation}
\end{example}

\begin{remark} \label{rm_mu_degenerate} Above we have assumed
that $S$ is an invertible endomorphism.
In fact, one can also define $\int_{\sim} \cdots d\mu $ if $S$ is not invertible
(in which case we call $d\mu$ ``degenerate'') by
setting $\int_{\sim} \cdots d\mu :=  \lim_{\eps \to 0}(\eps/\pi)^{n/2} \int \cdots e^{- \eps |x|^2} d\mu(x)$ where $n:= \dim(\ker(S))$.
This will be relevant in Sec. \ref{subsubsec4.1.1} and \ref{subsubsec4.3.1} below.
\end{remark}

Let us now go back to our infinite dimensional (informal) integral \eqref{eq_inf_int} above.
By applying a suitable limit argument  we will show
in Subsec. \ref{subsubsec3.3.2} below that
 the argument in Example \ref{ex2} can be used to evaluate the integral \eqref{eq_inf_int} above and that by doing so we obtain for all fixed
 $A^{\orth}_c \in \cA^{\orth}_c$,  $B \in \cB^{ess}_{reg}$, and $\eps < \eps_0$
 \begin{multline} \label{eq_Step1}
 \int_{\Check{\cA}^{\orth}}  \Tr_{\rho_1}\bigl(
 \Hol^{\eps}_{R_1}(\Check{A}^{\orth} + A^{\orth}_c, B)\bigr)
d\mu^{\orth}_B(\Check{A}^{\orth}) \\
 =  \Tr_{\rho_1}\bigl( \Hol^{\eps}_{R_1}(0 + A^{\orth}_c, B)) =  \Tr_{\rho_1}\bigl(
 \Hol^{\eps}_{R_1}(A^{\orth}_c, B))
\end{multline}
Of course, in order to arrive at Eq. \eqref{eq_Step1}
we need to verify a condition analogous to Eq. \eqref{eq_toy_model}
and  the analyticity \& growth assumption in Example \ref{ex2}.
We will do this in Sec. \ref{subsubsec3.3.2} below
by using the assumptions on $R_1 = \bar{R}_1$ made at the beginning of Sec. \ref{subsec3.3}
and  Condition \ref{ass1} for
 the auxiliary Riemannian metric ${\mathbf g}_{\Sigma}$ , cf.  Sec. \ref{subsubsec3.2.4} above.

\begin{remark} \label{rm_trivial_inner_int} Similar arguments will  play a crucial
role in Sec. \ref{subsec3.4} below and for the explicit evaluation of the expressions ``$T^{\eps}_{cl}(A^{\orth}_c, B)$'',  $cl  \in Cl_1(L,D)$, introduced in Sec. \ref{subsec3.5} below.
\end{remark}

\subsubsection{Justification of Eq. \eqref{eq_Step1} }
\label{subsubsec3.3.2}

As in Sec. \ref{subsubsec3.3.1} above
let $A^{\orth}_c \in \cA^{\orth}_c$, $B \in \cB^{ess}_{reg}$, and $\eps \in (0,\eps_0)$
be fixed.

\smallskip

Recall that  $ \rho:= \rho_1$ was a a representation of $G$ over a finite-dimensional complex vector space $V$.
Now observe that we have  (with $R_1 = R^{(s_0)}_1$, cf. Convention \ref{conv_s_0} above)
\begin{subequations}
 \begin{equation} \label{eq_Hol_vs_P}
 \Tr_{\rho}(\Hol^{\eps}_{R_1}(A^{\orth},B))
 = \Tr(\rho(\Hol^{\eps}_{R_1}(A^{\orth},B))) = \Tr(P^{\eps}_1(\rho;A^{\orth},B))
 \end{equation}
 where
$(P^{\eps}_s(\rho;A^{\orth},B))_{s \in [0,1]}$ is the unique solution of
\begin{equation}\tfrac{d}{ds} P^{\eps}_s(\rho;A^{\orth},B) =  P^{\eps}_s(\rho;A^{\orth},B) \cdot  \rho_*(D^{\eps}_s(A^{\orth},B)), \quad P^{\eps}_0(\rho;A^{\orth},B)=\id_V
\end{equation}
\end{subequations}
where $\cdot$ is the multiplication of $\End(V)$
and where $\rho_*: \cG \to \gl(V)$ is the derived representation of $\rho$
 and $D^{\eps}_s(A^{\orth},B)$ is defined by \eqref{eq_Deps_def} in the situation
  $R=R_1$. \par

Let us now expand $P^{\eps}_1(\rho;A^{\orth},B)$ in
 a Piccard-Lindeloeff series
\begin{multline} \label{eq_Pic_Lind}
P^{\eps}_1(\rho;A^{\orth},B) = \id_V + \int_0^1  \rho_*(D^{\eps}_s(A^{\orth},B)) ds_1
+ \int_0^1 \int_0^{s_2} \rho_*(D^{\eps}_{s_1}(A^{\orth},B))  \rho_*(D^{\eps}_{s_2}(A^{\orth},B))  ds_1 ds_2 \\
+ \int_0^1 \int_0^{s_3} \int_0^{s_2} \rho_*(D^{\eps}_{s_1}(A^{\orth},B))  \rho_*(D^{\eps}_{s_2}(A^{\orth},B))
  \rho_*(D^{\eps}_{s_3}(A^{\orth},B)) ds_1 ds_2 ds_3 + \ldots  \\
= \id_V  + \sum_{n=1}^{\infty} \int_{\triangle_n} \prod_{i=1}^n \bigl( \rho_*(D^{\eps}_{s_i}(A^{\orth},B)) \bigr) ds
\end{multline}
  where
$$ \triangle_n := \{s \in [0,1]^n \mid  s_1 \le s_2 \le \cdots  \le s_n\}$$

In view of Eq. \eqref{eq_Hol_vs_P},  Eq. \eqref{eq_Pic_Lind}, and  Eqs \eqref{eq_Deps_def} and \eqref{eq_Deps_def_prep} in Sec. \ref{subsubsec3.2.2} above
 we now see that, for each fixed $A^{\orth}_c$, $B$
   we can approximate\footnote{e.g., on the RHS of Eq. \eqref{eq_Pic_Lind} we can truncate the Lindeloef-Piccard series at the $n$th term and
   approximate each of the $n$ integrals in the truncated series by a Riemannian sum;
   moreover, on the RHS of Eq \eqref{eq_Deps_def} we also approximate the first $\int \cdots du$-integral by a Riemannian sum.}
$$\Check{\cA}^{\orth}  \ni \Check{A}^{\orth} \mapsto \Tr_{\rho}(\Hol^{\eps}_{R_1}(\Check{A}^{\orth}+A^{\orth}_c,B)) \in \bC $$
 by a  sequence $(F_n)_{n \in \bN}$ of functions $F_n:\Check{\cA}^{\orth} \to \bC$   (depending on  $A^{\orth}_c$, $B$)  of the form
 $$F_n(\Check{A}^{\orth}) = \Phi_n((\ll \phi^{\eps}_i, \Check{A}^{\orth}\gg)_{i \le d_n})$$
where $\ll \cdot, \cdot \gg:= \ll \cdot, \cdot \gg_{\Check{\cA}^{\orth}}$,
 where each $\Phi_n:\bR^{d_n} \to \bC$, $d_n \in \bN$, is a polynomial function  (depending on  $A^{\orth}_c$, $B$),  and each $\phi^{\eps}_i \in \Check{\cA}^{\orth}$, $i \le d_n$, is of
 the form
  \begin{equation} \phi^{\eps}_i = \pi_{\Check{\cA}^{\orth}}\bigl( T_{a_i} {\mathbf X}_{(l_{u_i})'_{\Sigma}(s_i)} \delta^{\eps}_{l_{u_i}(s_i)}\bigr)
 \end{equation}
  for some $s_i, u_i \in [0,1]$ and $a_i \le \dim(\ct)$ ($i \le d_n)$
  where $\pi_{\Check{\cA}^{\orth}}:\cA^{\orth} \to \Check{\cA}^{\orth}$ is the $\ll \cdot, \cdot \gg_{\cA^{\orth}}$-orthogonal projection, where
   $l_u$, $u \in [0,1]$, is the loop  $[0,1] \ni s \mapsto  R_1(i_{S^1}(s),u) \in S^2  \times S^1$, and  where we have set $(l_u)_{\Sigma}:= \pi_{\Sigma} \circ l_u$,
  cf. Convention \ref{conv_Sigma_statt_S2} above.
(Note that if $B \in \cB^{ess}_{reg} \backslash \cB_{reg}$ it is possible that
$\phi^{\eps}_i \notin \dom(C(B))$. Since $\dom(C(B))$ is dense in $(\Check{\cA}^{\orth}, \ll \cdot, \cdot \gg)$ we can in this case simply replace $\phi^{\eps}_i$ by an element $\dom(C(B))$ which is sufficiently close to $\phi^{\eps}_i$.)

\medskip

 Now  recall that above we have assumed that
   $R_1 = \bar{R}_1$ (cf. Convention \ref{conv_s_0} above) is a strictly canonical torus ribbon knot of standard type,  that  Condition \ref{ass1} is fulfilled,
  and that the number $s_0$ fixed in Sec. \ref{subsubsec3.2.6}  above
  is strictly smaller than $1$.  From these  assumptions it follows that
 for all sufficiently small $\eps > 0$ we have
 \begin{subequations}
  \begin{equation} \label{eq_self_covariance_0}\ll \phi^{\eps}_i, \star \phi^{\eps}_j \gg = 0 \quad \forall i,j \le d_n
   \end{equation}
 and therefore
 \begin{equation} \label{eq_self_covariance}
 \ll \phi^{\eps}_i, C(B) \phi^{\eps}_j \gg = 0\quad \forall i,j \le d_n
 \end{equation}
 \end{subequations}
 i.e., the analogue of Eq. \eqref{eq_toy_model} in Example \ref{ex2} above is fulfilled.
So, for each $n \in \bN$, we obtain an analogue of Eq. \eqref{eq_toy_model2}.
We now obtain  Eq. \eqref{eq_Step1} by sending $n \to \infty$
 and interchanging $\lim_{n \to \infty}$
with the integral functional $\int_{\Check{\cA}^{\orth}}  \cdots d\mu^{\orth}_B(\Check{A}^{\orth})$
 appearing in Eq. \eqref{eq_Step1}:
\begin{multline}
 \int_{\Check{\cA}^{\orth}}  \Tr_{\rho_1}\bigl(
 \Hol^{\eps}_{R_1}(\Check{A}^{\orth} + A^{\orth}_c, B)\bigr)
d\mu^{\orth}_B(\Check{A}^{\orth})
=  \int_{\Check{\cA}^{\orth}} \lim_{n \to \infty}
 \Phi_n((\ll \phi^{\eps}_i,\Check{A}^{\orth}\gg)_{i \le d_n})
d\mu^{\orth}_B(\Check{A}^{\orth}) \\
=  \lim_{n \to \infty}   \int_{\Check{\cA}^{\orth}}
 \Phi_n((\ll \phi^{\eps}_i, \Check{A}^{\orth}\gg)_{i \le d_n})
d\mu^{\orth}_B(\Check{A}^{\orth})
\overset{(*)}{=}  \lim_{n \to \infty} \Phi_n(\ll \phi^{\eps}_i, 0\gg)_{i \le d_n})
=   \Tr_{\rho_1}\bigl( \Hol^{\eps}_{R_1}(A^{\orth}_c, B))
\end{multline}
where in step $(*)$ we have applied the aforementioned
analogue\footnote{In fact, if we include one  more step where we use the pushforward
measure $\pi_*(d\mu^{\orth}_B)$ where
$\pi: \Check{\cA}^{\orth} \to V$ is the orthogonal projection onto the finite-dimensional
subspace $V$ of $\Check{\cA}^{\orth}$ which is spanned by $(\phi^{\eps}_i)_{i \le d_n}$
then we do not have to use an analogue of Eq. \eqref{eq_toy_model2} but can use
Eq. \eqref{eq_toy_model2} itself.}  of Eq. \eqref{eq_toy_model2}.

\begin{remark} \label{rm_Ha2_Anwendung1}
In the rigorous continuum framework
 (F2) described in Sec. \ref{subsec4.2} below
there is indeed a rigorous realization of the aforementioned limit argument
(similar to the argument given in  Proposition 6 in \cite{Ha2}). \par
In the simplicial framework (F1) described in Sec. \ref{subsec4.1}
the  limit argument can be avoided altogether and we can work with the original argument in Example \ref{ex2}.
\end{remark}

\subsubsection{Evaluation of the outer integral(s) in Eq. \eqref{eq2.48_regc}}
\label{subsubsec3.3.3}

By combining Eq. \eqref{eq2.48_regc} (in the special situation under consideration in the present
section) with Eq. \eqref{eq_Step1} we arrive at
\begin{multline}  \label{eq_Step2_Beginn} Z(S^2 \times S^1,L)
 \sim \lim_{\eps \to 0} \sum_{y \in I} \int_{\cB} \biggl[ \int_{\cA^{\orth}_c} \bigl\{ 1_{\cB^{ess}_{reg}}(B) \Det_{rig}(B)
  \exp\bigl( - 2\pi i k  \langle y, B(\sigma_0) \rangle \bigr)  \\
 \times   \Tr_{\rho_1}\bigl( \Hol^{\eps}_{R_1}( A^{\orth}_c, B)\bigr)\bigr\}
 \exp(i S_{CS}(A^{\orth}_c, B)) DA^{\orth}_c \biggr] DB
\end{multline}
where we have written the double integral as an iterated integral, cf. Remark \ref{rm_2.7}
in Sec. \ref{subsubsec2.3.1} above.\par

For simplicity let us now interchange, informally,
 in Eq. \eqref{eq_Step2_Beginn} the $\eps \to 0$-limit with the sum $\sum_y \cdots $ and the two integrals. (We  could avoid this informal interchange
 but  we would then have to work a bit harder, cf. Remark \ref{rm_3.3.4} below.)
Since  $\lim_{\eps \to 0} \Hol^{\eps}_{R_1}( A^{\orth}_c, B)= \Hol_{R_1}( A^{\orth}_c, B)$
we then  obtain
\begin{multline}  \label{eq_Step2_Beginn2} Z(S^2 \times S^1,L)
 \sim  \sum_{y \in I} \int_{\cB} \biggl[ \int_{\cA^{\orth}_c} \bigl\{ 1_{\cB^{ess}_{reg}}(B) \Det_{rig}(B)
  \exp\bigl( - 2\pi i k  \langle y, B(\sigma_0) \rangle \bigr)  \\
 \times   \Tr_{\rho_1}\bigl( \Hol_{R_1}( A^{\orth}_c, B)\bigr)\bigr\}
 \exp(i S_{CS}(A^{\orth}_c, B)) DA^{\orth}_c \biggr] DB
\end{multline}

 Let $D_{s}(A^{\orth}_c,B)$, $s \in [0,1]$,
 and  $l_u$, $u \in [0,1]$, be as in Sec. \ref{subsubsec3.2.2} above for $R = R_1$.
 (In particular, we have $l_u(s) = R_1(i_{S^1}(s),u)$ for all $s,u \in [0,1]$.)
  Observe that since $D_{s}(A^{\orth}_c,B) \in \ct$ we have
\begin{multline}  \label{eq_3.3.3neu_1}
  \Hol_{R_1}( A^{\orth}_c, B) = \exp\biggl(\int_0^1 D_{s}(A^{\orth}_c,B) ds\biggr) \\
= \exp\biggl(\int_0^1 \int_0^1 \bigl[ A^{\orth}_c(l'_u(s)) \bigr] ds du \biggr) \exp\biggl(\int_0^1 \int_0^1 \bigl[ (B dt)(l'_u(s)) \bigr] ds du \biggr)
\end{multline}

 Let $u \in [0,1]$  be fixed for a while.
 As in Sec. \ref{subsubsec3.2.2}  set $(l_u)_{\Sigma} := \pi_{\Sigma} \circ l_u$
 (cf.  Convention \ref{conv_Sigma_statt_S2} above).
 Recall that $R_1$ is a strictly canonical torus ribbon knot in $M = S^2 \times S^1$
 of standard type (cf. Definition \ref{def_3.3_2} above). This implies\footnote{In order to see this observe that if $R:=R_1$
 then $R_{S^2}(s,u)$ in the notation of Definition \ref{def_3.3_2}
 coincides with what now is denoted by $(l_u)_{\Sigma}(s)$.} that
  the restriction of $(l_u)_{\Sigma}:[0,1] \to S^2$ to the subinterval $[0,1/{\mathbf p}]$
   is a Jordan loop in $\Sigma = S^2$, which we will denote by ${\frak l}_u$.
   Another implication is that $(l_u)_{\Sigma}$ is the ${\mathbf p}$-fold concatenation of the loop ${\frak l}_u$ with itself.
Using this (for each  $u \in [0,1]$) and  Stokes' Theorem
we now  obtain for every $A^{\orth}_c \in \cA^{\orth}_c \cong \cA_{\Sigma,\ct}$
\begin{multline}  \label{eq_3.3.3neu_2}
\int_0^1 \int_0^1 \bigl[ A^{\orth}_c(l'_u(s)) \bigr] ds du
 = \int_0^1 \biggl[ \int_{l_u} A^{\orth}_c \biggr] du
 = \int_0^1 \biggl[ \int_{(l_u)_{\Sigma}} A^{\orth}_c \biggr] du  \\
= \int_0^1 \biggl[ {\mathbf p} \int_{{\frak l}_u} A^{\orth}_c \biggr] du
=  \int_0^1 \biggl[ {\mathbf p} \int_{\partial {\cal X}_u} A^{\orth}_c \biggr] du
=   \int_0^1 \biggl[ {\mathbf p} \int_{{\cal X}_u}   dA^{\orth}_c  \biggr] du
=  {\mathbf p} \int_{S^2}   dA^{\orth}_c    f_1
\end{multline}
 where  ${\cal X}_u$ is the connected component of $S^2 \backslash \arc({\frak l}_u)$ chosen such that the orientation of $\arc({\frak l}_u) = \partial {\cal X}_u$
induced by ${\cal X}_u$ is the same as the orientation induced by ${\frak l}_u$
 and where $f_1= f_{R_1}$ is the function $S^2 \to \bR$  given by
\begin{equation}
f_1(\sigma) = \int_0^1 1_{{\cal X}_u}(\sigma) du \quad \text{ for all $\sigma \in S^2$,}
\end{equation}
$1_{{\cal X}_u}$ being the indicator function of ${\cal X}_u$.
Next observe that
\begin{equation} \label{eq_3.36}
  \Tr_{\rho_1}(\exp(b)) = \sum_{\alpha \in \Lambda} m_{\lambda_1}(\alpha) e^{2 \pi i \langle \alpha, b \rangle}
  \quad \quad \forall b \in \ct
 \end{equation}
 where we use the notation of Appendix \ref{appB}  below.
 In particular,  $m_{\lambda_1}(\alpha)$  is the   multiplicity   of $\alpha \in \Lambda$
as a weight of $\rho_{1}$.  (Recall that $\lambda_1 \in \Lambda_+$ is the highest
weight of $\rho_1$.)  \par

 From Eq.  \eqref{eq_3.3.3neu_1}, Eq.  \eqref{eq_3.3.3neu_2},  Eq. \eqref{eq_3.36},  Remark \ref{rm_2.7} in Sec. \ref{subsubsec2.3.2} above, and the relation
 $\alpha\bigl(\int_{S^2}   dA^{\orth}_c    f_1\bigr) =
 \alpha\bigl(\int_{S^2}   A^{\orth}_c   \wedge df_1\bigr) = \ll A^{\orth}_c, \alpha \star df_1 \gg_{\cA^{\orth}} $  we therefore obtain
\begin{multline}    \Tr_{\rho_1}\bigl( \Hol_{R_1}( A^{\orth}_c, B)\bigr)
 \exp(i S_{CS}(A^{\orth}_c, B)) \\
   = \sum_{\alpha \in \Lambda} m_{\lambda_1}(\alpha)
  \biggl[ \exp\biggl(2 \pi i \alpha\biggl(\int_0^1 \int_0^1 \bigl[ (B dt)(l'_u(s)) \bigr] ds du \biggr)\biggr)  \\
  \times  \exp\biggl(2 \pi i  \ll  A^{\orth}_c, - k \star dB +
 \alpha {\mathbf p} (\star d  f_1)
  \gg_{\cA^{\orth}} \biggr) \biggr]
\end{multline}
Informally, we have
 \begin{multline} \label{eq_sec3.3.3_delta}
 \int \exp\bigl(2 \pi  i \ll  A^{\orth}_c, - k \star dB +   \alpha {\mathbf p} (\star d  f_1) \gg_{\cA^{\orth}} \bigr) DA^{\orth}_c \\
\sim \delta\bigl(  \star d( - k B +   \alpha {\mathbf p}   f_1) \bigr)
\sim \delta\bigl(  d(  B -  \tfrac{1}{k}  \alpha {\mathbf p}   f_1) \bigr)
\end{multline}
 where $\delta(\cdot)$ is the (informal) delta function on $\cA^{\orth}_c$.
From  Eq. \eqref{eq_Step2_Beginn2} and the last two equations\footnote{Recall that according to Remark \ref{rm_eigentlich_average} in Sec. \ref{subsec2.2}  above   the integral $\int \cdots DB$ on the RHS of Eq. \eqref{eq2.24} (and consequently also  on the RHS of Eq. \eqref{eq_Step2_Beginn})  should be interpreted as a suitable
 improper integral $\int^{\sim} \cdots DB$,  cf. Remark \ref{rm_sec2.2.1} above.
     This is why in Eq. \eqref{eq_Step2b_Ende} we use the   improper integral $\int^{\sim}_{\ct} \cdots db$ (cf.  Proposition \ref{prop2.1} in Sec. \ref{subsubsec2.2.1}).} we obtain
 \begin{multline} \label{eq_Step2b_Ende}  Z(S^2 \times S^1,L)
 \sim  \sum_{y \in I}    \sum_{\alpha \in \Lambda} m_{\lambda_1}(\alpha)
  \int^{\sim}_{\ct} db     \biggl[  1_{\cB^{ess}_{reg}}(B) \Det_{rig}(B)
  \exp\bigl( - 2\pi i k  \langle y, B(\sigma_0) \rangle \bigr) \\
  \times \exp\biggl(2 \pi i \alpha\biggl(\int_0^1 \int_0^1 \bigl[ (B dt)(l'_u(s)) \bigr] ds du \biggr)\biggr) \biggr]_{| B = b + \tfrac{1}{k} \alpha  {\mathbf p} f_1  }
\end{multline}

\begin{observation} \label{rm_X+_1} Let $R^1_{\Sigma}:= (R_1)_{\Sigma} := \pi_{\Sigma} \circ R_1$
(cf. Convention \ref{conv_Sigma_statt_S2}) and let
$X^{+}_1$ and $X^{-}_1$ be the two connected components
 of $S^2 \backslash \Image(R^1_{\Sigma}) = \Sigma \backslash \Image(R^1_{\Sigma})$.
 Here $X^{+}_1$ is determined by
   the condition $X^+_1 \subset {\cal X}_u$ for all $u \in [0,1]$
 where ${\cal X}_u$ is as above.
 (By contrast, $X^-_1 \cap {\cal X}_u = \emptyset$
  for every $u \in [0,1]$.)  Observe that we have $X^+_1 = {\cal X}_u$
either for $u=1$ (``Case I'') or for $u=0$ (``Case II'').
From the definitions it follows that  $f_1$ is given  explicitly by
\begin{subequations}  \label{eq_def_f1}
 \begin{equation} f_1(\sigma) = 1 \quad \text{ for }  \sigma \in X^+_1, \quad \quad f_1(\sigma) = 0 \quad \text{ for } \sigma \in X^-_1,
 \end{equation}
\begin{equation} \label{eq_def_f1_b}
f_1(\sigma)=
\begin{cases}
u_{\sigma} & \text{ in Case I } \\
1- u_{\sigma} & \text{ in Case II }
\end{cases} \quad \text{ for } \sigma \in \Image(R^1_{\Sigma})
\end{equation}
\end{subequations}
where  $u_{\sigma}$ is the unique $u \in [0,1]$ such that
$\sigma \in \arc({\frak l}_u) = \arc((l_u)_{\Sigma})$.
\end{observation}

Recall that above we
 interchanged   the $\eps \to 0$-limit in Eq. \eqref{eq_Step2_Beginn} with the sum $\sum_y \cdots $ and the two integrals.
This informal argument simplified the calculations above but it has an obvious disadvantage. The function $f_1$ is only piecewise smooth
a 1-form of the form $\alpha \star d f_1$ will in general not
be an element of $\cA_{\Sigma,\ct} \cong \cA^{\orth}_c$
and a map $B$ of the form $B  = b + \tfrac{1}{k}     \alpha {\mathbf p} f_1 $
will in general not be an element of $\cB$.
So strictly speaking,  the expressions $1_{\cB^{ess}_{reg}}(B)$ and
  $\Det_{rig}(B)$ for $B= b + \tfrac{1}{k}     \alpha {\mathbf p} f_1 $
  appearing on the RHS of Eq. \eqref{eq_Step2b_Ende} above are  not defined unless $\alpha \neq 0$ even though there are obvious candidates for such definitions, namely
  \begin{subequations}   \label{eq_subeq_sec3.3_0}
\begin{equation} \label{eq_subeq_sec3.3_0_a}
\Det_{rig}(B) =  \prod_{i=1}^2  \det\nolimits^{1/2}\bigl(1_{\ck}-\exp(\ad(B(Y_i)))_{|\ck}\bigr)^{\chi(Y_i)}
\end{equation}
and
\begin{equation}\label{eq_subeq_sec3.3_0_b} 1_{\cB_{reg}^{ess}}(B) =  \prod_{i=1}^2 1_{\ct_{reg}}(B(Y_i))
\end{equation}
\end{subequations}
where we  write $Y_1$ instead of $X^+_1$ and $Y_2$ instead of $X^-_1$
and where  $B(Y_i)$, $i=1,2$, is the (constant) value of
 $B = b + \tfrac{1}{k}     \alpha {\mathbf p} f_1 $ on $Y_i$. \par

 In the following remark we will sketch a more careful (informal) derivation of Eq. \eqref{eq_Step2b_Ende}  above which
 avoids the  aforementioned informal interchange of limit procedures
  and  where the ``candidates'' for $\Det_{reg}(B)$
and  $1_{\cB^{ess}_{reg}}(B)$, given by Eqs. \eqref{eq_subeq_sec3.3_0} above  appear automatically.

\begin{remark} \label{rm_3.3.4}
Here is an alternative derivation of Eq. \eqref{eq_Step2b_Ende}  above
which is preferable to the derivation above from a conceptual point of view  and which is relevant  for the rigorous
 implementation of the informal calculations in Sec. \ref{subsec3.3}
  within the rigorous continuum framework (F2) introduced in Sec. \ref{subsec4.2}   below. \par

Observe   that since $D^{\eps}_{s}(A^{\orth}_c,B) = D^{\eps}_{R_1,s}(A^{\orth}_c,B) \in \ct$
   (cf. Eq. \eqref{eq_Deps_def}  in Sec. \ref{subsubsec3.2.2} above) we obtain
(with $l_u$, $u \in [0,1]$, as above)
\begin{multline} \label{eq_3.35} \Hol^{\eps}_{R_1}( A^{\orth}_c, B) = \exp\biggl(\int_0^1 D^{\eps}_{s}(A^{\orth}_c,B) ds\biggr) \\
   = \exp\biggl( \sum_{a=1}^{\dim(\ct)}  T_a \ll  A^{\orth}_c, T_a j_{R_1}^{\eps} \gg_{\cA^{\orth}} \biggr) \exp\biggl(\int_0^1 \int_0^1 \bigl[ (B dt)(l'_u(s)) \bigr] ds du \biggr)
\end{multline}
where we have assumed, for convenience, that
the ortho-normal basis $(T_a)_{a \le \dim(\cG)}$ of $\cG$ fixed
in Sec. \ref{subsubsec3.2.2} above was chosen such that $T_a \in \ct$ for $a \le \dim(\ct)$
 and where we have set (cf. Convention \ref{conv_Sigma_statt_S2} above)
\begin{equation} \label{eq_def_j_R_1_eps}
j_{R_1}^{\eps}:=\int_0^1 \int_0^1  {\mathbf X}_{(l_u)'_{\Sigma}(s)} \delta^{\eps}_{(l_u)_{\Sigma}(s)} ds du \in \cA_{\Sigma,\ct} \cong \cA^{\orth}_c
 \end{equation}
 From Eq. \eqref{eq_3.35},  Eq. \eqref{eq_3.36}, and Remark \ref{rm_2.7} in Sec. \ref{subsubsec2.3.2} above
 and  $\sum_{a = 1}^{\dim(\ct)}  \langle \alpha, T_a \rangle  T_a =  \alpha$
  we therefore obtain
\begin{multline}  \label{eq_rm_sec3.3.3_1}  \Tr_{\rho_1}\bigl( \Hol^{\eps}_{R_1}( A^{\orth}_c, B)\bigr)
 \exp(i S_{CS}(A^{\orth}_c, B)) \\
   = \sum_{\alpha \in \Lambda} m_{\lambda_1}(\alpha)
  \biggl[ \exp\biggl(2 \pi i \alpha\biggl(\int_0^1 \int_0^1 \bigl[ (B dt)(l'_u(s)) \bigr] ds du \biggr)\biggr)  \\
  \times      \exp\biggl(2 \pi i  \ll  A^{\orth}_c, - k \star dB +  \alpha j_{R_1}^{\eps}
  \gg_{\cA^{\orth}} \biggr) \biggr]
\end{multline}
Moreover,  we have, informally
 \begin{equation}  \label{eq_rm_sec3.3.3_2}
 \int \exp\bigl(2 \pi  i \ll  A^{\orth}_c, - k \star dB +  \alpha j_{R_1}^{\eps} \gg_{\cA^{\orth}} \bigr) DA^{\orth}_c
\sim \delta\bigl( - \star dB +  \tfrac{1}{k} \alpha j_{R_1}^{\eps} \bigr)
\end{equation}
So far we have been working with a rather general choice
of Dirac families $\{(\delta^{\eps}_{\sigma})_{\eps < \eps_0} \mid \sigma \in \Sigma\}$,
cf. Sec. \ref{subsubsec3.2.2} above.
For the next argument it will be convenient to work with a restricted (but canonical) choice.
More precisely, we will assume  that  the family (of Dirac families)
  $((\delta^{\eps}_{\sigma})_{\eps < \eps_0})_{\sigma \in \Sigma}$
    is  ``translation invariant'' on $\Image(((R^{(s_0)}_1)_{\Sigma})_{|S^1 \times (0,1)})$
    when the latter set is embedded into $S^1 \times S^1$ in a suitable way\footnote{More
    precisely: in view of Condition \ref{ass1}
    in Sec. \ref{subsubsec3.2.4} we have
    $\Image(((R^{(s_0)}_1)_{\Sigma})_{|S^1 \times (0,1)})
    \subset  \Image((\bar{R}^1_{\Sigma})_{|S^1 \times (0,1)})  \cong S^1 \times (0,1) \cong S^1 \times (S^1 \backslash \{1\}) \subset S^1 \times S^1$. Here we have equipped each space
    with the ``obvious'' Riemannian metric.
    In particular,  $S^1 \times S^1$ is equipped with the product of ${\mathbf g}_{S^1}$
    with itself (cf. Sec. \ref{subsubsec3.2.4}). Since ${\mathbf g}_{S^1}$ is translation-invariant we also have a natural notion of translation invariance
    for families of Dirac families on $S^1 \times S^1$, which gives rise to a similar notion
    of translation invariance  for families of Dirac families
    on the Riemannian submanifold $\Image(((R^{(s_0)}_1)_{\Sigma})_{|S^1 \times (0,1)})$ of
    $S^1 \times S^1$.}. (This can always be achieved provided that  $ \eps_0 > 0$ in Sec. \ref{subsubsec3.2.2} was chosen  sufficiently small.)   One can show that then
\begin{equation} \label{eq_Image_star_d} j_{R_1}^{\eps} \in \Image(\star d)
\end{equation}
and
\begin{equation} \label{eq_lem_3.3}
 \lim_{\eps \to 0}  \bigl((\star d)^{-1} j_{R_1}^{\eps}\bigr)(\sigma)
= {\mathbf p} f_1(\sigma) + C \quad \text{uniformly in $\sigma$}
\end{equation}
 where  $C \in \bR$ is a constant (depending only on $R_1$) and $\star d$ is the restriction of  $\cB \ni B \mapsto \star d B \in \cA_{\Sigma,\ct}$
 onto the orthogonal complement $\cB'$ of $\cB_{c} \cong \ct$ (w.r.t. to the scalar product on $\cB$ induced by the Riemannian metric ${\mathbf g}_{\Sigma}$). \par

 By combining Eq. \eqref{eq_Step2_Beginn} with Eq.  \eqref{eq_rm_sec3.3.3_1}, Eq. \eqref{eq_rm_sec3.3.3_2}, and Eq. \eqref{eq_lem_3.3}
we arrive  at\footnote{\label{ft_f1+C}More precisely: we arrive at the
 modification of Eq. \eqref{eq_Step2b_Ende} above where
  $[ \cdots ]_{| B  = b + \tfrac{1}{k}     \alpha {\mathbf p} f_1 }$  is  replaced by $[ \cdots ]_{| B  = b + \tfrac{1}{k}     \alpha {\mathbf p} (f_1 + C)}$.
This constant $C$ can  be eliminated by performing, for each fixed $\alpha$,
the  change of variable $b \to b -  \tfrac{1}{k} \alpha C$ and by taking into account that
each of the four functions  appearing in Eqs. \eqref{eq5.51}  in  Sec. \ref{subsubsec3.3.5} below  is $I$-periodic.} Eq. \eqref{eq_Step2b_Ende} above
with $\Det_{reg}(B)$ and  $1_{\cB^{ess}_{reg}}(B)$  given by Eqs. \eqref{eq_subeq_sec3.3_0} above.
\end{remark}

\subsubsection{Some simplifications}
\label{subsubsec3.3.5}

Let   $B: S^2 \to \ct$ be of the form
$ B= b + b_1 f_1$  for some $b, b_1 \in \ct$
with $f_1$ as in Sec. \ref{subsubsec3.3.3} above.
From  Eq. \eqref{eq_def_f1_b} above it follows that either
$f_1((l_u)_{\Sigma}(s)) = u$ or $f_1((l_u)_{\Sigma}(s)) = 1-u$.
In particular,  $B((l_u)_{\Sigma}(s))$ is  independent of $s$.
  Observe also that
$\int_0^1 dt(l'_u(s)) ds = {\mathbf q}$ for every $u \in [0,1]$.
Taking this into account we obtain
\begin{multline} \label{eq_subeq_sec3.3}\int_0^1 \int_0^1 \bigl[ (B dt)(l'_u(s)) \bigr] ds du
= \int_0^1 \int_0^1 \bigl[B((l_u)_{\Sigma}(s))\cdot dt(l'_u(s)) \bigr] ds du \\
 =  {\mathbf q}  \int_0^1  B((l_u)_{\Sigma}(0)) du  =  {\mathbf q} (b + b_1/2)
 = {\mathbf q} \tfrac{B(Y_1) + B(Y_2)}{2}
  \end{multline}
where as in in Sec. \ref{subsubsec3.3.3} above we
have set $Y_1:= X^+_1$ and $Y_2 := X^-_1$ and where we
 write $B(Y_i)$, $i=1,2$, for the unique value of $B_{|Y_i}$. \par

Let us now assume without loss of generality\footnote{It is not difficult to see that
if $\sigma_0 \in Y_1$ we will get the same explicit
expression for $Z(S^2 \times S^1,L)$ as if $\sigma_0 \in Y_2$.}
that
 \begin{equation} \label{eq_conv1}  \sigma_0 \in Y_2
\end{equation}
Then $f_1(\sigma_0) = 0$ and $B(\sigma_0) = b$.
Combining  Eq. \eqref{eq_subeq_sec3.3} with  Eq. \eqref{eq_Step2b_Ende}
and Eqs \eqref{eq_subeq_sec3.3_0} in Sec. \ref{subsubsec3.3.3} above
 we arrive at
\begin{equation} \label{eq3.47}
 Z(S^2 \times S^1,L)  \sim  \sum_{\alpha \in \Lambda}
   m_{\lambda_1}(\alpha)    \sum_{y \in I}
  \int^{\sim}_{\ct}  e^{ - 2\pi  i k \langle y,  b \rangle}    F_{\alpha}(b) db
\end{equation}
where for $b \in \ct$ and $\alpha \in \Lambda$ we have set
 \begin{multline} \label{eq5.49}
 F_{\alpha}(b) :=  \biggl[ \bigl(  \prod_{i=1}^2 1_{\ct_{reg}}(B(Y_i)) \bigr) \bigl(  \exp(   \pi i {\mathbf q}   \langle \alpha,
 B(Y_1) + B(Y_2)  \rangle )   \\
 \times \prod_{i=1}^2  \det\nolimits^{1/2}\bigl(1_{\ck}-\exp(\ad(B(Y_i)))_{|\ck}\bigr)^{\chi(Y_i)}  \bigr) \biggr]_{| B  = b + \tfrac{1}{k}     \alpha {\mathbf p} f_1 }
\end{multline}
Now observe that the function  $F_{\alpha}(b)$ (and the whole integrand in Eq. \eqref{eq3.47} above) is $I$-invariant.
This follows from
\begin{subequations}  \label{eq5.51}
\begin{align}
e^{ 2\pi  i \eps \langle \alpha,  b + x \rangle} & = e^{ 2\pi  i \eps \langle \alpha,  b \rangle}
\quad \text{  for all $\alpha \in \Lambda$, $\eps \in \bZ$} \\
\det\nolimits^{1/2}(1_{\ck} - \exp(\ad(b + x))_{| \ck})  & =  \det\nolimits^{1/2}(1_{\ck} - \exp(\ad(b))_{| \ck})\\
1_{\ct_{reg}}(b+x) & = 1_{\ct_{reg}}(b) \\
 \exp\bigl( - 2\pi i k  \langle y, b + x \rangle \bigr) & =  \exp\bigl( - 2\pi i k  \langle y, b \rangle \bigr),
\end{align}
\end{subequations}
for all $b \in \ct$ and $x \in I$.
The first  equation  follows because  by definition $\Lambda$ is the lattice dual to $\Gamma = I$.  The second  equation follows\footnote{Since  $\det^{1/2}\nolimits(1_{\ck} - \exp(\ad(b))_{| \ck})$ is a square root of $\det(1_{\ck} - \exp(\ad(b))_{| \ck})$  and since   $\exp(\ad(b))=\Ad(\exp(b))$
 it is  immediately clear that
 $ \det^{1/2}\nolimits(1_{\ck} - \exp(\ad(b + x))_{| \ck})   = \pm  \det^{1/2}\nolimits(1_{\ck} - \exp(\ad(b))_{| \ck})$. The  argument above  is necessary in order to show that the sign appearing on the RHS of the last equation
 is ``$+$'' rather than ``$-$''.}
  from  Eq. \eqref{eq_appB_det1/2} in Appendix \ref{appB} below
 by taking into account that $(-1)^{ \sum_{\alpha \in \cR_+} \langle \alpha, x \rangle}
= (-1)^{ 2  \langle \rho, x \rangle} = 1$ for $x \in  \Gamma = I$
because  $\rho = \tfrac{1}{2} \sum_{\alpha \in \cR_+}$
is an element of the weight lattice $\Lambda$. The last two equations follow from
Remark \ref{rm_affineWeyl} in Appendix \ref{appB}.\par

Since the integrand in Eq. \eqref{eq3.47} is $I$-periodic we obtain
\begin{equation} \label{eq3.47b} Z(S^2 \times S^1,L)  \sim \sum_{\alpha \in \Lambda}
   m_{\lambda_1}(\alpha)    \sum_{y \in I}
  \int_{Q}  e^{ - 2\pi  i k \langle y,  b \rangle}    F_{\alpha}(b) db
\end{equation}
 where we have set
\begin{equation}  \label{eq5.54}
Q:= \{ \sum_i x_i e_i \mid x_i \in (0,1) \text{ for all $i \le m$}   \} \subset \ct,
\end{equation}
Here  $(e_i)_{i \le m}$ is an (arbitrary) fixed basis of $I$. \par

 By applying the Poisson summation formula at an informal level\footnote{In a rigorous treatment of this argument
  one needs to regularize the indicator functions $1_{\ct_{reg}}$ appearing
 on the RHS of Eq. \eqref{eq5.49} above, cf. Secs 3.7 and 5.4 in \cite{Ha9}.}
to the RHS of Eq. \eqref{eq3.47b} we obtain
\begin{align}  \label{eq5.63}
 Z(S^2 \times S^1,L) & \sim \sum_{\alpha \in \Lambda}
 m_{\lambda_1}(\alpha) \sum_{b \in \tfrac{1}{k} \Lambda}  1_Q(b)  F_{\alpha}(b) \nonumber \\
 & \overset{(*)}{\sim}   \sum_{\alpha_0,\alpha_1 \in \Lambda} m_{\lambda_1}(\alpha_1) 1_{k Q}(\alpha_0)
     F_{\alpha_1}(\tfrac{1}{k} \alpha_0)
\end{align}
 where in step $(*)$ we performed the change of variable
$b \to \alpha_0 := k b $  and wrote $\alpha_1$ instead of $\alpha$.
From Eq. \eqref{eq5.63} and Eq. \eqref{eq5.49} we obtain
(by taking into account that $\chi(Y_1) = \chi(Y_2) = 1$)
\begin{align}  \label{eq5.89} Z(S^2 \times S^1,L)  &  \sim
 \sum_{\alpha_0, \alpha_1  \in \Lambda} 1_{k Q}(\alpha_0)
  m_{\lambda_1}(\alpha_1)  \nonumber \\
&  \quad \quad \quad \times \biggl[ \biggl(  \prod_{i=1}^2 1_{\ct_{reg}}(B(Y_i)) \biggr)
\biggl( \prod_{i=1}^2 \det\nolimits^{1/2}\bigl(1_{\ck}-\exp(\ad(B(Y_i)))_{|\ck}\bigr)
  \biggr) \nonumber \\
&  \quad \quad \quad \times    \exp(   \pi i {\mathbf q}   \langle \alpha_1,
 B(Y_1) + B(Y_2)  \rangle )  \biggr]_{|
B  =  \tfrac{1}{k} ( \alpha_0 +    \alpha_1 {\mathbf p} f_1)}
\end{align}

\subsubsection{Rewriting Eq. \eqref{eq5.89} in quantum algebraic notation}
\label{subsubsec3.3.6}

As a preparation for  Sec. \ref{subsubsec3.3.8} below
we will now rewrite Eq. \eqref{eq5.89} using the quantum algebraic notation of Appendix \ref{appB.2}.

\smallskip

Recall that $\lambda_1 \in \Lambda_+$ is the highest
weight of $\rho_1$.
In the  following we will assume, for simplicity, that  $\lambda_1 \in \Lambda_+^k$ where $\Lambda_+^k$ is as in Appendix \ref{appB.2}.
For each $\alpha_0, \alpha_1 \in \Lambda$ we set
$$ B_{(\alpha_0,\alpha_1)} := \tfrac{1}{k}(\alpha_0 + \alpha_1 {\mathbf p} f_1)$$
we define
$\eta_{(\alpha_0,\alpha_1)}: \{Y_1, Y_2\} \to \Lambda$  by
\begin{equation}\eta_{(\alpha_0,\alpha_1)}(Y_i)= k B_{(\alpha_0,\alpha_1)}(Y_i) - \rho \quad \quad i =1,2
\end{equation}
Observe that for $\eta=\eta_{(\alpha_0,\alpha_1)}$ and $B= B_{(\alpha_0,\alpha_1)}$ we have, using the notation of Appendix \ref{appB.2} below  (and recalling Eq. \eqref{eq_conv1} in
Remark \ref{rm_3.3.4} above)
 \begin{subequations}
 \begin{equation}
 \alpha_0 = k B(Y_1) = \eta(Y_1)+\rho,
 \end{equation}
 \begin{equation}
 \alpha_1 = \tfrac{1}{\mathbf p} (k B(Y_1)- k B(Y_2)) = \tfrac{1}{\mathbf p} (\eta(Y_1)-\eta(Y_2))
 \end{equation}
\begin{align} \label{eq_prep_diagonal_argument}
 & \exp( \pi i {\mathbf q}   \langle \alpha_1,B(Y_1) + B(Y_2)\rangle \nonumber \\
 & \quad = \exp( \pi i {\mathbf q}   \langle  \tfrac{1}{\mathbf p} (\eta(Y_1)-\eta(Y_2)),\tfrac{1}{k}(\eta(Y_1)+\eta(Y_2) + 2 \rho)\rangle  = \theta_{\eta(Y_1)}^{\tfrac{\mathbf q}{\mathbf p}} \theta_{\eta(Y_2)}^{-\tfrac{\mathbf q}{\mathbf p}}
 \end{align}
 \begin{equation}
 \det\nolimits^{1/2}\bigl(1_{\ck}-\exp(\ad(B(Y_i)))_{|\ck}\bigr)) =  \det\nolimits^{1/2}\bigl(1_{\ck}-\exp(\ad(\tfrac{1}{k}(\eta(Y_i)+\rho))_{|\ck}\bigr)\bigr) \sim d_{\eta(Y_i)},
 \end{equation}
\end{subequations}
 In view of  the previous equations
 it is clear that we can rewrite Eq. \eqref{eq5.89} in the following form
\begin{align}  \label{eq5.89c}
Z(S^2 \times S^1,L) &   \sim
 \sum_{\alpha_0, \alpha_1  \in \Lambda} \biggl[
  m_{\lambda_1}\bigl(\tfrac{1}{\mathbf p} (\eta(Y_1)-\eta(Y_2))\bigr)
 1_{k Q}(\eta(Y_1)+\rho) \bigl( \prod_{i=1}^2  1_{\ct_{reg}}(\tfrac{1}{k}(\eta(Y_i)+\rho)\bigr) \bigr) \nonumber \\
 & \quad \quad \quad \times  d_{\eta(Y_1)} d_{\eta(Y_2)}      \theta_{\eta(Y_1)}^{\tfrac{\mathbf q}{\mathbf p}} \theta_{\eta(Y_2)}^{-\tfrac{\mathbf q}{\mathbf p}}  \biggr]_{|\eta  =  \eta_{( \alpha_0,\alpha_1)}}\nonumber \\
 & \overset{(*)}{\sim}
 \sum_{\eta_1, \eta_2  \in \Lambda } \biggl[
  m_{\lambda_1}\bigl(\tfrac{1}{\mathbf p} (\eta_1-\eta_2)\bigr)
1_{k (Q \cap \ct_{reg})}(\eta_1 + \rho)  1_{k \ct_{reg}}(\eta_2 + \rho)
d_{\eta_1} d_{\eta_2}      \theta_{\eta_1}^{\tfrac{\mathbf q}{\mathbf p}} \theta_{\eta_2}^{-\tfrac{\mathbf q}{\mathbf p}}  \biggr] \nonumber \\
& \sim  \sum_{\eta_1 \in (k (Q \cap \ct_{reg}) - \rho) \cap \Lambda,  \eta_2 \in
(k \ct_{reg} - \rho) \cap \Lambda} \biggl[
  m_{\lambda_1}\bigl(\tfrac{1}{\mathbf p} (\eta_1-\eta_2)\bigr)
d_{\eta_1} d_{\eta_2}      \theta_{\eta_1}^{\tfrac{\mathbf q}{\mathbf p}} \theta_{\eta_2}^{-\tfrac{\mathbf q}{\mathbf p}}  \biggr]
\end{align}
where in step $(*)$ we made the  change of variable
$(\alpha_0,\alpha_1) \to (\eta_1,\eta_2) := (\eta_{( \alpha_0,\alpha_1)}(Y_1), \eta_{( \alpha_0,\alpha_1)}(Y_2))$.
Let $P$ be the fundamental Weyl alcove (w.r.t. to the Weyl chamber $\CW$ fixed above),
cf. Eq. \eqref{eq_P_formula} in Appendix \ref{appB}.
Observe that the map
\begin{subequations}
\begin{equation}\cW_{\aff} \times P \ni (\tau,b) \mapsto \tau \cdot b \in \ct_{reg}
\end{equation}
is a well-defined bijection, cf. part (ii) of Remark \ref{rm_affineWeyl} in Appendix \ref{appB} below.
Moreover, there is a finite subset $W$ of $\cW_{\aff}$ such that
\begin{equation}W \times P \ni (\tau,b) \mapsto \tau \cdot b \in Q \cap \ct_{reg}
\end{equation}
\end{subequations}
is a bijection, too.
Clearly, these two bijections above  induce two other bijections
\begin{subequations} \label{eq_ast_bij}
\begin{equation}\cW_{\aff} \times (k P - \rho) \ni (\tau,b) \mapsto \tau \ast b \in k \ct_{reg} - \rho
\end{equation}
\begin{equation}W \times (k P - \rho) \ni (\tau,b) \mapsto \tau \ast b \in k (Q \cap \ct_{reg}) - \rho
\end{equation}
\end{subequations}
where $\ast: \cW_{\aff} \times \ct \to \ct$ is given by
\begin{equation} \label{eq_def_ast}
\tau \ast b = k \bigl( \tau \cdot \tfrac{1}{k} (b+\rho)\bigr) - \rho, \quad \quad
\text{for all $\tau \in \cW_{\aff}$ and $b \in \ct$}
\end{equation}

Observe that for all $\tau \in \cW_{\aff}$, $\eta,\eta_1,\eta_2 \in \Lambda$   we have (cf. Remark \ref{rm_Step6} below)
\begin{subequations} \label{eq_d+th_inv}
\begin{align}
\label{eq_d_inv}
 d_{\tau \ast \eta} & = (-1)^{\tau} d_{\eta},
 \end{align}
 \begin{equation} \label{eq_diagonal_inv}
  m_{\lambda_1}\bigl(\tfrac{1}{\mathbf p} (\tau \ast \eta_1- \tau \ast \eta_2)\bigr)
      \theta_{\tau \ast \eta_1}^{\tfrac{\mathbf q}{\mathbf p}} \theta_{\tau \ast \eta_2}^{-\tfrac{\mathbf q}{\mathbf p}} =
  m_{\lambda_1}\bigl(\tfrac{1}{\mathbf p} (\eta_1-  \eta_2)\bigr)
      \theta_{\eta_1}^{\tfrac{\mathbf q}{\mathbf p}} \theta_{\eta_2}^{-\tfrac{\mathbf q}{\mathbf p}}
 \end{equation}
 \end{subequations}

 \begin{remark}  \label{rm_Step6}
 (i)  In fact, we also have $\theta_{\tau \ast \eta} = \theta_{\eta}$
and in the special case where $\tau \in \cW$ we even have
$\theta_{\tau \ast \eta}^{\tfrac{\mathbf q}{\mathbf p}} = \theta_{\eta}^{\tfrac{\mathbf q}{\mathbf p}}$.
However, if ${\mathbf p} \neq \pm 1$ we cannot expect $\theta_{\tau \ast \eta}^{\tfrac{\mathbf q}{\mathbf p}} = \theta_{\eta}^{\tfrac{\mathbf q}{\mathbf p}}$ to hold
 for a general element $\tau$ of $\cW_{\aff}$.

\smallskip

(ii) According to Remark \ref{rm_affineWeyl} in Appendix \ref{appB} below the affine Weyl group
$\cW_{\aff}$ is generated by $\cW$ and the translations associated to the lattice
$\Gamma = I$ so it is enough to check Eq. \eqref{eq_d_inv} and Eq. \eqref{eq_diagonal_inv} for elements of $\cW$ and the aforementioned translations. If    $\tau \in \cW $  then $\tau \ast \eta = \tau \cdot \eta + \tau \cdot \rho - \rho$. On the other hand if $\tau$ is the translation by $y \in \Gamma$ we have
$\tau \ast \eta = \eta + k y$. Using this Eq. \eqref{eq_d_inv} follows from Eq. \eqref{eq_def_d}
  and Eq. \eqref{eq_def_S}
and   Eq. \eqref{eq_diagonal_inv} follows by taking into account\footnote{This is relevant only in the special
 case where $\tau$ is a translation by  $y \in \Gamma$.
 If $\tau \in \cW$
 then the validity of Eq. \eqref{eq_diagonal_inv} follows from the $\cW$-invariance of $m_{\lambda_1}(\cdot)$
 and the relations mentioned above.}  Eq. \eqref{eq_prep_diagonal_argument} above and Eq. \eqref{eq_mbar_def} in Appendix \ref{appB} below.
 \end{remark}

Combining  Eq.  \eqref{eq5.89c} with Eqs \eqref{eq_d+th_inv} we therefore obtain
\begin{align} \label{eq5.89d}
 & Z(S^2 \times S^1,L)  \nonumber \\
 &  \sim  \sum_{\eta_1, \eta_2 \in (k P - \rho) \cap \Lambda} \sum_{\tau_1 \in W, \tau_2 \in \cW_{\aff}}
\biggl[  m_{\lambda_1}\bigl(\tfrac{1}{\mathbf p} (\tau_1 \ast \eta_1- \tau_2 \ast \eta_2)\bigr)
d_{\tau_1 \ast \eta_1} d_{\tau_2 \ast \eta_2}      \theta_{\tau_1 \ast \eta_1}^{\tfrac{\mathbf q}{\mathbf p}} \theta_{\tau_2 \ast \eta_2}^{-\tfrac{\mathbf q}{\mathbf p}}  \biggr] \nonumber \\
& = \sum_{\eta_1, \eta_2 \in (k P - \rho) \cap \Lambda} \sum_{\tau_1 \in W, \tau_2 \in \cW_{\aff}}
\biggl[  m_{\lambda_1}\bigl(\tfrac{1}{\mathbf p} (\tau_1 \ast \eta_1- \tau_2 \ast \eta_2)\bigr)
(-1)^{\tau_1} (-1)^{\tau_2} d_{\eta_1} d_{\eta_2}      \theta_{\tau_1 \ast \eta_1}^{\tfrac{\mathbf q}{\mathbf p}} \theta_{\tau_2 \ast \eta_2}^{-\tfrac{\mathbf q}{\mathbf p}}  \biggr] \nonumber \\
& \overset{(*)}{=} \sum_{\eta_1, \eta_2 \in (k P - \rho) \cap \Lambda} \sum_{\tau_1 \in W, \tau \in \cW_{\aff}}
\biggl[ m_{\lambda_1}\bigl(\tfrac{1}{\mathbf p} ( \eta_1- \tau \ast \eta_2)\bigr)  (-1)^{\tau}
 d_{\eta_1} d_{\eta_2}      \theta_{\eta_1}^{\tfrac{\mathbf q}{\mathbf p}} \theta_{\tau \ast \eta_2}^{-\tfrac{\mathbf q}{\mathbf p}}  \biggr]  \nonumber \\
& \overset{(**)}{\sim}
 \sum_{\eta_1, \eta_2 \in \Lambda_+^k} \sum_{\tau \in \cW_{\aff}}
m^{\eta_1\eta_2}_{\lambda_1,\mathbf p}(\tau)
  d_{\eta_1} d_{\eta_2}      \theta_{\eta_1}^{\tfrac{\mathbf q}{\mathbf p}}  \theta_{\tau \ast \eta_2}^{-\tfrac{\mathbf q}{\mathbf p}}
\end{align}
where we have set for  $\lambda \in \Lambda_+$,  $\mu, \nu \in \Lambda$, $\mathbf p \in \bZ \backslash \{0\}$, and  $\tau \in \cW_{\aff}$
\begin{equation} \label{eq_def_plethysm}
  m^{\mu \nu }_{\lambda, \mathbf p}(\tau)  :=  (-1)^{\tau} m_{\lambda}\bigl(\tfrac{1}{\mathbf p} (\mu - \tau \ast \nu)\bigr) \in \bZ
\end{equation}
Above in step $(*)$ we applied Eq. \eqref{eq_diagonal_inv}
and made the change of  variable $\tau_2 \to \tau := \tau_1^{-1} \tau_2$
 and  in  step $(**)$ we have used Eq. \eqref{eq_rm_fund_Weyl_alcove}
 of Appendix \ref{appB}.

 \medskip

 Analogous (but considerably simpler)  computations for the empty  link $L = \emptyset$
 lead to
\begin{equation}\label{eq_Step6_fast_Ende}
Z(S^2 \times S^1) = Z(S^2 \times S^1,\emptyset) \sim \frac{1}{S^2_{00}}
\end{equation}
where the multiplicative (non-zero) constant represented by $\sim$ is the same
as that in Eq. \eqref{eq5.89d}
above. Combining Eq. \eqref{eq5.89d} and Eq. \eqref{eq_Step6_fast_Ende}   we conclude (cf. Eq. \eqref{eq_sec2.1_bracket} in Sec. \ref{subsec2.1})
\begin{equation} \label{eq_Step6_Ende}
\langle L \rangle  = \frac{Z(S^2 \times S^1,L)}{Z(S^2 \times S^1)}
= S^2_{00} \sum_{\eta_1, \eta_2 \in \Lambda_+^k} \sum_{\tau \in \cW_{\aff}} m^{\eta_1\eta_2}_{\lambda_1,\mathbf p}(\tau) \
d_{\eta_1} d_{\eta_2} \ \theta_{\eta_1}^{ \frac{{\mathbf q} }{{\mathbf p}}} \theta_{\tau \ast \eta_2}^{- \frac{{\mathbf q} }{{\mathbf p}}}
\end{equation}

\subsubsection{A useful generalization of Eq. \eqref{eq_Step6_Ende}}
\label{subsubsec3.3.7}

 As a preparation for Sec. \ref{subsubsec3.3.8} below let us now generalize Eq. \eqref{eq_Step6_Ende} above
 in a straightforward way.  Let $L=(R_1,R_2)$ be a 2-component link where $R_1$ is the torus ribbon knot above and
where $R_2:S^1 \times [0,1] \to S^2 \times S^1$ is a ``fiber ribbon'', i.e.
each of the loops  $R_2(\cdot,u): S^1 \to  S^2 \times S^1$, $u \in [0,1]$, is a fiber loop
in the sense of Sec. \ref{subsec3.1} above.  \par

Assume that $L$ is colored with $(\rho_1,\rho_2)$ and that
 $\lambda_1, \lambda_2 \in \Lambda_+^k$ where $\lambda_1, \lambda_2$ are the highest weights of
 $\rho_1$ and $\rho_2$.
By generalizing the computations  above in a straightforward way
(cf.  Sec. 5.5 in \cite{Ha9} for analogous computations
 within the simplicial setting of \cite{Ha9}) we obtain
\begin{equation} \label{eq_Step7_1} \langle L \rangle = S_{00} \sum_{\eta_1,\eta_2 \in \Lambda_+^k}
  \sum_{\tau \in \cW_{\aff}} m^{ \eta_1 \eta_2}_{\lambda_1,\mathbf p}(\tau)
d_{\eta_1} S_{\lambda_2 \eta_2}  \theta_{\eta_1}^{ \frac{{\mathbf q} }{{\mathbf p}}} \theta_{\tau \ast \eta_2}^{- \frac{{\mathbf q} }{{\mathbf p}}}
\end{equation}

\begin{remark} \label{rm_no_fiber_ribbon} Instead of using the original version of Eq. \eqref{eq_Step7_1}
we can also work with the variant of  Eq. \eqref{eq_Step7_1}
which is obtained by replacing in $L$ the ``fiber ribbon'' $R_2$ by a fiber loop $l_2:[0,1] \to S^2 \times S^1$, $l_2(s) = (\sigma_2,i_{S^1}(s))$ for all $s \in [0,1]$
(and for fixed $\sigma_2 \in S^2$).
 More precisely, we replace $L=(R_1,R_2)$ by the ``mixed'' ribbon/loop link
 $L=(R_1,l_2)$ and denote by
   $Z(S^2 \times S^1, L)$ and $\langle L \rangle$  the obvious path integrals.
Clearly, due to the ``mixing'' of ribbons and proper loops, from a conceptual point of view
 this variant of Eq. \eqref{eq_Step7_1} is   less natural
than its original version but  it has the  advantage of being somewhat
    easier to derive than  Eq. \eqref{eq_Step7_1}. Here is a sketch of this derivation: \par

Since   $ \Tr_{\rho_2}\bigl( \Hol_{l_2}(\Check{A}^{\orth} + A^{\orth}_c,   B)\bigr) = \Tr_{\rho_2}\bigl(\exp(B(\sigma_2))\bigr)$  an extra factor  $\Tr_{\rho_2}\bigl(\exp(B(\sigma_2))\bigr)$ will appear in the (obvious) modifications
 of the equations in Sec. \ref{subsubsec3.3.1}--\ref{subsubsec3.3.5},
 for example, in Eq. \eqref{eq_Step2_Beginn} and in Eq. \eqref{eq5.89}.
Let us now assume without loss of generality that the point $\sigma_2$ lies in
the region $Y_2$ (cf. the beginning of Sec. \ref{subsubsec3.3.5} above).
Then the extra factor $\Tr_{\rho_2}\bigl(\exp(B(\sigma_2))\bigr)$ in
(the modification of) Eq. \eqref{eq5.89}
   gives rise to an extra factor $\Tr_{\rho_2}\bigl(\exp(\tfrac{1}{k} (\eta_2 + \rho))\bigr) = \frac{S_{\eta_2 \lambda_2 }}{S_{ \eta_2 0}}$ in Eq. \eqref{eq5.89c},
  cf. Eq. \eqref{eq_Weyl_char} in Appendix \ref{appB}.
Since  apart from Eq. \eqref{eq_d+th_inv} we also
have $S_{(\tau \ast \eta) \lambda_2} = (-1)^{\tau} S_{ \eta \lambda_2}$
 we then obtain
from a computation completely analogous to the one in Eq. \eqref{eq5.89d} above
\begin{align}
  Z(S^2 \times S^1, L)   &  \sim
 \sum_{\eta_1, \eta_2 \in \Lambda_+^k} \sum_{\tau \in \cW_{\aff}}
  m^{ \eta_1 \eta_2}_{\lambda_1,\mathbf p}(\tau)
d_{\eta_1} d_{\eta_2} \frac{S_{\eta_2 \lambda_2 }}{S_{ \eta_2 0}} \theta_{\eta_1}^{ \frac{{\mathbf q} }{{\mathbf p}}} \theta_{\tau \ast \eta_2}^{- \frac{{\mathbf q} }{{\mathbf p}}}
\end{align}
Combining this with Eq. \eqref{eq_Step6_fast_Ende}  above
and taking into account that $d_{\eta_2} = \frac{S_{\eta_2 0}}{S_{0 0}}$ (cf. Eq. \eqref{eq_def_d} in Appendix \ref{appB}) we  arrive at Eq. \eqref{eq_Step7_1}.
\end{remark}

\subsubsection{Change of notation}
\label{subsubsec3.3.7b}

As a preparation for Sec. \ref{subsubsec3.3.8} below we will now modify our notation.
We will write
\begin{itemize}
\item $T_{{\mathbf p},{\mathbf q}}$ instead  of $R_1$,

\item $C$ instead of the fiber loop $l_2$ appearing in Remark \ref{rm_no_fiber_ribbon},

\item $\lambda$ instead of $\lambda_1$ and $\rho_{\lambda}$ instead of $\rho_1$,

\item $\alpha$ instead of $\lambda_2$ and $\rho_{\alpha}$ instead of $\rho_2$,

\item $Z(S^2 \times S^1, (T_{{\mathbf p},{\mathbf q}},\rho_{\lambda}), (C,\rho_{\alpha}))$
  instead of $Z(S^2 \times S^1, L)$,

\item   $\langle (T_{{\mathbf p},{\mathbf q}},\rho_{\lambda}), (C,\rho_{\alpha}) \rangle$ instead      of $\langle L \rangle$.
\end{itemize}

Using the new notation we can rewrite Eq. \eqref{eq_Step7_1}
(or rather, the variant of Eq. \eqref{eq_Step7_1} appearing in
Remark \ref{rm_no_fiber_ribbon}  above) as
\begin{equation} \label{eq_RT_WLO_spec} \langle (T_{{\mathbf p},{\mathbf q}},\rho_{\lambda}), (C,\rho_{\alpha}) \rangle = S_{00} \sum_{\eta_1,\eta_2 \in \Lambda_+^k}
  \sum_{\tau \in \cW_{\aff}} m^{ \eta_1 \eta_2}_{\lambda,\mathbf p}(\tau)
d_{\eta_1} S_{\alpha \eta_2}  \theta_{\eta_1}^{ \frac{{\mathbf q} }{{\mathbf p}}} \theta_{\tau \ast \eta_2}^{- \frac{{\mathbf q} }{{\mathbf p}}}
\end{equation}

\subsubsection{Derivation of the general Rosso-Jones formula for torus knots in $S^3$}
\label{subsubsec3.3.8}

Let us now combine Eq. \eqref{eq_RT_WLO_spec} above with Witten's  surgery formula
and derive the original Rosso-Jones formula (for general $G$) which is concerned with (colored)  torus knots in $S^3$. More precisely,
we will  use the following two informal arguments from  \cite{Wi}:
\begin{itemize}
\item $Z(S^2 \times S^1) = 1$, which implies
\begin{equation}  \label{eq_Z=1}
Z(S^2 \times S^1, (T_{{\mathbf p},{\mathbf q}},\rho_{\lambda}), (C,\rho_{\alpha}))
 = \langle (T_{{\mathbf p},{\mathbf q}},\rho_{\lambda}), (C,\rho_{\alpha}) \rangle
\end{equation}

\item  Witten's surgery formula\footnote{Here we use a notation which is very
similar to Witten's notation; note, however, that we write $(C,\rho_{\alpha})$ where Witten  writes $R_{\alpha}$.}:
\begin{equation} \label{eq_surgery_formula1}
Z(S^3,(\tilde{T}_{{\mathbf p},{\mathbf q}},\rho_{\lambda})) = \sum_{\alpha \in \Lambda_+^k} S_{\alpha 0} \ Z(S^2 \times S^1, (T_{{\mathbf p},{\mathbf q}},\rho_{\lambda}), (C,\rho_{\alpha}))
\end{equation}
where  $\tilde{T}_{{\mathbf p},{\mathbf q}}$ is the  torus ribbon knot in
$S^3$ which is obtained from\footnote{Note that up to equivalence and a change of framing
every  torus ribbon knot in $S^3$ can be obtained in this way.}
 $T_{{\mathbf p},{\mathbf q}}$
by performing the  surgery on $C$
which transforms $S^2 \times S^1$ into $S^3$ (cf. Fig. 16 on p. 389  in  \cite{Wi}).
\end{itemize}

 Combining  Eq. \eqref{eq_RT_WLO_spec} with Eq. \eqref{eq_Z=1} and
 Eq. \eqref{eq_surgery_formula1} (for every $\alpha \in \Lambda_+^k$) we obtain
\begin{align} \label{eq_RT_rewrite0}
 Z(S^3,(\tilde{T}_{{\mathbf p},{\mathbf q}},\rho_{\lambda}))
 & = \sum_{\alpha \in \Lambda_+^k} S_{\alpha 0} \biggl(S_{00} \sum_{\eta_1,\eta_2 \in \Lambda_+^k}
  \sum_{\tau \in \cW_{\aff}} m^{ \eta_1 \eta_2}_{\lambda,\mathbf p}(\tau)
d_{\eta_1} S_{\alpha \eta_2}  \theta_{\eta_1}^{ \frac{{\mathbf q} }{{\mathbf p}}} \theta_{\tau \ast \eta_2}^{- \frac{{\mathbf q} }{{\mathbf p}}}  \biggr) \nonumber\\
  & = S_{00} \sum_{\eta_1,\eta_2 \in \Lambda_+^k}  \sum_{\tau \in \cW_{\aff}}
   \biggl(\sum_{\alpha \in \Lambda_+^k} S_{\alpha 0}  S_{\alpha \eta_2} \biggr)
  m^{ \eta_1 \eta_2}_{\lambda,\mathbf p}(\tau)
d_{\eta_1} \theta_{\eta_1}^{ \frac{{\mathbf q} }{{\mathbf p}}} \theta_{\tau \ast \eta_2}^{- \frac{{\mathbf q} }{{\mathbf p}}}   \nonumber\\
   & \overset{(*)}{=} S_{00} \sum_{\eta_1,\eta_2 \in \Lambda_+^k}  \sum_{\tau \in \cW_{\aff}}  \bigl(C_{0 \eta_2} \bigr) m^{ \eta_1 \eta_2}_{\lambda,\mathbf p}(\tau)
d_{\eta_1} \theta_{\eta_1}^{ \frac{{\mathbf q} }{{\mathbf p}}} \theta_{\tau \ast \eta_2}^{- \frac{{\mathbf q} }{{\mathbf p}}}   \nonumber\\
  & \overset{(**)}{=} S_{00} \sum_{\eta_1\in \Lambda_+^k}  \sum_{\tau \in \cW_{\aff}}
  m^{ \eta_1 0}_{\lambda,\mathbf p}(\tau)
d_{\eta_1} \theta_{\eta_1}^{ \frac{{\mathbf q} }{{\mathbf p}}} \theta_{\tau \ast 0}^{- \frac{{\mathbf q} }{{\mathbf p}}}
\end{align}
Here  in Step $(*)$ we used $S^2 = C$  and the fact that $S$ is a symmetric matrix
and in Step $(**)$  we used $C_{0 \mu} = \delta_{\bar{0} \mu} = \delta_{0 \mu}$
(cf. Appendix \ref{appB} below). \par

For simplicity we will now assume that  $k$ is ``sufficiently large''.
(It can be shown  that  this restriction on $k$ can be dropped, i.e.
 we can actually derive Eq. \eqref{eq_my_RossoJones3}  for all $k > \cg$.)
 If $k$ is sufficiently large then
 the sum $\sum_{\tau \in \cW_{\aff}} \cdots $ appearing in the last term in Eq. \eqref{eq_RT_rewrite0}
can be replaced by $\sum_{\tau \in \cW} \cdots $.
Moreover, (for fixed $\lambda$, ${\mathbf p}$, and $\tau \in \cW$)  the coefficients  $m^{ \eta_1 0}_{\lambda,\mathbf p}(\tau)$
are non-zero only for a finite number of values of $\eta_1 \in \Lambda_+^k$.
So if $k$ is large enough we can replace the index set $\Lambda_+^k$ in Eq. \eqref{eq_RT_rewrite0} by $\Lambda_+$.
Making these two replacements, writing $\mu$ instead of $\eta_1$, and taking into account that for $\tau \in \cW$
we  have  $\theta_{\tau \ast 0}^{- \frac{{\mathbf q} }{{\mathbf p}}} =  \theta_{0}^{- \frac{{\mathbf q} }{{\mathbf p}}} = 1$ (cf. Remark \ref{rm_Step6} above) we arrive at
\begin{align} \label{eq_RT_rewrite}
Z(S^3,(\tilde{T}_{{\mathbf p},{\mathbf q}},\rho_{\lambda}))
& =  S_{00} \sum_{\mu \in \Lambda_+} \biggl( \sum_{\tau \in \cW}
 m^{ \mu 0}_{\lambda,\mathbf p}(\tau) \biggr) d_{\mu} \theta_{\mu}^{ \frac{{\mathbf q} }{{\mathbf p}}}
\end{align}
According to Eq \eqref{eq_def_plethysm} and Eq. \eqref{eq_def_ast} above
we have
\begin{equation}
\sum_{\tau \in \cW}
 m^{ \mu 0}_{\lambda,\mathbf p}(\tau) =  c^{\mu}_{\lambda,{\mathbf p}}
\end{equation}
where $c^{\mu}_{\lambda,{\mathbf p}}$ is defined by Eq. \eqref{eq_coef_Rosso_Jones} in Appendix \ref{appB} below.
 So we finally obtain
\begin{equation}  \label{eq_my_RossoJones3}
Z(S^3,(\tilde{T}_{{\mathbf p},{\mathbf q}},\rho_{\lambda}))  = S_{00}  \sum_{\mu \in \Lambda_+}
 c^{ \mu}_{\lambda,\mathbf p} d_{\mu} \theta_{\mu}^{ \frac{{\mathbf q} }{{\mathbf p}}},
\end{equation}
which is a version\footnote{By replacing $Z(S^3,(\tilde{T}_{{\mathbf p},{\mathbf q}},\rho_{\lambda}))$
with the Reshetikhin-Turaev invariant $RT(S^3,(\tilde{T}_{{\mathbf p},{\mathbf q}},\rho_{\lambda}))$
one obtains a rigorous version of Eq. \eqref{eq_my_RossoJones3}
which -- as is shown in Appendix A in \cite{Ha9} --
is equivalent to the original version  of the Rosso-Jones formula.
(Observe that the original version of the  Rosso-Jones formula deals with unframed torus knots while
in the present paper we are working with ribbon torus knots).} of the Rosso-Jones formula,
cf. \cite{RoJo} and Eq. (10) in  \cite{GaMo}.

\begin{remark} \label{rm_3.3.9_I}  In Appendix \ref{appB} below we define
the coefficients  $(c^{\mu}_{\lambda,{\mathbf p}})_{\mu, \lambda \in \Lambda_+}$ appearing in the Rosso-Jones formula by  formula \eqref{eq_coef_Rosso_Jones}.
In fact, these coefficients $(c^{\mu}_{\lambda,{\mathbf p}})_{\mu, \lambda \in \Lambda_+}$
are usually defined using a different formula,
cf. Eq. (9) in  \cite{GaMo} and Eq. (7) in Sec. 2 in  \cite{GaVu}.
So  Eq. \eqref{eq_coef_Rosso_Jones} of  Appendix \ref{appB} mentioned above
is then not a definition but an identity, which was discovered only recently in \cite{GaVu}
(cf. Lemma 2.1 in \cite{GaVu}) and rediscovered in \cite{Ha9}.
\end{remark}

To my knowledge there are three other approaches for the informal
evaluation of $Z(S^3,K)$ for colored torus knots  $K$ in $S^3$, firstly the ``knot operator'' approach by  \cite{LabRam} and then the two path integral approaches by \cite{BeaWi,Bea} and \cite{BlTh5}.
In the following two remarks we comment on these three approaches.

\begin{remark}\label{rm_3.3.9_III}
The ``knot operator'' approach by \cite{LabRam}, mentioned above,
 was applied in the series\footnote{The first five of these papers deal with the following special cases:
 1.~$G=SU(2)$ and $K$  is colored with the defining representation of $G=SU(2)$
 (cf. \cite{LabLlaRam});
 2.~$G=SU(2)$ and arbitrary knot colors (cf. \cite{IsLabRam});
 3.~$G=SU(N)$ and $K$ is colored with the defining representation of $G=SU(N)$ (cf. \cite{LabMa1});
 4.~$G=SO(N)$ and $K$ is colored with the defining representation of $G=SO(N)$ (cf.  \cite{LabPer});
 5.~$G=SU(N)$ and arbitrary knot colors (cf. \cite{LabMa2}).
The case of arbitrary simple, simply-connected, compact Lie group $G$ and  arbitrary knot colors
was then covered in \cite{Ste}.} of papers \cite{LabLlaRam,IsLabRam,LabMa1,LabMa2,LabPer,Ste}
  to the explicit   evaluation of  $Z(S^3,K)$ for colored torus knots  $K$ in $S^3$
and equivalence with the  Rosso-Jones formula was shown.
Note that the knot operator approach
 uses the  Hamiltonian formulation of Chern-Simons theory.
Accordingly, it involves  only very few genuine path integral arguments (i.e. arguments which deal directly/explicitly with the CS path integral).
\end{remark}

\begin{remark}\label{rm_3.3.9_II}  The two path integral approaches by \cite{Bea} and \cite{BlTh5} rely on the observation  by Moser (cf. Ref. ``[85]'' in \cite{Bea})
that a knot $K$ in $S^3$ is a torus knot iff
it can be  represented as a ``fiber loop''\footnote{``Seifert loop'' in the terminology of \cite{Bea}.} in the sense of Sec. \ref{sec1} above
by considering $S^3$  as Seifert fibered space in a suitable way.
 By combining this  observation\footnote{This observation plays
 a  crucial role both in \cite{Bea} and in \cite{BlTh5}.
 Accordingly, since  general knots/links cannot be represented as fiber loops/links,
 the computation of $Z(M,L)$ for general $L$  is not considered in \cite{Bea} or \cite{BlTh5}.}
with results from orbifold theory and the theory of moduli spaces,
$Z(S^3,K)$ was evaluated explicitly  in \cite{Bea} for colored torus knots  $K$ in $S^3$
  using  non-Abelian localization.
 For the special case $G =  SU(2)$ it was verified in \cite{Bea} that the explicit  expression obtained for $Z(S^3,K)$ agrees with the corresponding expression in the Rosso-Jones formula.
(To my knowledge this has not yet been verified  for general $G$
 even though it should not be too difficult to do so by
 using formula \eqref{eq_coef_Rosso_Jones} of  Appendix \ref{appB} below.)\par

 Recall from the introduction that in \cite{BlTh4,BlTh5}
torus gauge fixing was generalized first to the case where $M$ is a non-trivial $S^1$-bundle and
 later to the case where $M$ is a Seifert fibered space and this can be used
 to obtain an explicit evaluation of  $Z(M)$ and $Z(M,L)$, where $L$ is a fiber link in $M$.
 As observed above,  torus knots in $S^3$  can be  represented as  ``fiber loops''
when $S^3$ is considered  as Seifert fibered space in a suitable way.
 By doing so the approach in \cite{BlTh5} allows the explicit  evaluation of
  $Z(S^3,K)$ for colored torus knots $K$ in $S^3$ and it can be expected that this evaluation will
  lead to expressions which are equivalent to those  in \cite{Bea} (which, as mentioned above,  in the case $G=SU(2)$
 were shown to be equivalent to those in the Rosso-Jones formula).
 In contrast to the treatment in \cite{Bea} the treatment in \cite{BlTh5}
 does not require any arguments involving moduli spaces
 but some  technical results from orbifold theory  are  necessary.
 (In particular,   a version of  Hodge decomposition for orbifolds, due to Baily,
  the Gysin sequence for $S^1$-bundles over 2-dimensional orbifolds,
 and  the  Riemann-Roch-Kawasaki index theorem  for  orbifolds are used in \cite{BlTh5},
 cf. references ``[1]'', ``[11]'', and ``[13]'' in  \cite{BlTh5}.)

\smallskip

The evaluation of $Z(S^3,K)$ (with $K= (\tilde{T}_{{\mathbf p},{\mathbf q}},\rho_{\lambda})$) which we have given above  has the advantage of avoiding  both the use of moduli space arguments and orbifold theory.
On the other hand it has the disadvantage of having to rely on the use of Witten's surgery formula (which was derived in \cite{Wi} using  the Hamiltonian formulation of Chern-Simons theory).
 So in contrast to the evaluations   in \cite{Bea} and \cite{BlTh5}
our evaluation of $Z(S^3,K)$ is not a pure path integral evaluation.
It would be desirable to find a way to eliminate Witten's surgery argument
and, indeed,  by combining the ideas/methods in the present paper with those in \cite{BlTh4} this is probably possible, cf. Sec. \ref{subsec5.1} below.
\end{remark}

\subsection{Special case III. Links  in $M = \Sigma \times S^1$ without ``double points''}
\label{subsec3.4}

In Sec. \ref{subsec3.5} below we will consider the case of general (strictly admissible) colored  ribbon links $L = ((R_1, R_2, \ldots, R_m), (\rho_1, \rho_2, \ldots, \rho_m))$,
 $m \in \bN$,
 in $M = \Sigma \times S^1$,   where
$Z(\Sigma \times S^1,L)$ is expected to be given by\footnote{Recall from Sec. \ref{subsec2.4}
above that we expect $Z(\Sigma \times S^1,L) \sim RT(\Sigma \times S^1,L)$
and according to Eq. \eqref{eq_RT_vs_|L|} in  Appendix \ref{appC} we have  $RT(\Sigma \times S^1,L) \sim |L|$.}
\begin{equation} \label{eq_Z(M,L) sim_shadow}
Z(\Sigma \times S^1,L) \sim RT(\Sigma \times S^1,L) \sim |L|= \sum_{\eta \in col(L)}
|L|_1^{\eta}\,|L|_2^\eta\,|L|_3^\eta\,|L|_4^\eta
\end{equation}
with   $|L|_i^{\eta}$, $i \le 4$, $\eta \in \col(L)$,  as
 in Appendix \ref{appC} below
and with  $\col(L)$ denoting the set of maps $Y(L) \to \Lambda_+^k$ where
$$Y(L) = \{Y_0, Y_1, \ldots, Y_{m'}\}, \text{ $m' \in \bN$ },$$
 is  the set of connected components of
$ \Sigma \backslash ( \bigcup_{i=1}^m \Image(R^i_{\Sigma}))  $.
(Recall from Sec. \ref{subsubsec3.2.1} that $R^i_{\Sigma} := (R_i)_{\Sigma} := \pi_{\Sigma} \circ R_i$.)

\smallskip

As a preparation for Sec. \ref{subsec3.5} (and in particular, in order to show how major building blocks of the ``shadow invariant'' $|L|$ appear naturally)
we will now consider the following simplified situation
where  $L = ((R_1, R_2, \ldots, R_m), (\rho_1, \rho_2, \ldots, \rho_m))$,
 $m \in \bN$,  fulfills the following two conditions:
   \begin{description}
   \item[(C1)] The maps $R^i_{\Sigma}$ neither intersect each other nor themselves\footnote{I.e. these maps have pairwise disjoint
   images and each $R^i_{\Sigma}:S^1 \times [0,1] \to \Sigma$ is an embedding.}
    \item[(C2)] Each of the maps $R^i_{\Sigma}$, $i \le m$ is null-homotopic.
  \end{description}
In this special situation  Eq. \eqref{eq_Z(M,L) sim_shadow} reduces to (cf. Eqs. \eqref{eqA.5} in Appendix \ref{appC}  and  Remark \ref{rm_Sec3.4.1} below)
 \begin{multline} \label{eqA.8}
Z(\Sigma \times S^1,L) \sim |L| = \sum_{\eta \in col(L)}
|L|_1^{\eta}\,|L|_2^\eta\,|L|_3^\eta  \\
=  \sum_{\eta\in col(L)} \biggl( \prod_{i=1}^m N_{\lambda_i \eta(Y^-_{i})}^{\eta(Y^+_{i})} \biggr)
 \biggl(  \prod_{Y \in Y(L)}
(d_{\eta(Y)})^{\chi(Y)}  (\theta_{\eta(Y)})^{\gleam(Y)} \biggr)
\end{multline}
where $\lambda_i \in \Lambda_+$ is the highest weight of $\rho_i$
 and where $\gleam(Y)$ is given by Eq. \eqref{eqA.1} below.
(For simplicity, we will assume in the following that $\lambda_i \in \Lambda_+^k$.)

\medskip

 Let $L^0=(l_1,l_2, \ldots, l_m)$ be the proper link associated to $L$,
  cf. Definition \ref{def_3.2_2} above.
  Let us set $l^i_{\Sigma} := (l_i)_{\Sigma} := \pi_{\Sigma} \circ l_i$
 and $l^i_{S^1} := (l_i)_{S^1}:= \pi_{S^1} \circ l_i$.

\begin{remark} \label{rm_Sec3.4.1}
(i) In the special case where $L$ fulfills conditions (C1) and (C2) above
  the set $V(L^0)$ is empty and
the set $E_{loop}(L^0)$ coincides with $E(L^0)$, so  Eq. \eqref{eqA.4}  then
 indeed  reduces to\footnote{Observe that  $Y^{\pm}_{i}= Y^{\pm}_{e}$
  with $e = l^i_{\Sigma}$   where $Y^{\pm}_{e}$
is as in Appendix \ref{appC}.} Eq. \eqref{eqA.8} above.

\smallskip

(ii) Observe also that in the special case where $L$ fulfills conditions (C1) and (C2)
  we have $m' = m$  and  there is $Y \in Y(L) \cong Y(L^0)$ such that
\begin{subequations} \label{eq_chi_Y}
\begin{equation}\chi(Y) = 2 -2g -\# \{ j \le m \mid
\arc(l^j_{\Sigma}) \subset \partial Y\}
\end{equation}
where $g$ is the genus of $\Sigma$ while for all the other $Y \in Y(L)$
we have
\begin{equation}\chi(Y) = 2 -\# \{ j \le m \mid
\arc(l^j_{\Sigma}) \subset \partial Y\}
\end{equation}
\end{subequations}
 Moreover,  for every $Y \in Y(L) \cong Y(L^0)$   we have the explicit formula
\begin{equation} \label{eqA.1}
\gleam(Y) = \sum_{i \text{ with } \arc(l^i_{\Sigma}) \subset
\partial Y}  \wind(l^i_{S^1}) \cdot \sgn(Y;l^i_{\Sigma}) \in \bZ
\end{equation}
where $ \wind(l^i_{S^1})$ is the winding number of the loop
 $l^i_{S^1}$
 and where $ \sgn(Y;l^i_{\Sigma})$
 is given by
 \begin{equation} \label{eqA.2}
  \sgn(Y; l^{i}_{\Sigma}):=
\begin{cases} 1 & \text{ if  $Y \subset \bar{X}^+_i$ }\\
-1 & \text{ if  $Y \subset \bar{X}^-_i$ }\\
\end{cases}
\end{equation}
where $\bar{X}^{+}_i$ and $\bar{X}^{-}_i$  are the two connected components
of $\Sigma \backslash \Image(l^i_{\Sigma})$ (cf. Condition (C2) above). More precisely,
$\bar{X}^{+}_i$ is determined by the condition
 that the orientation of $\Image(l^i_{\Sigma}) = \partial \bar{X}^{+}_i$
induced by $\bar{X}^{+}_i$ coincides with the orientation induced by $l^i_{\Sigma}$.
\end{remark}

In the rest of this section we will now derive Eq. \eqref{eqA.8}
by evaluating the RHS of Eq. \eqref{eq2.48_regc}
explicitly for the special $L$ described  above.

\subsubsection{Explicit evaluation of $Z(\Sigma \times S^1,L)$}
\label{subsubsec3.4.1}

Let us first evaluate the RHS of Eq. \eqref{eq2.48_regc}
explicitly for the special $L$ described at the beginning of Sec. \ref{subsec3.4} above.
In Sec. \ref{subsubsec3.4.2} below we will then rewrite the explicit expression
on the RHS of Eq. \eqref{eq3.93} in a suitable way and by doing so we will obtain Eq. \eqref{eqA.8}.

\medskip

We begin by evaluating the inner integral on the RHS of Eq. \eqref{eq2.48_regc}.
  In a completely analogous way as in Secs \ref{subsubsec3.3.1}--\ref{subsubsec3.3.2}
   we obtain (cf.     Remark \ref{rm_trivial_inner_int} above)
   with $R_i = R_i^{(s_0)}$ (cf.  Convention \ref{conv_s_0} in Sec. \ref{subsubsec3.2.6} above)
\begin{equation} \label{eq_inner_int_sec3.4}
 \int_{\Check{\cA}^{\orth}} \bigl(\prod_i  \Tr_{\rho_i}\bigl(
 \Hol^{\eps}_{R_i}(\Check{A}^{\orth} + A^{\orth}_c, B)\bigr) \bigr)
d\mu^{\orth}_B(\Check{A}^{\orth})  =  \prod_i \Tr_{\rho_i}\bigl(
 \Hol^{\eps}_{R_i}(A^{\orth}_c, B))
\end{equation}
so the RHS of Eq. \eqref{eq2.48_regc} simplifies and we obtain
\begin{multline}  \label{eq3.92}
 Z(S^2 \times S^1,L)
 \sim \lim_{\eps \to 0} \sum_{y \in I} \int_{\cB} \biggl[ \int_{\cA^{\orth}_c} \bigl\{ 1_{\cB^{ess}_{reg}}(B) \Det_{rig}(B)
  \exp\bigl( - 2\pi i k  \langle y, B(\sigma_0) \rangle \bigr)  \\
 \times \bigr(\prod_i  \Tr_{\rho_i}\bigl( \Hol^{\eps}_{R_i}( A^{\orth}_c, B)\bigr) \bigr) \bigr\}
 \exp(i S_{CS}(A^{\orth}_c, B)) DA^{\orth}_c \biggr] DB
\end{multline}

For each $i \le m$  let
$X^+_i$ and $X^-_i$ be the two connected components  of
 $\Sigma \backslash \Image(R^i_{\Sigma})$ (cf. Condition (C2) above) where $X^+_i$ is given by
 $X^+_i \subset \bar{X}^+_i$ with  $ \bar{X}^+_i$ as in Remark \ref{rm_Sec3.4.1} above. \par

 Moreover, let $f_i:=f_{R_i}$ be the function  $\Sigma \to [0,1]$
 defined in an analogous way as the function $f_1$ in Sec. \ref{subsubsec3.3.3} above.
 Instead of working with $f_i$ it will be more convenient to work with
 the functions  $\bar{f}_i: \Sigma \to [0,1]$  given by
 $\bar{f}_i = f_i + C_i$
 where $C_i \in \bR$ is chosen such that  $\bar{f}_i(\sigma_0) = 0$. \par

Finally, for each $i \le m$ we denote by\footnote{Note that if $m=1$ we have $Y(L)=\{Y_0,Y_1\}$ and $Y^+_1 = X^+_1$ and $Y^-_1 = X^-_1$.
This probably makes it easier  to compare the notation here with the one in Sec. \ref{subsubsec3.3.3}.}
 $Y^+_i$ (or $Y^-_i$, respectively) the unique $Y \in Y(L)$ which is contained
 in $X^+_i$ (or $X^-_i$, respectively) and has a common boundary
 with $\Image(R^i_{\Sigma})$. \par

By an obvious modification of the arguments in Sec. \ref{subsubsec3.3.3} above
 we then obtain
  \begin{multline}  \label{eq_sec3.4_step_fun} Z(\Sigma \times S^1,L) \\
 \sim  \sum_{y \in I}    \sum_{\alpha_1,\alpha_2, \ldots,\alpha_m \in \Lambda}
 \bigl(\prod_j m_{\lambda_j}(\alpha_j) \bigr)
  \int_{\ct}^{\sim} db     \biggl[  1_{\cB^{ess}_{reg}}(B) \Det_{rig}(B)
  \exp\bigl( - 2\pi i k  \langle y, B(\sigma_0) \rangle \bigr) \\
  \times \prod_j
  \exp\biggl(2 \pi i \alpha_j\biggl(\int_0^1 \int_0^1  \bigl[ (B dt)((l^j_u)'(s)) \bigr] ds du \biggr)\biggr) \biggr]_{| B = b + \tfrac{1}{k} \sum_i \alpha_i  f_i }
\end{multline}
where for each $j \le m$ and $u \in [0,1]$ we have set
\begin{equation} \label{eq_def_lju}  l^j_u:= (l_j)_u := R_j(\cdot,u)
\end{equation}
Note that according to Footnote \ref{ft_f1+C} in Sec. \ref{subsubsec3.3.3} above the value of the RHS of Eq. \eqref{eq_sec3.4_step_fun} above does not change if we replace each $f_i$
by $\bar{f}_i$. We will do this in the following but will
simply write $f_i$ instead of $\bar{f}_i$.

\smallskip

By modifying the calculations in Sec. \ref{subsubsec3.3.5} in a straightforward way we obtain
\begin{align} \label{eq3.93}  Z(\Sigma \times S^1,L)  &  \sim
 \sum_{\alpha_0, \alpha_1, \ldots, \alpha_m  \in \Lambda} 1_{k Q}(\alpha_0)
 \bigl(\prod_{j=1}^m m_{\lambda_j}(\alpha_j) \bigr)  \nonumber \\
&  \quad \quad \quad \times \biggl[ \biggl(  \prod_{i=0}^m 1_{\ct_{reg}}(B(Y_i)) \biggr)
\biggl( \prod_{i=0}^m \det\nolimits^{1/2}\bigl(1_{\ck}-\exp(\ad(B(Y_i)))_{|\ck}\bigr)^{\chi(Y_i)}
  \biggr) \nonumber \\
&  \quad \quad \quad \times  \biggl( \prod_{j=1}^m  \exp(   \pi i \wind(l^j_{S^1})  \langle \alpha_j,
 B(Y^+_j) + B(Y^-_j)  \rangle )  \biggr) \biggr]_{|
B  =  \tfrac{1}{k} ( \alpha_0 +    \sum_i \alpha_i f_i)}
\end{align}
where we have used that $\wind((l^j_u)_{S^1}) =\wind((l^j_{1/2})_{S^1})$
(cf. \eqref{eq_def_lju} above)
and $l^j_{1/2} = l_j$ (cf.  the beginning of Remark \ref{rm_Sec3.4.1}
and Definition \ref{def_3.2_2}).  \par
Here and below we use the following convention:
If $B= \tfrac{1}{k} ( \alpha_0 +    \sum_i \alpha_i f_i)$
we write $B(Y_j)$ for the unique value of the (constant) map $B_{| Y_j}$.
Similarly, we will write    $f_i(Y_j)$  for the unique value of the (constant) function $(f_i)_{|Y_j}$.

\subsubsection{Rewriting Eq. \eqref{eq3.93} in quantum algebraic notation}
\label{subsubsec3.4.2}

 For each  sequence $(\alpha_i)_{0 \le i \le m}$ of elements of $\Lambda$ we set
$$B_{(\alpha_i)_i} := \tfrac{1}{k} \bigl( \alpha_0 +   \sum_{j=1}^m  \alpha_j \cdot   f_j  \bigr)$$
and introduce the map  $\eta_{(\alpha_i)_i}: Y(L) = \{Y_0,Y_1, \ldots,
Y_{m}\} \to \Lambda $ given
 by
\begin{equation}
\eta_{(\alpha_i)_i}(Y):= k B_{(\alpha_i)_i}(Y) - \rho = \alpha_0 +   \sum_{i=1}^m  \alpha_i \cdot   f_i(Y)
  - \rho \quad \quad \forall Y \in Y(L)\end{equation}

Then with $B= B_{(\alpha_i)_i}$ and $\eta = \eta_{(\alpha_i)_i}$
we have
\begin{subequations} \label{eq_3.4_subeqs}
\begin{equation} \label{eq_alpha_j} \alpha_j =   \eta(Y^+_j)    -  \eta(Y^-_j) \quad \text{ for  $1 \le j \le m$ },
\end{equation}
\begin{equation}\det\nolimits^{1/2}\bigl(1_{\ck}-\exp(\ad(B(Y_i)))_{|\ck}\bigr) \sim d_{\eta(Y_i)},
\end{equation}
\begin{align}
& \prod_j \exp(   \pi i \ \wind(l^j_{S^1})  \langle \alpha_j,
 B(Y^+_j) + B(Y^-_j)  \rangle )  \nonumber \\
 & =   \prod_{j} \exp\left( \tfrac{\pi i}{k} \wind(l^j_{S^1})
   \bigl[ \langle \eta(Y^+_j), \eta(Y^+_j) + 2\rho \rangle -
  \langle \eta(Y^-_j), \eta(Y^-_j) + 2\rho \rangle \bigr] \right) \nonumber \\
     &  =  \prod_{Y \in Y(L)} \exp\left(   \tfrac{\pi i}{k}  \biggl(\sum\nolimits_{j \text{ with } \arc(l^j_{\Sigma}) \subset \partial Y}  \wind(l^j_{S^1})   \sgn(Y;l^j_{\Sigma})\biggr) \langle \eta(Y), \eta(Y) + 2\rho \rangle \right) \nonumber \\
&  \overset{(*)}{=}  \prod_{Y \in Y(L)}  \exp\left(   \tfrac{\pi i}{k}
\ \gleam(Y) \cdot \langle \eta(Y), \eta(Y) + 2\rho \rangle  \right) \nonumber \\
& = \left( \prod\nolimits_{Y \in Y(L)} \bigl(\theta_{\eta(Y)}\bigr)^{
\gleam(Y)} \right)
\end{align}
 where in step $(*)$ we used Eq. \eqref{eqA.1} above.

\smallskip

Let us now assume without loss of generality that $\sigma_0 \in Y_0$.
Then we have
 \begin{equation} \label{eq_alpha_0} \alpha_0 = \eta_{(\alpha_i)_i}(Y_0) + \rho
 \end{equation}
\end{subequations}
Using Eqs \eqref{eq_3.4_subeqs} we see that we can rewrite Eq. \eqref{eq3.93} as
\begin{multline} \label{eq3.96}
Z(\Sigma \times S^1,L)
\sim    \sum_{(\alpha_i)_i \in \Lambda^{m+1}} \biggl[
1_{kQ - \rho}(\eta(Y_0)) \prod_{Y \in Y(L)} 1_{k \ct_{reg} - \rho}(\eta(Y))\\
 \times    \biggl(\prod_{j=1}^m  m_{\lambda_j}(\eta(Y^+_j)    -  \eta(Y^-_j))\biggr) \\
  \times \biggl( \prod_{Y \in Y(L)} ( d_{\eta(Y)})^{\chi(Y)} (\theta_{\eta(Y)})^{ \gleam(Y)}\biggr) \biggr]_{|\eta = \eta_{(\alpha_i)_i}}
  \end{multline}

  Recall that $\col(L)$ denotes the set of
maps  $\eta: Y(L) = \{Y_0,Y_1, \ldots, Y_{m}\} \to  \Lambda_+^k$. In the
following let  $\col'(L)$  denote the set of maps
$\eta : \{Y_0,Y_1, \ldots, Y_{m}\} \to \Lambda  \cap (k \ct_{\reg} -
\rho) $

 \begin{observation}  \label{obs_lemma2}
  The map
   $$\Phi:  \{ (\alpha_i)_{0 \le i \le m} \in \Lambda^{m+1} \mid
   \eta_{(\alpha_i)_i}(Y_t) \in (k \ct_{reg} - \rho) \text{ for all $t \in \{0,1,\ldots,m\}$}
     \} \to \col'(L)$$
given by $\Phi((\alpha_i)_{0 \le i \le m}) = \eta_{(\alpha_i)_i}$
is a  well-defined bijection.
(That $\Phi$ is  well-defined and surjective is easy to see.  That  $\Phi$ is also injective follows from Eq. \eqref{eq_alpha_j} and   Eq. \eqref{eq_alpha_0} above.)
  \end{observation}

Combining Eq. \eqref{eq3.96} with  Observation \ref{obs_lemma2} we obtain\footnote{Here we  use
$ 1_{kQ - \rho}(\eta(Y_0))  1_{k \ct_{reg} - \rho}(\eta(Y_0))
 =  1_{k(Q \cap \ct_{reg}) - \rho}(\eta(Y_0))$.}
\begin{multline} \label{eq70}
Z(\Sigma \times S^1,L)
 \sim  \sum_{\eta \in \col'(L)} \biggl[ 1_{k(Q \cap \ct_{reg}) - \rho}(\eta(Y_0))
  \biggl(\prod_{j=1}^m  m_{\lambda_j}(\eta(Y^+_j)    -  \eta(Y^-_j))\biggr) \\
  \times    \biggl(\prod_{Y \in Y(L)} (d_{\eta(Y)})^{\chi(Y)}  (\theta_{\eta(Y)})^{ \gleam(Y)} \biggr)  \biggr]
\end{multline}

Let $(\cW_{\aff})^{Y(L)}$ denote  the space of maps
 $\tau: \{Y_0,Y_1, \ldots, Y_{m}\} \to \cW_{\aff}$.
We can then rewrite Eq.   \eqref{eq70} as
  \begin{align} \label{eq_change_to_tau_tilde}
& Z(\Sigma \times S^1, L) \nonumber \\
& \overset{(*)}{\sim} \sum_{\eta \in \col(L)}
\sum_{{\tau} \in (\cW_{\aff})^{Y(L)}, \tau(Y_0) \in W}  \biggl[   \biggl(\prod_{j=1}^m  m_{\lambda_j}(\tau(Y^+_j) \ast \eta(Y^+_j)    - \tau(Y^-_j) \ast \eta(Y^-_j))\biggr) \nonumber \\
 & \quad \times    \biggl(\prod_{Y \in Y(L)} (d_{\tau(Y) \ast \eta(Y)})^{\chi(Y)}  (\theta_{\tau(Y) \ast \eta(Y)})^{ \gleam(Y)} \biggr)  \biggr] \nonumber \\
&  \overset{(**)}{\sim}  \sum_{\eta \in \col(L)}
\sum_{{\tau} \in (\cW_{\aff})^{Y(L)}, \tau(Y_0) \in W  } \biggl[
 \biggl(\prod_{j=1}^m  m_{\lambda_j}\bigl({\tau}(Y^+_j) \ast \eta(Y^+_j)    -  {\tau}(Y^-_j) \ast \eta(Y^-_j)\bigr)\biggr)  \nonumber \\
& \quad \quad \times \biggl( \prod_{Y \in Y(L)} \bigl((-1)^{{\tau}(Y)}\bigr)^{\chi(Y)}\biggr) \biggl( \prod_{Y \in Y(L)} (d_{\eta(Y)})^{\chi(Y)} (\theta_{\eta(Y)})^{ \gleam(Y)} \biggr) \biggr] \nonumber \\
&  \overset{(+)}{=}  \sum_{\eta \in \col(L)}
\sum_{\tilde{\tau} \in (\cW_{\aff})^{m}, \tilde{\tau}_0 \in W  } \biggl[
 \biggl(\prod_{j=1}^m  m_{\lambda_j}\bigl( \eta(Y^+_j)    -  \tilde{\tau}_j \ast \eta(Y^-_j)\bigr)\biggr)  \nonumber \\
& \quad \quad \times \biggl( \prod_{j=1}^m (-1)^{\tilde{\tau}_j} \biggr) \biggl( \prod_{Y \in Y(L)} (d_{\eta(Y)})^{\chi(Y)} (\theta_{\eta(Y)})^{ \gleam(Y)} \biggr) \biggr]
\end{align}
with $W \subset \cW_{\aff}$  as in Sec. \ref{subsubsec3.3.6}
and where in step $(*)$ we used the two bijections \eqref{eq_ast_bij},
 in step $(**)$ we used Eqs \eqref{eq_d_inv} and
the first relation in part (i) of Remark \ref{rm_Step6} above,
and in step $(+)$ we used that
$$ m_{\lambda_j}\bigl({\tau}(Y^+_j) \ast \eta(Y^+_j)    -  {\tau}(Y^-_j) \ast \eta(Y^-_j)\bigr)
= m_{\lambda_j}(\eta(Y^+_j)    - ({\tau}(Y^+_j)^{-1} \cdot {\tau}(Y^-_j)) \ast \eta(Y^-_j)\bigr)$$
and then made the change of variable\footnote{Observe that the map  $\tau \to \tilde{\tau}$ introduced above is a (well-defined) bijection
from $\{ \tau \in \cW_{\aff}^{Y(L)} \mid \tau(Y_0) \in W \}$ onto $W \times (\cW_{\aff})^m$. }
 $\tau = (\tau_0,\tau_1,\tau_2, \ldots \tau_m) \to \tilde{\tau}=(\tilde{\tau}_0,\tilde{\tau}_1, \ldots \tilde{\tau}_m)$ given by
$\tilde{\tau}_0 = \tau(Y_0)$ and
 $\tilde{\tau}_j := {\tau}(Y^+_j)^{-1} \cdot {\tau}(Y^-_j)$, for $j \le m$,
and took into account that
\begin{multline*}  \prod_{Y \in Y(L)} \bigl((-1)^{{\tau}(Y)}\bigr)^{\chi(Y)} \overset{(++)}{=}
 \prod_Y \bigl((-1)^{\tau(Y)}\bigr)^{\# \{ j \le m \mid \arc(l^j_{\Sigma}) \subset \partial Y\}} \\
   = \prod_{j=1}^m (-1)^{\tau(Y^+_j)} (-1)^{\tau(Y^-_j)}
  =  \prod_{j=1}^m (-1)^{\tilde{\tau}_j}
 \end{multline*}
Here step $(++)$ follows from Eqs. \eqref{eq_chi_Y} in Remark \ref{rm_Sec3.4.1} above.
 By combining Eq. \eqref{eq_change_to_tau_tilde}
with the relation
 $$\sum_{\tilde{\tau}_j \in \cW_{\aff}}  (-1)^{\tilde{\tau}_j}   m_{\lambda_j}\bigl(  \eta(Y^+_j)    - \tilde{\tau}_j \ast \eta(Y^-_j) \bigr)  =
  N_{\lambda_j \eta(Y^-_j) }^{\eta(Y^+_j)}  \quad \quad \text{ for all $j \le m$}$$
(cf.    Eq. \eqref{eq_quantum_racah} in Appendix \ref{appB.2})
we  finally obtain
\begin{multline*}
 Z(\Sigma \times S^1,L) \\
  \sim  \sum_{\eta \in \col(L)}\biggl[  \biggl(\prod_{j=1}^m N_{\lambda_j  \eta(Y^-_j)  }^{ \eta(Y^+_j)} \biggr)  \biggl( \prod_{Y \in Y(L)} (d_{\eta(Y)})^{\chi(Y)} \biggr)
    \biggl( \prod_{Y \in Y(L)} (\theta_{\eta(Y)})^{ \gleam(Y)} \biggr) \biggr]
\end{multline*}

\subsection{The case of ``generic'' links  in $M = \Sigma \times S^1$}
\label{subsec3.5}

We will finally consider the case of ``generic''\footnote{Note that in view of Remark \ref{rm_adm_links_gen} and Remark \ref{rm_many_adm_links} below, ``admissible'' ribbon links (in the sense of Definition \ref{def_3.5_3} below) can be considered
to be  ``generic''. Moreover, in the special case $\Sigma \cong S^2$ also ``strictly admissible'' ribbon links (in the sense of Definition \ref{def_3.5_4} below) are ``generic''.} ribbon links $L$ in $M = \Sigma \times S^1$.
In contrast to  Sec. \ref{subsec3.3} and Sec. \ref{subsec3.4}
where for the special ribbon links $L$ considered there we gave a complete evaluation of $Z(\Sigma \times S^1,L)$ we will now only sketch the overall strategy for evaluating $Z(\Sigma \times S^1,L)$. It should be noted, though,
 that with some extra work one can indeed obtain an explicit combinatorial formula for
$Z(\Sigma \times S^1,L)$ also for  general ``strictly admissible'' $L$
(cf. Definition \ref{def_3.5_4} below).
The difference with respect to the treatment of  the three special cases mentioned
above is that  we do not verify here that the explicit expressions obtained for $Z(\Sigma \times S^1,L)$ for  general strictly admissible $L$ agrees with those in $RT(\Sigma \times S^1,L)$, cf. the paragraph after Eq. \eqref{eq_sec3.5.2b0} in Sec. \ref{subsubsec3.5.2} below.

\subsubsection{Admissible  links and (strictly) admissible ribbon links in $M = \Sigma \times S^1$}
\label{subsubsec3.5.1}

As a preparation for Definition \ref{def_3.5_2} and Definition \ref{def_3.5_3} below
 we first introduce several definitions for proper links
 $L=(l_1,l_2, \ldots, l_m)$, $m \in \bN$,  in $M=\Sigma \times S^1$. \par

 For given $L$ we call $p \in \Sigma$ a ``double point'' (resp. a ``triple point'') of $L$
if the intersection of $\pi_{\Sigma}^{-1}(\{p\})$ with the union of the images of $l_1, l_2, \ldots l_m$
contains at least two (or at least three, respectively) elements. We set
\begin{equation}V(L) := \{p \in \Sigma \mid \text{ $p$ is a double point of $L$} \}
\end{equation}
 Moreover, we will denote by   $E(L)$ the set of
curves in $\Sigma$ into which the loops $l^1_{\Sigma}, l^2_{\Sigma},
\ldots, l^m_{\Sigma}$ (defined again by $l^i_{\Sigma} := (l_i)_{\Sigma} := \pi_{\Sigma} \circ l_i$) are decomposed when being ``cut''  in the
points of $V(L)$.

\begin{definition} \label{def_3.5_0} Let $L=(l_1,l_2, \ldots, l_m)$, $m \in \bN$, be a (smooth) link in $M=\Sigma \times S^1$.
We  call $L$ admissible iff
\begin{itemize}

\item There are no triple points of $L$,
\item The projected loops $l^1_{\Sigma}, l^2_{\Sigma},
\ldots, l^m_{\Sigma}$ only have transversal intersections\footnote{\label{ft_transversal}More precisely: for each $p \in V(L)$ the two corresponding tangent vectors in $T_p\Sigma$ (which are  given by $(l^i_{\Sigma})'(\bar{t})$ and $(l^{j}_{\Sigma})'(\bar{u})$  where  $\bar{t}, \bar{u} \in [0,1]$, $i, j \le n$,   such that
$p= l^i_{\Sigma}(\bar{t}) = l^{j}_{\Sigma}(\bar{u})$) are not parallel to each other.}.
(In particular, $V(L)$ is a finite set.)
\item Each $l^i_{\Sigma}$, $i \le m$, is an immersion
(i.e. none of the tangent vectors $(l^i_{\Sigma})'(t)$, $t \in [0,1]$
vanishes).
\end{itemize}
\end{definition}

\begin{remark} \label{rm_adm_links_gen} Observe that every (smooth) link $L$ in  $M=\Sigma \times S^1$
is  equivalent (i.e. isotopic) to an admissible one.
\end{remark}

A decomposition of a link  $L=(l_1,l_2, \ldots, l_m)$, $m \in \bN$,
  in $M=\Sigma \times S^1$ is a tuple $D=(D_i)_{i \le m}$ where
each $D_i$ is a finite decomposition\footnote{More precisely: each $D_i$
 is a finite sequence $(t_j)_{j \le n}$, $n \in \bN$, of elements of $[0,1]$
 fulfilling $0 = t_1 < t_2 < \cdots < t_{n} = 1$.
 Clearly, $D_i$ induces a sequence $([t_j,t_{j+1}])_{j \le n-1}$
 of subintervals of $[0,1]$.}  of $\dom(l_i)=[0,1]$ into subintervals.
 Using $D_i$ the loop $l_i$, $i \le m$, decomposes into finitely many (sub)curves.
 We will denote this set of curves by $\cC(l_i,D_i)$ and we set
 $\cC(L,D):= \bigcup_i \cC(l_i,D_i)$.
For each $c \in \cC(L,D)$ we set  $c_{\Sigma} := \pi_{\Sigma} \circ c$.
By identifying $\dom(c)$ with $[0,1]$ in the obvious way we can consider
each $c \in \cC(L,D)$ as a map $[0,1] \to \Sigma \times S^1$
and each $c_{\Sigma}$ as a map $[0,1] \to \Sigma$.

\begin{definition} \label{def_3.5_1}
\begin{enumerate}
\item  A decomposition $D$ of $L$
is called a ``cluster decomposition'' of $L$ iff the following two conditions are fulfilled

\begin{itemize}
\item For every $c \in \cC(L,D)$ the projected curve
$c_{\Sigma}$ is a smooth embedding and none of the two endpoints of $c_{\Sigma}$
lies in $V(L)$

\item For every $c \in \cC(L,D)$ we have either of the following: \par

    Case 1: For every $c'  \in \cC(L,D)$, $c \neq c'$,
            the two curves $c_{\Sigma}$ and $c'_{\Sigma}$
              may have an endpoint in common
             but they  do not intersect ``properly''\footnote{More precisely,  $c_{\Sigma}(s) = c'_{\Sigma}(s')$
            can only occur for some $s,s' \in [0,1]$
            iff $s = 0 \wedge s'=1$ or $s = 1 \wedge s'=0$.}.\par

    Case 2: There is exactly  one $c'  \in \cC(L,D)$, $c \neq c'$, such that the two
curves $c_{\Sigma}$ and $c'_{\Sigma}$ intersect properly. Moreover,
$c_{\Sigma}$ and $c'_{\Sigma}$ intersect  in exactly one point $p \in \Sigma$
and they intersect transversally (cf. Footnote \ref{ft_transversal} above).
\end{itemize}

\item  A ``1-cluster'' of $L$ (induced by $D$) is a set of the form $\{c\}$ where   $c \in \cC(L,D)$ is as in  Case 1 above.

\item  A ``2-cluster''
  of $L$ (induced by $D$) is a set of the form $\{c,c'\}$ where $c, c' \in \cC(L,D)$ are as in Case 2 above.
\end{enumerate}
\end{definition}

\begin{observation} \label{obs_sec3.5} A (smooth) link $L$ in $M=\Sigma \times S^1$
is admissible iff it possesses a cluster decomposition.
\end{observation}

Let us now go back to the case of ribbon links in $M=\Sigma \times S^1$.
 \smallskip

Let   $L=(R_1,R_2, \ldots, R_m)$, $m \in \bN$, be a horizontal ribbon link
in $M=\Sigma \times S^1$. \par

We say that $p \in \Sigma$ is a double point of $L$ iff  the intersection of $\pi_{\Sigma}^{-1}(\{p\})$ with $\bigcup_{i=1}^m \Image(R_i)$
contains at least two elements.
A decomposition of $L$  is a tuple $D=(D_i)_{i \le m}$ where
each $D_i$ is a finite decomposition
 of $[0,1]$ into subintervals $(I_i^j)_j$.
 Via $D_i$ the (closed) ribbon $R_i$, $i \le m$, decomposes into
 finitely many ``pieces'', i.e. maps
 of the form  $S:I_i^j \times [0,1] \to \Sigma \times S^1$.
  We will denote this set of maps by $\cS(R_i,D_i)$. Moreover, we set
 $\cS(L,D):= \bigcup_i \cS(R_i,D_i)$ and we fix an (arbitrary)  order relation on $\cS(L,D)$
 (which will be kept fixed in the following).
Finally, for each  $S \in \cS(L,D)$ we set
   $S_{\Sigma} := \pi_{\Sigma} \circ S$. By  identifying each  $I_i^j$ with $[0,1]$
    in the obvious way we can  consider each $S$ as a map $[0,1] \times [0,1] \to \Sigma \times S^1$    and  each  $S_{\Sigma}$ as a map $[0,1] \times [0,1] \to \Sigma$.

\begin{definition}\label{def_can_intersection}
 We say that two  smooth maps
$S^i_{\Sigma}:[0,1] \times [0,1] \to \Sigma$, $i \in \{1,2\}$,
``intersect transversally'' iff -- using suitable reparametrizations and a suitable chart of $\Sigma$
--  we can transform''\footnote{More precisely:
$S^i_{\Sigma}$, $i \in \{1,2\}$, intersect transversally
iff it is possible to find  smooth reparametrizations $\psi_i: [0,1] \times [0,1] \to [0,1] \times [0,1]$
and an open chart $(U,\varphi)$ of $\Sigma$ such that each $\Image(S^i_{\Sigma})$, $i \in \{1,2\}$,
is contained in $U$ and such that
$S^i_{can} = \varphi \circ S^i_{\Sigma} \circ \psi_i$ for $i \in \{1,2\}$.} $(S^1_{\Sigma},S^2_{\Sigma})$  into $(S^1_{can},S^2_{can})$
where $S^i_{can}: [0,1] \times [0,1] \to \bR^2$, $i \in \{1,2\}$, are given by
$S^1_{can}(s,u) = (2s -1, u - 1/2)$ and $S^2_{can}(s,u) = (u - 1/2, 2s -1)$
 for all $s, u \in [0,1]$.
\end{definition}

\begin{definition} \label{def_3.5_2}
\begin{enumerate}
\item  A decomposition $D$ of $L$
is called a ``cluster decomposition'' of $L$ iff the following two conditions are fulfilled

\begin{itemize}
\item For every $S \in \cS(L,D)$ the projected map
$S_{\Sigma}$ is injective and no endpoint of $S_{\Sigma}$
(i.e. none of the points $p \in \{ S_{\Sigma}(s,u) \mid s \in \{0,1\}, u \in [0,1]\}$) is a double point of $L$.

\item For every $S \in \cS(L,D)$ we have either of the following: \par

    Case 1: For every $S'  \in \cS(L,D)$, $S \neq S'$,
            the two maps $S_{\Sigma}$ and $S'_{\Sigma}$ ``do not intersect properly''\footnote{More precisely:  $S_{\Sigma}(s,u)= S'_{\Sigma}(s',u')$ for $s,s',u,u' \in [0,1]$ can only occur when $s = 0 \wedge s'=1$ or $s = 1 \wedge s'=0$.}. \par

    Case 2: There is exactly  one $S'  \in \cS(L,D)$, $S \neq S'$, such that the two
maps $S_{\Sigma}$ and $S'_{\Sigma}$ intersect properly. Moreover,  $S_{\Sigma}$ and $S'_{\Sigma}$ intersect transversally  in the sense of Definition \ref{def_can_intersection} above.
\end{itemize}

\item  A 1-cluster of $L$ induced by $D$ is a set $\{S\}$ where $S \in \cS(L,D)$ is as in Case 1.

\item A 2-cluster of $L$ induced by $D$ is a set $\{S_1,S_2\}$ where $S_1, S_2 \in \cS(L,D)$
are as in Case 2.
\end{enumerate}
\end{definition}

In view of the order relation on $\cS(L,D)$ fixed above
 we can consider a 2-cluster as an ordered pair $(S_1,S_2)$. (This will convenient below).
 We will denote the set of 1-clusters (and 2-clusters, respectively) of $L$ induced by $D$
 by $Cl_1(L,D)$ (and   $Cl_2(L,D)$, respectively).
  We set  $Cl(L,D) := Cl_1(L,D) \cup Cl_2(L,D)$. The order relation on $\cS(L,D)$
  fixed above   induces an order relation on $Cl(L,D)$.

\medskip

Observation \ref{obs_sec3.5} above motivates the following
   definition.

\begin{definition} \label{def_3.5_3}
An admissible  ribbon link in $M=\Sigma \times S^1$
is a  horizontal ribbon link  in $M = \Sigma \times S^1$
which  possesses a cluster decomposition.
\end{definition}

\begin{remark} \label{rm_many_adm_links} The set of admissible ribbon links is very large.
Indeed, every admissible proper link  when equipped with a
horizontal framing leads to an admissible ribbon link. More precisely, we have
the following: \par

Let $L_{pr}$ be a (proper) link in $M = \Sigma \times S^1$ which is admissible in the sense of Definition \ref{def_3.5_0} above.
Let $L$ be a  horizontal ribbon link such that $L^0 = L_{pr}$, cf. Definition \ref{def_3.2_1} and Definition \ref{def_3.2_2} in Sec. \ref{subsec3.2}.
Now let $L^{(s)} = (R_1^{(s)}, R_2^{(s)}, \ldots R_m^{(s)})$, for $s \in (0,1]$,
where each $R_i^{(s)}$ is given as in  Sec. \ref{subsubsec3.2.1} above.
Then if $s$ is  sufficiently small,
$L^{(s)}$ will be an admissible ribbon link.
\end{remark}

\begin{definition} \label{def_3.5_4}
 An admissible  ribbon link
   $L=(R_1,R_2, \ldots, R_m)$, $m \in \bN$,  in $M=\Sigma \times S^1$
 is called ``strictly admissible'' iff
\begin{itemize}
 \item Each  $R^i_{\Sigma}$, $i \le m$, is null-homotopic, and

 \item It is possible to choose the auxiliary Riemannian metric ${\mathbf g}_{\Sigma}$ in Sec. \ref{subsec3.2}
  such that  Condition \ref{ass1} in Sec. \ref{subsubsec3.2.4} can be fulfilled for $L$.
\end{itemize}
 (Note that every admissible ribbon link $L$ can be reparametrized so that
Condition \ref{ass1} in Sec. \ref{subsubsec3.2.4} can be fulfilled).
\end{definition}

\subsubsection{Evaluation of $Z(\Sigma \times S^1,L)$ for general (strictly admissible) $L$: sketch}
\label{subsubsec3.5.2}

Let  $L = ((R_1, R_2, \ldots, R_m), (\rho_1, \rho_2, \ldots, \rho_m))$,
 $m \in \bN$,
 be an arbitrary colored, strictly admissible ribbon link in $M = \Sigma \times S^1$.
For the evaluation of $Z(\Sigma \times S^1,L)$
 we will essentially follow the same steps as in  Sec. \ref{subsec3.3} and Sec. \ref{subsec3.4} above:

 \smallskip

\noindent
{\bf Step 1:} We evaluate  (for each fixed $A^{\orth}_c \in \cA^{\orth}_c$, $B \in \cB$,
and $\eps < \eps_0$) the inner integral
\begin{equation} \label{eq_I_eps_L_Def}
I^{\eps}_L(A^{\orth}_c, B):= \int_{\Check{\cA}^{\orth}} \left( \prod_i  \Tr_{\rho_i}\bigl(
 \Hol^{\eps}_{R_i}(\Check{A}^{\orth} + A^{\orth}_c, B)\bigr) \right)
d\mu^{\orth}_B(\Check{A}^{\orth})
\end{equation}
in Eq.  \eqref{eq2.48_regc} (with $R_i = R_i^{(s_0)}$, cf.
Convention \ref{conv_s_0} in Sec. \ref{subsubsec3.2.6} above). \par

Observe that in contrast to  the situation in the two special cases II and III
in Sec. \ref{subsec3.3} and Sec.  \ref{subsec3.4} above  Step 1 is now no longer trivial.
Still, using the following procedure it is possible to perform Step 1. \par

\begin{itemize}
\item Let us fix a cluster decomposition $D$ of $L$.

\smallskip

Recall that each color $\rho_i$, $i \le m$, is  an irreducible representation
$\rho_i: G \to \Aut(V_i)$ where
$V_i$ is a finite-dimensional, complex vector space.
For every $S \in \cS(L,D)$ we set $\rho_S := \rho_i$ and $V(S) := V_i$
where $i \le m$ is the unique index such that
$S \in \cS(R_i,D_i)$ (i.e. such that $S$ is a ``piece  of the closed ribbon $R_i$'').
Moreover, if $cl = \{S\} \in Cl_1(L,D)$ we will write $V(cl)$ instead of $V(S)$
and if $cl = (S_1,S_2) \in Cl_2(L,D)$ we will write $V_i(cl)$ instead of $V(S_i)$, $i = 1,2$.

\smallskip

For each    $\eps>0$, $A^{\orth} \in \cA^{\orth}$,  $B \in \cB$, and $cl \in  Cl(L,D)$
we define
\begin{equation} \cP^{\eps}_{cl}(A^{\orth},B) \in
\begin{cases}
\End(V(cl))   & \text{ if $cl  \in Cl_1(L,D)$} \\
 \End(V_1(cl)) \otimes \End(V_2(cl))
     & \text{ if $cl  \in Cl_2(L,D)$}
\end{cases}
\end{equation}
by
\begin{equation} \label{eq_Sec3.5_P-Def}
\cP^{\eps}_{cl}(A^{\orth},B):=
\begin{cases}
  \rho_{S_1}( P^{\eps}_{S_1}( A^{\orth}, B))  & \text{ if $cl = \{S_1\} \in Cl_1(L,D)$} \\
   \rho_{S_1}(P^{\eps}_{S_1}( A^{\orth}, B)) \otimes  \rho_{S_2}(P^{\eps}_{S_2}( A^{\orth}, B))
   & \text{ if $cl = (S_1,S_2) \in Cl_2(L,D)$}
\end{cases}
\end{equation}
  where we use the notation of Remark \ref{rm_gen_notation}
  in Sec. \ref{subsec3.2}. \par

 It is not difficult to see
  that there is a linear form $\beta_{L,D}$ on
  \begin{equation} \label{eq_tensor_prod} \bigl( \otimes_{cl \in Cl_1(L,D)} \End(V(cl))  \bigr) \otimes
  \bigl( \otimes_{cl \in Cl_2(L,D)} \bigl( \End(V_1(cl)) \otimes \End(V_2(cl)) \bigr) \bigr)
  \end{equation}
 such that for all $\eps>0$ we have
\begin{equation} \label{eqinle4}
\prod_i \Tr_{\rho_i} (\Hol_{R_i}^{\eps}(A^{\orth}, B)) =
\prod_i \Tr_{\rho_i} (P_{R_i}^{\eps}(A^{\orth}, B))
 = \beta_{L,D}\bigl( \otimes_{cl \in Cl(L,D)} \cP_{cl}^{\eps}(A^{\orth},B) \bigr)
 \end{equation}
(See also Eq. \eqref{eq_appD_beta_term} in Appendix \ref{appD}  below for an explicit
formula for  $\beta_{L,D}$.)
Above  we interpret
``$\otimes_{cl \in Cl(L,D)} \cP_{cl}^{\eps}(A^{\orth},B)$'' on the RHS of
Eq. \eqref{eqinle4} as an element of the tensor product  \eqref{eq_tensor_prod} in
the obvious way.

\item Let us introduce the short
 notation\footnote{We write $\Phi^{\orth}_{B}(f)$ both for functions $f: \Check{\cA}^{\orth} \to \bC$ and
 for vector valued maps $f: \Check{\cA}^{\orth} \to V$ where $V$ is a real vector space.}
$\Phi^{\orth}_{B}:= \int_{\Check{\cA}^{\orth}} \cdots
d\mu^{\orth}_B(\Check{A}^{\orth})$ and let us denote by
$\cP_{cl}^{\eps}(\cdot + A^{\orth}_c, B)$ (for $A^{\orth}_c \in \cA^{\orth}_c$, $B \in \cB$) the map
$$\Check{\cA}^{\orth} \ni \Check{A}^{\orth} \mapsto \cP_{cl}^{\eps}(\Check{A}^{\orth} + A^{\orth}_c, B)
\in \begin{cases}
\End(V(cl))   & \text{ if $cl  \in Cl_1(L,D)$} \\
 \End(V_1(cl)) \otimes \End(V_2(cl))
     & \text{ if $cl  \in Cl_2(L,D)$}
\end{cases}$$

 The key observation now is that
the family of maps $(\cP_{cl}^{\eps}(\cdot + A^{\orth}_c, B))_{cl \in Cl(L,D)}$
has the following ``independence'' property w.r.t. the functional  $\Phi^{\orth}_{B}$\footnote{Cf. Eq. (5.35) in \cite{Ha4} and Eq. (6.3) in  \cite{Ha2} which are closely related (rigorous) results. In fact,
the derivation of Eq. \eqref{eqinle5} is a slightly more complicated than the derivation of Eq. (5.35) in \cite{Ha4}.  We need to use Condition \ref{ass1} above and also the assumption that $s_0 < 1$, cf. Sec. \ref{subsubsec3.2.6}.}:
\begin{equation} \label{eqinle5}
  \Phi^{\orth}_{B} \bigl( \bigl( \otimes_{cl \in Cl(L,D)} \cP_{cl}^{\eps}(\cdot + A^{\orth}_c, B) \bigr) \bigr)
 =  \otimes_{cl \in Cl(L,D)}   \Phi^{\orth}_{B}\bigl(  \cP_{cl}^{\eps}(\cdot + A^{\orth}_c, B)  \bigr)
\end{equation}
 for all sufficiently small $\eps>0$.
From Eq. \eqref{eqinle4} and Eq. \eqref{eqinle5} and the linearity of $\beta_{L,D}$ we obtain
\begin{equation}  \label{eq_pre_statesum}
  I^{\eps}_L(A^{\orth}_c, B) =  \Phi^{\orth}_{B} \bigl(\beta_{L,D} \circ \bigl( \otimes_{cl \in Cl(L,D)} \cP_{cl}^{\eps}(\cdot + A^{\orth}_c,B) \bigr) \bigr)
  = \beta_{L,D}\bigl( \otimes_{cl \in Cl(L,D)}  T^{\eps}_{cl}(A^{\orth}_c, B)  \bigl)
\end{equation}
 for all sufficiently small $\eps>0$ where we have set
\begin{equation} \label{eq_def_Rcl}
 T^{\eps}_{cl}(A^{\orth}_c,  B) :=     \Phi^{\orth}_{B}\bigl(  \cP_{cl}^{\eps}(\cdot + A^{\orth}_c,  B) \bigl)
\end{equation}

\item We evaluate $T^{\eps}_{cl}(A^{\orth}_c, B)$  for $cl  \in Cl_1(L,D)$. This is  easy.
Using the same argument as in Sec. \ref{subsubsec3.3.1} and  \ref{subsubsec3.3.2}
(cf. Remark \ref{rm_trivial_inner_int} in Sec. \ref{subsubsec3.3.1})
we obtain for  $cl = \{S\} \in Cl_1(L,D)$
\begin{multline} \label{eq_T_cl1_expl}
T^{\eps}_{cl}(A^{\orth}_c, B) =  \int_{\Check{\cA}^{\orth}}  \cP_{cl}^{\eps}(\Check{A}^{\orth}+ A^{\orth}_c,
 B) d\mu^{\orth}_B(\Check{A}^{\orth})
=  \int_{\Check{\cA}^{\orth}}  \rho_S(P_{S}^{\eps}(\Check{A}^{\orth} + A^{\orth}_c,
 B)) d\mu^{\orth}_B(\Check{A}^{\orth}) \\
= \rho_S(P_{S}^{\eps}(0 + A^{\orth}_c,  B)) =  \rho_S(P_{S}^{\eps}(A^{\orth}_c, B)) = \exp\bigl(\int_0^1  (\rho_S)_* (D_{S,s}^{\eps}(A^{\orth}_c,  B)) ds\bigr) \end{multline}
where we use the notation of Remark \ref{rm_gen_notation} in Sec. \ref{subsec3.2}.

\item Next, we evaluate $T^{\eps}_{cl}(A^{\orth}_c, B)$ for $cl \in Cl_2(L,D)$.

In contrast to the  case $cl \in Cl_1(L,D)$, the evaluation of $T^{\eps}_{cl}(A^{\orth}_c, B)$ for $cl \in Cl_2(L,D)$ is now non-trivial. However, using Eq. \eqref{eq_Pic_Lind}  and the well-known formula
for the moments of Gaussian measures
 it is  not too difficult to write down an explicit formula  involving a (rather complicated) infinite series of powers of $1/k$,
similarly to the derivation of Eq. (6.7) in \cite{Ha2}.
(It should be noted  that, in contrast to  Eq. (6.7) in \cite{Ha2}, the coefficients in front of each power of $1/k$ will now also depend on $k$).

\item Finally, by combining Eq. \eqref{eq2.48_regc} with Eq. \eqref{eq_pre_statesum} we obtain
\begin{multline} \label{eq_step1_ende}
Z(\Sigma \times S^1,L ) \sim \lim_{\eps \to 0} \sum_{y \in I}  \int_{\cA^{\orth}_c \times \cB}
  \biggl\{ 1_{\cB^{ess}_{reg}}(B) \Det_{rig}(B)  \exp\bigl( - 2\pi i k  \langle y, B(\sigma_0) \rangle \bigr) \\
  \times  \beta_{L,D}\bigl( \otimes_{cl \in Cl(L,D)}  T^{\eps}_{cl}(A^{\orth}_c, B)  \bigl) \biggr\}
 \exp(i S_{CS}(A^{\orth}_c, B)) (DA^{\orth}_c \otimes DB)
\end{multline}

 Let us now rewrite Eq. \eqref{eq_step1_ende} in such a way that its RHS becomes
more similar to the RHS of Eq. \eqref{eqA.4} in Appendix \ref{appC}
 (with the explicit expressions \eqref{eqA.5}   inserted into it).
 As in Appendix \ref{appC} let $L^0$  be  the proper link
 associated to $L$ (cf. Definition \ref{def_3.2_2}  in Sec. \ref{subsubsec3.2.1})
 and  set
$$V(L) := V(L^0),  \quad E(L) := E(L^0) $$
Observe that there is an obvious 1-1-correspondence between
$Cl_2(L,D)$ and $V(L)$. Moreover, we can assume without loss of generality\footnote{This can, of course, always be achieved by ``merging'' every ``chain'' of  consecutive 1-clusters  to a single 1-cluster.}
that the cluster decomposition $D$ was chosen to be ``minimal'' in the sense
that every 1-cluster $cl \in Cl_1(L,D)$ ``connects'' two different 2-clusters.
Then there is also a 1-1-correspondence between
$Cl_1(L,D)$ and $E(L)$. For $x \in V(L)$ (and $e \in E(L)$, respectively)
 let $cl(x) \in Cl_2(L,D)$ (or  $cl(e) \in Cl_1(L,D)$, respectively) denote the corresponding
 2-cluster (1-cluster).
Accordingly, we can rewrite Eq. \eqref{eq_step1_ende} as
\begin{multline} \label{eq_step1_ende_rewritten}
Z(\Sigma \times S^1, L) \sim \lim_{\eps \to 0} \sum_{y \in I}  \int_{\cA^{\orth}_c \times \cB}
  \biggl\{ 1_{\cB^{ess}_{reg}}(B) \Det_{rig}(B)  \exp\bigl( - 2\pi i k  \langle y, B(\sigma_0) \rangle \bigr) \\
  \times  \beta_{L,D}\bigl(  \bigl( \otimes_{e \in E(L)}  T^{\eps}_{e}(A^{\orth}_c, B)  \otimes \bigl( \otimes_{x \in V(L)}  T^{\eps}_{x}(A^{\orth}_c, B)  \bigr) \bigl) \biggr\} \\
  \times  \exp(i S_{CS}(A^{\orth}_c, B)) (DA^{\orth}_c \otimes DB)
\end{multline}
where, for $x \in V(L)$ and $e \in E(L)$, we have set
 \begin{equation} \label{eq_def_Teps_e}
 T^{\eps}_{e}(A^{\orth}_c, B):= T^{\eps}_{cl(e)}(A^{\orth}_c, B) \quad
 \text{ and } \quad T^{\eps}_{x}(A^{\orth}_c, B):= T^{\eps}_{cl(x)}(A^{\orth}_c, B).
 \end{equation}
\end{itemize}

\noindent {\bf Steps 2--4:} We perform in Eq. \eqref{eq_step1_ende_rewritten} above
     the  integral, the  sum $\sum_y$, and the $\eps \to 0$-limit,
     cf Appendix \ref{appD} below
     (and the paragraph after Eq. \eqref{eq_sec3.5.2b} below).

\smallskip

\noindent {\bf Step 5:} We rewrite the algebraic expression
 obtained after performing  Steps 2--4 in  quantum algebraic notation.

\medskip

From Eq. \eqref{eq_T_cl1_expl} we easily obtain
an explicit (and ``closed'') expression for  $T^{\eps}_{e}(A^{\orth}_c, B)$.
By contrast, even though each  $T^{\eps}_{x}(A^{\orth}_c, B)$, $x \in V(L)$,
 can be written  explicitly as suitable infinite sums (cf. the paragraph
 after Eq. \eqref{eq_T_cl1_expl} above)
  so far we do not have a closed expression for $T^{\eps}_{x}(A^{\orth}_c, B)$. \par

Since the integral on the RHS of Eq. \eqref{eq_step1_ende_rewritten}
only involves the Abelian fields $A^{\orth}_c$ and $B$
and since the informal complex measure
$\exp(i S_{CS}(A^{\orth}_c, B)) (DA^{\orth}_c \otimes DB)$
is of Gauss-type one would expect that the evaluation of the integral
is straightforward. \par

There is, however, one potential complication: the informal measure
$\exp(i S_{CS}(A^{\orth}_c, B)) (DA^{\orth}_c \otimes DB)$ is ``degenerate''.
More precisely, the kernel of (the symmetric bilinear form associated
to) the quadratic form  $S_{CS}(A^{\orth}_c, B)$ on $\cA^{\orth}_c \oplus \cB
\cong  \cA_{\Sigma,\ct} \oplus \cB$ is
 $\cA_{closed} \oplus \cB_c$ where $\cB_c:= \{ B \in \cB \mid B \text{ is constant }\}$
and $\cA_{closed}:= \{ \alpha  \in  \cA_{\Sigma,\ct} \mid d\alpha = 0\}$.
On the other hand, if it turns out that  the function
\begin{equation}  \label{eq_F_Q1_def}
F^{\eps}_L(A^{\orth}_c, B) :=  \beta_{L,D}\bigl(  \bigl( \otimes_{e \in E(L)}  T^{\eps}_{e}(A^{\orth}_c, B)  \otimes \bigl( \otimes_{x \in V(L)}  T^{\eps}_{x}(A^{\orth}_c, B)  \bigr) \bigl)
\end{equation}
fulfills
\begin{equation} \label{eq_(Q1)}  F^{\eps}_L( A_{closed} + A_{coex}, B)
= F^{\eps}_L(A_{coex}, B)
  \end{equation}
 for all $B \in \cB$,  $A_{closed} \in  \cA_{closed}$, and $A_{coex} \in
 \cA_{coex} := \{\star d f \mid f \in \Omega^0(\Sigma,\ct)\} \in \cA_{\Sigma,\ct}$
 then the aforementioned complication does not have any serious
 consequences.
 And indeed, for the special links $L$  considered in  Sec. \ref{subsec3.4}
above, where $V(L) = \emptyset$ and each $R^i_{\Sigma}$ is null-homotopic,
it is easy to check\footnote{Since $V(L) = \emptyset$ in Sec. \ref{subsec3.4}
this follows easily from the aforementioned explicit expression  for  $T^{\eps}_{e}(A^{\orth}_c, B) $, cf.   Eq. \eqref{eq_T_cl1_expl} above.}  that \eqref{eq_(Q1)} is indeed fulfilled  provided that we use  the ``canonical'' choice for the
family (of Dirac families) $((\delta^{\eps}_{\sigma})_{\eps < \eps_0})_{\sigma \in \Sigma}$ as in Remark \ref{rm_3.3.4} above.
 This is the deeper reason why we did not get any problems in Sec. \ref{subsec3.4}
 above  when evaluating $Z(\Sigma \times S^1, L)$.  Recall that by doing so we arrived at
 \begin{equation} \label{eq_sec3.5.2b0}
Z(\Sigma \times S^1,L) \sim |L|= \sum_{\eta \in col(L)}
|L|_1^{\eta}\,|L|_2^\eta\,|L|_3^\eta
\end{equation}
where the factors $|L|_i^{\eta}$, $i \le 3$, are as in Sec. \ref{subsec3.4} above.  \par

 The obvious  two questions now are whether Eq. \eqref{eq_(Q1)}
 also holds for general strictly admissible $L$
and whether  we obtain indeed
\begin{equation} \label{eq_sec3.5.2b}
Z(\Sigma \times S^1,L) \sim |L|= \sum_{\eta \in col(L)}
|L|_1^{\eta}\,|L|_2^\eta\,|L|_3^\eta\,|L|_4^\eta
\end{equation}
after carrying out Steps 2--5 above.
(Here  $|L|_4^{\eta}$ is given by
$|L|_4^{\eta} = \contr_{D(L)} \bigl( \otimes_{x \in V(L)} T(x,\eta) \bigr)$
with $\contr_{D(L)}$ and $T(x,\eta)$ as in Appendix \ref{appC}.) \par

It should be possible to verify each of these two questions
``perturbatively'' by using the aforementioned explicit infinite series for
$T^{\eps}_{x}(A^{\orth}_c, B)$, $x \in V(L)$,
cf. the ``bullet point'' after Eq. \eqref{eq_T_cl1_expl} above.
 Note that since at present  we do not have a closed formula for
 $T^{\eps}_{x}(A^{\orth}_c, B)$ or $T_{x}(A^{\orth}_c, B) := \lim_{\eps \to 0} T^{\eps}_{x}(A^{\orth}_c, B)$  we cannot yet give a completely explicit non-perturbative treatment of these two questions (in contrast to the situation in Sec. \ref{subsec3.3} and  Sec. \ref{subsec3.4} where for the special class of links $L$ considered there a non-perturbative treatment was possible). On the other hand, as we will show
in Appendix \ref{appD} below, even without having a closed formula for
$T_{x}(A^{\orth}_c, B)$ we can still get quite far with a non-perturbative
 evaluation of $Z(\Sigma \times S^1,L)$.
 In particular, we will  make it plausible that
 also for general strictly admissible $L$ we have a good chance of obtaining
 Eq. \eqref{eq_sec3.5.2b}, cf. Remark \ref{rm_appD_not_expl} in Appendix \ref{appD}.

\begin{remark} \label{rm_KiRe} In order to clarify if Eq. \eqref{eq_sec3.5.2b} holds
for general strictly admissible $L$ it may be
useful to do consider first the special case $G=SU(2)$
and to restrict one's attention to those
ribbon links $L$ in $\Sigma \times S^1$
which stay inside $\Sigma \times (S^1 \backslash \{1\})$.
In this special case we can use the formulas in \cite{KiRe}
 which allow us to rewrite the RHS of Eq. \eqref{eq_sec3.5.2b}  using ``$R$-matrices''
 instead of quantum $6j$-symbols.
\end{remark}

\begin{remark}
 The strategy used in Step 1 is similar to the strategy used
in  Sec. 6 in \cite{Ha2} (see also  \cite{CCFM,Lab} and \cite{ASen}) for the evaluation of the Chern-Simons path integral $Z(\bR^3,L)$ on $\bR^3$  in the axial gauge
 where $L$ is  a colored framed link in $\bR^3$.
 However, as was observed in  \cite{Ha2,CCFM,Lab}
the Chern-Simons path integral $Z(\bR^3,L)$
 in the axial gauge is  problematic\footnote{In \cite{Ha2} we left it
 open what the  deeper reason
 for this is. One explanation could be that  axial gauge fixing is  so ``singular''
 that we can in general not expect meaningful results when applying it.
 Alternatively, the problems with $Z(\bR^3,L)$
 could indicate  that something  is ``wrong'' with the Chern-Simons path integral $Z(M,L)$ when $M$ is a  non-compact 3-manifold.}: \par
 It turns out  that   the values of $Z(\bR^3,L)$ do not agree with those expected in the standard literature   even for links $L$ without
 double points\footnote{When  considering the Chern-Simons path integral $Z(\bR^3,L)$ on $\bR^3$  in the axial gauge  it is unclear why quantum groups (or rather, the corresponding R-matrices)     should enter the computations.
    Note that a quantum group $U_q(\cG_{\bC})$, $q \in \bC \backslash \{-1,0,1\}$, is obtained
    from the classical enveloping algebra $U(\cG_{\bC})$
    by a deformation process that involves a fixed Cartan subalgebra $\ct$ of $\cG$.
    But such a Cartan subalgebra does not play any role
    when considering the Chern-Simons path integral $Z(\bR^3,L)$ in the axial gauge.
By contrast, when considering the Chern-Simons path integral $Z(\Sigma \times S^1,L)$
in the torus gauge a Cartan subalgebra $\ct$ plays an important  role right from the beginning.}.
 However,  the explicit expressions obtained for $Z(\bR^3,L)$
 are surprisingly close to the correct expressions, cf. Sec. 6 and Sec. 7 in \cite{Ha2}.
 For example, for a restricted class of the ``loop smearing''
 regularization procedure which we use, the values of $Z(\bR^3,L)$
 are invariant under  Reidemeister I and Reidemeister II moves.
 Moreover, in the special case\footnote{In Sec. 7 in \cite{Ha2} we actually considered the non-simply connected  group $G=SO(N)$. This is, however, equivalent
 to the situation where $G=Spin(N)$ and where each link color $\rho_i$
 comes from a $SO(N)$-representation.} $G=Spin(N)$   one ``almost recovers''
  Kauffman's  state models for the HOMFLY  polynomial at special values.
  (It is easy to imagine that by studying this special case
  one could have rediscovered Kauffman's state models.
 I emphasize that we really use $G=Spin(N)$ here  even though the
 HOMFLY polynomial is usually associated to  the groups  $G=SU(N)$, $N \ge 2$.)

 \smallskip

Because of this I am optimistic  that   we will  indeed obtain Eq. \eqref{eq_sec3.5.2b} for general strictly admissible ribbon links $L$ when evaluating $Z(\Sigma \times S^1,L)$ given by Eq. \eqref{eq2.48_regc} or Eq. \eqref{eq2.48_reg}  (or by a suitable modification of Eq. \eqref{eq2.48_reg},
e.g. the modification sketched in Remark \ref{rm_BF_necessary} below).
\end{remark}

\begin{remark} \label{rm_BF_necessary}
It may useful to perform what in \cite{Ha7b} was called the
 ``Transition to the BF-theoretic setting''. This amounts to applying a suitable
 linear change of variable to the CS path integral on $M= \Sigma \times S^1$
 with group $G \times G$ and parameters $(k,-k)$, cf. Sec. 7 in \cite{Ha7b} and Appendix D in  \cite{Ha7b}.
 Note that as long as one works with proper loop holonomies
 the formulas which we obtain after performing
 the aforementioned linear change of variable   will be equivalent
 to the original formulas.
 However, once we switch to ribbon holonomies this  no longer needs to be the case
 and, accordingly,  the aforementioned change of variable can be expected to lead to something new.
 In particular, the change of variable discussed in  Appendix D in  \cite{Ha7b},
 which involves  a complex structure on $\cG \oplus \cG$,
  could be interesting, since it allows
   the complexification $\cG_{\bC}$ of $\cG$ to enter the picture.
  This reduces the gap between the CS path integral approach  and the algebraic approach in \cite{ReTu1,ReTu2,turaev} where complex Lie algebras play an essential role.
\end{remark}

\section{Rigorous realization $Z^{t.g.f}_{rig}(\Sigma \times S^1,L)$
of $Z^{t.g.f}(\Sigma \times S^1,L)$.}
\label{sec4}

 Sec. \ref{sec3} was dedicated to what in Sec. \ref{sec1} we called ``Problem (P1)''.
Now we will focus on Problem (P2), i.e. the problem of
making rigorous sense of the torus gauge fixed CS path integral $Z^{t.g.f}(\Sigma \times S^1,L)$ given by Eq. \eqref{eq_Ztgf_def} (or, alternatively, the RHS of Eq. \eqref{eq2.48_ribbon} above).
This  is desirable for the following reasons:
\begin{itemize}

\item[(R1)] In order to have a chance of making progress regarding the open problems in Quantum Topology mentioned in  Sec.  \ref{subsec5.2} below
 we need to solve both Problem (P1) and Problem (P2) of Sec. \ref{sec1}.

\item[(R2)]  The study of the informal CS path integral gives rise to an interesting ``paradox'':
  On the one hand the study of the informal CS path integral
      leads to very deep mathematics but, from the purist's point of view,
       since it is only informal it does not contain
          any real mathematics at all.
          Obviously, once we have a rigorous realization
          $Z^{t.g.f}_{rig}(\Sigma \times S^1,L)$ of $Z^{t.g.f}(\Sigma \times S^1,L)$
          this paradox is resolved for manifolds of the form $M=\Sigma \times S^1$
           (cf.  Remark \ref{rm_sec4.7} in Sec. \ref{subsec4.6}  below for additional  comments).

\item[(R3)] Torus gauge fixing is quite ``singular''\footnote{One manifestation of
this ``singularity'' is the ``instability'' phenomenon described in Sec. \ref{subsec4.6} below.} so it would be good to have a rigorous definition and evaluation
 of the torus gauge fixed CS path integral  $Z^{t.g.f}(\Sigma \times S^1,L)$.
\end{itemize}

\smallskip

In Sec. \ref{subsec4.1}--\ref{subsec4.3}
 we will sketch three different approaches/frameworks (F1), (F2), and (F3)
for obtaining a rigorous realization $Z^{t.g.f}_{rig}(\Sigma \times S^1,L)$
of $Z^{t.g.f}(\Sigma \times S^1,L)$.

 \smallskip

For each of these three approaches/frameworks  we proceed in the following way:
 \begin{itemize}
 \item First we replace the spaces $\Check{\cA}^{\orth}$, $\cA^{\orth}_c$, and $\cB$ by
       suitable modifications\footnote{\label{ft_Frechet_space}When working with the original spaces of smooth 1-forms and maps
       $\Check{\cA}^{\orth}$, $\cA^{\orth}_c$, and $\cB$
        the informal integral functionals given by Eqs \eqref{eq4_int_functionals} below  cannot be defined in a satisfactory way.  This is related to the well-known fact in measure theory that
       on most infinite-dimensional topological vector spaces $E$
       a (non-trivial) cylinder set measure  does not extend to a true measure.
       An important exception is the case where $E$ is the dual of a nuclear Frechet space, and it is precisely this exception which plays a crucial for approach (F2).}  $\Check{\cA}^{\orth}_{mod}$, $(\cA^{\orth}_c)_{mod}$, and $\cB_{mod}$
        (e.g. finite-dimensional analogues like in framework (F1) and (F3)
       or suitably extended  spaces like in framework (F2)).

 \item Then we find a rigorous realization of the informal integral
functionals\footnote{In frameworks (F1) and (F3) this boils down
to finding a rigorous realization of the expressions $S_{CS}(\Check{A}^{\orth},B)$ and $S_{CS}(A^{\orth}_c,B)$ for $\Check{A}^{\orth} \in \Check{\cA}^{\orth}_{mod}$,  $A^{\orth}_c \in (\cA^{\orth}_c)_{mod}$, and $B \in \cB_{mod}$.
(Recall that $ d\mu^{\orth}_B(\Check{A}^{\orth})$ is the normalization of $\exp(i  S_{CS}( \hat{A}^{\orth}, B)) D\hat{A}^{\orth}$, cf. Eq. \eqref{eq_def_mu_B} in Sec. \ref{subsubsec2.3.1} above.)
By contrast, in framework (F2) we construct the two integral functional $\Phi^{\orth}_B$
and $\Psi$ directly, without trying to give a separate meaning to the expressions
$S_{CS}(\Check{A}^{\orth},B)$ and $S_{CS}(A^{\orth}_c,B)$.
Moreover, in framework (F2)  we will actually not define $\Phi^{\orth}_{B}$
 for all $B \in \cB_{mod}$ but only $\Phi^{\orth}_{B_r}$
 where $B_r$ is a  suitable regularization of  $B \in \cB_{mod}$.}
 associated to the two  informal (``Gauss-type'') complex measures appearing in  Eq. \eqref{eq2.48}, i.e. of the
 functionals\footnote{\label{ft_another_funct}In fact, for reasons explained in Sec. \ref{subsec4.5} below,
 in Framework (F2) it will not be the  informal integral functional
 $  \int_{\cA^{\orth}_c \times \cB} \cdots  \exp(i S_{CS}(A^{\orth}_c, B)) (DA^{\orth}_c \otimes DB)$ for which we will find a rigorous realization but
 another (closely related) informal integral functional.}
  \begin{subequations}  \label{eq4_int_functionals}
  \begin{align} \Phi^{\orth}_B & :=
  \int \cdots d\mu^{\orth}_B(\Check{A}^{\orth}), \quad \quad \quad  B \in \cB_{mod}\\
  \Psi & := \int \cdots  \exp(i S_{CS}(A^{\orth}_c, B)) (DA^{\orth}_c \otimes DB)
\end{align}
\end{subequations}
These rigorous realizations $\Phi^{\orth}_B$ and $\Psi$
 must have reasonably large domains $\dom(\Phi^{\orth}_B) \subset Fun(\Check{\cA}^{\orth}_{mod},\bC)$ and $\dom(\Psi) \subset Fun((\cA^{\orth}_c)_{mod} \times \cB_{mod},\bC)$. (Here  $Fun(X,\bC) = \bC^X$.)

\item Next, we find, for each $\Check{A}^{\orth} \in \Check{\cA}^{\orth}_{mod}$,  $A^{\orth}_c \in (\cA^{\orth}_c)_{mod}$, and $B \in \cB_{mod}$
a rigorous analogue (or regularized version)  of the  expression  (this is the choice  for framework (F1))
\begin{subequations}
 \begin{multline} \label{eq_F_informal2}
F(\Check{A}^{\orth},A^{\orth}_c,B):= \\
\exp\bigl( - 2\pi i k  \langle y, B(\sigma_0) \rangle \bigr) 1_{\cB_{reg}}(B) \Det_{rig}(B)  \left( \prod_i  \Tr_{\rho_i}\bigl(
 \Hol_{R_i}(\Check{A}^{\orth} + A^{\orth}_c, B)\bigr) \right)
\end{multline}
appearing in Eq. \eqref{eq2.48_ribbon} or,
 in the case of framework (F3), of the  expression
 \begin{multline} \label{eq_F_informal1}
F(\Check{A}^{\orth},A^{\orth}_c,B):= \\
\exp\bigl( - 2\pi i k  \langle y, B(\sigma_0) \rangle \bigr) 1_{\cB^{ess}_{reg}}(B) \Det_{rig}(B)  \left( \prod_i  \Tr_{\rho_i}\bigl(
 \Hol_{R_i}(\Check{A}^{\orth} + A^{\orth}_c, B)\bigr) \right)
\end{multline}
or,  in the case of framework (F2),  of the  expression (for each fixed $\eps < \eps_0$ and $s \in (0,1)$)
 \begin{multline} \label{eq_F_informal}
F(\Check{A}^{\orth},A^{\orth}_c,B):= \\
\exp\bigl( - 2\pi i k  \langle y, B(\sigma_0) \rangle \bigr) 1_{\cB^{ess}_{reg}}(B) \Det_{rig}(B)  \left( \prod_i  \Tr_{\rho_i}\bigl(
 \Hol^{\eps}_{R^{(s)}_i}(\Check{A}^{\orth} + A^{\orth}_c, B)\bigr) \right)
\end{multline}
\end{subequations}
appearing on the RHS of Eq. \eqref{eq_Ztgf_def} above.
Doing so we obtain a function $F_{rig}: \Check{\cA}^{\orth}_{mod} \times (\cA^{\orth}_c)_{mod} \times  \cB_{mod}  \to \bC$.
The function $F_{rig}$ must be defined/constructed in such a way that
\begin{subequations}
 \begin{equation} \label{eq4.4a} F_{rig}(\cdot,A^{\orth}_c, B) \in \dom(\Phi^{\orth}_B)
 \end{equation}
for all  $A^{\orth}_c \in (\cA^{\orth}_c)_{mod},
 B \in \cB_{mod}$ and also
\begin{equation} \label{eq4.4b}  [(\cA^{\orth}_c)_{mod} \times \cB_{mod} \ni (A^{\orth}_c, B) \mapsto \Phi^{\orth}_B(F_{rig}(\cdot,A^{\orth}_c, B)) \in \bC] \in \dom(\Psi)
\end{equation}
\end{subequations}

\item The final definition for $Z^{t.g.f}_{rig}(\Sigma \times S^1,L)$
  is now obtained by combining the first three points above in the obvious way
  (according to the RHS of Eq. \eqref{eq_Ztgf_def}  or the RHS of Eq. \eqref{eq2.48_ribbon} above)  and by adding suitable limits for the elimination of
  the regularization parameters which are involved in the definition of $F_{rig}$.
  In the case of framework (F3), we also have to perform a continuum limit.
\end{itemize}

\subsection{The  simplicial framework (F1)}
\label{subsec4.1}

A concrete/full ``implementation'' of framework (F1) was given in \cite{Ha9} (which improves
the original implementation in \cite{Ha7a}).

\subsubsection{Outline}
\label{subsubsec4.1.1}

\begin{itemize}

\item The spaces $\Check{\cA}^{\orth}_{mod}$, $(\cA^{\orth}_c)_{mod}$, and $\cB_{mod}$ are chosen to
be (finite-dimensional) simplicial analogues\footnote{For every $p \in \{0,1,2\}$ and every real vector space $V$
 the space of $V$-valued $p$-cochains $C^p(K,V)$ for some fixed  finite smooth triangulation (or polyhedral cell decomposition)
 $K$ of $\Sigma$ is a  simplicial analogue of the space of $V$-valued $p$-forms $\Omega^p(\Sigma,V)$.} of the continuum spaces $\Check{\cA}^{\orth}$, $\cA^{\orth}_c$, and $\cB$.

\item The two informal integral functionals \eqref{eq4_int_functionals}
appearing above are defined as follows: \par

First we introduce  natural simplicial analogues $S^{disc}_{CS}(\Check{A}^{\orth},B)$ and $S^{disc}_{CS}(A^{\orth}_c,B)$
 of the functions $S_{CS}(\Check{A}^{\orth},B)$ and $S_{CS}(A^{\orth}_c,B)$.
 This gives us two  well-defined complex measures $\exp(i S_{CS}(\Check{A}^{\orth},B)) D\Check{A}^{\orth}$
 and  $\exp(i S_{CS}(A^{\orth}_c,B)) DA_c^{\orth} \otimes DB$,
where $D\Check{A}^{\orth}$, $DA_c^{\orth}$, and $DB$ are the (normalized) Lebesgue measures on the finite dimensional spaces  $\Check{\cA}^{\orth}_{mod}$,  $(\cA^{\orth}_c)_{mod}$, and $\cB_{mod}$
(equipped with a natural scalar product).
By normalizing the complex measure $\exp(i S_{CS}(\Check{A}^{\orth},B)) D\Check{A}^{\orth}$
we obtain a simplicial analogue of $d\mu^{\orth}_B$.
The simplicial analogues of the two integral functionals \eqref{eq4_int_functionals} above can now be realized rigorously as regularized integrals
in the same way as  the functional $\int_{\sim} \cdots d\mu$
appearing in Sec. \ref{subsubsec3.3.1} above. (For the second integral
functional we need to use the definition in  Remark \ref{rm_mu_degenerate} in Sec.  \ref{subsubsec3.3.1}).

\item The rigorous function $F_{rig}$ is obtained by constructing
a  natural ``simplicial analogue''   of the continuum expression Eq.  \eqref{eq_F_informal2} above (cf. Sec. \ref{subsec4.5} below for more details
  and  Sec. 3 in \cite{Ha9} for full details).

\item Depending on the concrete implementation that we use
a limit for the elimination of one regularization parameter may be  necessary
(cf. Sec. \ref{subsec4.5} below).

\end{itemize}

\subsubsection{Comments}
\label{subsubsec4.1.2}

Plus points:

\begin{itemize}
\item  Framework (F1) is very simple,  even elementary. As a result rigorous proofs are easy to obtain,
cf. Theorems 5.7 and 5.8 in \cite{Ha9} and Theorem 3.5 in \cite{Ha7b} for a rigorous version and proof of the informal results in Sec. \ref{subsec3.3} and Sec. \ref{subsec3.4} above.

\item So far (F1) is the only rigorous approach/framework
 which was carried out completely in the three special situations
covered in Sec. \ref{subsec3.1}, Sec. \ref{subsec3.3}, and Sec. \ref{subsec3.4} above.
\end{itemize}

\noindent
Drawbacks:

\begin{itemize}
\item If one wants to have a reasonable chance of  obtaining a rigorous treatment
for the case of general $L$ (i.e. of generalizing Theorem 3.5 in \cite{Ha7b} and Theorems 5.7 and 5.8 in \cite{Ha9} to the case of general admissible ribbon links $L$,
cf. Sec. \ref{subsec3.5})
   within (F1)  it seems to be necessary to make the transition to the $BF$-theoretic setting,    cf. Remark \ref{rm_BF_necessary} in Sec. \ref{subsubsec3.5.2} above and Sec. 7 in \cite{Ha7b}.

\item In view of the observations in Appendix D in \cite{Ha7a} it seems very unlikely that a suitable  ``(approximative) unfixing the gauge''-procedure (cf. the last paragraph of  Sec. \ref{subsec5.2}  below)
 can be implemented within (F1).
 \end{itemize}

\subsection{The continuum framework (F2)}
\label{subsec4.2}

This approach/framework was inspired by
 \cite{ASen}, which (to my knowledge) was the first paper
 to study, for non-Abelian $G$,
 the rigorous realization of $Z(M,L)$ for any manifold $M$,
 namely the non-compact manifold $M=\bR^3$. \par

A concrete implementation of framework (F2)  was given in \cite{Ha6b}
(using many ideas from   \cite{ASen}),  cf. Sec. \ref{subsec4.5} below for more details.

\subsubsection{Outline}
\label{subsubsec4.2.1}

\begin{itemize}

\item The spaces $\Check{\cA}^{\orth}_{mod}$, $(\cA^{\orth}_c)_{mod}$, and $\cB_{mod}$ are chosen to
be  considerably larger than $\Check{\cA}^{\orth}$, $\cA^{\orth}_c$, and $\cB$.
They consist of distributional elements, for example, we
take  $\Check{\cA}^{\orth}_{mod}:= \cN'$ (with the weak topology) where  $\cN$ is the nuclear Frechet space obtained from $\Check{\cA}^{\orth}$ by equipping that space with a suitable family of semi-norms, cf. Footnote \ref{ft_Frechet_space} above.

\item The framework of White Noise Analysis (WNA) is used for the rigorous
     construction of the integral functionals $\Psi$ and $\Phi^{\orth}_B$ (or, rather
  $\Phi^{\orth}_{B^{r}}$ where $B^{r}$ is a suitable regularization of
     $B \in \cB_{mod}$).
     (We refer to  \cite{ASen,ASen2} for a  summary of the constructions of WNA which are relevant here.)
         For example,
         $\Phi^{\orth}_{B^{r}}$  can be realized rigorously as  a continuous linear functional $\Phi^{\orth}_{B^{r}}: (\cN) \to \bC$   where  $(\cN)$ is a suitable subspace  of $L^2(\cN',d\mu_{can})$,
     $d\mu_{can}$ being the canonical Gaussian Borel measure on $\cN' = \Check{\cA}^{\orth}_{mod}$. Regarding the implementation of $\Psi$
         we refer the reader to Sec. \ref{subsec4.5} below
         (cf. also Footnote \ref{ft_another_funct} above).

\item We need to construct $F_{rig}$ such that  for each fixed $A^{\orth}_c \in (\cA^{\orth}_c)_{mod}$ and $B \in \cB_{mod}$
 the function  $F_{rig}(\cdot,A^{\orth}_c,B)$  is an element of $(\cN)$.
 (This takes care of Eq. \eqref{eq4.4a} above.)
    Since we are working with regularized holonomies  this is easy.
    What remains to be done is to prove  Eq. \eqref{eq4.4b} above,
   or, rather, the analogue  of Eq. \eqref{eq4.4b} obtained after introducing suitable
    regularizations of the two terms  $1_{\cB^{ess}_{reg}}(B)$ and $ \Det_{rig}(B)$, cf. Sec. \ref{subsec4.5} below.

\item Three limits  for the elimination of the regularization
parameters\footnote{Apart from the parameters $\eps$ and $s$ appearing in Eq. \eqref{eq_F_informal} in the concrete implementation given in \cite{Ha6b} there is a third parameter $n$  which is used for the regularization $B^r$ of $B$
mentioned above and  the regularization of
 the two terms  $1_{\cB^{ess}_{reg}}(B)$ and $ \Det_{rig}(B)$, cf. Sec. \ref{subsec4.5} below.}
involved  are necessary.
\end{itemize}

\subsubsection{Comments}
\label{subsubsec4.2.2}

Plus points:

\begin{itemize}
\item  Framework (F2) is closest\footnote{And in fact,  in Sec. \ref{sec3}
        we (indirectly) referred to Framework (F2) several times in order to justify
       some of the informal arguments appearing there, cf., e.g., Remark \ref{rm_Ha2_Anwendung1} in Sec. \ref{subsubsec3.3.2} and Eq. \eqref{eqinle5} in Sec. \ref{subsubsec3.5.2}.} to the informal treatment in Sec. \ref{sec3}.
       If the informal calculations sketched in Sec. \ref{subsec3.5} lead to the correct result for $Z(\Sigma \times S^1,L)$    for general admissible $L$ then it is almost certain that Framework (F2) will allow us to make these computations rigorous.
        (In particular, a ``transition to the BF-theoretic setting''  will not be necessary, unless   it is necessary already for the informal treatment, cf.
         Remark \ref{rm_BF_necessary} in Sec. \ref{subsubsec3.5.2} above).

\item Framework (F2) provides the most direct way for resolving the ``paradox'' mentioned in (R2) above.
(Also when using framework (F1) and (F3) this paradox can be resolved but the argument
    is then less direct, cf. Remark \ref{rm_sec4.7} in Sec. \ref{subsec4.6}  below.)
\end{itemize}

\noindent
Drawbacks:

\begin{itemize}
\item  Framework (F2) is quite technical. Partly because of this,  full proofs
    have not been given yet.

\item It seems unlikely that  ``(approximative) unfixing the gauge'', cf. the last paragraph of
 Sec. \ref{subsec5.2}  below, can be
  implemented within Framework (F2).
\end{itemize}

\subsection{The ``mixed'' framework (F3)}
\label{subsec4.3}

 Framework  (F3) was briefly sketched in Sec. 3.10 in \cite{Ha9}
 but a  concrete implementation  has not been given yet. \par

 The basic idea of Framework  (F3) is to
  combine constructions from the simplicial and the continuum setting.
 By doing so we can combine some of the advantages of the simplicial setting (like finite dimensional realizations of the space
 $\Check{\cA}^{\orth}_{mod}$, $(\cA^{\orth}_c)_{mod}$, and $\cB_{mod}$) with some of the advantages of the continuum setting
 (like a very natural definition for the rigorous realization
 $F_{rig}(\Check{A}^{\orth},A^{\orth}_c,B)$ of the function \eqref{eq_F_informal1} above).

\subsubsection{Outline}
\label{subsubsec4.3.1}

\begin{itemize}

\item The spaces $\Check{\cA}^{\orth}_{mod}$, $(\cA^{\orth}_c)_{mod}$, and $\cB_{mod}$ are obtained
by embedding suitable  simplicial spaces
into the spaces $\Check{\cA}^{\orth}_{pw}$, $(\cA^{\orth}_c)_{pw}$, and $\cB_{pw}$
which are the extensions of   $\Check{\cA}^{\orth}$, $\cA^{\orth}_c$, and $\cB$
containing piecewise smooth elements.

\item The two informal integral functionals \eqref{eq4_int_functionals}
 are defined in an analogous way as in (F1).
 The advantage now is that the functions $S_{CS}(\Check{A}^{\orth},B)$ and $S_{CS}(A^{\orth}_c,B)$ for $\Check{A}^{\orth} \in \Check{\cA}^{\orth}_{mod}$,  $A^{\orth}_c \in (\cA^{\orth}_c)_{mod}$, and $B \in \cB_{mod}$
 are given canonically.

\item The definition of $F_{rig}(\Check{A}^{\orth},A^{\orth}_c,B)$ is as follows:
 the last two factors on the RHS of Eq. \eqref{eq_F_informal1} can be defined
 in a canonical way. For the factor $\Det_{rig}(B)$ on the RHS of
  Eq. \eqref{eq_F_informal1}
 we have several options: we can use again the ``continuum'' definition given by Eqs \eqref{eq_Det_rig_B} in Sec. \ref{subsubsec2.3.2}
 or we use the simplicial definition mentioned in Remark \ref{rm_Det_disc} above.
 The realization of $1_{\cB^{ess}_{reg}}$ requires a suitable regularization.

 \item  The regularization parameters have to be eliminated using suitable limits.
  Moreover, we also have to perform a continuum limit.
\end{itemize}

\subsubsection{Comments}
\label{subsubsec4.3.2}

Plus points:
 \begin{itemize}
  \item Framework (F3) is not as simple as the simplicial framework (F1) but still quite simple and not very technical.

 \item  I consider the chances that framework (F3) will allow a successful treatment
   of general $L$ to be fairly good (but not as good as when working with framework (F2)). In particular,  for framework (F3) a ``transition to the BF-theoretic setting'' will not be necessary, unless  it is necessary already for the informal treatment, cf. Remark \ref{rm_BF_necessary}  in Sec. \ref{subsubsec3.5.2} above.

  \item In framework (F3) the chances for a successful implementation of a suitable ``(approximative) unfixing the gauge'' procedure (cf. the last paragraph of
 Sec. \ref{subsec5.2}  below) seem to be  fairly good.
  \end{itemize}

\noindent
Drawbacks:
\begin{itemize}
 \item The basic idea in (F3) of combining/mixing
  constructions from the simplicial and the continuum setting
    makes (F3)   less natural/elegant than a purely simplicial framework or a pure continuum framework.

  \item As mentioned above, a continuum limit
  is  necessary in (F3).

  \end{itemize}

\subsection{Comparison of the three frameworks}
\label{subsec4.4}

\begin{center}
\begin{tabular}{|  c | c | p{1.8cm} | p{3.1cm} |  p{2.0cm} |  p{2.3cm} |}
\hline
Framework  &
    Simplicity & Continuum limit necessary? & Concrete implementation given?/ Carried out for all 3 special cases? &  Chances for a successful treatment of  general $L$  &
    Chances for a successful implementation of  ``unfixing of the gauge'' \\ \hline
  (F1) &    very simple  & no         & yes / yes & negligible & negligible \\  \hline
 (F2)  &   quite technical & no    & yes / no & very good     & low  \\  \hline
 (F3)  &   rather simple & yes      & no / no &  fair & fair  \\ \hline
\end{tabular}
\end{center}

Comments:

\begin{itemize}

\item The ``three special cases'' referred to in the fourth column are those considered
 in Sec. \ref{subsec3.1}, Sec. \ref{subsec3.3}, and Sec. \ref{subsec3.4} above.

\item ``Chances for a successful treatment of  general $L$'':
 This refers to the chances of being able to find a rigorous derivation of Eq. \eqref{eq_sec3.5.2b} above for general  (strictly admissible) $L$  within the original framework, i.e.  without having to
  make a serious modification of the approach/framework
   (provided that Eq. \eqref{eq_sec3.5.2b} can be derived on an informal level).
   For framework (F1) these chances seem to be zero  unless we  make  the transition   to the BF-theoretic setting, cf.  Sec. \ref{subsec4.1} above.

\item For ``unfixing of the gauge'' see  the last paragraph of
 Sec. \ref{subsec5.2}  below.

\end{itemize}

\subsection{Some remarks regarding the concrete implementation of (F1), (F2), and (F3)}
 \label{subsec4.5}

In \cite{Ha9} and \cite{Ha6b} we have given concrete
implementations of the frameworks (F1) and (F2),  respectively.
Here are some brief remarks regarding these implementations
(plus a remark regarding the implementation of (F3)):

\begin{itemize}
\item[(F1)] We first fix a (sufficiently fine) finite (polyhedral) cell decomposition $\cK$ of $\Sigma$. For technical reasons we then construct from
    $\cK$   another cell decomposition $q\cK$
which is finer than  $\cK$ but coarser than the barycentric subdivision $b\cK$ of $\cK$.
Using $q\cK$ we then obtain simplicial analogues
$\Check{\cA}^{\orth}(q\cK)$, $\cA^{\orth}_c(q\cK)$, and $\cB(q\cK)$
of the three spaces  $\Check{\cA}^{\orth}$, $\cA^{\orth}_c$, and $\cB$.
The spaces $\Check{\cA}^{\orth}_{mod}$, $(\cA^{\orth}_c)_{mod}$, and $\cB_{mod}$
 are then defined as suitable subspaces of
$\Check{\cA}^{\orth}(q\cK)$, $\cA^{\orth}_c(q\cK)$, and $\cB(q\cK)$
 (for reasons explained in Sec. 3  in \cite{Ha9}).
The definitions of the expressions $S^{disc}_{CS}(\Check{A}^{\orth},B)$ and $S^{disc}_{CS}(A^{\orth}_c,B)$ mentioned in Sec. \ref{subsubsec4.1.1} above are very natural.
The definition of the first three of  the four factors appearing
 on the RHS of Eq. \eqref{eq_F_informal2}  are also  very natural\footnote{The simplicial analogue for the first factor is obvious. As a simplicial analogue of $\Det(B)$
 we use $\Det^{disc}(B)$, mentioned in  Remark \ref{rm_Det_disc} in Sec. \ref{subsec3.2}.
 The simplicial analogue of the factor $1_{\cB_{reg}}(B)$ is essentially straightforward.
  The only point worth mentioning is that, for technical reasons  (cf. Secs 3.7 and 5.4 in \cite{Ha9}) the simplicial analogue of   $1_{\cB_{reg}}(B)$  requires  a suitable regularization.}.  The simplicial analogues for the terms $\Hol_{R_i}(A)$ appearing in
 in the fourth factor in  Eq. \eqref{eq_F_informal2}
 used in \cite{Ha9} are also  natural but in contrast to the implementation
 of $\Hol_{R_i}(A)$ in frameworks (F2) and (F3) definitely not canonical.
 In fact, among a number of possible definitions of $\Hol_{R_i}(A)$
 which would all be  natural
 one has to  choose the one definition that gives rise to the correct values for
  $Z^{t.g.f}_{rig}(\Sigma \times S^1,L)$. (This is an illustration
  of the ``instability'' phenomenon which we discuss in  Sec. \ref{subsec4.6} below.)
Finally, for reasons  explained in Sec. 3.11 in \cite{Ha9}
we need to apply a suitable regularization (cf. (M2) in Sec. 3.11 in \cite{Ha9}).

\item[(F2)]   In Sec. \ref{subsubsec4.2.1} we already explained roughly
 how the space $\Check{\cA}^{\orth}_{mod}$
 and the functional(s) $\Phi^{\orth}_{B^{r}}$ are defined in framework (F2). \par

The definition/realization of the functional $\Psi$ is somewhat trickier.
For reasons explained now $\Psi$ will not be constructed as a rigorous realization of the informal integral functional  $  \int_{\cA^{\orth}_c \times \cB} \cdots  \exp(i S_{CS}(A^{\orth}_c, B)) (DA^{\orth}_c \otimes DB)$ but  of a closely related (informal) integral functional.
Recall from Sec. \ref{subsubsec3.5.2} above that
in the special case considered in Sec. \ref{subsec3.4} we have Eq. \eqref{eq_(Q1)}
 with $F^{\eps}_L(A^{\orth}_c, B)$ as in Eq. \eqref{eq_F_Q1_def}  (provided that
  we use the ``canonical'' choice for the family of Dirac families $((\delta^{\eps}_{\sigma})_{\eps < \eps_0})_{\sigma \in \Sigma}$ as in
   Remark \ref{rm_3.3.4} above) and that we expect
  Eq. \eqref{eq_(Q1)} to hold also for general $L$.
   If this is indeed the case
  then the expression inside the
  $  \int_{\cA^{\orth}_c \times \cB} \cdots  \exp(i S_{CS}(A^{\orth}_c, B)) (DA^{\orth}_c \otimes DB)$-integral on the RHS of Eq. \eqref{eq_Ztgf_def}
  only depends on the $A_{coex}$-component of
  $A^{\orth}_c = A_{closed} + A_{coex}$, which means that the integration
  over the $A_{closed}$-component is trivial and can be carried out right away.
  After doing so we are left with  an (informal) integral functional
  of the form $  \int_{\cA_{coex} \times \cB} \cdots  \exp(i S_{CS}(A_{coex}, B)) (DA_{coex} \otimes DB)$.
  For technical reasons we also make use of  the decomposition
  $\cB = \cB' \oplus \cB_c$ with $\cB'$ and $\cB_c$ as in
  Remark \ref{rm_3.3.4} above.
  The rigorous functional  $\Psi$ mentioned above is constructed
    as a rigorous realization of the   (informal) integral functional
  $  \int_{\cA_{coex} \times \cB'} \cdots  \exp(i S_{CS}(A_{coex}, B')) (DA_{coex} \otimes DB')$
  using the framework of WNA.

  \smallskip

Finally, note that  for making sure that Eq. \eqref{eq4.4b} holds
we need to regularize the two factors $1_{\cB^{ess}_{reg}}(B)$
and $ \Det_{rig}(B)$ appearing in Eq. \eqref{eq_F_informal} above in a suitable way.
In Appendix B in \cite{Ha6b} we give a concrete
suggestion for such regularizations, which are, however,
 not very elegant\footnote{They rely on choosing a cell decomposition of $\Sigma$
which is not in the spirit of a continuum approach.}.

\item[(F3)] A concrete implementation of (F3) has not been written up explicitly yet,
            but due to the observation above that almost all constructions in (F3)
            are ``canonical'' it should be straightforward to do so.
\end{itemize}

\subsection{Some comments on the ``instability'' of $Z^{t.g.f.}(\Sigma \times S^1,L)$ }
\label{subsec4.6}

The informal  torus gauge fixed CS path integral   $Z^{t.g.f.}(\Sigma \times S^1,L)$
shows a certain degrees of ``instability'' in the sense that
the explicit values of
certain candidates for a rigorous realization $Z^{t.g.f.}_{rig}(\Sigma \times S^1,L)$
of $Z^{t.g.f.}(\Sigma \times S^1,L)$ depend in a delicate way on the details
of this realization.
This is most obvious when using the simplicial framework (F1)
 for  obtaining  $Z^{t.g.f.}_{rig}(\Sigma \times S^1,L)$.
$Z^{t.g.f.}_{rig}(\Sigma \times S^1,L)$
 can then be considered as a kind of ``lattice  regularization'' of the informal
 expression  $Z^{t.g.f.}(\Sigma \times S^1,L)$.
   Usually, when one  works with a lattice regularization in Quantum  Field Theory
 one has to perform a suitable continuum limit.
 As we saw above, in the simplicial framework (F1) such a continuum limit is
 actually not necessary,
  which is a great  advantage of the simplicial approach.
 (Of course, if we want to apply a continuum limit anyway, this is still possible  but the limit will turn out to be trivial.)
  On the other hand there is  a price to pay:  in contrast to the standard  situation in QFT where the continuum limit is usually independent of the lattice regularization
   the value of $Z^{t.g.f.}_{rig}(\Sigma \times S^1,L)$
   will depend on the details of the lattice regularization\footnote{Cf. the discussion
   regarding the  possible definitions of $\Hol_{R_i}(A)$ in the concrete implementation
   of framework (F1) sketched in    Sec. \ref{subsec4.5} for a concrete example.}
  and even if we choose to apply a  continuum limit anyway   this dependence on the details will not disappear.
  As a result,  only for a distinguished subclass of  lattice regularizations
  the expression $Z^{t.g.f.}_{rig}(\Sigma \times S^1,L)$ will have the correct value.

  \smallskip

  In the following remark we give one possible
  (but somewhat speculative) interpretation of the instability phenomenon   mentioned above.

\begin{remark} \label{rm_sec4.7}
Suppose that\footnote{Observe that from what we said in Sec. \ref{subsec3.5}
there are good chances that Eq. \eqref{eq_sec3.5.2b} in Sec. \ref{subsec3.5} above holds
for general strictly admissible $L$.
If so then the rigorous implementation
within framework (F2) above is almost certainly possible,
and there are fair chances that the rigorous implementation
is also be possible  within  Framework (F3).
Finally, there seem to be fair chances that ``(approximative) unfixing of the gauge'' is
possible within Framework (F3).}
\begin{itemize}
\item[1.]  Eq. \eqref{eq_sec3.5.2b} holds
for general strictly admissible $L$,
i.e. the informal evaluation of $Z^{t.g.f}(\Sigma \times S^1,L)$ leads to the correct result.

\item[2.]  Using   framework (F3) above (or a suitable variant of (F3))
it is possible to obtain a rigorous realization $Z^{t.g.f}_{rig}(\Sigma \times S^1,L)$ of $Z^{t.g.f}(\Sigma \times S^1,L)$.

\item[3.] The rigorous framework which we use
allows  ``(approximative) unfixing of the gauge''
 (cf. the last paragraph of Sec. \ref{subsec5.2}  below)
 and therefore allows us to relate the rigorous expression
 $Z^{t.g.f}_{rig}(\Sigma \times S^1,L)$  to (the informal expression) $Z(\Sigma \times S^1,L)$  in a suitable way.

\end{itemize}

If all this is the case then  both $Z^{t.g.f.}(\Sigma \times S^1,L)$
 and  $Z(\Sigma \times S^1,L)$
 can be considered as idealized (informal) continuum limits\footnote{In some sense the non-gauge fixed  CS theory  would be  an effective theory.
 In contrast to what is normally the case with effective theories
 this effective theory would have a rather special property:
 The expression  $Z(\Sigma \times S^1,L)$
   would not just be an approximation of the ``primary''/``real'' expression
  $Z^{t.g.f.}_{rig}(\Sigma \times S^1,L)$ but the informal perturbative
  evaluation of $Z(\Sigma \times S^1,L)$ would
  lead exactly to the same values as the evaluation of
  $Z^{t.g.f.}_{rig}(\Sigma \times S^1,L)$.}
  of the rigorously defined expression $Z^{t.g.f.}_{rig}(\Sigma \times S^1,L)$.
  In particular, it would then be natural to consider $Z^{t.g.f.}_{rig}(\Sigma \times S^1,L)$ as  ``primary'' and $Z^{t.g.f.}(\Sigma \times S^1,L)$  and  $Z(\Sigma \times S^1,L)$ as ``secondary''.
  Now observe that it is quite possible
  that    the definition of $Z^{t.g.f.}_{rig}(\Sigma \times S^1,L)$  will involve
  certain technical features (like, e.g., a distinguished class of lattice   ``regularizations''\footnote{If we consider $Z^{t.g.f.}_{rig}(\Sigma \times S^1,L)$ to be primary
  the word  ``regularization'' is actually not really appropriate anymore.})
  which will no longer be visible after the  (idealized) continuum limit
  has been carried out.
  The ``instability phenomenon'' mentioned above  is created ``artificially''
  by changing one's mind about  what is primary and what is secondary.
  It is only when one considers $Z^{t.g.f.}(\Sigma \times S^1,L)$
 (or  $Z(\Sigma \times S^1,L)$) as ``primary'' and $Z^{t.g.f.}_{rig}(\Sigma \times S^1,L)$
 as ``secondary'' (namely as a kind of regularization)
 that one feels the need to explain why $Z^{t.g.f.}_{rig}(\Sigma \times S^1,L)$
  involves a  distinguished subclass of  lattice regularizations.
\end{remark}

\section{Outlook}
\label{sec5}

\subsection{Generalization to the case where $M$ is the total space of a non-trivial $S^1$-bundle (or a more general Seifert fibered space)}
\label{subsec5.1}

As mentioned in Sec. \ref{sec1} the results in \cite{BlTh1} on the CS path integral
 on manifolds of the form $M=\Sigma \times S^1$   were generalized
in \cite{BlTh4} to non-trivial $S^1$-bundle spaces $M$ and in \cite{BlTh5} to Seifert fibered spaces $M$. It is natural to try to combine the ideas/methods in the present paper with those in  \cite{BlTh4,BlTh5} and to try to extend the results in \cite{BlTh4,BlTh5}
to general links $L$ in the aforementioned 3-manifolds $M$.
(Recall from the introduction  that the only links $L$  considered in \cite{BlTh4,BlTh5} are fiber links.)  \par

For the applications to the open problems (OP1) and (OP3) mentioned in Sec. \ref{subsec5.2} below it is sufficient to consider the special case where $M=S^3$ (which can be considered as a non-trivial $S^1$-bundle via the Hopf fibration).
As a first step one should therefore study whether by combining the ideas/methods in  \cite{BlTh4}
with those in Secs \ref{sec2}--\ref{sec3} of  the present paper
one can ``define'' and evaluate a suitable torus gauge fixed CS path integral
 $Z^{t.g.f}(S^3,L)$ explicitly for general colored links $L$
in $S^3$ (such that the explicit value of $Z^{t.g.f}(S^3,L)$ coincides
with $RT(S^3,L)$)  and whether one can obtain a rigorous realization $Z^{t.g.f}_{rig}(S^3,L)$
 of $Z^{t.g.f}(S^3,L)$ by adapting the frameworks (F1)--(F3) in Sec. \ref{sec4} above
 in a suitable way (or by using another framework). \par

If this can be done successfully the next step would be to
introduce/define $Z^{t.g.f}(M,L)$
 and $Z^{t.g.f}_{rig}(M,L)$ for general colored links $L$ in all those
   Seifert fibered spaces $M$ considered in \cite{BlTh5}.

\subsection{Potential applications to Quantum Topology}
\label{subsec5.2}

Several open conjectures in  Quantum Topology  can be ``proven'' on an informal level
 by assuming the
equivalence between $RT(M,L)$  and $Z(M,L)$
and by applying informal path integral methods for the perturbative evaluation of
$Z(M,L)$. This is, e.g., the case for the following open problems (OP1), (OP2), and\footnote{In the case of (OP3)  the informal approach of \cite{Guk,Wi10}
does not quite give a full informal ``proof'' but goes a long way towards such a proof.}
(OP3):

\begin{description}

\item[(OP1)] As is shown in \cite{GMM,BN,AxSi1,AxSi2,BoTa,AlFr,Thu} the informal CS path integral $Z^{L.g.f}(S^3,L)$ in the Lorenz gauge
     can be evaluated on a perturbative level for general colored, framed links $L$ in $S^3$.
    More precisely,   $\langle L \rangle^{L.g.f} = Z^{L.g.f}(S^3,L)/Z^{L.g.f}(S^3)$ can be expanded
     as an asymptotic series of powers of $1/k$ and
     the coefficients in this series involve complicated analytic expressions
      called ``configuration space integrals''.
    In view of the expected equivalence between  $Z(S^3,L)$ and $RT(S^3,L)$
    one  therefore arrives at the conjecture that
    these configuration space integrals also appear (with the same
    coefficients as predicted by \cite{GMM,BN,AxSi1,AxSi2,BoTa,AlFr,Thu})
    when expanding  $RT_{norm}(S^3,L)$  as an asymptotic series of  powers of $1/k$.
    To my knowledge, to date this conjecture has been proven only
    up to order $6$, cf. Sec. 7 in \cite{Les} and Sec. 3.2 in \cite{BN04}.

\item[(OP2)]    By using the equivalence of $RT(M)$ and
the original (= non-gauge fixed) CS path integral
      $Z(M)$ and by applying standard techniques from asymptotic analysis
      (including  the stationary phase method)
            to the (informal) evaluation of $Z(M)$ as $k \to \infty$
        one arrives at the so-called ``Perturbative Expansion Conjecture'',
        which relates $RT(M)$ to geometric/topological concepts
 like moduli spaces, Reidemeister torsion, spectral flow, and the Chern-Simons invariant, cf. \cite{And1}.

\item[(OP3)] The so-called ``Volume Conjecture'', which relates
the colored Jones polynomial of a (hyperbolic) knot $K$ in $S^3$
to the hyperbolic volume $vol_{hyp}(S^3 \backslash K)$
 of the knot complement  $S^3 \backslash K$,
 is one of the most important open problems
in Knot Theory. So far, it has only been verified
 for a small number of special knots $K$.
There is, however, a very promising path integral approach
(cf. \cite{Guk,Wi10})  which goes a long way towards giving an informal ``proof''
of the volume conjecture for general $K$ by using arguments
involving  the CS path integral for $G = SU(2)$ and $M=S^3$ on the one hand,
the CS path integral for $G = SL(2)$ and  $M = S^3 \backslash K$ on the other hand,
and  suitable analytic continuation arguments relating the two types
of CS path integrals.
\end{description}

If it were possible to obtain  rigorous realizations  $Z^{L.g.f.}_{rig}(S^3,L)$,
 $Z_{rig}(M)$, and $Z_{rig}(S^3,L)$ of $Z^{L.g.f.}(S^3,L)$,
 $Z(M)$, and $Z(S^3,L)$  then  one could hope to be able to turn the informal path integral ``proofs'' mentioned above into rigorous proofs.  \par

Now, as we have seen in Sec. \ref{sec3} and Sec. \ref{sec4} above,
for the manifolds of the form  $M=\Sigma \times S^1$
there  are good chances that the informal evaluation
of the torus gauge fixed CS path integral
$Z^{t.g.f}(M,L)$  (given by the RHS of Eq. \eqref{eq2.48_ribbon}
or Eq. \eqref{eq_Ztgf_def} above) leads to the same values as $RT(M,L)$
and that one can  obtain a rigorous realization
$Z^{t.g.f}_{rig}(M,L)$ of $Z^{t.g.f}(M,L)$.
As we explained in  Sec. \ref{subsec5.1}  above,
 it is probably possible to  generalize  our approach to the case
 where $M$ is  a non-trivial $S^1$-bundle,
 which would lead, in particular, to a rigorous realization
 $Z^{t.g.f}_{rig}(S^3,L)$ of the suitably ``defined''
 informal expression $Z^{t.g.f}(S^3,L)$.
 In a future version of the present paper  I plan to include an additional part of the appendix
 where I sketch the possible implementation of a  suitable ``(approximative)
 unfixing of the gauge''-procedure and an ``(approximative) changing the gauge''-procedure
  and how this could be exploited for  relating    $Z^{t.g.f}_{rig}(S^3,L)$
   directly  to the expressions appearing in the  (informal) perturbative evaluation
  of  $Z(S^3,L)$ and $Z^{L.g.f.}(S^3,L)$.

\bigskip

\noindent
{\it Acknowledgements:} It is a pleasure for me to thank  Sergio Albeverio for several stylistic suggestions
and for pointing out an inaccuracy
in  Sec. \ref{subsubsec3.5.2},
and  Thierry L{\'e}vy for a useful feedback on Remark \ref{rm_sec2.2.1} above.  \par

Since this review paper is partly based on my papers \cite{Ha3b,Ha4,Ha7a,Ha7b}
I take the opportunity  to thank again everyone who contributed
 with useful suggestions and comments   to these papers.
 In particular,  I want to thank  Dietmar Arlt,  Laurent Freidel,
Benjamin Himpel,   Jean-Claude Zambrini, and the (anonymous) referee of the paper \cite{Ha4}.
Moreover, I want to thank Sebastian de Haro  for pointing out the reference \cite{Tu2}
to me and for his collaboration in \cite{HaHa}.
 Finally, it is a great pleasure for me to
  thank  Sergio Albeverio and  Ambar N. Sengupta for
 their support and encouragement over  the past years
and for arousing my interest in the rigorous study of the non-Abelian
 Chern-Simons path integral which was initiated in \cite{ASen}.
 \cite{ASen} was the original and, apart from \cite{Wi} and \cite{BlTh1}, the most important
  inspiration for my work on the Chern-Simons path integral.

 \renewcommand{\thesection}{\Alph{section}}
\setcounter{section}{0}

\section{Summary of Lie theoretic and quantum algebraic notation}
\label{appB}

In the present part of the Appendix we list the Lie theoretic and quantum algebraic notation
used in the paper and we also recall some useful identities.

\subsection{Lie theoretic notation}
\label{appB.1}

Recall that in Sec. \ref{sec2} we fixed a simple, simply-connected compact
Lie group $G$ with Lie algebra $\cG$ and a maximal torus $T$ of $G$
with Lie algebra $\ct$. We set

\begin{itemize}

\item $G_{reg}  := \{g \in G \mid g \text{ is regular}\}$, cf. Remark \ref{rm_appB_neu} below,

\item $ \cG_{reg} := \exp^{-1}(G_{reg})$,

\item  $ T_{reg}  := T \cap G_{reg}$,

\item  $\ct_{reg}  := \exp^{-1}(T_{reg}) = \ct \cap \cG_{reg}$.

\end{itemize}

\begin{remark} \label{rm_appB_neu}
(i) An element $g$ of $G$ is called ``regular'' iff it is contained
in exactly one maximal torus of $G$, cf. Sec. 3 in Chap. IV in \cite{Br_tD}.

\smallskip

\noindent
(ii) The connected components of $\ct_{reg}$ are called
 ``Weyl alcoves''.
\end{remark}

Using the scalar product   $\langle \cdot, \cdot \rangle$
on $\cG$ which we fixed in Sec. \ref{sec2} above
we now make  the obvious identification $\ct \cong \ct^*$.

\begin{itemize}
\item $\cR \subset \ct^*$  denotes the set of real roots associated to ($\cG, \ct)$

\item $\Check{\cR}$ denotes the set of  real coroots, i.e. $\Check{\cR} := \{\Check{\alpha} \mid \alpha \in \cR\} \subset \ct$
       where $\Check{\alpha}: = \frac{ 2 \alpha}{\langle \alpha,  \alpha \rangle}$.

\item $\Gamma  \subset \ct$  denotes the lattice generated by the set of real coroots.

\item $\Lambda \subset \ct^*$ denotes the real weight lattice associated to $(\cG,\ct)$,
i.e. the lattice in $\ct$ which is dual to $\Gamma$.

 \item $\cW$ denotes the Weyl group associated to $(\cG,\ct)$.
       For $\tau \in \cW$ we  denote by $(-1)^{\tau}$ the determinant
       of $\tau \in \Aut(\ct)$.

\item  $\cW_{\aff}$  denotes the affine Weyl group associated to $(\cG,\ct)$, i.e.
  the group of isometries of $\ct \cong \ct^*$
   generated by the orthogonal reflections on the hyperplanes $H_{\alpha,k}$, $\alpha \in \cR$, $k \in \bZ$, where  $H_{\alpha,k}:= \alpha^{-1}(k)$.

\item $I := \ker(\exp_{| \ct})$.
   \end{itemize}

\begin{remark} \label{rm_affineWeyl}
(i) From the assumption that $G$ is simply-connected it follows that $\Gamma = I$
 (cf. Theorem 7.1 in Chap. V in \cite{Br_tD}).

\smallskip

(ii)   $\cW_{\aff}$ is  generated by    $\cW$ and  the translations $T_x: \ct \ni b \mapsto  b + x \in \ct$, $x \in \Gamma = I$. In fact, $\cW_{\aff}$ is the semi-direct product
of $\cW$ and the  group $\{T_x \mid x \in \Gamma\} \cong \Gamma$, so we can make the identification $\cW_{\aff} \cong \cW \times \Gamma$
(cf. Proposition 7.9 in Chap. V in \cite{Br_tD}).
 For $\tau = (\sigma,x) \in \cW_{\aff}  \cong \cW \times \Gamma$ we write  $(-1)^{\tau}$ instead of $(-1)^{\sigma}$.

\smallskip

(iii) We have
 $\ct_{\reg} = \ct \backslash \bigcup_{\alpha \in \cR, k \in \bZ} H_{\alpha,k}$
 where $H_{\alpha,k}$, $\alpha \in \cR$, $k \in \bZ$ is as above.

\smallskip

(iv) Every $\tau \in  \cW_{\aff}$ leaves $\ct_{reg}$ invariant
      and transfers each Weyl alcove into another  Weyl alcove.
 Accordingly, $\cW_{\aff}$ acts on the set of Weyl alcoves. One can show that this action is  free and transitive (cf. Proposition 7.10 in Chap. V in \cite{Br_tD}).

\smallskip

(v) $\forall x,y \in \Gamma: \langle x, y \rangle \in \bZ$, and
$\forall x \in \Gamma: \langle x, x \rangle \in 2\bZ$.
 In order to see this it is enough to show that
\begin{equation} \label{eq_CartanMatrix} \langle \Check{\alpha}, \Check{\beta} \rangle \in \bZ \quad \text{ for all coroots  $\Check{\alpha}, \Check{\beta}$}
\end{equation}
According to the general theory of semi-simple Lie algebras we have
 $2\tfrac{\langle \Check{\alpha}, \Check{\beta} \rangle}{\langle \Check{\alpha}, \Check{\alpha} \rangle}
\in \bZ$. Moreover,  there are at most two different (co)roots lengths
and  the quotient between the square lengths of the long and short coroots is either 1, 2, or 3.
Since  the normalization of $\langle
\cdot,  \cdot \rangle$ was chosen such that
$\langle \Check{\alpha}, \Check{\alpha} \rangle = 2$ holds
if $\Check{\alpha}$ is a short coroot we therefore have
$\langle \Check{\alpha}, \Check{\alpha} \rangle/2 \in \{1,2,3\}$
and \eqref{eq_CartanMatrix} follows.
\end{remark}

Let us now also fix a Weyl chamber $\CW$.

\begin{itemize}
\item $\cR_+$ denotes the set of positive (real) roots
associated to $(\cG,\ct)$ and $\CW$.

 \item  $\Lambda_+$ denotes the set of dominant (real) weights
 associated to $(\cG,\ct)$ and $\CW$.

\item  $\rho$  denotes the half-sum of the positive (real) roots.
\item $\theta$  denotes the unique  long (real) root in  $\overline{\CW}$.
\item We set $ \cg:= 1 + \langle \theta,\rho \rangle$. (Note that $\cg$ is the dual Coxeter number of $\cG$.)

\item  The fundamental Weyl alcove  is  the unique Weyl alcove $P$ which is contained in the Weyl chamber $\CW$ fixed above and which has $0 \in \ct$ on its  boundary. $P$  is given explicitly by
\begin{equation} \label{eq_P_formula} P =   \{b  \in \CW \mid \langle b,\theta \rangle < 1 \}.
\end{equation}

\item We have
\begin{equation} \label{eq_appB_det}
\det(1_{\ck}-\exp(\ad(b))_{|\ck}) =  \prod_{\alpha \in \cR+}  4 \sin^2(\pi  \alpha(b)) \quad b \in \ct
\end{equation}
In Sec. \ref{subsec3.2} above we introduced
the  ``square root''
 $\det^{1/2}(1_{\ck}-\exp(\ad(\cdot))_{|\ck}):\ct \to \bR$ of
 $\det(1_{\ck}-\exp(\ad(\cdot))_{|\ck}): \ct \to \bR$ by setting
\begin{equation} \label{eq_appB_det1/2}
\det\nolimits^{1/2}(1_{\ck}-\exp(\ad(b))_{|\ck}) :=  \prod_{\alpha \in \cR+}  2 \sin(\pi  \alpha(b))
\quad \forall b \in \ct
\end{equation}

\item For $\lambda \in  \Lambda_+$ let $\lambda^* \in
\Lambda_+$ denote the weight conjugated to $\lambda$ and
$\bar{\lambda} \in \Lambda_+ $ the weight conjugated to $\lambda$
``after applying a shift by $\rho$''. More precisely, $\bar{\lambda}$
is given by $\bar{\lambda} + \rho = (\lambda + \rho)^*$.

\end{itemize}

For every $\lambda \in \Lambda_+$ we denote by $\rho_{\lambda}$
the (up to equivalence) unique
irreducible, finite-dimensional, complex representation of $G$
with highest weight $\lambda$.
For every $\mu \in \Lambda$ we will denote by
$m_{\lambda}(\mu)$  the   multiplicity   of $\mu$
as a weight in $\rho_{\lambda}$.
It will be convenient to introduce $  \bar{m}_{\lambda}: \ct \to \bZ$  by
 \begin{equation}\label{eq_mbar_def}
  \bar{m}_{\lambda}(b) =
\begin{cases} m_{\lambda}(b) & \text{ if } b \in \Lambda\\
0 & \text{ otherwise }
\end{cases}
\end{equation}
Instead of  $\bar{m}_{\lambda}$ we often write  $m_{\lambda}$.

\subsection{Quantum algebraic notation}
\label{appB.2}

Recall that in Sec. \ref{sec2} above we fixed $k \in \bN$.  We set
\begin{equation} \Lambda_+^k :=  \{ \lambda \in \Lambda_+  \mid  \langle \lambda + \rho ,\theta \rangle < k \}
=  \{ \lambda \in \Lambda_+  \mid  \langle \lambda,\theta \rangle\leq k - \cg\}
\end{equation}
In Sec. \ref{sec3} the following formula proved to be very useful
 \begin{equation} \label{eq_rm_fund_Weyl_alcove}
\Lambda^{k}_+ =  \Lambda  \cap (k P - \rho)
\end{equation}
where $P$ is the fundamental Weyl alcove, cf. Eq. \eqref{eq_P_formula} above.

\begin{remark}
In order to see that Eq. \eqref{eq_rm_fund_Weyl_alcove} holds
observe first that according to Eq. \eqref{eq_P_formula} above we have
$$ (k P - \rho)
  =     \{k b - \rho  \mid b \in \CW \text{ and } \langle b ,\theta \rangle < 1 \}
 =      \{\bar{b} \in \ct \mid \bar{b} + \rho \in \CW \text{ and } \langle \bar{b} + \rho,\theta\rangle < k \}
$$
and therefore
$$ \Lambda  \cap (k P - \rho)
 =    \{\lambda \in \Lambda  \mid \lambda + \rho \in \CW \text{ and } \langle \lambda + \rho,\theta\rangle < k \} \overset{(*)}{=}  \{\lambda \in \Lambda \cap \overline{\CW}  \mid \langle \lambda + \rho,\theta\rangle < k \}  = \Lambda^{k}_+$$
Here in step $(*)$ follows because for each $\lambda \in  \Lambda$,
    $\lambda + \rho $ is in the open Weyl chamber $\CW$ iff $\lambda$
    is in the closure $\overline{\CW}$ (cf. the last remark   in Sec. V.4 in \cite{Br_tD}).
    \end{remark}

Let $C$ and $S$  be the $\Lambda_+^k \times \Lambda_+^k$ matrices with complex entries    given by
\begin{subequations} \label{eq_def_C+S}
\begin{align}
  C_{\lambda \mu} & := \delta_{\lambda \bar{\mu}}, \\
\label{eq_def_S} S_{\lambda \mu} & :={i^{\# \cR _{+}}\over k^{\dim(\ct)/2}}  \frac{1}{|\Lambda/\Gamma|^{1/2}}
\sum_{\tau \in \cW} (-1)^{\tau} e^{- {2\pi i\over k} \langle \lambda + \rho , \tau \cdot
(\mu + \rho) \rangle }
\end{align}
\end{subequations}
for all $\lambda, \mu \in \Lambda_+^k$
where $\# \cR _{+}$ is the number of elements of $\cR _{+}$
and $|\Lambda/\Gamma|$ is the index of $\Gamma$ in $\Lambda$.
(Note that according to Proposition 7.16 in Chap. V in \cite{Br_tD}
 $|\Lambda/\Gamma|$ coincides with the order of the center of $G$.)
We have
\begin{equation} \label{eq_S2=C} S^2 = C
\end{equation}

\begin{remark} \label{rm_mod_cat} The matrix $S$ introduced above coincides with the  $S$-matrix of the modular category associated
   to  $U_q(\cG_{\bC})$    with $q:= \exp( \tfrac{2 \pi i}{k})$ (cf. Sec. 1.4 in Chap. II in \cite{turaev})
   or, equivalently, the  $S$-matrix of the WZW model associated to $\Sigma$, $G$,
       and the level $\bar{k} := k - \cg$.
       Moreover, $\Lambda_+^k$ is the set of dominant weights which are ``integrable
       at level $\bar{k}$''.
 \end{remark}

 Let $\theta_{\lambda}$ and $d_{\lambda}$ for  $\lambda \in \Lambda$
 be  given by\footnote{\label{ft_warning}For $r \in \bQ$ we will write $\theta_{\lambda}^r$ instead of
$e^{r \cdot\frac{\pi i}{k} \langle \lambda,\lambda+2\rho \rangle}$.
Note that this notation is somewhat dangerous since $\theta_{\lambda_1} = \theta_{\lambda_2}$
does, of course, in general not imply $\theta_{\lambda_1}^r = \theta_{\lambda_2}^r$.}
\begin{subequations} \label{eq_def_th+d}
\begin{align}
\label{eq_def_th}
 \theta_{\lambda}  & := e^{\frac{\pi i}{k} \langle \lambda,\lambda+2\rho \rangle}\\
 \label{eq_def_d}
d_{\lambda} & := \frac{S_{\lambda 0}}{S_{00}}
\overset{(*)}{=} \prod_{\alpha \in \cR_+}
\frac{\sin(\frac{\pi}{k} \langle \lambda+\rho,\alpha \rangle) }{\sin(\frac{\pi}{k} \langle \rho,\alpha \rangle)}
\end{align}
\end{subequations}
where in Eq. \eqref{eq_def_d} we have  generalized the definition of $S_{\lambda \mu}$ to the situation
of general $\lambda, \mu \in \Lambda$ using again Eq. \eqref{eq_def_S}
and where step $(*)$ follows from the Weyl denominator formula (cf., e.g.,
 part (iii) in Theorem 1.7 in Chap. VI in \cite{Br_tD}).

\smallskip

From Eq. \eqref{eq_appB_det1/2} and  Eq. \eqref{eq_def_d} we obtain for  $\lambda \in  \Lambda_+^k$
\begin{equation} \label{eq_S_lamba0}
\det\nolimits^{1/2}(1_{\ck}-\exp(\ad((\lambda+\rho)/k))_{| \ck}) \sim S_{\lambda 0}
\end{equation}
where $\sim$ denotes equality up to a multiplicative constant which
is independent of $\lambda$.
Moreover, according to Weyl's character formula we have for all $\lambda, \mu \in  \Lambda_+^k$:
 \begin{equation} \label{eq_Weyl_char}
 \Tr_{\rho_{\lambda}}(\exp((\mu + \rho)/k)) = \tfrac{S_{\mu \lambda}}{S_{\mu 0}}
\end{equation}
For  $\lambda, \mu, \nu \in \Lambda_+^k$ we set\footnote{The notation $N_{\lambda \mu \nu}$ is motivated by Eq. \eqref{eq_Verlinde_conj} in Sec. \ref{subsec3.1}.}
\begin{subequations}
 \begin{equation}  \label{eq_def_Verlinde} N_{\lambda \mu \nu}:=
 \sum_{\alpha \in \Lambda_+^k} \frac{ S_{\alpha \lambda}  S_{\alpha \mu}  S_{\alpha \nu} }{S_{\alpha 0}}
 \end{equation}
 \begin{equation} \label{eq_fusion_rules} N_{\lambda \mu}^{\nu}:= N_{\lambda \mu \bar{\nu}}
 \end{equation}
 \end{subequations}
The numbers  $N_{\lambda \mu \nu}$,  $\lambda, \mu, \nu \in \Lambda_+^k$,
 are called
``Verlinde numbers'' and  $N_{\lambda \nu}^{\mu}$, $\lambda, \mu, \nu \in \Lambda_+^k$,
  are  the so-called ``fusion coefficients''.  \par

Observe that the following identity holds (called
  the ``quantum Racah formula'' in \cite{Saw})
\begin{equation}  \label{eq_quantum_racah}
N_{\lambda \nu}^{\mu} = \sum_{\tau \in \cW_{\aff}} (-1)^{\tau} m_{\lambda}\bigl(\mu - \tau \ast \nu\bigr)
\end{equation}
for all $\lambda \in \Lambda_+$,  $\mu, \nu \in \Lambda$ where $\ast: \cW_{\aff} \times \ct \to \ct$  is as in Eq. \eqref{eq_def_ast} above.

\smallskip

Finally, we set for all $\lambda \in \Lambda_+$,  $\mu  \in \Lambda$, and
$\mathbf p \in \bZ \backslash \{0\}$ (cf. Remark \ref{rm_3.3.9_I} in Sec. \ref{subsubsec3.3.8})
\begin{equation}  \label{eq_coef_Rosso_Jones}
c_{\lambda, \mathbf p}^{\mu} := \sum_{\tau \in \cW} (-1)^{\tau} m_{\lambda}\bigl(\tfrac{1}{\mathbf p} (\mu - \tau \cdot \rho + \rho)\bigr)
\end{equation}

\section{Some technical details for Sec. \ref{sec2}}
\label{appA}

We will now fill in some of the technical details omitted in Sec. \ref{sec2} above.
Observe that Sec. \ref{appA.1}, Sec. \ref{appA.3}, and Sec. \ref{appA.4} are rigorous
while in  Sec. \ref{appA.2},  Sec. \ref{appA.5}, and Sec. \ref{appA.6}
 we use several informal arguments.
(Secs \ref{appA.2} and \ref{appA.6} are new,
Secs \ref{appA.3}--\ref{appA.4} are based on \cite{Ha3c}
and Sec. \ref{appA.5} is based both on \cite{Ha3c} and on Appendix B of \cite{Ha7a}.)

\medskip

\subsection{Proof of Proposition \ref{prop2.1}}
\label{appA.1}

{\bf First proof:} Let $f$, $dg$, $\ck$, and $1_{\ck}$  be as in Proposition \ref{prop2.1}.
From Theorem 1.11 in Chap. IV in \cite{Br_tD} we obtain
\begin{equation}  \int_G f(g) dg   \sim  \int_T f(t) \det(1_{\ck} - \Ad(t^{-1})_{|\ck}) dt
\end{equation}
where $dt$ is the  normalized Haar measure on $T$.
Eq. \eqref{eq_WeylInt}  in Proposition \ref{prop2.1} now follows from
$\Ad(\exp(b)^{-1}) = \Ad(\exp(-b)) = \exp(\ad(-b))$, $b \in \ct$,
and the relation  $\det(1_{\ck} - \exp(\ad(-b)))_{|\ck} = \det(1_{\ck} - \exp(\ad(b)))_{|\ck}$,
cf. Eq. \eqref{eq_appB_det} in Appendix \ref{appB} above.

\smallskip

Here is a second proof of Eq. \eqref{eq_WeylInt},
which is   more complicated  than the first one
but which  is a useful preparation for Appendix \ref{appA.2} below.

\smallskip


\smallskip

\noindent {\bf Second proof:} Recall the notation $G_{reg}$ and $\ct_{reg}$  introduced in Appendix \ref{appB.1} above.
Let $P \subset \ct_{reg}$ be a fixed Weyl alcove. We will use the following three observations:
\begin{itemize}

\item The map $q: P \times G/T \ni  (b,\bar{g}) \mapsto  \bar{g} \exp(b) \bar{g}^{-1}  \in G_{reg}$  is a (well-defined) diffeomorphism.
  (Here we set, for each  $\bar{g} \in G/T$ and $t \in T$,
$ \bar{g} t \bar{g}^{-1} := g t g^{-1} \in G$
where $g$ is an arbitrary  element of $G$ fulfilling $gT = \bar{g}$.)
 In order to see this observe first that, according to
  Prop. 7.11 in Sec. V.7 in \cite{Br_tD}
$q$  is a  (connected) smooth covering.
 Moreover, according to Lemma 7.3 in Sec. V.7 in \cite{Br_tD}
 the assumption that $G$ is simply-connected implies that also  $G_{reg}$ is simply-connected. These two observations imply that  $q$ is  a
trivial, connected, smooth covering, i.e.  a diffeomorphism.

\item From Proposition 1.8 in Chap. IV in \cite{Br_tD}
it follows\footnote{Note that Proposition 1.8 in Chap. IV in \cite{Br_tD}
is stated in terms of volume forms on the relevant manifolds. By rewriting it in terms
of Borel measures we obtain Eq. \eqref{eq_transformed_measure}. } that the pushforward measure $(q^{-1})_*(dg)$ of $dg$ under $q^{-1}$ is given by
\begin{equation} \label{eq_transformed_measure} (q^{-1})_*(dg) \sim   \det(1_{\ck} - \exp(\ad(b))_{| \ck}) db \otimes d\bar{g}
\end{equation}
where $d\bar{g}$ is the normalized left-invariant Borel measure on
$G/T$  and where ``$\otimes$'' denotes the product of two measures.

\item  $G \backslash G_{reg}$ is a zero-set w.r.t. to the Haar measure $dg$ on $G$.
This follows, e.g., from the observation in  the proof of
 Lemma 7.3 in Sec. V.7  in \cite{Br_tD} that $G \backslash G_{reg}$
 is contained in a submanifold $N$ of $G$
 with codimension  at least $3$ (in the sense explained there).

\end{itemize}

 From the three points above we now obtain  for every conjugation-invariant
 continuous function $f:G \to \bC$
\begin{align} \label{eq_WeylInt_reg}
\int_G f(g) dg  & =  \int_{G_{reg}} f(g) dg \nonumber \\
& = \int (f \circ q) \ (q^{-1})_* (dg) \nonumber \\
& \sim \int_{P \times G/T} f(q(b,\bar{g})) \
 \det(1_{\ck} - \exp(\ad(b))_{| \ck}) db \otimes d\bar{g} \nonumber \\
 & = \int_{P} \int_{G/T} f(\bar{g} \exp(b) \bar{g}^{-1})
 \det(1_{\ck} - \exp(\ad(b))_{| \ck})  d\bar{g} db \nonumber \\
& = \int_{P} f(\exp(b)) \det(1_{\ck} - \exp(\ad(b))_{| \ck}) db
\end{align}
Finally, observe that the multiplicative constant hidden in the symbol ``$\sim$''
is independent of the Weyl alcove $P$ chosen above.
Since $\ct_{reg}$ is the disjoint union of all Weyl alcoves we obtain from
Eq. \eqref{eq_WeylInt_reg} by a trivial ``averaging'' procedure
$$\int_G f(g) dg \sim  \int^{\sim} 1_{\ct_{reg}}(b) f(\exp(b))
 \det(1_{\ck} - \exp(\ad(b))_{| \ck}) db$$
 where $\int^{\sim} \cdots db$ is as in Proposition \ref{prop2.1} in Sec. \ref{subsubsec2.2.1}.
   Since the complement of $\ct_{reg}$ in $\ct$ is a $db$-zero zet
 we arrive (again) at Eq. \eqref{eq_WeylInt}.

\subsection{Derivation of Eq.  \eqref{eq2.24}}
\label{appA.2}

Our approach  for deriving Eq. \eqref{eq2.24} in Sec. \ref{subsubsec2.2.2}
will be similar (but not totally analogous) to the derivation of Eq. \eqref{eq_AppA1_lang} in Sec. \ref{subsubsec2.2.1} above.

\begin{remark} \label{rm_appA.2}
Here are  two aspects of the derivation of Eq. \eqref{eq2.24}
which are not analogous to the derivation of  Eq. \eqref{eq_AppA1_lang} above
and which are responsible for the  differences  between the RHS of Eq.   \eqref{eq2.24}
and the RHS of (the incorrect) Eq. \eqref{eq2.18} in Sec. \ref{subsubsec2.2.2}:
\begin{itemize}
\item[1.] While the standard left-action of $G$ on $G/T$ is transitive
the left-action  of $\G_{\Sigma} = C^{\infty}(\Sigma,G)$ on $C^{\infty}(\Sigma,G/T)$
 defined in Sec. \ref{subsubsec2.2.2} above  is not transitive if $\Sigma$ is compact.
   In other words: The orbit space $C^{\infty}(\Sigma,G/T)/\G_{\Sigma}$ is then
   non-trivial (cf. part (iii) of Proposition \ref{prop_n(cl)} in Sec. \ref{subsec2.3}).

\item[2.] While the complement of $\ct_{reg}$ in $\ct$ is  a zero set w.r.t. the measure $db$       one cannot argue at an informal level that\footnote{Observe that   $\codim(\ct \backslash \ct_{reg}) = 1$ so a generic smooth map $\Sigma \to \ct$
      will in general not remain inside $\ct_{reg}$.}
 $C^{\infty}(\Sigma,\ct) \backslash C^{\infty}(\Sigma,\ct_{reg})$ is a zero-set w.r.t. the (informal) measure $DB$. Accordingly,  we cannot argue at an informal level that in Eq. \eqref{eq2.24} we can  replace $\int_{C^{\infty}(\Sigma,\ct_{reg})} \cdots DB$
by $\int_{C^{\infty}(\Sigma,\ct)} \cdots DB$.
\end{itemize}
\end{remark}

Let $\chi: \cA \to \bC$ be a (not necessarily $\G$-invariant) function.
Using the observation that $\cA = \cA^{\orth} \oplus \cA^{||}$
(cf.  Eq. \eqref{eq_cA_decomp} in Sec. \ref{subsubsec2.2.2})
 let us first rewrite the LHS of Eq.  \eqref{eq2.24} as
\begin{equation} \label{eq_appA.2_a}
\int_{\cA} \chi(A) DA
  =  \int_{\cA^{||}}  \chi_{red}(A^{||})  DA^{||}
  \end{equation}
 where  $DA^{||}$ is the (informal) Lebesgue measure on $\cA^{||}$
and where we have introduced the function
$\chi_{red}: \cA^{||} \to \bC$ by
\begin{equation} \label{eq_def_chi_red} \chi_{red}(A^{||}) := \int_{\cA^{\orth}} \chi(A^{\orth} + A^{||}) DA^{\orth} \quad \forall A^{||} \in \cA^{||},
 \end{equation}
 $DA^{\orth}$ being the informal Lebesgue measure on $\cA^{\orth}$.
Now consider the $\G$-action on $\cA^{||}$ given by
$$ ( A^{||} \cdot {\Omega})(\sigma) = A^{||}(\sigma) \cdot {\Omega}(\sigma) \quad \forall \sigma \in \Sigma$$
where we have made the  obvious identifications
\begin{subequations}
\begin{align}
\cA^{||} & \cong C^{\infty}(\Sigma,\cA_{S^1}) \\
\G & \cong C^{\infty}(\Sigma,\G_{S^1})
\end{align}
\end{subequations}
where  the two spaces $C^{\infty}(\Sigma,\cA_{S^1})$ and $C^{\infty}(\Sigma,\G_{S^1})$ are defined in a totally analogous way as the space $C^{\infty}(S^1,\cA_{\Sigma})$ in Sec. \ref{subsec2.3} above.

\begin{observation} \label{obs_appA.2} If $\chi$ is $\G$-invariant then
$\chi_{red}$ is $\G$-invariant as well.
\end{observation}
{\it ``Proof'':} If $\chi: \cA \to \bC$ is $\G$-invariant
 we have, informally, for all $A^{||} \in \cA^{||}$ and $\Omega \in \G$
 \begin{align} \label{eq_appA.2_Ende}
\chi_{red}(A^{||} \cdot \Omega) & =   \chi_{red}( {\Omega}^{-1} A^{||} {\Omega}  + ({\Omega}^{-1} d {\Omega})^{||} ) =
  \int_{\cA^{\orth}} \chi(A^{\orth} +
  {\Omega}^{-1} A^{||} {\Omega}  + ({\Omega}^{-1} d {\Omega})^{||}  ) DA^{\orth} \nonumber \\
& \overset{(*)}{=}
 \int_{\cA^{\orth}} \chi({\Omega}^{-1} A^{\orth} {\Omega}  +
({\Omega}^{-1} d {\Omega})^{\orth} +
  {\Omega}^{-1} A^{||} {\Omega}  + ({\Omega}^{-1} d {\Omega})^{||}  ) DA^{\orth} \nonumber \\
& =  \int_{\cA^{\orth}} \chi(( A^{\orth} + A^{||}) \cdot \Omega) DA^{\orth}
=  \int_{\cA^{\orth}} \chi( A^{\orth} + A^{||}) DA^{\orth} = \chi_{red}(A^{||})
\end{align}
where $({\Omega}^{-1} d {\Omega})^{\orth}$ and
$({\Omega}^{-1} d {\Omega})^{||}$
are the components of ${\Omega}^{-1} d {\Omega} \in \cA_{\Sigma} \subset \cA^{\orth}$ w.r.t. the decomposition \eqref{eq_cA_decomp} in Sec. \ref{subsec2.2}
and where in step $(*)$ we have used the
 change of variable\footnote{This change of variable argument
was inspired by a similar argument used in \cite{BlTh3},
cf. Eqs. (6.5) and (6.6) in Sec. 6 of \cite{BlTh3}.}
$A^{\orth} \to   {\Omega}^{-1} A^{\orth} {\Omega}  +
({\Omega}^{-1} d {\Omega})^{\orth}$
and have taken into account that $DA^{\orth}$ is invariant under
this affine transformation whose linear part has, informally, determinant  $1$.
(Note that since $G$ is compact and connected we have
 $\det(\Ad(g))=1$ for all $g \in G$.)  \hfill{$\square$}

\bigskip

From now on we will assume that $\chi$ is $\G$-invariant.
According to Observation \ref{obs_appA.2} above $\chi_{red}$ is then $\G$-invariant as well.
In particular, $\chi_{red}$ is $\tilde{\G}$-invariant where we have
set  $\tilde{\G}   := \{ \Omega \in \G \mid \forall \sigma \in \Sigma:
\Omega(\sigma,1) = 1 \}$. As in Sec. \ref{subsubsec2.2.1}
 it follows that there is a  $\bar{\chi}_{red}: C^{\infty}(\Sigma,G) \to \bC$ such that
 \begin{equation}\chi_{red} = \bar{\chi}_{red} \circ \hat{p}
 \end{equation}
 where  $\hat{p}  :  C^{\infty}(\Sigma, \cA_{S^1}) \to C^{\infty}(\Sigma, G)$
is given by $(\hat{p}(A^{||}))(\sigma)= p(A^{||}(\sigma))$
 for all $A^{||} \in C^{\infty}(\Sigma, \cA_{S^1})$ and  $\sigma \in \Sigma$.
(Here $p$ is as in  Sec. \ref{subsubsec2.2.1}.)
Taking this into account we obtain the following  analogue
of Eq. \eqref{eq_sec2.2.1a} in Sec. \ref{subsubsec2.2.1}
\begin{multline}   \label{eq_appA.2_b}
  \int_{\cA^{||}}  \chi_{red}(A^{||})  DA^{||}
 = \int_{ C^{\infty}(\Sigma,\cA_{S^1})}
 \chi_{red}(A^{||})  DA^{||}  \\
 = \int_{ C^{\infty}(\Sigma,\cA_{S^1})}
 \bar{\chi}_{red}(\hat{p}(A^{||}))  DA^{||}
 \sim \int_{C^{\infty}(\Sigma,G)} \bar{\chi}_{red}(\Omega) D\Omega
 \end{multline}
 where  $D\Omega$ is the (informal) normalized Haar measure  on $C^{\infty}(\Sigma,G) = \G_{\Sigma}$.

 \medskip

Since   $\chi_{red}$ is not only $\tilde{\G}$-invariant but even
 $\G$-invariant we can conclude
 that the function $\bar{\chi}_{red}$ is
 conjugation invariant, i.e. invariant
 under the $\G_{\Sigma}$-action on itself by conjugation.
 We will exploit this in a similar way as we exploited the conjugation invariance
 of the function $f$ appearing in Eq. \eqref{eq_WeylInt_reg} in Appendix \ref{appA.1}
 above. Before we do this we need some preparations.

 \smallskip

 First observe that since, by assumption,  $\dim(\Sigma)=2$,
 and since $G \backslash G_{reg}$
 is contained in a submanifold of $G$
 with codimension  at least $3$ (cf. Appendix \ref{appA.1} above)
  a ``generic'' map  $\Omega:\Sigma \to G$ will remain inside $G_{reg}$ so, informally, the set $C^{\infty}(\Sigma,G) \backslash C^{\infty}(\Sigma,G_{reg})$
can be considered as a  zero-set w.r.t. $D\Omega$. \par

Let $q: P \times G/T \to G_{reg}$ be the diffeomorphism introduced in
Appendix \ref{appA.1} (where $P$ is the Weyl alcove fixed there).
Let  $\hat{q}  :  C^{\infty}(\Sigma, P) \times   C^{\infty}(\Sigma, G/T)  \to C^{\infty}(\Sigma, G_{reg})$
be given by
$\hat{q}(B,\bar{g})(\sigma)= q(B(\sigma),\bar{g}(\sigma))$
 for all $B \in C^{\infty}(\Sigma, P)$, $\bar{g} \in C^{\infty}(\Sigma, G/T)$, and  $\sigma \in \Sigma$.  As $q$ is a diffeomorphism  we conclude that $\hat{q}$ is a bijection.
In analogy with Eq. \eqref{eq_WeylInt_reg} in Appendix \ref{appA.1} above
we now obtain, informally,
\begin{subequations}\label{eq_appA.2_pre_all}
 \begin{multline} \label{eq_appA.2_c_pre}
 \int_{C^{\infty}(\Sigma,G)} \bar{\chi}_{red}(\Omega) D\Omega  = \int_{C^{\infty}(\Sigma,G_{reg})} \bar{\chi}_{red}(\Omega) D\Omega \\
  = \int_{C^{\infty}(\Sigma, P) \times   C^{\infty}(\Sigma, G/T)} (\bar{\chi}_{red} \circ \hat{q}) \ (\hat{q}^{-1})_* D\Omega
\end{multline}
 Observe that Eq. \eqref{eq_transformed_measure} in Appendix \ref{appA.1}
 above now suggests that we have, informally,
 $$ (\hat{q}^{-1})_* D\Omega
 \sim \det\bigl(1_{\ck}-\exp(\ad(B)_{| \ck})\bigr) DB \otimes D\bar{g}
$$
 and therefore
 \begin{multline} \label{eq_appA.2_c_pre2}
 \int_{C^{\infty}(\Sigma, P) \times   C^{\infty}(\Sigma, G/T)} (\bar{\chi}_{red} \circ \hat{q}) \ (\hat{q}^{-1})_* D\Omega \\
  \sim \int_{C^{\infty}(\Sigma, P)}  \int_{C^{\infty}(\Sigma, G/T)}
\bar{\chi}_{red}(\hat{q}(B,\bar{g})) \   \det\bigl(1_{\ck}-\exp(\ad(B)_{| \ck})\bigr)  D\bar{g} DB
\end{multline}
 where $D\bar{g} $ is the (informal) $\G_{\Sigma}$-invariant measure
 on $C^{\infty}(\Sigma, G/T)$
 normalized such that every $\G_{\Sigma}$-orbit
 has volume $1$.

\smallskip

Similarly to the second part of Eq. \eqref{eq_WeylInt_reg} in Appendix \ref{appA.1} above
we  obtain
 \begin{align} \label{eq_appA.2_c}
&\int_{C^{\infty}(\Sigma, P)}  \int_{C^{\infty}(\Sigma, G/T)}
\bar{\chi}_{red}(\hat{q}(B,\bar{g})) \   \det\bigl(1_{\ck}-\exp(\ad(B)_{| \ck})\bigr)  D\bar{g} DB  \nonumber \\
&  = \int_{C^{\infty}(\Sigma, P)} \biggl[ \int_{C^{\infty}(\Sigma, G/T)}
\bar{\chi}_{red}(\bar{g} \ \exp(B) \ \bar{g}^{-1}) D\bar{g} \biggr]  \det\bigl(1_{\ck}-\exp(\ad(B)_{| \ck})\bigr)  DB  \nonumber \\
  & \overset{(*)}{\sim}  \sum_{\cl \in C^{\infty}(\Sigma,G/T)/ \G_{\Sigma}}
 \int_{C^{\infty}(\Sigma,P)} \biggl[ \bar{\chi}_{red}(\bar{g}_{\cl}
  \exp(B)  \bar{g}_{\cl}^{-1} ) \biggr]   \det\bigl(1_{\ck}-\exp(\ad(B)_{| \ck})\bigr)  DB
 \end{align}
\end{subequations}
where  $(\bar{g}_{\cl})_{\cl \in  C^{\infty}(\Sigma,G/T)/ \G_{\Sigma})}$, $\bar{g}_{\cl} \in C^{\infty}(\Sigma,G/T)$, is
 an arbitrary     system
 of representatives of $C^{\infty}(\Sigma,G/T)/ \G_{\Sigma}$
and where, for $t \in C^{\infty}(\Sigma, T)$ and
$\bar{g} \in C^{\infty}(\Sigma, G/T)$, we have denoted by $\bar{g} t\bar{g}^{-1}$
the element of $C^{\infty}(\Sigma, G)$ given by
$(\bar{g} t\bar{g}^{-1})(\sigma):= \bar{g}(\sigma) t(\sigma) \bar{g}(\sigma)^{-1}$ for all $\sigma \in \Sigma$. \par

Note that in  step $(*)$ we used that, as a result of the conjugation invariance of
$\bar{\chi}_{red}$, for each fixed $B$ the function $C^{\infty}(\Sigma, G/T) \ni \bar{g} \mapsto \bar{\chi}_{red}(\bar{g} \ \exp(B) \ \bar{g}^{-1}) \in \bC$
 is constant on each $\G_{\Sigma}$-orbit.

 \bigskip

By combining Eq. \eqref{eq_appA.2_a},  Eq. \eqref{eq_appA.2_b}, Eq. \eqref{eq_appA.2_c_pre},
Eq. \eqref{eq_appA.2_c_pre2},  and Eq. \eqref{eq_appA.2_c} and by taking into account that
$$\bar{\chi}_{red}(\bar{g}_{\cl}
  \exp(B)  \bar{g}_{\cl}^{-1} )
  = \bar{\chi}_{red}(\exp(\bar{g}_{\cl} B
    \bar{g}_{\cl}^{-1})) = \bar{\chi}_{red}(\hat{p}(\bar{g}_{\cl} B
    \bar{g}_{\cl}^{-1} dt)) = \chi_{red}(\bar{g}_{\cl} B
    \bar{g}_{\cl}^{-1} dt)$$
    we now obtain
\begin{multline} \label{eq_appA.2_ende0}
\int_{\cA} \chi(A) DA \sim \sum_{\cl \in C^{\infty}(\Sigma,G/T)/ \G_{\Sigma}}
 \int_{C^{\infty}(\Sigma,P)} \chi_{red}(\bar{g}_{\cl}
  B  \bar{g}_{\cl}^{-1} dt) \det\bigl(1_{\ck}-\exp(\ad(B)_{| \ck})\bigr)  DB
\end{multline}
Observe that the multiplicative constant implicit in $\sim$
is independent of the choice of the Weyl alcove $P$.
Since $\cB_{reg} = C^{\infty}(\Sigma,\ct_{reg})$ is the disjoint union
of $\{ C^{\infty}(\Sigma,P') \mid P' \text{ is a Weyl alcove of } \ct\}$
we now obtain from Eq. \eqref{eq_appA.2_ende0}
by a trivial averaging procedure  analogous to the one used in the last paragraph
 in Appendix \ref{appA.1}
\begin{multline} \label{eq_appB.2_Ende}
\int_{\cA} \chi(A) DA \sim \sum_{\cl \in C^{\infty}(\Sigma,G/T)/ \G_{\Sigma}}
 \int^{\sim} 1_{\cB_{reg}}(B) \chi_{red}(\bar{g}_{\cl}
  B  \bar{g}_{\cl}^{-1} dt) \det\bigl(1_{\ck}-\exp(\ad(B)_{| \ck})\bigr)  DB
\end{multline}
if $\int^{\sim} \cdots DB$ is a suitable (informal) ``improper integral'', cf. Remark \ref{rm_sec2.2.1} and part (ii) of Remark \ref{rm_eigentlich_average} in Sec. \ref{subsec2.2} above. Combining Eq. \eqref{eq_appB.2_Ende} with  Eq. \eqref{eq_def_chi_red}  above
we now arrive at Eq.  \eqref{eq2.24}.

\subsection{Derivation of Eq. \eqref{eq_S_CS_change_of_gauge}}
\label{appA.3}

Recall from Sec. \ref{subsubsec2.1.1} above
 that  if $\alpha$ or $\beta$ is a 0-form we write $\alpha \beta$
instead of $\alpha \wedge \beta$.

\medskip

Let $\cA_{\Sigma \backslash \{\sigma_0\}}:= \Omega^1(\Sigma \backslash \{\sigma_0\},\cG)$
and let $C^{\infty}(S^1, \cA_{\Sigma \backslash \{\sigma_0\}})$ be defined in a completely
analogous way as the space $C^{\infty}(S^1, \cA_{\Sigma})$ in Sec. \ref{subsec2.3} above.
For $A^{\orth} \in C^{\infty}(S^1, \cA_{\Sigma \backslash \{\sigma_0\}})$
 and $B \in C^{\infty}(\Sigma \backslash \{\sigma_0\},\ct)$
let us set
\begin{multline} \label{eq_A1_0} S'_{CS}(A^{\orth} + B dt)
 :=   - k\pi  \int_{S^1} \lim_{\eps \to 0} \biggl[  \int_{\Sigma \backslash  B_{\eps}(\sigma_0)}  \Tr\bigl(A^{\orth}(t) \wedge \bigl(\partial/\partial t + \ad(B) \bigr) \cdot  A^{\orth}(t)\bigr) \\
  -  2  \Tr\bigl(d(A^{\orth}(t)) B\bigr)  \biggr] dt
\end{multline}
if the limit exists. (Above $d$ is the differential for differential forms on $\Sigma$.)
Observe that if we consider $\cA^{\orth} \cong C^{\infty}(S^1, \cA_{\Sigma})$
as a subspace of $C^{\infty}(S^1, \cA_{\Sigma \backslash \{\sigma_0\}})$
and  $\cB= C^{\infty}(\Sigma,\ct)$ as a subspace of $C^{\infty}(\Sigma \backslash \{\sigma_0\},\ct)$
in the obvious way then  from Stokes' Theorem  it follows that
$S'_{CS}(A^{\orth} + B dt ) = S_{CS}(A^{\orth} + B dt )$
it $A^{\orth} \in \cA^{\orth}$ and $B \in \cB$.
Moreover, we also have
 \begin{equation}\label{eq_A1_1}
 S'_{CS}(A^{\orth} + \Omega_{\cl} B  \Omega_{\cl}^{-1} dt ) =
  S_{CS}(A^{\orth} + \Omega_{\cl} B  \Omega_{\cl}^{-1} dt )
 \end{equation}
 where $ S_{CS}(A^{\orth} + \Omega_{\cl} B  \Omega_{\cl}^{-1} dt )$ is given as in Sec. \ref{subsec2.3}.
On the other hand, in general we have
$$S'_{CS}(A^{\orth} \cdot \Omega_{\cl} +   B dt) \neq S_{CS}(A^{\orth} \cdot \Omega_{\cl} +   B dt)$$
where $S_{CS}(A^{\orth} \cdot \Omega_{\cl} +   B dt)$ is given as in Sec. \ref{subsec2.3}
and where ``$\cdot$''  refers to the obvious $\G_{\Sigma \backslash \{\sigma_0\}}$
action on $C^{\infty}(S^1, \cA_{\Sigma \backslash \{\sigma_0\}})$.
In order to see this observe
 \begin{align} \label{eq_A1_2}
&  S'_{CS}(A^{\orth} \cdot \Omega_{\cl} +   B dt) - S_{CS}(A^{\orth}
\cdot \Omega_{\cl} +   B dt) \nonumber \\
   & = 2 \pi k  \lim_{\eps \to 0} \int_{\Sigma \backslash  B_{\eps}(\sigma_0)}
 \bigl( \Tr(d(A^{\orth}_c \cdot \Omega_{\cl})  B) -  \Tr((A^{\orth}_c \cdot \Omega_{\cl}) \wedge dB ) \bigr)  \nonumber \\
  & = 2 \pi k  \lim_{\eps \to 0} \int_{\Sigma \backslash  B_{\eps}(\sigma_0)}
 d(\Tr((A^{\orth}_c \cdot \Omega_{\cl} )  B))   \overset{(*)}{=} 2 \pi k  \lim_{\eps \to 0} \int_{\partial B_{\eps}(\sigma_0)}
 \Tr((A^{\orth}_c \cdot \Omega_{\cl})  B)    \nonumber \\
    & = 2 \pi k \biggl[ \lim_{\eps \to 0} \int_{\partial B_{\eps}(\sigma_0)}
 \Tr(  \Omega^{-1}_{\cl} A^{\orth}_c  \Omega_{\cl}  B)  +  \lim_{\eps \to 0} \int_{\partial B_{\eps}(\sigma_0)}
 \Tr(  \Omega^{-1}_{\cl} d \Omega_{\cl}  B)  \biggr] \nonumber \\
&  \overset{(+)}{=}  2 \pi k \biggl[ 0 + \Tr\bigl( \bigl(\lim_{\eps \to 0} \int_{\partial B_{\eps}(\sigma_0)}  \Omega_{\cl}^{-1} d\Omega_{\cl}\bigr)  B(\sigma_0) \bigr) \biggr]  =    - 2 \pi k \langle n(\cl),  B(\sigma_0) \rangle
 \end{align}
 where we have set
 $A^{\orth}_c := \int \pi_{\ct}(A^{\orth}(t)) dt \in \cA_{\Sigma,\ct}$ and
 where in step $(*)$ in the last line we have used Stokes' Theorem\footnote{\label{ft_opposite_orientation}Accordingly, the orientation
 on $\partial B_{\eps}(\sigma_0)$ is the orientation induced by
 the orientated surface $\Sigma \backslash B_{\eps}(\sigma_0)$,
 i.e. the orientation opposite to the orientation
 induced by the orientated surface $\overline{B_{\eps}(\sigma_0)}$.}.
 (We also used Eq. \eqref{eq_scal_prod_normalization} and Eq. \eqref{eq_def_ncl}.) \par
 Above step $(+)$  holds for all $B$ provided that the lift  $\Omega_{\cl}$ of $\bar{g}_{\Sigma}$   chosen
 in Sec. \ref{subsec2.3}  does not oscillate/vary ``too wildly''\footnote{On the other hand step $(+)$ holds
 even for wildly oscillating $\Omega_{\cl}$ if $B$ is locally constant around
 $\sigma_0$, which is the only case which will be relevant
 later, cf. Remark \ref{rm_comm0} in Appendix \ref{appA.5} below.}
 around $\sigma_0$. (Such a choice of $\Omega_{\cl}$ is always possible.)  \par

Eq. \eqref{eq_S_CS_change_of_gauge} now follows by combining Eq. \eqref{eq_A1_1} and Eq. \eqref{eq_A1_2}
with\footnote{As Eq. \eqref{eq_A1_3} is simpler than Eq. \eqref{eq_S_CS_change_of_gauge}
the reader may wonder  why in Sec. \ref{subsec2.3} we did not work with $S'_{CS}(\cdot)$
instead of $S_{CS}(\cdot)$. The reason for this is explained in Remark \ref{rm_AppA.3}
in Sec. \ref{subsec2.3} above.}
 \begin{equation}\label{eq_A1_3}
  S'_{CS}(A^{\orth} \cdot \Omega_{\cl} +   B dt) =
   S'_{CS}(A^{\orth} + \Omega_{\cl} B  \Omega_{\cl}^{-1} dt)
  \end{equation}

\noindent
{\bf Proof of Eq.  \eqref{eq_A1_3}:} Let $B \in \cB \subset C^{\infty}(\Sigma \backslash \{\sigma_0\},\ct)$
 and $A^{\orth} \in \cA^{\orth}  \cong C^{\infty}(S^1, \cA_{\Sigma}) \subset C^{\infty}(S^1, \cA_{\Sigma \backslash \{\sigma_0\}})$.  Observe that
\begin{equation}
(A^{\orth} \cdot \Omega_{\cl})(t) = A^{\orth}(t) \cdot \Omega_{\cl}
\end{equation}
where the ``$\cdot$'' on the RHS  refers to the obvious $\G_{\Sigma \backslash \{\sigma_0\}}$ action on $\cA_{\Sigma \backslash \{\sigma_0\}}$.
Taking this into account  we obtain
\begin{align*} & S'_{CS}(A^{\orth} \cdot \Omega_{\cl} + B dt ) \\
& =  -  k\pi \lim_{\eps \to 0}  \int_{S^1} \biggl[  \int_{\Sigma \backslash   B_{\eps}(\sigma_0)}
\bigl[ \Tr\bigl((A^{\orth}(t) \cdot \Omega_{\cl}) \wedge \bigl(\partial/\partial t + \ad(B) \bigr) \cdot
 (A^{\orth}(t) \cdot \Omega_{\cl})\bigr) \\
&  \quad -  2  \Tr(d(A^{\orth}(t) \cdot \Omega_{\cl})  B) \bigr]  \biggr] dt \\
& \overset{(*)}{=} - k\pi \lim_{\eps \to 0}  \int_{S^1} \biggl[  \int_{\Sigma \backslash   B_{\eps}(\sigma_0)}
\bigl[\Tr\bigl(A^{\orth}(t)  \wedge \bigl(\partial/\partial t + \ad(\Omega_{\cl} B \Omega_{\cl}^{-1}) \bigr) \cdot
 A^{\orth}(t)\bigr) -2 \Tr(d(A^{\orth}(t)) \Omega_{\cl} B \Omega_{\cl}^{-1})) \\
& \quad +  \Tr(    d \Omega_{\cl}  \Omega_{\cl}^{-1}\wedge \tfrac{\partial}{\partial t}   A^{\orth}(t))  \bigr]  \biggr] dt \\
& = S'_{CS}(A^{\orth}  +  \Omega_{\cl} B  \Omega_{\cl}^{-1}dt )
  - k\pi \lim_{\eps \to 0} \int_{S^1} \frac{d}{dt} \biggl[ \int_{\Sigma \backslash B_{\eps}(\sigma_0)}
 \Tr(  d\Omega_{\cl}  \Omega_{\cl}^{-1}\wedge    A^{\orth}(t) )\biggr] dt \\
& =  S'_{CS}(A^{\orth}  +  \Omega_{\cl} B  \Omega_{\cl}^{-1}dt ) -  k\pi \lim_{\eps \to 0} \bigl[ 0 \bigr] = S'_{CS}(A^{\orth}  +  \Omega_{\cl} B  \Omega_{\cl}^{-1}dt )
\end{align*}
Here step $(*)$ follows -- setting $\Omega:= \Omega_{\cl}$ --
from
\begin{align*}
& \Tr\bigl((A^{\orth}(t) \cdot \Omega) \wedge \bigl(\partial/\partial t + \ad(B) \bigr) \cdot
 (A^{\orth}(t) \cdot \Omega)\bigr) \\
& =  \Tr(\Omega^{-1} A^{\orth}(t) \Omega \wedge
 ( \tfrac{\partial}{\partial t} + \ad(B)) \cdot \Omega^{-1} A^{\orth}(t) \Omega)  \\
& \quad  +
  \bigl\{ \Tr(  \Omega^{-1} A^{\orth}(t) \Omega \wedge
  \ad(B)  \cdot \Omega^{-1}  d \Omega )  +
  \Tr(   \Omega^{-1}  d \Omega \wedge ( \tfrac{\partial}{\partial t} + \ad(B))  \cdot
   \Omega^{-1} A^{\orth}(t) \Omega)   \bigr\}\\
& \quad +  \Tr( \Omega^{-1}  d \Omega \wedge \ad(B )  \cdot
 \Omega^{-1}  d \Omega) \\
& =  \Tr\bigl(A^{\orth}(t)  \wedge \bigl(\partial/\partial t + \ad(\Omega B \Omega^{-1}) \bigr) \cdot
 A^{\orth}(t)\bigr) \\
& \quad  +
  \bigl\{2  \Tr(   A^{\orth}(t)  \wedge  \ad(\Omega B \Omega^{-1})  \cdot   d \Omega \ \Omega^{-1})
  +   \Tr(    d \Omega  \Omega^{-1}\wedge \tfrac{\partial}{\partial t}   A^{\orth}(t) )  \bigr\}\\
 & \quad - 2 \Tr(d\Omega  \Omega^{-1} \wedge  d\Omega  B  \Omega^{-1})
\end{align*}
 and
\begin{align*}
&  \Tr(d(A^{\orth}(t) \cdot \Omega)  B)  \\
 & = \Tr( d(\Omega^{-1} A^{\orth}(t) \Omega)  B )
 +  \Tr(d(\Omega^{-1}  d \Omega)   B )\\
 & = \Tr( (- \Omega^{-1} d\Omega \Omega^{-1} ) \wedge A^{\orth}(t) \Omega  B)
   +  \Tr(\Omega^{-1} d(A^{\orth}(t)) \Omega B )  -
 \Tr( \Omega^{-1} A^{\orth}(t) \wedge d\Omega  B)  \\
 & \quad +   \Tr(d(\Omega^{-1}  d \Omega)  B) \\
 & = \Tr(\Omega^{-1} d(A^{\orth}(t)) \Omega B ))  +
   \Tr(   A^{\orth}(t)  \wedge  (\ad(\Omega B \Omega^{-1})  \cdot   d \Omega \ \Omega^{-1})) \\
 & \quad   \Tr(- ( \Omega^{-1} d\Omega  \wedge   \Omega^{-1} d\Omega) B)\\
  & = \Tr(d(A^{\orth}(t)) \Omega B \Omega^{-1})  +
   \Tr(   A^{\orth}(t)  \wedge  (\ad(\Omega B \Omega^{-1})  \cdot
    d \Omega \ \Omega^{-1})) \\
 & \quad -  \Tr(d\Omega  \Omega^{-1} \wedge  d\Omega B  \Omega^{-1}  )
 \end{align*}
\hfill $\square$

\subsection{Proof of Proposition \ref{prop_n(cl)}}
\label{appA.4}

\noindent {\bf Proof of part (i):} Let $\cl$, $\bar{g}_{\cl}$, and $ \Omega_{\cl}$ be as in Sec. \ref{subsec2.3}
and set  $\bar{g} := \bar{g}_{\cl}$ and $\Omega := \Omega_{\cl}$.
Let $s$ be a (smooth) local section  of the bundle
   $\pi_{G/T}: G \to G/T$ such that $\bar{g}(\sigma_0) \in \dom(s)$
   and let $U$ be an  open neighborhood of  $\sigma_0$
fulfilling $\Image(\bar{g}_{|U }) \subset \dom(s)$.
   Then  $\Omega_0:= s \circ \bar{g}_{|U } \in C^{\infty}(U,G)$    and
   there is a (unique) $t \in C^{\infty}(U \backslash \{\sigma_0\}, T)$
   such that   $  \Omega_{|U \backslash \{\sigma_0\}} = (\Omega_0)_{|U \backslash \{\sigma_0\}} \cdot t$.
  For sufficiently small $\eps > 0$ we have
   $\overline{B_{\eps}(\sigma_0)} \subset U $   and\footnote{Recall that  $\pi_{\ct}:\cG \to \ct$ is the orthogonal projection
   w.r.t. the $\Ad$-invariant scalar product $\langle \cdot,\cdot \rangle$ on $\cG$.
   From this it follows that $\pi_{\ct}(t^{-1}at) = \pi_{\ct}(a)$ for all $a \in \cG$ and $t \in T$.}
\begin{equation} \label{eq_appA.4} \int_{\partial B_{\eps}(\sigma_0)} \pi_{\ct} (\Omega^{-1} d\Omega)
  =  \int_{\partial B_{\eps}(\sigma_0)} \pi_{\ct}\bigl(
    \Omega_0^{-1} d\Omega_0 \bigr)  +  \int_{\partial B_{\eps}(\sigma_0)} t^{-1} dt
    \end{equation}
    Since $d\Omega_0$  is bounded\footnote{Here the notion ``bounded''
     is defined in terms of the Riemannian metric $\mathbf g$, cf.  Footnote \ref{ft_bounded} below. }  on  the  (compact) set   $\overline{B_{\eps}(\sigma_0)}$      we  have
     $ \lim_{\eps \to 0} \int_{\partial B_{\eps}(\sigma_0)} \pi_{\ct}\bigl(
    \Omega_0^{-1} d\Omega_0 \bigr) = 0$.
On the other hand, $\int_{\partial B_{\eps}(\sigma_0)} t^{-1} dt$
   is independent\footnote{Observe that in contrast to the situation in
  the proof of part (ii) below we cannot conclude $\int_{\partial B_{\eps}(\sigma_0)} t^{-1} dt = 0 $
  as $t$ is only defined on $U \backslash \{\sigma_0\}$ and not on all of $\Sigma \backslash \{\sigma_0\}$.} of $\eps$. (This follows, e.g., from Stokes' theorem and
   $d  (t^{-1} dt) = - t^{-1} dt \wedge t^{-1} dt = 0$.)
  We conclude  that $\lim_{\eps \to 0}  \int_{\partial B_{\eps}(\sigma_0)} \pi_{\ct} (\Omega^{-1} d\Omega)$   exists.    \hfill $\square$

 \smallskip

\noindent {\bf Proof of part (ii):} Let $\cl$  and $ \Omega_{\cl}$ be as in Sec. \ref{subsec2.3}.
Moreover, let $\Omega$ be an arbitrary element of $\G_{\Sigma} = C^{\infty}(\Sigma,G)$
  and $t$ an arbitrary element of  $C^{\infty}(\Sigma \backslash \{\sigma_0\}, T)$.
  Finally, let us   set $\Omega'_{\cl} := \Omega \Omega_{\cl} t$. The assertion then  follows easily from
$$\lim_{\eps \to 0} \int_{\partial B_{\eps}(\sigma_0)}  \pi_{\ct}\bigl( (\Omega'_{\cl})^{-1} d\Omega'_{\cl}\bigr)
\overset{(*)}{=}  \lim_{\eps \to 0} \int_{\partial B_{\eps}(\sigma_0)}  \pi_{\ct}\bigl( (\Omega_{\cl} t)^{-1} d(\Omega_{\cl} t)\bigr) \overset{(**)}{=}  \lim_{\eps \to 0} \int_{\partial B_{\eps}(\sigma_0)}  \pi_{\ct}\bigl( \Omega_{\cl}^{-1} d\Omega_{\cl}\bigr)$$
Here step $(*)$ follows from a short computation taking into account that
 $\Omega$ is defined an all of $\Sigma$ (and $d\Omega$ is therefore bounded)
and step $(**)$ follows from an argument analogous to the one in Eq. \eqref{eq_appA.4} above and by taking into account that
   $\int_{\partial B_{\eps}(\sigma_0)} t^{-1} dt = - \int_{\Sigma \backslash B_{\eps}(\sigma_0)} t^{-1} dt \wedge t^{-1} dt =  0$  by Stokes' theorem.
   \hfill $\square$

 \bigskip

Let $ \cl \in C^{\infty}(\Sigma,G/T)/\G_{\Sigma}$. It is not difficult to see that
$\cl$ has a representative $\bar{g}_{\cl} \in C^{\infty}(\Sigma,G/T)$ which is constant on a neighborhood $U$
of $\sigma_0$ taking only the value $T \in G/T$ there.
Observe that every lift $ \Omega_{\cl} \in C^{\infty}(\Sigma \backslash \{\sigma_0\},G)$ of
$(\bar{g}_{\cl})_{|\Sigma \backslash \{\sigma_0\}}$  will then only take values in $T$
on $U \backslash \{\sigma_0\}$.  We will call  a $\bar{g}_{\cl} \in C^{\infty}(\Sigma,G/T)$ and a $\Omega_{\cl}$  with the properties just described
``a standard representative of $\cl$'' and ``a standard lift associated to $\cl$'', respectively.

 \begin{observation}  \label{obs_appA.4_1}
 Let $\cl\in C^{\infty}(\Sigma,G/T)/\G_{\Sigma}$ and let
 $\Omega_{\cl}$ be a standard lift associated to $\cl$. Then for sufficiently small
  $\eps >0$ we have
   $t:= (\Omega_{\cl})_{| \partial B_{\eps}(\sigma_0)} \in C^{\infty}(\partial B_{\eps}(\sigma_0),T)$ and
$$ \text{ $n(\cl)= \int_{\partial B_{\eps}(\sigma_0)} t^{-1} dt$} $$
 Moreover, every  $\Omega \in C^{\infty}(\Sigma \backslash \{\sigma_0\},G)$ which on a neighborhood of $\sigma_0$
    only takes values in $T$ is a standard lift associated to some $\cl\in C^{\infty}(\Sigma,G/T)/\G_{\Sigma}$.

\end{observation}

\begin{observation} \label{lem2} Let $\psi:  C^{\infty}(S^1,T) \ni t \mapsto \int_{S^1} t^{-1} dt \in \ct$.
Then we have:
\begin{itemize}
\item  $\Image(\psi) = I=  \ker(\exp_{| \ct})$
\item $\psi(t_1)= \psi(t_2)$ for $t_1,t_2 \in C^{\infty}(S^1,T)$ implies
that $t_1$ and $t_2$ are homotopic.
\end{itemize}
\end{observation}

\noindent {\bf Proof of part (iii):}
 From Observation \ref{obs_appA.4_1} and Observation \ref{lem2}  it follows immediately that  $\Image(n) \subset I$.
From Observation \ref{obs_appA.4_1} and Observation \ref{lem2}
 it will  also follow that  $\Image(n) \supset I$ provided that we can show that
for all sufficiently small $\eps > 0$ every smooth map
 $t: \partial B_{\eps}(\sigma_0) \to T$  can be extended
 smoothly to a map $\Omega:\Sigma \backslash \{\sigma_0\} \to G$
  which  on $\overline{B_{\eps}(\sigma_0)} \backslash \{\sigma_0\}$
takes only values in $T$.   But this follows easily from the assumption that  $G$ is simply-connected.

\smallskip

It remains to be shown that  $n$ is injective.
 Let $\cl_1,\cl_2 \in C^{\infty}(\Sigma,G/T)/\G_{\Sigma}$ with $n(\cl_1)=n(\cl_2)$.
 For $i=1,2$ let
   $\bar{g}_{i}$ be a standard representative of $\cl_i$ and
 $\Omega_i$ be   lift of $(\bar{g}_{\cl})_{|\Sigma \backslash \{\sigma_0\}}$
 (and hence  a standard lift associated to $\cl_i$).
  For sufficiently small $\eps >0$
 we then have   $t_i := (\Omega_i)_{|B_{\eps}(\sigma_0) \backslash \{\sigma_0\}} \in C^{\infty}(B_{\eps}(\sigma_0) \backslash \{\sigma_0\} ,T)$ and $t:= t_1  t^{-1}_2 \in C^{\infty}(B_{\eps}(\sigma_0) \backslash \{\sigma_0\} ,T)$.
  Observation \ref{obs_appA.4_1} then implies that
\begin{align*}
\int_{\partial B_{\eps}(\sigma_0)} t^{-1} dt & = \int_{\partial B_{\eps}(\sigma_0)} t_1^{-1} dt_1 - \int_{\partial B_{\eps}(\sigma_0)} t_2^{-1} dt_2  = n(\cl_1) - n(\cl_2) = 0
\end{align*}
According to  Observation  \ref{lem2}  $t_{|\partial B_{\eps}(\sigma_0)}$ is
null-homotopic, which  implies that $t:B_{\eps}(\sigma_0) \backslash \{\sigma_0\} \to T$  can  be extended smoothly
to a  map $\overline{t}: \Sigma \backslash \{\sigma_0\} \to T$.
Clearly, $\Omega:= \Omega_1  \overline{t}^{-1} \Omega_2^{-1}
 \in \G_{\Sigma \backslash \{\sigma_0\}}$
 is locally constant around $\sigma_0$
 and can therefore  be extended  to an element
$\overline{\Omega}$ of $\G_{\Sigma}$ in a trivial way.
 So we finally obtain $\overline{\Omega} \bar{g}_2= \bar{g}_1$ and therefore   $\cl_1= \cl_2$.
 \hfill $\square$.

\begin{remark}
 In view of the relation $C^{\infty}(\Sigma,G/T)/\G_{\Sigma} \cong [\Sigma,G/T]$  (cf. Remark \ref{rm2.1} above) and the relation
  $  n(\cl) = \lim_{\eps \to 0} \int_{\partial B_{\eps}(\sigma_0)}  \pi_{\ct}\bigl( \Omega_{\cl}^{-1} d\Omega_{\cl}\bigr) = \pm \int_{\Sigma \backslash \{\sigma_0\}} \pi_{\ct}\bigl( \Omega_{\cl}^{-1} d\Omega_{\cl} \wedge \Omega_{\cl}^{-1} d\Omega_{\cl} \bigr)$
 where we have set
 $ \int_{\Sigma \backslash \{\sigma_0\}} \pi_{\ct}\bigl( \Omega_{\cl}^{-1} d\Omega_{\cl} \wedge \Omega_{\cl}^{-1} d\Omega_{\cl} \bigr) :=  \lim_{\eps \to 0} \int_{\Sigma \backslash  B_{\eps}(\sigma_0)} \pi_{\ct}\bigl( \Omega_{\cl}^{-1} d\Omega_{\cl} \wedge \Omega_{\cl}^{-1} d\Omega_{\cl} \bigr)$
 it is clear that  Proposition \ref{prop_n(cl)} is very closely related
 to the argument  in  Sec. 5 in  \cite{BlTh3} (cf., in particular Eq. (5.5) in \cite{BlTh3}),
 which involves  the Kirillov-Kostant symplectic forms on
   the regular coadjoint orbits of $G$ (each of which can be identified with $G/T$) and the winding numbers of their pull-backs on $\Sigma$
 via  smooth maps   $\bar{g} : \Sigma \to G/T$.
 By elaborating this argument in a suitable way it should not be difficult
 to obtain an alternative (but less direct) proof of Proposition \ref{prop_n(cl)}.
\end{remark}

\subsection{Justification of the change of variable $A^{\orth} \to A^{\orth} \cdot \Omega_{\cl}^{-1}$ in Eq. \eqref{eq_change_of_variable}}
\label{appA.5}

 Let $\cl \in C^{\infty}(\Sigma,G/T)/\G_{\Sigma}$ and $B \in \cB$ be fixed.
 In the following we will justify  the change of variable
 $A^{\orth} \to A^{\orth} \cdot \Omega_{\cl}^{-1}$ or,
 equivalently,  $ A^{\orth} \cdot \Omega_{\cl} \to A^{\orth}$
appearing in  step $(*)$ in Eq. \eqref{eq_change_of_variable} in Sec. \ref{subsec2.3} above.
 First observe that in view of
 $ A^{\orth} \cdot \Omega_{\cl} =  \Omega_{\cl}^{-1} A^{\orth} \Omega_{\cl} +
  \Omega^{-1}_{\cl} d \Omega_{\cl}  $
   the change of variable  $ A^{\orth} \cdot \Omega_{\cl} \to A^{\orth}$
   in  step $(*)$ in Eq. \eqref{eq_change_of_variable}
   can be realized by performing the following  two   changes of variable
   one after the other:
 \begin{description}
 \item[(CoV1)]  $\Omega_{\cl}^{-1} A^{\orth} \Omega_{\cl} \to A^{\orth} $,
 \item[(CoV2)] $A^{\orth} + \Omega^{-1}_{\cl} d \Omega_{\cl} \to A^{\orth} $.
 \end{description}
 It is enough to justify each of these two changes of variable separately.

 \medskip

 \noindent
 {\bf Justification of (CoV1):} It is a standard procedure in Constructive Quantum Field Theory (CQFT)
  for making rigorous sense of a given informal path integral expression to
    extend  the ``path space'' in a suitable way, the implicit assumption being that this does not change     the value of the corresponding path integral, cf. Remark \ref{rm_extended_spaces_CQFT} below.  \par

  In order to justify (CoV1) we can do something similar.
  We replace the space $\cA^{\orth} \cong C^{\infty}(S^1,\cA_{\Sigma})$ appearing
  in the second line in Eq. \eqref{eq_change_of_variable}
  by the extended space
  $$\overline{\cA^{\orth}}:= C^{\infty}(S^1,\overline{\cA_{\Sigma}}) $$
  where $\overline{\cA_{\Sigma}}$ is a suitable extension of the space
  $\cA_{\Sigma}$. For example, we can choose $\overline{\cA_{\Sigma}}$
  to be the space of $\cG$-valued 1-forms $A_c$ on $\Sigma$ which are
   smooth on $\Sigma \backslash \{\sigma_0\}$
          and   bounded in a suitable sense\footnote{\label{ft_bounded}For example, $\|\cdot\|_{\infty}$-bounded
  where $\|\cdot\|_{\infty}$ is the norm given by
  $\|A\|_{\infty} := \sup_{\sigma \in \Sigma \backslash \{\sigma_0\}}
  \|A_{\sigma}\|_{\sigma,{\mathbf g}}$ where
   $\|\cdot\|_{\sigma,{\mathbf g}}$ is the norm on
  $\Hom(T_{\sigma} \Sigma, \cG)$ induced
  by   the  Riemannian metric $\mathbf g$ on $\Sigma$
  and  $\langle \cdot, \cdot \rangle$.}.
  (Observe that the integrand in the second line in Eq. \eqref{eq_change_of_variable}
    makes sense for every $A^{\orth} \in \overline{\cA^{\orth}}$.) \par

  Now the justification of the change of variable (CoV1) is straightforward since $A^{\orth} \to \Omega_{\cl} A^{\orth} \Omega_{\cl}^{-1}$ is a well-defined linear transformation of $\overline{\cA^{\orth}}$ whose determinant  equals  $1$, informally.
  (Recall that  since $G$ is compact and connected we have
 $\det(\Ad(g))=1$ for all $g \in G$.)
  After carrying out (CoV1)   we replace $\overline{\cA^{\orth}}$
  again by $\cA^{\orth}$ (assuming again that by doing so the value of the path integral does not change). By doing so we arrive at the expression
  \begin{multline} \label{eq_after_CoV1}
   \exp\bigl( - 2 \pi i k  \langle  n(\cl), B(\sigma_0)\rangle \bigr)
  \times   \int_{\cA^{\orth}}
   \left( \prod_{i=1}^m  \Tr_{\rho_i}(\Hol_{l_i}(A^{\orth}
     + A_{\sing}(\cl)   +
 B dt)) \right)  \\
 \times \exp(iS_{CS}(A^{\orth} +  A_{\sing}(\cl) +
 B dt)) DA^{\orth}
 \end{multline}
  where we have set $ A_{\sing}(\cl) :=  \Omega^{-1}_{\cl} d \Omega_{\cl}$.

\begin{remark} \label{rm_extended_spaces_CQFT}
The extended spaces used in CQFT are usually  rather large, for example
they often consist of  spaces   of distributions (or distributional forms) while the original spaces usually consist of smooth functions/forms.
By contrast, the replacement   $\cA_{\Sigma} \to \overline{\cA_{\Sigma}}$
used for the change of variable (CoV1) above  is quite modest. \par

Another difference  with respect to the standard procedure in CQFT is that here we only extend the space    $\cA^{\orth}$ ``temporarily'', i.e. even though we initially make the replacement $\cA^{\orth} \to \overline{\cA^{\orth}}$ we later go back\footnote{On the other hand,   in   the ``continuum  framework'' (F2) for making rigorous sense of $Z^{t.g.f}(\Sigma \times S^1,L)$
 (cf.  Sec. \ref{subsubsec4.2.1} above)
  we later replace the space  $\cA^{\orth}$  again by a modified space which  then is
 a large extension of $\cA^{\orth}$.} from   $\overline{\cA^{\orth}}$ to $\cA^{\orth}$, assuming again that    this will not change the value of the corresponding path integral.
\end{remark}

Note that the change of variable (CoV1) above, even though it is not a well-defined   transformation of $\cA^{\orth}$,
 does not involve any singularities but only points of discontinuity. This made it easy to use a ``temporary extension of space''-argument
(cf.  Remark \ref{rm_extended_spaces_CQFT} above)  for the justification of (CoV1).
 By contrast, the 1-form  $A_{\sing}(\cl) =  \Omega^{-1}_{\cl} d \Omega_{\cl}$ appearing in the  change of variable  (CoV2)
   has a singularity in the point $\sigma_0$
    and the map  $A^{\orth} \to A^{\orth} - A_{\sing}(\cl)$ is therefore
      far from being a well-defined transformation of $\cA^{\orth}$.
     Accordingly,     we can   not be sure that the  naive change of variable (CoV2) will lead to
    the correct result. (Remark \ref{rm_AppA.3} in Sec. \ref{subsec2.3}  above
    and Remark \ref{rm_comm1} below illustrate how things could go wrong.)
    One could still find a way to justify the change of variable  (CoV2)
   with the help of a suitable ``extension of space''-argument\footnote{In contrast to the
   2-form $d  A_{\sing}(\cl)$, appearing in  the change of variable (CoV2)'
   in Footnote \ref{ft_CoV2'} below, the 1-form  $A_{\sing}(\cl)$ appearing in
   (CoV2) is locally integrable.}
    but it is safer to use the following  more careful argument.

\medskip

 \noindent
 {\bf Justification of (CoV2):}    Let $U \subset \Sigma$ be an open neighborhood of $\sigma_0$  such that
  $$U \subset \Sigma \backslash (\bigcup_{j=1}^m \arc(l^j_{\Sigma})).$$
   Moreover,
  choose a pair $(V,\varphi)$ where  $V$ is an  open neighborhood of $\sigma_0$ with $\overline{V} \subset U$
 and where $\varphi$ is a smooth function $\Sigma \to [0,1]$ fulfilling
$$\varphi \equiv 1 \quad \text{ on $V$} \quad \quad \text{ and } \quad \quad  \varphi \equiv 0 \quad \text{on $\Sigma \backslash U$}$$
From the assumptions above it follows that
for all $A^{\orth} \in \cA^{\orth}$ we have
\begin{equation}  \label{eq_appB'_2a0} \Tr_{\rho_i}\bigl( \Hol_{l_i}\bigl(A^{\orth} + A_{\sing}(\cl) + Bdt\bigr)\bigr)  \\
=\Tr_{\rho_i}\bigl( \Hol_{l_i}\bigl(A^{\orth} + (1-\varphi) A_{\sing}(\cl) + Bdt\bigr) \bigr)
\end{equation}
We can consider
 $(1-\varphi) A_{\sing}(\cl) \in \cA_{\Sigma \backslash \{\sigma_0\}}$  as an element of  $\cA_{\Sigma} \subset \cA^{\orth}$ (by trivially extending $(1-\varphi) A_{\sing}(\cl)$ in the point $\sigma_0$).
Accordingly, we obtain
\begin{align}  \label{eq_appB'_2a}
& \int_{\cA^{\orth}}   \left( \prod_{i=1}^m  \Tr_{\rho_i}(\Hol_{l_i}( A^{\orth} + A_{\sing}(\cl) +
 B dt)) \right)    \exp(iS_{CS}( A^{\orth} + A_{\sing}(\cl) +B dt))   DA^{\orth} \nonumber \\
& = \int_{\cA^{\orth}}   \left( \prod_{i=1}^m  \Tr_{\rho_i}(\Hol_{l_i}( A^{\orth} + (1- \varphi) A_{\sing}(\cl) +
 B dt)) \right)    \exp(iS_{CS}( A^{\orth} + A_{\sing}(\cl) +B dt))   DA^{\orth}  \nonumber \\
& \overset{(*)}{=} \int_{\cA^{\orth}}   \left( \prod_{i=1}^m  \Tr_{\rho_i}(\Hol_{l_i}( A^{\orth}  +
 B dt)) \right)    \exp(iS_{CS}( A^{\orth}  + \varphi A_{\sing}(\cl) +B dt))   DA^{\orth}  \nonumber \\
 & =  \biggl[ \int_{\cA^{\orth}} \prod_i  \Tr_{\rho_i}\bigl( \Hol_{l_i}(A^{\orth} + Bdt)\bigr)    \exp(i  S_{CS}( A^{\orth} +B dt))  DA^{\orth} \biggr]  \times \nonumber\\
  & \hspace{8cm} \times  \exp\biggl( i  2 \pi k \int_{\Sigma \backslash
\{\sigma_0\}} \! \! \Tr\bigl(\varphi A_{\sing}(\cl) \wedge dB\bigr)\biggr)
\end{align}
where in step $(*)$ we applied
 the informal change of variable $A^{\orth} \to A^{\orth} - (1-\varphi) A_{\sing}(\cl)$
(which is now justified since  $(1-\varphi) A_{\sing}(\cl)$ is an element of  $\cA^{\orth}$,
cf. the paragraph after Eq. \eqref{eq_appB'_2a0} above). \par

Observe that Eq. \eqref{eq_appB'_2a} holds for all $U$, $V$, and $\varphi$ satisfying the assumption above.
Moreover, the term $ \int_{\Sigma \backslash
\{\sigma_0\}} \! \! \Tr\bigl(\varphi A_{\sing}(\cl) \wedge  dB\bigr)
= \int_{\Sigma \backslash
\{\sigma_0\}} \! \! \Tr\bigl(A_{\sing}(\cl) \wedge \varphi dB\bigr)$
can be made arbitrarily small by choosing $\supp(U)$ small enough.
From this we conclude that
\begin{multline}  \label{eq_appB'_2b}
 \int_{\cA^{\orth}}   \left( \prod_{i=1}^m  \Tr_{\rho_i}(\Hol_{l_i}( A^{\orth} + A_{\sing}(\cl) +
 B dt)) \right)    \exp(iS_{CS}( A^{\orth} + A_{\sing}(\cl) +B dt))   DA^{\orth}  \\
=  \int_{\cA^{\orth}} \prod_{i=1}^m  \Tr_{\rho_i}\bigl( \Hol_{l_i}(A^{\orth} + Bdt)\bigr)    \exp(i  S_{CS}( A^{\orth} +B dt))  DA^{\orth}
\end{multline}

(Note that this is exactly the formula which one would obtain by performing the
 naive change of variable (CoV2).)
Applying Eq. \eqref{eq_appB'_2b}
to the expression \eqref{eq_after_CoV1} above
we then obtain the last expression in Eq. \eqref{eq_change_of_variable}.

\begin{remark} \label{rm_comm0}
  Observe that Eq. \eqref{eq_appB'_2a} and Eq. \eqref{eq_appB'_2b} taken together seem to imply that
 \begin{equation} \label{eq_mock_contra} \exp\biggl( i  2 \pi k \int_{\Sigma \backslash
\{\sigma_0\}} \! \! \Tr\bigl(\varphi A_{\sing}(\cl) \wedge dB\bigr)\biggr) = 1,
\end{equation}
which would be a contradiction.
However, a closer look shows that  Eq. \eqref{eq_appB'_2a} and Eq. \eqref{eq_appB'_2b} together
in fact only imply that  Eq. \eqref{eq_mock_contra} holds {\it unless} the integral on the RHS of Eq. \eqref{eq_appB'_2b} vanishes.
In Appendix B.2 in \cite{Ha7a} it is shown (on an informal level) that the RHS of Eq. \eqref{eq_appB'_2b}
indeed vanishes unless\footnote{Observe that there is an obvious complication here:
Those $B$ which have this property will actually not be smooth (unless they are constant)
and will therefore not be well-defined elements of $\cB$. This complication can be
eliminated if we advance  the switch from loops to closed ribbons (and regularized
 ribbon holonomies) from Secs \ref{subsubsec3.2.1} and \ref{subsubsec3.2.3} to Sec. \ref{subsec2.3}. More precisely,
 the introduction of (smeared) ribbon holonomies (which we consider as a regularization
of the original loop holonomies) must take place after we have applied
Eq. \eqref{eq_Tr_Hol_change_of_gauge} since in Eq. \eqref{eq_Tr_Hol_change_of_gauge} we use an argument based on $\G$-invariance, which is only valid for loop holonomies
(cf. the second bullet point after Eq. \eqref{eq2.48_ribbon}  in Sec. \ref{subsubsec3.2.1}).}
 $B$ is constant on each connected component $Y$ of
$\Sigma \backslash ( \bigcup_{j=1}^m \arc(l^j_{\Sigma}))$.
But in this case Eq. \eqref{eq_mock_contra} does hold so there is no contradiction.
\end{remark}

\begin{remark} \label{rm_comm1}
  In Sec. \ref{subsec2.3}, instead of working with the expression $S_{CS}(A^{\orth}+B dt)$ given by Eq. \eqref{eq_S_CS_improper} in Sec. \ref{subsec2.3},
we could have decided to work with the expression $S'_{CS}(A^{\orth}+B dt)$ given by Eq. \eqref{eq_A1_0}.
Then instead of  Eq. \eqref{eq_S_CS_change_of_gauge} in Sec. \ref{subsec2.3} we would use Eq. \eqref{eq_A1_3} above and, accordingly, the first equation in  Eq. \eqref{eq_change_of_variable} in Sec. \ref{subsec2.3} would read
 \begin{subequations}
\begin{multline}  \label{eq_comm1a}
\int_{\cA^{\orth}}
   \left( \prod_{i=1}^m  \Tr_{\rho_i}(\Hol_{l_i}(A^{\orth} +
  \bar{g}_{\cl} B  \bar{g}_{\cl}^{-1} dt)) \right)   \exp(iS_{CS}(A^{\orth} +
  \bar{g}_{\cl} B  \bar{g}_{\cl}^{-1} dt))   DA^{\orth} \\
= \int_{\cA^{\orth}}
   \left( \prod_{i=1}^m  \Tr_{\rho_i}(\Hol_{l_i}(A^{\orth} \cdot \Omega_{\cl} +
 B dt)) \right)   \exp(iS'_{CS}(A^{\orth} \cdot  \Omega_{\cl} +
 B dt)) DA^{\orth}
 \end{multline}
 Now, using the   change of variable (CoV1)
 (which can  again be justified in a similar way as above)
 we can rewrite the last expression in the previous equation as
\begin{align} \label{eq_comm1b}
 & \int_{\cA^{\orth}}
   \left( \prod_{i=1}^m  \Tr_{\rho_i}(\Hol_{l_i}(A^{\orth} \cdot \Omega_{\cl} +
 B dt)) \right)   \exp(iS'_{CS}(A^{\orth} \cdot  \Omega_{\cl} +
 B dt)) DA^{\orth} \nonumber \\
 & =  \int_{\cA^{\orth}}   \left( \prod_{i=1}^m  \Tr_{\rho_i}(\Hol_{l_i}( A^{\orth} + A_{\sing}(\cl) +
 B dt)) \right)    \exp(iS'_{CS}( A^{\orth} + A_{\sing}(\cl) +B dt))   DA^{\orth}
\end{align}
Observe that if in the last expression we  performed the change of variable $A^{\orth} + \Omega^{-1}_{\cl} d \Omega_{\cl} \to A^{\orth} $, which we will call (CoV2)'
(in order to distinguish it from the analogous change of variable (CoV2)
appearing above in the main text)  in a naive way we would  arrive at
\begin{align*}
& \int_{\cA^{\orth}}   \left( \prod_{i=1}^m  \Tr_{\rho_i}(\Hol_{l_i}( A^{\orth} + A_{\sing}(\cl) +
 B dt)) \right)    \exp(iS'_{CS}( A^{\orth} + A_{\sing}(\cl) +B dt))   DA^{\orth} \\
 & =  \int_{\cA^{\orth}}   \left( \prod_{i=1}^m  \Tr_{\rho_i}(\Hol_{l_i}( A^{\orth}  +
 B dt)) \right)    \exp(iS'_{CS}( A^{\orth}  +B dt))   DA^{\orth}
\end{align*}
which is not correct\footnote{\label{ft_CoV2'}It is interesting to analyze what would happen if we tried  to use an ``extension of space''-argument in order to justify the  (incorrect) change of variable  (CoV2)'.  The replacement
 $ S'_{CS}( A^{\orth} + A_{\sing}(\cl) +B dt) \to S'_{CS}( A^{\orth} +B dt)$
 involves a replacement $ dA^{\orth} + d A_{\sing}(\cl) \to dA^{\orth}$.
 Now observe that  the differential $dA_{\sing}(\cl)$ of $A_{\sing}(\cl)$ is not locally integrable around $\sigma_0$ w.r.t.
 the measure $d\mu_{\mathbf g}$, cf. Sec. \ref{subsubsec3.2.2}. This rules out   any reasonable  ``extension of space''-argument
for justifying this change of variable  (CoV2)'.  In particular, we cannot even use the large extended spaces
 containing distributional forms mentioned in  Remark \ref{rm_extended_spaces_CQFT} above.}.
  In fact, according to the  careful (informal) argument
 in\footnote{Note that in  \cite{Ha7a}
  the notation $A_{\sing}(\cl)$ is used for the 1-form $\pi_{\ct}(\Omega^{-1}_{\cl} d \Omega_{\cl})$ rather than $\Omega^{-1}_{\cl} d \Omega_{\cl}$
     but it is obvious how to modify the argument there.}
 Appendix B.3 in \cite{Ha7a}    we have
\begin{align}  \label{eq_comm1c}
& \int_{\cA^{\orth}}   \left( \prod_{i=1}^m  \Tr_{\rho_i}(\Hol_{l_i}( A^{\orth} + A_{\sing}(\cl) +
 B dt)) \right)    \exp(iS'_{CS}( A^{\orth} + A_{\sing}(\cl) +B dt))   DA^{\orth} \nonumber \\
 & = \exp\bigl( - 2 \pi i k  \langle  n(\cl), B(\sigma_0)\rangle \bigr)
 \int_{\cA^{\orth}}   \left( \prod_{i=1}^m  \Tr_{\rho_i}(\Hol_{l_i}( A^{\orth}  +
 B dt)) \right)    \exp(iS'_{CS}( A^{\orth}  +B dt))   DA^{\orth}
\end{align}
\end{subequations}
From Eqs. \eqref{eq_comm1a}, \eqref{eq_comm1b}, and \eqref{eq_comm1c}
 and $S'_{CS}( A^{\orth}  +B dt) = S_{CS}( A^{\orth}  +B dt)$
we  now obtain again  Eq. \eqref{eq_change_of_variable}.
\end{remark}

\subsection{Justification of the replacement $\cB_{reg} \to \cB_{reg}^{ess}$ in Sec.
\ref{subsubsec3.2.3}}
\label{appA.6}

Let us now justify the replacement $\cB_{reg} \to \cB_{reg}^{ess}$ in Sec.
\ref{subsubsec3.2.3}. More precisely, we will show that
in Eq. \eqref{eq2.24} in Sec. \ref{subsec2.2} we can replace $\cB_{reg}$ by $\cB_{reg}^{ess}$.
(This then implies that also in Sec. \ref{subsec2.3} we can replace
  $\cB_{reg}$ by $\cB_{reg}^{ess}$ everywhere.)
In order to do so we will modify/extend some of the arguments in Appendix \ref{appA.2}. \par

Recall from Appendix \ref{appA.2} that the
 diffeomorphism  $q: P \times G/T \ni (b, \bar{g}) \mapsto   \bar{g} \exp(b) \bar{g}^{-1} \in G_{reg}$ induces a  bijection
$\hat{q}: C^{\infty}(\Sigma,P) \times C^{\infty}(\Sigma,G/T) \to
C^{\infty}(\Sigma,G_{reg})$.
Similarly, the  smooth
map  $q':\ct \times G/T \ni (b, \bar{g}) \mapsto   \bar{g} \exp(b) \bar{g}^{-1} \in G$   induces an injection
$$\hat{q}': \{B \in \cB_{reg}^{ess} \mid B(\sigma_0) \in P\} \times C^{\infty}(\Sigma,G/T) \to C^{\infty}(\Sigma,G)$$

By modifying  Eqs \eqref{eq_appA.2_pre_all} in Appendix \ref{appA.2} above
in a suitable way we obtain
\begin{multline} \int_{C^{\infty}(\Sigma,G)} \bar{\chi}_{red}(\Omega) D\Omega  \\
  \sim  \int_{\{B \in \cB_{reg}^{ess} \mid B(\sigma_0) \in P\}} \biggl[ \sum_{\cl \in C^{\infty}(\Sigma,G/T)/ \G_{\Sigma}} \bar{\chi}_{red}(\bar{g}_{\cl}
  \exp(B)  \bar{g}_{\cl}^{-1} ) \biggr]
     \det\bigl(1_{\ck}-\exp(\ad(B)_{| \ck})\bigr)  DB
\end{multline}
which then leads to
 \begin{multline} \label{eq_last_A.6}
  \int_{\cA} \chi(A) DA
  \sim \sum_{\cl \in C^{\infty}(\Sigma,G/T)/ \G_{\Sigma}}
 \int_{\{B \in \cB_{reg}^{ess} \mid B(\sigma_0) \in P\}} \biggl[ \int_{\cA^{\orth}} \chi(A^{\orth} + (\bar{g}_{\cl}  B  \bar{g}_{\cl}^{-1}) dt)  DA^{\orth} \biggr]  \\
 \times \det\bigl(1_{\ck}-\exp(\ad(B)_{| \ck})\bigr)  DB
\end{multline}
By applying  Eq. \eqref{eq_last_A.6}   to all different choices of $P$
(and using that $\ct \backslash \ct_{reg}$ is a $db$-zero set)
we arrive at the version of Eq. \eqref{eq2.24} in Sec. \ref{subsec2.2} where
$\cB_{reg}$ is replaced by $\cB^{ess}_{reg}$.

\section{The shadow invariant $|L|$}
\label{appC}

The algebraic approach to the quantum invariants by Reshetikhin and Turaev, which works with
 quantum group representations and surgery operations, can be reformulated leading
 to the so-called ``shadow world'' approach (cf. \cite{KiRe},
\cite{Tu2}, and part II of \cite{turaev}) which also works with quantum group representations but eliminates  the use of surgery operations.
Let us briefly recall the definition of the
shadow invariant in the  situation relevant for us, i.e. for the base manifold
$M=\Sigma \times S^1$.

\smallskip

Let $L= (l_1, l_2, \ldots, l_m)$, $m \in \bN$,  be an admissible  link
in  $M= \Sigma \times S^1$ (cf. Definition \ref{def_3.5_0})
such that each $l^i_{\Sigma}:S^1 \to \Sigma$, $i \le m$, is null-homotopic.
We will  assume that
each $l_i$, $i \le m$, is equipped with a  horizontal framing, cf.
Definition \ref{def_3.2_2} in Sec. \ref{subsubsec3.2.1} above. \par

 As in Sec. \ref{subsec3.5} we will denote by
  $V(L)$  the set of double points of $L$ (i.e. the set of those $p \in \Sigma$ where the loops
 $l^i_{\Sigma}$, $i \le m$, cross themselves or each other).
 Moreover, we will denote by   $E(L)$ the set of
curves in $\Sigma$ into which the loops $l^1_{\Sigma}, l^2_{\Sigma},
\ldots, l^m_{\Sigma}$ are decomposed when being ``cut''  in the
points of $V(L)$.   \par

From the assumption that $L$ is admissible it follows
that there are only finitely many  connected components $Y_0, Y_1, Y_2, \ldots, Y_{m'}$, $m' \in \bN$ (``faces'')
 of  $\Sigma \backslash ( \bigcup_i \arc(l^i_{\Sigma}))$. We  set
 \begin{equation} \label{eq_appC_YL} Y(L):= \{ Y_0, Y_1, Y_2, \ldots, Y_{m'} \}.
 \end{equation}
 By  $col(L)$ we denote the set of all maps $\eta: Y(L) \to \Lambda^k_+$ (= ``area
colorings''). \par

As explained in \cite{Tu2} one can associate to each face $Y \in Y(L)$
 in a natural way a  half integer called the ``gleam'' of $Y$
 (notation: $\gleam(Y)$).

\smallskip

From now on we will assume that each loop $l_i$ in the link $L$  is equipped with a ``color'' $\rho_i$,
i.e. an irreducible finite-dimensional complex representation of $G$.
By   $\gamma_i \in \Lambda_+$ we denote the highest weight of
 $\rho_i$ and we set     $\gamma(e):= \gamma_{i(e)}$
 for each $e \in E(L)$  where $i(e)$ denotes the unique
    index $i \le m$  such that  $\arc(e) \subset \arc(l_i)$.\par

We can now define the ``shadow invariant'' $|L| $
associated to the pair $(\cG,k)$ and  the colored, horizontally   framed   link $L$
 by (cf. Remark \ref{rm_Turaev_Vgl} below)
 \begin{equation} \label{eqA.4}
|L|:=  \sum_{\eta \in col(L)}
|L|_1^{\eta}\,|L|_2^\eta\,|L|_3^\eta\,|L|_4^\eta
\end{equation}
  with
  \begin{subequations} \label{eqA.5}
\begin{align} |L|_1^{\eta}&=\prod_{Y \in Y(L)} (d_{\eta(Y)})^{\chi(Y)}\\
|L|_2^{\eta} &= \prod_{Y \in Y(L)}  (\theta_{\eta(Y)})^{\gleam(Y)}\\
\label{eq_XL3}  |L|_3^{\eta} &= \prod_{e \in E_{loop}(L)} N^{\eta(Y^+_e)} _{\gamma(e)\eta(Y^-_e) } \\
\label{eq_XL4}
|L|_4^\eta&=  \contr_{D(L)} \bigl( \otimes_{x \in V(L)} T(x,\eta) \bigr)
  \end{align}
 \end{subequations}
  Here  $d_{\lambda}$, $\theta_{\lambda}$, $N_{ \mu \nu}^{\lambda}$ (for $\lambda, \mu, \nu \in \Lambda^k_+$) are as in Appendix \ref{appB} above
  and $E_{loop}(L)$ is the subset of those $e \in E(L)$ which are  loops.
  Moreover,   $Y^+_e$ (or $Y^-_e$, respectively) denotes the  unique face $Y$
  such that $\arc(e) \subset \partial Y$ and,
  additionally,  the orientation on $\arc(e)$ induced by the standard orientation on $\arc(l_{i(e)})$
coincides with (or is opposite to, respectively) the orientation induced on  $e \subset \partial Y$
by the orientation on  $Y$.  \par

Each factor  $T(x,\eta)$ appearing in $|L|_4^{\eta}$
is an element of a certain finite-dimensional complex vector space $W(x,\eta)$.
The  definitions of both $W(x,\eta)$ and $T(x, \eta)$ involve six elements of $\Lambda_+^k$,
 firstly,  the values $\eta(Y_i(x))$ for the four faces $Y_i(x) \in Y(L)$, $i \le 4$, having
$x$ on their boundary and, secondly, the
highest weights $\gamma_1(x)$ and $\gamma_2(x)$
of the two colors $\rho_i$ and $\rho_j$
associated to the two loops $l_{i}$ and $l_{j}$
whose $\pi_{\Sigma}$-projections intersect in $x$.
Moreover, the definition of $T(x,\eta)$  involves  the so-called (normalized) ``quantum 6j-symbols'' associated to the quantum group $U_q(\cG_{\bC})$
 where $q:= \exp( \tfrac{2 \pi i}{k})$ (cf.   Chap. VI and Chap. XI in \cite{turaev}).
 Finally, $D(L)$ is the graph $(V(L),E_{red}(L))$
where $E_{red}(L)  := E(L) \backslash E_{loop}(L)$ and
$\contr_{D(L)}: \otimes_{x \in V(L)} W(x,\eta) \to \bC$ is a suitable
linear functional which depends on the structure of  $D(L)$.
(See the last paragraph of Remark \ref{rm_Turaev_Derive} below
for additional comments regarding  $T(x,\eta)$ and  $contr_{D(L)}$.)

\begin{remark} \label{rm_Turaev_Vgl}
The definition of $|L|$ above  generalizes  the definition of the ``shadow invariant'' in \cite{Tu2}. More precisely, the invariant defined in \cite{Tu2}
is the special case of $|L|$ where $\cG= su(2)$.
By combining the ideas in Sec. 5 in \cite{Tu2} with those in Chapters VI, VIII  and XI in \cite{turaev}
it should not be too difficult to show that $|L|$ is indeed
a topological invariant also if  $\cG$ is the Lie algebra of a general simple, (simply-connected) compact Lie group $G$. However, to my knowledge such a proof has not yet been written
down anywhere\footnote{In the special case where the set $E_{loop}(L)$ is empty the RHS of Eq. \eqref{eqA.4} above
 coincides with the RHS of   Eq. (25) in \cite{PoRe} where the topological invariance of the RHS
 of their Eq. (25) is mentioned but not proven. In \cite{PoRe} it is also mentioned
 that their formula (25) follows from the general results in  \cite{turaev}. Eq. \eqref{eq_rem_C.2 main} below makes this  explicit.}.

 \smallskip

Instead of proving the topological invariance of $|L|$ directly we will prove it indirectly
by expressing $|L|$ in terms of the topological invariant  $|M,\Gamma|$ defined  in Sec. X.7.3 in \cite{turaev}. More precisely, we have
 \begin{equation} \label{eq_rem_C.2 main} | \Sigma \times S^1, L^*| \sim |  CY(\Sigma \times S^1,L^*)| \sim |L|
 \end{equation}
 where
 \begin{itemize}
 \item $L^*$ denotes the framed, colored link in $\Sigma \times S^1$ obtained from the framed, colored  link $L= (l_1, l_2, \ldots, l_m)$, $i \le m$, fixed above
   by replacing each color $\rho_i$ of $L$  with the dual color $\rho_i^*$,

\item the first ``$|\cdot|$'' refers to the invariant  $|M,\Gamma|$ defined  by\footnote{The first equation in Sec. X.7.3 in \cite{turaev}
reads (using the notation of \cite{turaev}) $|M,\Gamma| = \cD^{2(b_2(M) - b_3(M))} \dim'(\Gamma)^{-1} |CY(M,\Gamma)| \in H(\Gamma)$.
 The two observations important for us are that
 in the special case where  $\Gamma$ is a link we have, firstly, $H(\Gamma) = \bC$ and, secondly,  the factor $\dim'(\Gamma)$ is then trivial.
In order to understand the first point note that the notation $H(\Gamma)$ in Sec. X.7.3
 is a short notation for $H(\Gamma,\lambda)$, defined as in Sec. X.1.1 in \cite{turaev},
 where  $\lambda$ is the coloring of   $\Gamma$ and note also that  in the special case where  $\Gamma$ is a link
it does not contain  any vertices. The second point  follows from Eq. (5.5b) in Sec. X.5.5 in \cite{turaev} by taking into account that
 the boundary $\partial X$ of $X = CY(M,\Gamma) $ only has ``circle 1-strata'' in the terminology of Sec. VI.4.1 in \cite{turaev} if $\Gamma = L^*$,
 cf. the fourth bullet point in the list  in point (i) below. }   the first equation in Sec. X.7.3 in \cite{turaev},

\item ``$CY(\Sigma \times S^1,L^*)$'' is as explained in part (i) of the present remark,

\item the second   ``$|\cdot|$'' is the map described in part (ii) of the present remark,

 \item  ``$\sim$'' denotes equality   up to a multiplicative constant  independent of\footnote{By contrast, this ``constant'' may depend
 on $\Sigma$, $G$, and $k$.} $L$.
\end{itemize}

\noindent
Note that, apart from showing the topological invariance of $|L|$, Eq. \eqref{eq_rem_C.2 main}
will have the additional benefit of allowing us to find  the explicit relation between $|L|$ and the Reshetikhin-Turaev invariant
$RT(\Sigma \times S^1,L)$, cf. Eq. \eqref{eq_RT_vs_|L|} below.

\smallskip

We will now sketch   the definition/construction of $CY(\Sigma \times S^1,L^*)$
and the second map $|\cdot|$ appearing in Eq. \eqref{eq_rem_C.2 main} above.
In Remark \ref{rm_Turaev_Derive} below we will then sketch how one can verify directly (for an important special case)
that the second ``$\sim$'' in Eq. \eqref{eq_rem_C.2 main} indeed holds.
Note that the treatment here is definitely not self-contained but we hope that it will still be helpful for the reader, in particular
because it will probably allow the reader to  navigate  more quickly through the  sections of  \cite{turaev} which are relevant for us.

\medskip

\noindent
{\bf (i)}  The notation $CY(\Sigma \times S^1,L^*)$ appearing above refers to a ``shadowed 2-polyhedron'' (in the sense of
  Sec. VIII.1.2 in \cite{turaev}) which is  constructed as follows:

\begin{itemize}
\item  First we  choose a suitable ``skeleton''  of $M = \Sigma \times S^1$, i.e. an orientable ``simple 2-polyhedron''\footnote{Cf. Sec. VIII.1.1 in \cite{turaev}.}   $X_0$   contained in $M$ (and with  $\partial X_0 = \emptyset$) such that $M \backslash X_0$ is the disjoint union of open 3-balls,
  cf.  Sec. IX.2.1 in \cite{turaev}.
Note that we cannot choose  $X_0 =  \Sigma \times \{1\}$ but
we can choose $X_0$ such that $X_0 \supset  \Sigma \times \{1\} \cong \Sigma$ and such that $X_0$ has no ``vertices'' and only ``circle 1-strata''
in the terminology of \cite{turaev}.

 \item Once we have $X_0$ we can construct  from the framed  link $L^*$ a ``shadowed\footnote{The adjective ``shadowed'' refers to the family of ``gleams'', i.e. the half integers which are associated canonically to   the ``regions'' of $(l^i_{X_0})_{i \le m}$ in $X_0$, where the notion of ``region''
     is defined as in Sec. VIII.3.1 in \cite{turaev}.}
  system of loops'' $(l^i_{X_0})_{i \le m}$ in $X_0$
       in the sense of    Sec. VIII.3.1 in \cite{turaev}.   Each loop  $l^i_{X_0}$
       is obtained from the corresponding loop $l_i$ of $L^*$ by a kind of ``projection'' into $X_0$.
       The details of this construction are explained in Sec. IX.3.2 in \cite{turaev}.

 \item  The shadowed 2-polyhedron $CY(\Sigma \times S^1,L^*)$ appearing above
   (= ``the cylinder over the shadowed system of loops'' $(l^i_{X_0})_{i \le m}$ in the terminology of \cite{turaev})
     is constructed from the shadowed system of loops $(l^i_{X_0})_{i \le m}$ in $X_0$
      by attaching an annulus $\bar{Z}_i \cong S^1 \times [0,1]$, $i \le m$, along each of the
   loops $l^i_{X_0}$ in a suitable way.    This construction is alluded to  at the end of Sec. IX.3.3 in \cite{turaev} but the details
   of this construction are  actually given earlier\footnote{Or later, namely  in Sec. IX.8.3 in \cite{turaev} where the general definition of $CY(M,\Gamma)$
      is given. In the special situation where $\Gamma$ is a link (and where $M = \Sigma \times S^1$)
  it is more convenient to follow Sec. VIII.3.2 in \cite{turaev}.},
  namely in  Sec. VIII.3.2 in \cite{turaev}.

  \item   The   boundary   $\partial X$ of $X = CY(\Sigma \times S^1,L^*)$ is the disjoint union of m copies of $S^1$.
  (This is the result of the aforementioned  attachment of   $m$ annuli $\bar{Z}_i$).
  In the terminology  of Sec. VI.4.1 in \cite{turaev} we say that  $\partial X$
   consists only of   ``circle 1-strata'' .
   We equip each of the $m$ connected components of $\partial X \cong \coprod_{i=1}^m S^1$
   with the corresponding color $\rho_i^*$ coming from the colored link $L^*$.
   By doing so we obtain a ``coloring'' $\lambda$   of $\partial X$ in the sense of Sec. X.1.2 in \cite{turaev}.

\end{itemize}

\medskip

\noindent
{\bf (ii)}  The second map $|\cdot|$ appearing in Eq. \eqref{eq_rem_C.2 main} above is the map  defined in Sec. X.1.2 in \cite{turaev}.
 More precisely: What is denoted above by $|X|$ with $X= CY(\Sigma \times S^1,L^*)$
  is actually a short notation for $|X,\lambda|$   defined by  the last equation in Sec. X.1.2 in \cite{turaev}, which after dropping a
  factor independent of $L$ reads
  \begin{equation} \label{eq_def_Xlambda}
 |X,\lambda| \sim \sum_{\varphi \in col(X), \partial \varphi = \lambda}
  |X|_1^{\varphi} |X|_2^{\varphi} |X|_3^{\varphi} |X|_4^{\varphi}  |X|_5^{\varphi}
  \end{equation}
   where $\lambda$ is the coloring of $\partial X$  induced by   $L^*$ (cf. the fourth bullet point of part (i) below),
   where $col(X)$ is the set of ``colorings'' of $X$
  and where  $|X|_i^{\varphi}$ for $i \le 5$  and  $\varphi \in col(X)$
  is defined as in Sec.  X.1.2 in \cite{turaev}. (We will give some more details
  regarding $col(X)$ and  the terms $|X|_i^{\varphi}$  in Remark \ref{rm_Turaev_Derive} below.)

\end{remark}

\begin{remark}  \label{rm_Turaev_Derive}
For the convenience of the reader  we will now sketch  (for the special case where $L \subset \Sigma \times (S^1 \backslash \{1\})$)
a proof for the assertion above that   we have indeed
\begin{equation} \label{eq_rem_C.2 main2} |CY(\Sigma \times S^1,L^*)| \sim |L|.
\end{equation}
(The details for the general case will be postponed to a later version of the present paper.)
We will use the notation introduced in Remark \ref{rm_Turaev_Vgl} above.

\medskip

\noindent
{\bf (i)} In the special case where $L \subset \Sigma \times (S^1 \backslash \{1\})$ we can choose\footnote{By contrast, for general $L$ in $\Sigma \times S^1$ this is not the case,          which complicates matters considerably.} the skeleton $X_0$ above
         such that the projected loops $l^i_{X_0}$ do not meet the (circle) 1-strata of $X_0$, which implies
        that there is a 1-1-correspondence between the vertices of  $X = CY(\Sigma \times S^1,L^*)$
          and our set  $V(L)$.

\medskip

\noindent
{\bf (ii)} By contrast, as  $X_0$ cannot be chosen to be equal to $\Sigma \times \{1\}$ the
``regions'' of the shadowed system of loops $(l^i_{X_0})_{i \le m}$ in $X_0$
  (in the sense of Sec. VIII.3.1 in  \cite{turaev}) are not in 1-1-correspondence with the elements of our set $Y(L)$.
   Moreover, as a result of the aforementioned attachment of $m$ annuli $\bar{Z}_i$,   the set of ``regions'' $Reg(X)$   of $X = CY(\Sigma \times S^1,L^*)$ (in the sense of Sec. VIII.1.1 in  \cite{turaev}) is even larger.
   There are $m$ regions $Z_i \cong S^1 \times (0,1)$ of $X$, coming from the annuli  $\bar{Z}_i$, $i \le m$,
   which do not correspond to any of the elements of  $Y(L)$. \par

   This means that, while the area colorings $\eta$ appearing in the formula for $|L|$ (cf. Eq. \eqref{eqA.4} above)
   are elements of $(\Lambda^k_+)^{Y(L)} \cong (\Lambda^k_+)^n$ where $n:= \# Y(L)$, the
   colorings $\varphi$ appearing in Eq. \eqref{eq_def_Xlambda} above
   for $X = CY(\Sigma \times S^1,L^*)$   can be considered as elements of $(\Lambda^k_+)^{n' + m}$  where   $n' > n$ is the number of regions of the shadowed system of loops $(l^i_{X_0})_{i \le m}$ in $X_0$. \par

   A second difference is that, while the sum in Eq. \eqref{eqA.4} above is over all $\eta$,
   the sum in Eq. \eqref{eq_def_Xlambda} above is only
   over those  $\varphi$ obeying the relevant boundary condition\footnote{\label{ft_boundary_cond}Note that the value of  $\varphi$ on each $Z_i$ is fixed by  the boundary condition $\partial \varphi = \lambda$    appearing under the $\sum$-sign in Eq. \eqref{eq_def_Xlambda} above.}.
  So in the case of Eq. \eqref{eqA.4} above we have a sum over $(\Lambda^k_+)^n$ while in Eq. \eqref{eq_def_Xlambda}
  we have a sum over $(\Lambda^k_+)^{n'}$.   \par

Fortunately, it is quite easy to reduce the sum over $(\Lambda^k_+)^{n'}$ to a sum over  $(\Lambda^k_+)^n$
by summing out  $n' - n$ suitably chosen components of  $(\Lambda^k_+)^{n'}$
 and by taking into account that
 $\sum_{\alpha} N_{\alpha\beta\gamma} d_{\alpha} = d_{\beta} d_{\gamma}$ and $\sum_{\alpha} S_{\alpha 0}^2 = 1$.
 (Note that  $h^{\alpha\beta\gamma}$ and $\dim(\alpha)$  in the notation of \cite{turaev}
coincide with our $N_{\alpha\beta\gamma}$  and $d_{\alpha}$.)
The details of this argument will be given in later version of the present paper. \par

After summing out the aforementioned $n' - n$ components of $(\Lambda^k_+)^{n'}$ we then arrive at a  sum over
 $(\Lambda^k_+)^n$, which can be identified  with the sum over  all those ``reduced'' colorings $\varphi : Y(L) \cup \{Z_i \mid i \le m\} \to \Lambda^k_+$
which fulfill the  boundary condition mentioned in Footnote \ref{ft_boundary_cond} above.
 The important point is that the value of $|X,\lambda|$
is  again given by Eq. \eqref{eq_def_Xlambda} above where now each $|X|_i^{\varphi}$, $i \le 5$, is reexpressed in terms of the ``reduced'' $\varphi$
and where the sum over $\varphi$ is reinterpreted as the sum over the set of ``reduced'' colorings $\varphi$.

\medskip

\noindent
{\bf (iii)} In the following let us fix  $\varphi : Y(L) \cup \{Z_i \mid i \le m\} \to \Lambda^k_+$ and let $\eta$ denote the area coloring $Y(L) \to \Lambda^k_+$ corresponding to $\varphi$, i.e. $\eta = \varphi_{| Y(L)}$.
We will now compare  the five factors
$|X|_1^{\varphi}$, $|X|_2^{\varphi}$,  $|X|_3^{\varphi}$, $|X|_4^{\varphi}$, and $|X|_5^{\varphi}$
   appearing in Sec. X.1.2 in \cite{turaev} (reexpressed in terms of the ``reduced'' $\varphi$)
  with the four factors  $|L|_1^{\eta}$, $|L|_2^{\eta}$, $|L|_3^{\eta}$, and $|L|_4^{\eta}$
 appearing in our Eq. \eqref{eqA.4}.

 \begin{itemize}

\item  $|X|_1^{\varphi}$   is trivial for $X= CY(\Sigma \times S^1,L^*)$ since in this case  $\partial X$
consists of circle 1-strata  and therefore does not contain any vertices (cf. the last paragraph in Remark \ref{rm_Turaev_Vgl} above).

\item  In order to see that the term $|X|_2^{\varphi}$ leads to our term $|L|_1^{\eta}$
one  uses the relation $\chi(Z_i)= \chi(S^1 \times (0,1)) = 0$.

\item In order to see that the term $|X|_3^{\varphi}$ gives rise
to our term  $|L|_2^{\eta}$  note  that  $(v'_{i})^2$ in the notation of \cite{turaev}
  corresponds to our $\theta_\lambda$ if $\lambda = i$
and that the gleams of the $m$ regions $Z_i \cong S^1 \times (0,1)$  equal  the framing numbers
   of the $m$ loops $l_i$ in $L$, cf. Sec. VIII.3.2 in  \cite{turaev}.
   In the special case where $L$ is horizontally  framed (which we have assumed above) all these framing numbers are equal to $0$.

\item In order to see that the term $|X|_4^{\varphi}$ gives rise to our term
  $|L|_3^{\eta}$  recall first  that $h^{ijk}$ in the notation of \cite{turaev}
    coincides with our $N_{ijk}$ and that we have $N_{ij}^k = N_{ij\bar{k}}$.
    Moreover,  one should note that the asymmetric treatment of the two faces $Y^+_e$ and $Y^-_e$
   in each factor $N^{\eta(Y^+_e)} _{\gamma(e)\eta(Y^-_e)}$   on the RHS of Eq. \eqref{eq_XL3} above
    is mirrored in the definition of the term  $|X|_4^{\varphi}$ in \cite{turaev}
    even though at first look\footnote{The point to note is that when each face $Y \in Y(L)$ is equipped with the orientation induced by
    the orientation on $\Sigma$ (as we have assumed implicitly above)
      then $Y^+_e$ and $Y^-_e$  will induce different orientations on the bounding edge $e$.
    By contrast, in each factor $h_{\varphi}(g)$  (where $g$ is our $e$)
     the orientation of the three relevant regions, i.e.  $Y^+_e$, $Y^-_e$, and $Z_{i(e)}$  is supposed to induce the same orientation on $e$.
    So under the correspondence between $\varphi$ and $\eta$ which we are using   $\varphi(Y^+_e)$ (or $\varphi(Y^-_e)$, respectively)
    must be replaced by $\varphi(\bar{Y}^+_e) = \overline{\varphi(Y^+_e)}$   (or $\overline{\varphi(Y^-_e)}$, respectively)
    where $\bar{Y}^{\pm}_e$ denotes the face $Y^{\pm}_e$ when equipped with the opposite orientation and
    where $\bar{\lambda}$ for $\lambda = \varphi(Y^{\pm}_e)$
    is the dual weight.}  each factor $h_{\varphi}(g)$
    of  $|X|_4^{\varphi}$ seems to have a symmetric definition.

\item Finally note that   the second factor
of the tensor product appearing in the formula for $|X|_5^{\varphi}$ is trivial
 since  $\partial X$ only consists of circle 1-strata  and therefore does not contain any vertices.
 So the expression for $|X|_5^{\varphi}$  reduces
 to ``$cntr(\otimes_x |x|_{\varphi}) = cntr(\otimes_x |\Gamma^{\varphi}_x|)$'' in the notation of Sec. X.1.2 in \cite{turaev}.
  Note that ``$contr$'' in the notation of \cite{turaev} coincides with what we denote by $contr_{D(L)}$
  and $|\Gamma^{\varphi}_x|$, $x \in V(L)$, coincides
  with what we denote by $T(x,\eta)$   (where $\eta$ and $\varphi$ are related as described above).
 \end{itemize}

\end{remark}

Let us now go back to the case of ribbon links.
  Let $L$ be a general strictly admissible ribbon link in
$M = \Sigma \times S^1$ and let $L^0$ be the proper (framed) link associated to $L$,
cf. Definition \ref{def_3.2_1} in Sec. \ref{subsubsec3.2.1}.
Observe that $L^0$ is then automatically admissible and horizontally framed.
We will write $|L|$ instead of $|L^0|$ and
 $|L|^{\eta}_i$ instead of $|L^0|^{\eta}_i$.
 Moreover, we will the identify the set $Y(L^0)$ (defined at the beginning of Appendix \ref{appC})  with the set $Y(L)$ defined as
 in Sec. \ref{subsec3.4}  above.
According to  Theorem 7.3.1 in Sec. X.7.3 in \cite{turaev}
and Remark \ref{rm_Turaev_Vgl} above  we have
\begin{equation} \label{eq_RT_vs_|L|}
RT(\Sigma \times S^1,L) \sim |\Sigma \times S^1,L^*| \sim  \ |L|.
\end{equation}
(Observe that we use the notation $RT(M,L)$  for what in \cite{turaev} is denoted
by $\tau(M,L)$.)

\section{Performing Steps 2--4 in Sec. \ref{subsubsec3.5.2}}
\label{appD}

\noindent
{\it Note:} The present part of the appendix  is somewhat speculative and
will not be included in the print version of the present paper.

\medskip

We will now sketch the strategy for performing Steps 2--4 in Sec. \ref{subsubsec3.5.2}.
(At the end of Appendix \ref{appD} we will also make some comments regarding Step 5.)
In contrast to the calculations  in Sec. \ref{sec3} above
  we do this not  as a preparation for a rigorous treatment but
 simply  in order to make it plausible that
 also for general strictly admissible $L$ we have a good chance of obtaining
 \begin{equation} \label{eq_D.0}
   Z(\Sigma \times S^1,L) \sim |L|= \sum_{\eta \in col(L)}
|L|_1^{\eta}\,|L|_2^\eta\,|L|_3^\eta\,|L|_4^\eta
\end{equation}
Accordingly, in the present part of the appendix
we take the liberty to use several somewhat
 sloppy  (informal) arguments (cf., e.g., Footnotes \ref{ft_sloppy_3}, \ref{ft_sloppy_2} and \ref{ft_sloppy_4}  below).

\smallskip

As in Sec. \ref{subsubsec3.3.3} above let us
(informally) interchange the $\eps \to 0$-limit
 with the integral and the $\sum_y$-sum.
By doing so we arrive  at the following variant of  Eq. \eqref{eq_step1_ende_rewritten}
\begin{multline} \label{eq_step1_ohne_eps}
Z(\Sigma \times S^1, L) \sim  \sum_{y \in I}  \int_{\cA^{\orth}_c \times \cB}
  \biggl\{ 1_{\cB^{ess}_{reg}}(B) \Det_{rig}(B)  \exp\bigl( - 2\pi i k  \langle y, B(\sigma_0) \rangle \bigr) \\
  \times  \beta_{L,D}\bigl(  \bigl( \otimes_{e \in E(L)}  T_{e}(A^{\orth}_c, B)  \otimes \bigl( \otimes_{x \in V(L)}  T_{x}(A^{\orth}_c, B)  \bigr) \bigl) \biggr\} \\
  \times  \exp(i S_{CS}(A^{\orth}_c, B)) (DA^{\orth}_c \otimes DB)
\end{multline}
where we have set
\begin{equation} \label{eq_def_Te}
T_{e}(A^{\orth}_c, B) := \lim_{\eps \to 0} T^{\eps}_{e}(A^{\orth}_c, B), \quad \text{ and }
\quad T_{x}(A^{\orth}_c, B) := \lim_{\eps \to 0} T^{\eps}_{x}(A^{\orth}_c, B)
\end{equation}
Let us now  rewrite Eq. \eqref{eq_step1_ohne_eps} as
an iterated integral\footnote{\label{ft_sloppy_3}In fact,
instead of working with $\int \cdots DA_{\Sigma}$ and $\int \cdots DB$
 it would have
several conceptual advantages to  work with
suitable improper  integrals $ \int^{\sim} \cdots  DA_{\Sigma}$
and $\int^{\sim} \cdots  DB$ as in Remark \ref{rm_sec2.2.1}
and Remark \ref{rm_eigentlich_average} above, cf. Footnote \ref{ft_sloppy_4} and Footnote \ref{ft_sloppy_5} below.}
\begin{equation} \label{eq_D.2}
Z(\Sigma \times S^1, L) \sim  \sum_{y \in I}  \int_{\cB}
  \biggl\{ 1_{\cB^{ess}_{reg}}(B) \Det_{rig}(B)  \exp\bigl( - 2\pi i k  \langle y, B(\sigma_0) \rangle \bigr) F_{L,D}(B)  \biggr\}   DB
\end{equation}
 with
\begin{equation} \label{eq_D.3} F_{L,D}(B) := \int_{\cA_{\Sigma,\ct}}
 \beta_{L,D}\bigl(  \bigl( \otimes_{e \in E(L)}  T_{e}(A_{\Sigma}, B)  \otimes \bigl( \otimes_{x \in V(L)}  T_{x}(A_{\Sigma}, B)  \bigr) \bigl) \bigr) \exp(i S_{CS}(A_{\Sigma}, B)) DA_{\Sigma}
\end{equation}
where $DA_{\Sigma}$ is the informal Lebesgue measure on
$\cA_{\Sigma,\ct} \cong  \cA^{\orth}_c$.

\medskip

Recall that the ribbon link $L = (R_1,R_2, \ldots, R_m)$ fixed in Sec. \ref{subsubsec3.5.2} above  is  colored, i.e.  each $R_j$
is equipped with an irreducible representation
$\rho_j: G \to \Aut(V_j)$ where $V_j$ is a finite-dimensional complex vector space.

\begin{convention} \label{conv_appD_1}
(i) For each $e \in E(L)$ we denote by $S(e)$
  the unique (``open'') ribbon $[0,1] \times [0,1] \to \Sigma \times S^1$
  given by  $cl(e) = \{S(e)\}$. Similarly, for each $x \in V(L)$ we denote
  by $S_1(x)$ and $S_2(x)$ the two (``open'') ribbons $[0,1] \times [0,1] \to \Sigma \times S^1$   given by  $cl(x) = (S_1(x),S_2(x))$.

\smallskip

(ii) For $e \in E(L)$ and $x \in V(L)$
   we denote by $j(e)$ (or $j_1(x)$ or $j_2(x)$, respectively)
      the unique index  $j \le m$,   such that $S(e)$ (or $S_1(x)$ or $S_2(x)$, respectively) ``is a piece''  of the closed ribbon $R_j$.
\end{convention}

From  Eq. \eqref{eq_T_cl1_expl} above we obtain  the following explicit formula
 for\footnote{\label{ft_appD_177}Recall from the paragraph before
  Remark \ref{rm_KiRe}  in Sec. \ref{subsubsec3.5.2} above that
 we can also obtain an explicit formula for  $T_{x}(A_{\Sigma}, B)$, $x \in V(L)$,
 as an infinite sum of powers of $1/k$ (whose coefficients also depend on $k$)
  but we do not yet have a closed formula for this sum.}  $T_{e}(A_{\Sigma}, B) \in \Aut(V_{j(e)})$,  $e \in E(L)$:
\begin{align} \label{eq_D.4} T_{e}(A_{\Sigma}, B) & = \lim_{\eps \to 0}
 \exp\biggl(\int_0^1  (\rho_{j(e)})_* (D_{S(e),s}^{\eps}(A_{\Sigma},  B)) ds\biggr) \nonumber \\
& = \exp\biggl(\int_0^1 \int_0^1 \bigl[ (\rho_{j(e)})_*\bigl(A_{\Sigma}((c_{e,u})'_{\Sigma}(s))) + (\rho_{j(e)})_*((B  dt)(c'_{e,u}(s))\bigr) \bigr]  ds du \biggr) \nonumber \\
&  = \rho_{j(e)}\biggl(\exp\biggl(\int_0^1 \int_0^1 A_{\Sigma}((c_{e,u})'_{\Sigma}(s))
 ds du +  \int_0^1 \int_0^1 (B  dt)(c'_{e,u}(s)) ds du\biggr)\biggr)
\end{align}
  where $c_{e,u}: [0,1] \to \Sigma \times S^1$, $u \in [0,1]$, are the curves
given by $c_{e,u}(s) = S(e)(s,u)$ for all $s \in [0,1]$.

\medskip

It will be  convenient to fix, for  each $j \le m$,
 a basis $(v^j_a)_{a \le \dim(V_j)}$ of $V_j$ such
 that  $(\rho_j)_{|T}$ leaves each of the 1-dimensional subspaces $\bC v^j_a$, $a \le \dim(V_j)$ invariant.
The real  weight corresponding to $v^j_a$ will be denoted by
$\alpha^j_a $. In other words, $\alpha^j_a \in \Lambda$ is given by
\begin{equation} \label{eq_appD_weights}
\rho_j(\exp(b)) v^j_a = e^{2 \pi i \langle \alpha^j_a, b \rangle} v^j_a
\quad  \forall b \in \ct
\end{equation}

Observe that Eq. \eqref{eq_D.4} and Eq. \eqref{eq_appD_weights} imply
\begin{equation} \label{eq_D.6}
(T_{e}(A_{\Sigma}, B))_{a_e a'_e}
= \delta_{a_e a'_e}
e^{2 \pi i \langle \alpha_{a_e}, \int_0^1 \int_0^1 A_{\Sigma}((c_{e,u})'_{\Sigma}(s)) du ds \rangle} e^{2 \pi i \langle \alpha_{a_e}, \int_0^1 \int_0^1 (B  dt)(c'_{e,u}(s)) ds du \rangle}
\end{equation}
where we have written $ \alpha_{a_e}$ instead of  $\alpha^{j(e)}_{a_e}$.

\begin{convention} \label{conv_appD_2}
(i) For each $j \le m$ we set $\cI_j:= \{1,2,\ldots, \dim(V_j)\}$.

 \smallskip

 (ii) We set $(\cI):= \times_{e \in E(L)} \cI_{j(e)}$.

\smallskip

(iii) For an element $(a(e))_{e \in E(L)}$ of $(\cI)$
we will usually use the short notation $(a)$.
\end{convention}

\begin{convention} \label{conv_appD_3}
For every $x \in V(L)$ we denote by    $e_1(x)$ and $e_2(x)$
 the two elements of $E(L)$ having $x$ as the endpoint
 and we denote by  $e_3(x)$ and $e_4(x)$
 the two elements of $E(L)$ having $x$ as the starting point.
 (Observe that $e_1(x)$, $e_2(x)$, $e_3(x)$, and $e_4(x)$
 are not necessarily distinct).
The enumeration is such that $e_1(x)$ and $e_3(x)$ lie in the
open ribbon $S_1(x)$ while $e_2(x)$ and $e_4(x)$ lie in the
open ribbon $S_2(x)$, cf. Convention \ref{conv_appD_1} above.
\end{convention}

From the characterization of the linear form $\beta_{L,D}$ which we introduced in Sec. \ref{subsubsec3.5.2} above and from
Eq. \eqref{eq_D.6}  we conclude that
\begin{multline} \label{eq_appD_beta_term} \beta_{L,D}\bigl(  \bigl( \otimes_{e \in E(L)}  T_{e}(A^{\orth}_c, B)  \otimes \bigl( \otimes_{x \in V(L)}  T_{x}(A^{\orth}_c, B)  \bigr) \bigl)\\
 = \sum_{(a) \in (\cI)} \biggl[ \prod_{x \in V(L)}  (T_{x}(A_{\Sigma}, B))^{a(e_1(x)) a(e_2(x))}_{a(e_3(x)) a(e_4(x))} \\
 \times   \prod_{e \in E(L)} (T_{e}(A_{\Sigma}, B))_{a(e) a(e)} \biggr]
 \end{multline}
   where  $(T_{e}(A_{\Sigma}, B))_{a a'}$ are the components of $T_{e}(A_{\Sigma}, B)$
  w.r.t. to the basis  $(v^{j}_a)_{a \le \dim(V_j)}$ with $j = j(e)$
  and where
  $(T_{x}(A_{\Sigma}, B))^{a b}_{a' b'}$
  are the components of $T_{x}(A_{\Sigma}, B)$
  w.r.t. to the  basis $(v^{j_1}_a  \otimes v^{j_2}_b)_{a,b}$
  (with $j_1 = j_1(x)$, $j_2 = j_2(x)$), i.e. given by
  $T_{x}(A_{\Sigma}, B) \cdot (v^{j_1}_{a'}  \otimes v^{j_2}_{b'})
   = \sum_{a,b} (T_{x}(A_{\Sigma}, B))^{a b}_{a' b'}  (v^{j_1}_{a}  \otimes v^{j_2}_{b}$).

  \medskip

Clearly, from Eq. \eqref{eq_D.3} and Eq. \eqref{eq_appD_beta_term} we obtain
\begin{equation}\label{eq_D.8}
F_{L,D}(B)= \sum_{(a) \in (\cI)}  F_{(a)}(B)
\end{equation}
where we have set, for each  $(a) \in (\cI)$
\begin{multline} \label{eq_D.9}
F_{(a)}(B) :=
\int_{\cA_{\Sigma,\ct}} \biggl[ \prod_{x \in V(L)}
  (T_{x}(A_{\Sigma}, B))^{a(e_1(x)) a(e_2(x))}_{a(e_3(x)) a(e_4(x))} \\
 \times \prod_e (T_{e}(A_{\Sigma}, B))_{a(e) a(e)}
    \exp(i S_{CS}(A_{\Sigma}, B)) \biggr] DA_{\Sigma}
\end{multline}

The following remark will play an important role later.

\begin{remark}  \label{rm_appD_1}
(i) We say that a functional
 $F: \Omega^p(\Sigma,V) \to \bC$, $p \in \{0,1\}$,  has the ``loop decomposition property'' iff  for  every (piecewise) smoothly embedded
 circle $C \subset \Sigma$ which is contractible in $\Sigma$
 there are functionals $F_{K_j}:\Omega^p(K_j,V) \to \bC$, $j=1,2$,
  such that for all $\omega \in \Omega^p(\Sigma,V)$ we have
 \begin{equation}  F(\omega)  =
     F_{K_1}(\omega_{|K_1}) \times F_{K_2}(\omega_{| K_2})
\end{equation}
where $K_1$ and $K_2$ are the two closed subsets of $\Sigma$ obtained
 as the closure of the two connected components of $\Sigma \backslash C$. \par
 If  $F$ has the ``loop decomposition property'' in the sense above we obtain,
 informally\footnote{Note that we could rewrite  Eq. \eqref{eq_D.11_rem}
 in a more symmetric way  similarly to the last equation on page 633  in \cite{Lev}.
 Even though Eq. \eqref{eq_D.11_rem} is less elegant than its symmetric version
 it is more convenient for the calculations below.},
\begin{align} \label{eq_D.11_rem}
\int_{\Omega^p(\Sigma,V)} F(\omega) D\omega
& =  \int_{\Omega^p(K_1,V)} F_{K_1}(\omega_1) \biggl[ \int_{\Omega^p(K_2,V)}
  F_{K_2}(\omega_2) \delta_{[(\omega_2)_{|C} = (\omega_1)_{|C}]} D\omega_2 \biggr] D\omega_1
\end{align}
where $D\omega$, $D\omega_i$, $i=1,2$,  are the
obvious informal Lebesgue measures on $\Omega^p(\Sigma,V)$, $\Omega^p(K_i,V)$, $i \le 2$,
and where  $\delta_{[(\omega_2)_{|C} = (\omega_1)_{|C}]} D\omega_2$ is
 (a somewhat sloppy  notation for)   the  (informal) measure on $\{\omega_2 \in \Omega^p(K_2,V) \mid  (\omega_2)_{|C} = (\omega_1)_{|C} \}$  obtained by ``disintegration'' or ``conditioning''\footnote{\label{ft_sloppy_4}In measure theory the intuitive notion of restricting a  measure $d\mu$  on a measurable space $X$
  to a $d\mu$-zero-subset of $X$ can be implemented rigorously in many cases (and, in particular, in situations which are analogous to the situation here) with the help of the so-called disintegration theorem
  or --- in the case where $d\mu$ is a probability measure (which is relevant  also for
  the   application in the present remark, provided that we reinterpret
  $\int \cdots D\omega_2$
   as a suitable improper integral $\int^{\sim} \cdots D\omega_2$
   as in  Remark \ref{rm_sec2.2.1} in Sec. \ref{subsubsec2.2.1} above) ---
  with the help of the rigorous notion of conditional expectations/measures.}.

\smallskip

(ii) The definition of the notion ``loop decomposition property'' and
 the (informal) Eq. \eqref{eq_D.11_rem} can be generalized in an obvious way
to the situation where instead of one embedded circle $C$ we have
a finite family $(C_i)_{i \le n}$ of non-intersecting,
(piecewise) smoothly embedded circles which are contractible in $\Sigma$.
\end{remark}

 For each $x \in V(L)$ we   set $D_x := \Image(\pi_{\Sigma} \circ S_1(x)) \cap \Image(\pi_{\Sigma} \circ S_2(x)) \subset \Sigma$,
   cf. Convention \ref{conv_appD_1} above. Moreover, by
         $\Sigma'$ we denote the closure of $\Sigma \backslash \bigl( \bigcup_{x \in V(L)} D_x  \bigr)$ (in $\Sigma$).\par

From the definitions above (cf., in particular,  Eq. \eqref{eq_def_Te} above
and Eq. \eqref{eq_def_Teps_e}, Eq. \eqref{eq_def_Rcl},
and Eq. \eqref{eq_Sec3.5_P-Def} in Sec. \ref{subsubsec3.5.2}) we see  that for each fixed
$A_{\Sigma} \in \cA_{\Sigma, \ct}$  each of the factors
$(T_{x}(A_{\Sigma}, B))^{a(e_1(x)) a(e_2(x))}_{a(e_3(x)) a(e_4(x))}$
in Eq. \eqref{eq_D.9} above only depends on $A_{D_x}:= (A_{\Sigma})_{| D_x}$.
Similarly, each factor $(T_{e}(A_{\Sigma}, B))_{a(e) a(e)}$
only depends on $A_{\Sigma'}:= (A_{\Sigma})_{| \Sigma'}$.
On the other hand, we have
$$\exp(i S_{CS}(A_{\Sigma}, B)) = \exp(i S_{CS}(A_{\Sigma'}, B))
\prod_{x \in V(L)} \exp(i S_{CS}(A_{D_x}, B))$$
where we have set
$S_{CS}(A_{\Sigma'}, B):= 2 \pi k \int_{\Sigma'} \Tr(d A_{\Sigma'} \cdot  B)$
 and $S_{CS}(A_{D_x}, B):= 2 \pi k \int_{D_x} \Tr(d A_{D_x} \cdot B)$.
Taking this into account (as well as the relation $\partial D_x \cong S^1$,  $x \in V(L)$)
we obtain from the generalization of Eq. \eqref{eq_D.11_rem}  which is
  mentioned in part (ii) of Remark \ref{rm_appD_1}
  above\footnote{\label{ft_sloppy_2}This argument is a bit sloppy. The integral $\int \cdots  DA_{\Sigma'}$ in Eq. \eqref{eq_D.10} should actually be a ``conditioned integral''  as in  Remark \ref{rm_appD_1}. For example, in the special case where $V(L)$ consists of only one element $x$ we can take  $K_1 = D_x$ and $K_2 = \Sigma'$ and apply (the original version of) Eq. \eqref{eq_D.11_rem}, which would lead to an expression involving a
  $\int \cdots \delta_{[(A_{\Sigma'})_{|\partial \Sigma'} =
(A_{D_x})_{|\partial \Sigma'}]} DA_{\Sigma'}$-integral.
Strictly speaking we should therefore rewrite some of the
$\int \cdots DA_{\Sigma'}$-integrals
appearing  below as conditioned integrals in a suitable way. Doing so  would, however, not lead to anything  new, i.e. we would again arrive at Eq. \eqref{eq_D.10_b}  below.  }
\begin{multline} \label{eq_D.10} F_{(a)}(B)  =
\biggl[ \int_{\cA_{\Sigma',\ct}}\prod_e (T_{e}(A_{\Sigma'}, B))_{a(e) a(e)}
    \exp(i S_{CS}(A_{\Sigma'}, B))  DA_{\Sigma'} \biggr]  \biggl[\prod_{x \in V(L)} J_{x}(B) \biggr]
\end{multline}
where
\begin{equation} \label{eq_J_x(B)-def} J_{x,(a)}(B) := \int_{\cA_{D_x},\ct}    (T_{x}(A_{D_x}, B))^{a(e_1(x))
a(e_2(x))}_{a(e_3(x)) a(e_4(x))}
    \exp(i S_{CS}(A_{D_x}, B)) DA_{D_x}
\end{equation}
and where we have set  $\cA_{\Sigma',\ct} := \Omega^1(\Sigma',\ct)$ and
 $\cA_{D_x,\ct} := \Omega^1(D_x,\ct)$.

\smallskip

We will call $(a) \in (\cI)$ admissible if there is a
 $f_{(a)}: \Sigma' \to \ct$ such that
 \begin{itemize}
 \item  $f_{(a)}$ is constant on each\footnote{From the definition of $Y(L)$ and
 of $\Sigma'$ it follows each $Y \in Y(L)$ is contained in $\Sigma'$.}  $Y \in Y(L)$,
\item $f_{(a)}$ is continuous,
\item For each $e \in E(L)$ the restriction of $f_{(a)}$
to  $\Image(\pi_{\Sigma} \circ S(e)) \cong [0,1] \times [0,1]$ is given by
$$f_{(a)}((s,u))=
\begin{cases} \alpha_{a(e)} \cdot u + const & \text{ in Case I}\\
 \alpha_{a(e)} \cdot (1-u) + const & \text{ in Case II}
\end{cases} \quad \text{ for all $(s,u) \in [0,1] \times [0,1]$}$$
   where ``Case I'' and ``Case II'' are
  defined in a similar way as the two cases in Observation \ref{rm_X+_1} in Sec. \ref{subsubsec3.3.3}   above.
\end{itemize}
 If $f_{(a)}$ exists it is uniquely determined up to an additive constant, which
 we will fix by demanding that
 \begin{equation} \label{eq_D.13}
 f_{(a)}(\sigma_0) = 0
 \end{equation}

\begin{example} \label{eq_appD} In the special situation on Sec. \ref{subsec3.4}
 where $V(L) = \emptyset$ and
 $E(L) = \{l_1,l_2, \ldots, l_m\}$ every $(a) \in (\cI)$ is admissible
and  we have $f_{(a)} = \sum_{i =1}^m \alpha_i f_i$ where
$\alpha_i := \alpha^{j(l_i)}_{a(l_i)}$ and $f_i$ is as in Sec. \ref{subsec3.4} above.
\end{example}

Observe that for $(a) \in  (\cI)_{adm} := \{ (a) \in (\cI) \mid (a) \text{ is admissible}\}$  we have
\begin{multline} \sum_{e \in E(L)} \biggl\langle  \alpha_{a(e)} , \int_0^1 \int_0^1 A_{\Sigma'}((c_{e,u})'_{\Sigma}(s))  ds du \biggr\rangle \\
= \sum_{e \in E(L)} \biggl\langle  \alpha_{a(e)} , \int_0^1  \int_{(c_{e,u})_{\Sigma}}  A_{\Sigma'} du  \biggr\rangle =  \ll A_{\Sigma'},  \star d f_{(a)} \gg_{\cA_{\Sigma',\ct}}
\end{multline}
for $A_{\Sigma'} \in \cA_{\Sigma',\ct}$
where $\ll \cdot, \cdot \gg_{\cA_{\Sigma',\ct}}$
is the scalar product on $\cA_{\Sigma',\ct}$ induced by ${\mathbf g_{\Sigma}}$.
Using this as well as
$$ S_{CS}(A_{\Sigma'},B)  =   - 2 \pi k  \ll   A_{\Sigma'},
 \star dB \gg_{\cA_{\Sigma',\ct}}$$ and  Eq. \eqref{eq_D.6}  we obtain
for $(a) \in (\cI)_{adm}$
 \begin{multline}
\int_{\cA_{\Sigma',\ct}} \prod_e (T_{e}(A_{\Sigma'}, B))_{a(e) a(e)}
    \exp(i S_{CS}(A_{\Sigma'}, B)) DA_{\Sigma'} \\
 = \delta\bigl(d\bigl(B_{| \Sigma'} - \tfrac{1}{k} f_{(a)}\bigr)\bigr) \biggl(\prod_e e^{2 \pi i \langle \alpha_{a(e)}, \int_0^1 \int_0^1 (B  dt)(c'_{e,u}(s)) ds du \rangle} \biggr)
\end{multline}
On the other hand,  for $(a) \notin (\cI)_{adm}$ we have
$$\int_{\cA_{\Sigma',\ct}} \prod_e (T_{e}(A_{\Sigma'}, B))_{a(e) a(e)}
    \exp(i S_{CS}(A_{\Sigma'}, B)) DA_{\Sigma'}  = 0$$

 Combining the last two equations with Eq. \eqref{eq_D.10} we obtain
 \begin{subequations}  \label{eq_D.10_b}
\begin{equation}
 F_{(a)}(B) =  \delta\bigl(d\bigl(B_{| \Sigma'} - \tfrac{1}{k}  f_{(a)}\bigr)\bigr) \biggl(\prod_e e^{2 \pi i \langle \alpha_{a(e)}, \int_0^1 \int_0^1 (B  dt)(c'_{e,u}(s)) ds du \rangle} \biggr)  \prod_{x \in V(L)} J_{x,(a)}(B)
\end{equation}
if $(a) \in (\cI)_{adm}$ and
\begin{equation}
 F_{(a)}(B) = 0
\end{equation} if $(a) \notin (\cI)_{adm}$.
\end{subequations}
Combining this with Eq. \eqref{eq_D.2} and Eq. \eqref{eq_D.8}
 we now obtain
\begin{multline} \label{eq_D.15}
Z(\Sigma \times S^1, L)
 \sim  \sum_{y \in I}  \sum_{(a) \in (\cI)_{adm}} \int_{\cB}
  \biggl\{  \delta\bigl(d\bigl(B_{| \Sigma'} - \tfrac{1}{k}  f_{(a)}\bigr)\bigr) 1_{\cB^{ess}_{reg}}(B) \Det_{rig}(B)  \exp\bigl( - 2\pi i k  \langle y, B(\sigma_0) \rangle \bigr)  \\
   \times \biggl(\prod_e e^{2 \pi i \langle \alpha_{a(e)}, \int_0^1 \int_0^1 (B  dt)(c'_{e,u}(s)) ds du \rangle} \biggr)  \prod_{x \in V(L)} J_{x,(a)}(B) \biggr\}   DB
\end{multline}
Now observe that  each factor $J_{x,(a)}(B)$  is actually
 a function of $B_{|D_x}$.    On the other hand, the factors
 $ \exp\bigl( - 2\pi i k  \langle y, B(\sigma_0) \rangle \bigr)$,
 $\int_0^1 \int_0^1 (B  dt)(c'_{e,u}(s)) ds du$, and  $\delta\bigl(d\bigl(B_{| \Sigma'} - \tfrac{1}{k}  f_{(a)}\bigr)\bigr)$
only depend on $ B_{| \Sigma'}$. From  Eq. \eqref{eq_explicit_formula_Detreg}
and Observation \ref{obs_ass1} in Sec. \ref{subsubsec3.2.5} above
it follows that the same is true for
the factor $\Det_{rig}(B)$. (Accordingly, we will sometimes write
 $\Det_{rig}(B_{| \Sigma'})$ instead of $\Det_{rig}(B)$ in the following.)
 Finally, observe that we have
$1_{\cB^{ess}_{reg}}(B) = 1_{\cB^{ess}_{reg}(\Sigma')}(B_{|\Sigma'}) \prod_{x \in V(L)}
1_{\cB^{ess}_{reg}(D_x)}(B_{|D_x})$ where we have set
 $\cB^{ess}_{reg}(K):= \{ B \in C^{\infty}(K,\ct) \mid \forall \sigma \in K:
  \alpha \in \cR: [ B_{\alpha}(\sigma) \in \bZ
  \ \Rightarrow \ dB_{\alpha}(\sigma) \neq 0 ]\}$
for $K = \Sigma'$ or $K = D_x$, $x \in V(L)$.  \par

Consequently, for each fixed $(a) \in (\cI)_{adm}$
the integrand on the RHS of Eq. \eqref{eq_D.15} has the ``loop decomposition property'' in the sense of  Remark \ref{rm_appD_1} above and by
applying (in step $(*)$) part (ii) of Remark \ref{rm_appD_1}  we obtain\footnote{\label{ft_sloppy_5}Note  that in contrast to the situation in Eq. \eqref{eq_D.16} in Remark \ref{rm_appD_1} above, where $F_{K_1}(\omega_1)$ is a proper function,   we now have a delta-function expression
   $ \delta\bigl(d\bigl(B_{| \Sigma'} - \tfrac{1}{k}  f_{(a)}\bigr)\bigr)$.
   It is therefore not totally clear if the application
   of the argument in Remark \ref{rm_appD_1} is really justified.
    This complication can be  ``defused'' somewhat
 if (as mentioned in  Footnote \ref{ft_sloppy_3} above)
 we work with a suitable (informal) improper  integral $\int^{\sim} \cdots DA_{\Sigma}$
 as in Remark \ref{rm_sec2.2.1}
   instead of working with the original informal integral $\int \cdots DA_{\Sigma}$.
  By doing so the informal expressions
   $\delta^{(\eps)}\bigl(d\bigl(B_{| \Sigma'} - \tfrac{1}{k}  f_{(a)}\bigr)\bigr)$, $\eps >0$ appear instead of the delta-function
         $\delta\bigl(d\bigl(B_{| \Sigma'} - \tfrac{1}{k}  f_{(a)}\bigr)\bigr)$.
  Here  $\delta^{(\eps)}:  \cB_{\Sigma'} \to \bR$  is given, informally, by
 $\delta^{(\eps)}(B) =   \exp(-\tfrac{1}{4 \eps} \|B\|^2)/  \int \exp(-\tfrac{1}{4 \eps} \|B\|^2) DB$. }
\begin{align} \label{eq_D.16}
& \int_{\cB}
  \biggl\{ \delta\bigl(d\bigl(B_{| \Sigma'} - \tfrac{1}{k}  f_{(a)}\bigr)\bigr) 1_{\cB^{ess}_{reg}}(B) \Det_{rig}(B)  \exp\bigl( - 2\pi i k  \langle y, B(\sigma_0) \rangle \bigr)  \nonumber\\
  & \quad \quad \times \bigl(\prod_e e^{2 \pi i \langle \alpha_{a(e)}, \int_0^1 \int_0^1 (B  dt)(c'_{e,u}(s)) ds du \rangle} \bigr)  \prod_{x \in V(L)} J_{x,(a)}(B)  \biggr\}   DB
   \nonumber\\
& \overset{(*)}{=} \int_{\cB_{\Sigma'}} \biggl\{
   \bigl[ \delta\bigl(d\bigl(B_{\Sigma'} - \tfrac{1}{k}  f_{(a)}\bigr)\bigr) 1_{\cB^{ess}_{reg}(\Sigma')}(B_{\Sigma'}) \Det_{rig}(B_{\Sigma'}) \nonumber  \\
   & \quad \quad \quad \quad \times \exp\bigl( - 2\pi i k  \langle y, B_{\Sigma'}(\sigma_0) \rangle \bigr)
       \bigl(\prod_e e^{2 \pi i \langle \alpha_{a(e)}, \int_0^1 \int_0^1 (B_{\Sigma'}  dt)(c'_{e,u}(s)) ds du \rangle} \bigr) \bigr]  \nonumber \\
        & \quad \quad \times  \biggl[ \prod_{x \in V(L)}
    \int_{\cB_{D_x}}  J_{x,(a)}(B_{D_x}) 1_{\cB^{ess}_{reg}(D_x)}(B_{D_x})
    \delta_{[(B_{D_x})_{| \partial D_x} = (B_{\Sigma'})_{| \partial D_x}]} DB_{D_x}  \biggr] \biggr\}  DB_{\Sigma'}   \nonumber\\
& \overset{(**)}{=} \int^{\sim}_{\ct} \ db  \ e^{- 2\pi i k  \langle y, b  \rangle}   \biggl[ 1_{\cB^{ess}_{reg}(\Sigma')}(B_{\Sigma'}) \Det_{rig}(B_{\Sigma'})  \times \nonumber\\
   & \quad  \quad  \times \bigl(\prod_e e^{2 \pi i \langle \alpha_{a(e)}, \int_0^1 \int_0^1 (B_{\Sigma'}  dt)(c'_{e,u}(s)) ds du \rangle} \bigr)
    \bigl(  \prod_{x \in V(L)} \cJ_{x,(a)}((B_{\Sigma'})_{| \partial D_x}) \bigr) \biggr]_{| B_{\Sigma'} = b+ \tfrac{1}{k} f_{(a)} }
\end{align}
where we have set $\cB_{\Sigma'}:= C^{\infty}(\Sigma',\ct)$ and
 $\cB_{D_x} := C^{\infty}(D_x,\ct)$,
 where $DB_{\Sigma'}$ is the (informal) Lebesgue measure on $\cB_{\Sigma'}$
 and $DB_{D_x}$ is the (informal) Lebesgue measure on $\cB_{D_x}$
 and where we have set for $b_{\partial D_x}:\partial D_x \to \ct$
 \begin{equation} \label{eq_cJ_x(B)-def}
 \cJ_{x,(a)}(b_{\partial D_x}) :=  \int_{\cB_{D_x}} J_{x,(a)}(B_{D_x}) 1_{\cB^{ess}_{reg}(D_x)}(B_{D_x})
   \delta_{[(B_{D_x})_{| \partial D_x} = b_{\partial D_x}]} DB_{D_x}
 \end{equation}
   The notation $ \delta_{[(B_{D_x})_{| \partial D_x} = b_{\partial D_x}]} DB_{D_x}$ here
   and the notation  $\delta_{[(B_{D_x})_{| \partial D_x} = (B_{\Sigma'})_{| \partial D_x}]} DB_{D_x} $ in Eq. \eqref{eq_D.16} above  is
    explained in Remark \ref{rm_appD_1} above.\par

   In step $(**)$   we used  Eq. \eqref{eq_D.13} above.

\smallskip

Let us set
$\cF_{(a)}(b):=  F(B_{\Sigma'})_{| B_{\Sigma'} = b+ \tfrac{1}{k} f_{(a)}}$
where $F(B_{\Sigma'})$ is the expression appearing inside $[\cdots]$
in the last two lines of Eq. \eqref{eq_D.16} above.
It is plausible to expect that $\ct \ni b \mapsto \cF_{(a)}(b) \in \bC$ is $I$-periodic  (for each $(a) \in (\cI)$).
 If this is the case then, by using analogous arguments
 as in Sec. \ref{subsubsec3.3.5} above (and, in particular,  the Poisson summation formula),
we obtain from Eq. \eqref{eq_D.15} and Eq. \eqref{eq_D.16}
\begin{multline} \label{eq_D.17}
Z(\Sigma \times S^1, L)
 \sim    \sum_{(a) \in (\cI)_{adm}} \sum_{\alpha_0 \in \Lambda} \biggl[
  1_{k Q}(\alpha_0) 1_{\cB^{ess}_{reg}(\Sigma')}(B) \Det_{rig}(B) \\
  \times \bigl(\prod_{e \in E(L)} e^{2 \pi i \langle \alpha_{a(e)}, \int_0^1 \int_0^1 (B  dt)(c'_{e,u}(s)) ds du \rangle} \bigr)
 \bigl(  \prod_{x \in V(L)} \cJ_{x,(a)}(B_{| \partial D_x}) \bigr)
\biggr]_{| B =  B_{(a),\alpha_0}}
\end{multline}
 with $Q$ given by Eq. \eqref{eq5.54} in Sec. \ref{subsubsec3.3.5} above
and where we have set $B_{(a),\alpha_0}:= \tfrac{1}{k}( \alpha_0 +   f_{(a)})$
 for each $\alpha_0 \in \Lambda \cap (k Q)$ and $(a) \in (\cI)$.

\begin{remark} \label{rm_appD_not_expl}
Recall from the paragraph before Remark \ref{rm_KiRe} in Sec. \ref{subsubsec3.5.2}
 above that
even though we can write $T_x(A_{\Sigma},B)$, $x \in V(L)$, explicitly
as an infinite series of powers of $1/k$ (whose coefficients also depend on $k$) we do not yet have a closed formula
for $T_x(A_{\Sigma},B)$ and therefore neither for $\cJ_{x,(a)}((B_{(a),\alpha_0})_{| \partial D_x})$.
This means that we have not yet carried out Steps 2--4 of Sec. \ref{subsubsec3.5.2}
completely.  Anyway, Eq. \eqref{eq_D.17} above is explicit enough in order to allow us to make some comments  regarding Step 5 and, in particular,
to make it plausible that also for general strictly admissible $L$ we have a good chance of
arriving at Eq. \eqref{eq_D.0}.
\end{remark}

On the RHS of Eq. \eqref{eq_D.17} we can now make the change of variable
$ B_{(a),\alpha_0} \to  \eta_{(a),\alpha_0}$ where,
for each $\alpha_0 \in \Lambda \cap (k Q)$ and $(a) \in (\cI)$, we
have set $\eta_{(a),\alpha_0}:= k B_{(a),\alpha_0} - \rho =  \alpha_0 +   f_{(a)} - \rho$.
(In view of Example \ref{eq_appD} above this  essentially  generalizes the
change of variable used  in Sec. \ref{subsubsec3.4.2}  above).
Now observe that each $\eta_{(a),\alpha_0}$
  takes values in $\Lambda$ and is constant on each $Y \in Y(L)$.
Moreover, we can express $(a)$ and $\alpha_0$ by the values
of $(\eta(Y))_{Y \in Y(L)}$ with $\eta = \eta_{(a),\alpha_0}$.
Accordingly, we  can rewrite the RHS of Eq. \eqref{eq_D.17}
as a  sum over maps $\eta: Y(L) \to \Lambda$.

\begin{observation} \label{obs_appD}
Let $\cJ_{x}(\eta)$ be the expression by which the factor
 $\cJ_{x,(a)}((B_{(a),\alpha_0})_{| \partial D_x})$, $x \in V(L)$, gets replaced
 when rewriting  the RHS of Eq. \eqref{eq_D.17}
as a  sum over maps $\eta: Y(L) \to \Lambda$ as explained above.
Then $\cJ_{x}(\eta)$ only depends\footnote{In order to see this recall
Eq. \eqref{eq_cJ_x(B)-def} and Eq. \eqref{eq_J_x(B)-def} above
and note that the RHS of Eq. \eqref{eq_J_x(B)-def}
only depends on $(a) \in (\cI)$ via the four components
$a(e_i(x))$, $i =1,2,3,4$.}
 on the  values of $\eta(Y)$ for the four faces $Y \in Y(L)$
having $D_x$ on its boundary
and on the representations $\rho_{j_1(x)}$ and $\rho_{j_2(x)}$,
cf. Convention \ref{conv_appD_1} above,
 i.e., in other words, on the colors
of the two loops $l_i$ and $l_j$ (contained in the proper link $L^0$ associated to
$L$) whose $\pi_{\Sigma}$-projections intersect in $x$. (This observation will be useful
below, cf. the last paragraph before Remark \ref{rm_appD_Ende}.)
\end{observation}

It is not difficult to see that
those $\eta: Y(L) \to \Lambda$
which do not take values in $ \Lambda \cap (k \ct_{reg} - \rho)$
do not contribute to the sum mentioned above.
(This is a consequence of the presence of the factor $1_{\cB^{ess}_{reg}(\Sigma')}(B)$
on the RHS of Eq. \eqref{eq_D.17}).
Accordingly,  we  can rewrite the RHS of Eq. \eqref{eq_D.17}
as a  sum over maps $\eta: Y(L) \to \Lambda \cap (k \ct_{reg} - \rho)$,
and, by using the bijections \eqref{eq_ast_bij} in Sec. \ref{subsubsec3.3.6} above,
we can rewrite the RHS of Eq. \eqref{eq_D.17}
 as sum over all maps $Y(L) \to \Lambda_+^k$,  i.e. as a sum of the form $\sum_{\eta \in col(L)} \cdots $.
 (Observe that apart from the latter sum
 we also have a sum $\sum_{\tau \in (\cW_{\aff})^{Y(L)}} \cdots $
 as in Eq. \eqref{eq_change_to_tau_tilde} in Sec. \ref{subsubsec3.4.2} above.)

 \medskip

We have now seen how the sum $\sum_{\eta \in col(L)} \cdots$
 on the RHS of Eq. \eqref{eq_D.0} above arises
when rewriting the RHS of Eq. \eqref{eq_D.17}
in the way explained above.
Let us now make some comments
regarding the  four factors  $|L|^{\eta}_1$, $|L|^{\eta}_2$,  $|L|^{\eta}_3$,
and  $|L|^{\eta}_4$ on the RHS of Eq. \eqref{eq_D.0}.
Recall that in Sec. \ref{subsubsec3.4.2} above we showed already
in the special case $V(L) = \emptyset$ how the
 first three  factors  $|L|^{\eta}_1$, $|L|^{\eta}_2$,  $|L|^{\eta}_3$
 appear in the heuristic evaluation of  $Z(\Sigma \times S^1,L)$.
 It is  easy to believe that this will also be the case
 in the situation $V(L) \neq \emptyset$, cf. Remark \ref{rm_appD_Ende} below. \par

By far the most interesting and complicated factor on the RHS of Eq. \eqref{eq_D.0}
is the factor $|L|^{\eta}_4$.
At the moment it is totally open whether we really obtain this factor after
rewriting the RHS of Eq. \eqref{eq_D.17} in the way explained above.
However, the following observation gives reason for  optimism.  \par

 From the definition of  $|L|^{\eta}_4$ in  Eq. \eqref{eq_XL4}
 we see that  $|L|^{\eta}_4$ is a linear combination of products
 of matrix elements\footnote{W.r.t. to a suitable choice of basis of the vector space $W(x,\eta)$  mentioned in Appendix \ref{appC} above.} of  $T(x,\eta)$.
Note that this is  analogous to the situation on the RHS of Eq. \eqref{eq_D.17}
where we have the  products   $\prod_{x \in V(L)} \cJ_{x,(a)}((B_{(a),\alpha_0})_{| \partial D_x})$ which later lead to products  $\prod_{x \in V(L)} \cJ_{x}(\eta)$
with $\cJ_{x}(\eta)$ as in Observation \ref{obs_appD} above.
Moreover, as we observed in Observation \ref{obs_appD}  the factors
$\cJ_{x}(\eta)$ have
 several properties that are analogous\footnote{Note that we cannot expect a complete analogy at this stage.  According to Remark \ref{rm_appD_Ende}  below
  we expect  the factors $\cJ_{x}(\eta)$ to contribute
 not only to the factor  $|L|^{\eta}_4$  but also to the factors
 $|L|^{\eta}_1$ and  $|L|^{\eta}_2$ on the RHS of Eq. \eqref{eq_D.0}.
 Note also  that after rewriting the RHS of Eq. \eqref{eq_D.17} in the way explained above   we not only have a  sum $\sum_{\eta \in col(L)} \cdots$
  but also a sum   $\sum_{\tau \in (\cW_{\aff})^{Y(L)}} \cdots $.} to the
 properties of  $T(x,\eta)$  which are mentioned in the second paragraph after Eq. \eqref{eq_XL4} in Appendix \ref{appC} above.

\begin{remark} \label{rm_appD_Ende} Above I mentioned that
also  in the situation $V(L) \neq \emptyset$ it is plausible to expect
that the three  factors  $|L|^{\eta}_1$, $|L|^{\eta}_2$,  $|L|^{\eta}_3$
 appear during the heuristic evaluation of  $Z(\Sigma \times S^1,L)$.
In fact, for the factor $|L|^{\eta}_3$
this is definitely the case. Regarding the factor $|L|^{\eta}_2$ note that the
formula for $\gleam(Y)$ in Eq. \eqref{eqA.1} in Sec. \ref{subsec3.4} is now not sufficient
anymore. The general formula for $\gleam(Y)$  contains contributions
coming form the vertices $x \in V(L)$. I expect that
 these contributions will come from the factors $\cJ_x(\eta)$ mentioned in
Observation \ref{obs_appD} above.
The factors $|L|^{\eta}_1$ are somewhat problematic.
Observe that for general strictly admissible links $L$
Eq. \eqref{eq_Gauss_bonnet_gen} in Sec. \ref{subsubsec3.2.5}
is in general not applicable as the faces $Y \in Y(L)$ in general
will have ``corners'', which are associated to the vertices $x \in V(L)$.
We will therefore have to use a generalization of  Eq. \eqref{eq_Gauss_bonnet_gen}
where on the RHS of the generalization of
Eq. \eqref{eq_Gauss_bonnet_gen}  extra terms  involving  the angles in the corners of $Y$
appear, cf.  Footnote \ref{ft_gen_Gauss_Bonnet}
in Sec. \ref{subsubsec3.2.5}.   It is conceivable  that these extra terms
also come from the $\cJ_x(\eta)$-factors. In fact, if it turns out that
 Condition \ref{ass1} in Sec. \ref{subsubsec3.2.4} on the auxiliary Riemannian metric
  ${\mathbf g}_{\Sigma}$ can be dropped (cf. Remark \ref{rm_sec3.2.7} in Sec. \ref{subsubsec3.2.6})
  then there are good chances that  this is indeed the case.
  If Condition \ref{ass1} cannot be dropped we  still have the alternative
  of using  a different approach for defining $\Det_{rig}(B)$,
   for example, the approach sketched in Remark \ref{rm_Det_disc} in Sec. \ref{subsubsec3.2.5} above where the  Gauss-Bonnet formula is irrelevant.
\end{remark}

\end{document}